%% file: arxiv_version.tex
\documentclass[a4paper,UKenglish,cleveref, autoref]{lipics-v2019}

\title{Engineering Negative Cycle Canceling for Wind Farm Cabling}

\author{Sascha Gritzbach}{Karlsruhe Institute of Technology, Germany}{}{https://orcid.org/0000-0002-2835-392X}{}

\author{Torsten Ueckerdt}{Karlsruhe Institute of Technology, Germany}{}{https://orcid.org/0000-0002-0645-9715}{}

\author{Dorothea Wagner}{Karlsruhe Institute of Technology, Germany}{}{https://orcid.org/0000-0003-4728-7013}{}

\author{Franziska Wegner}{Karlsruhe Institute of Technology, Germany}{}{https://orcid.org/0000-0001-5146-4145}{}

\author{Matthias Wolf}{Karlsruhe Institute of Technology, Germany}{all authors: firstname.secondname@kit.edu}{https://orcid.org/0000-0003-1411-6330}{}

\authorrunning{S. Gritzbach, T. Ueckerdt, D. Wagner, F. Wegner, and M. Wolf}

\Copyright{Sascha Gritzbach, Torsten Ueckerdt, Dorothea Wagner, Franziska Wegner, and Matthias Wolf}



\category{}

\relatedversion{A preliminary version of this work was published in the Proceedings of the 27th~Annual European Symposium on Algorithms (ESA'19). This work became part of the dissertation of the first author.}

\supplement{}

\funding{This work was funded (in part) by the Helmholtz Program Storage and Cross-linked Infrastructures, Topic 6
Superconductivity, Networks and System Integration, by the Helmholtz future topic Energy Systems
Integration, and by the German Research Foundation (DFG) as part of the Research Training Group GRK~2153: Energy Status Data -- Informatics Methods for its Collection, Analysis and Exploitation.}

\acknowledgements{We thank our colleagues Lukas Barth and Marcel Radermacher for valuable discussions and Lukas Barth for providing a code interface for simultaneous \milp-solver experiments.}

\nolinenumbers 

\hideLIPIcs  

\EventEditors{Michael A. Bender, Ola Svensson, and Grzegorz Herman}
\EventNoEds{3}
\EventLongTitle{27th Annual European Symposium on Algorithms (ESA 2019)}
\EventShortTitle{ESA 2019}
\EventAcronym{ESA}
\EventYear{2019}
\EventDate{September 9--11, 2019}
\EventLocation{Munich/Garching, Germany}
\EventLogo{}
\SeriesVolume{144}
\ArticleNo{52}

\usepackage{mathtools}
\usepackage{booktabs}
\usepackage{makecell,multirow}
\usepackage{colortbl}
\usepackage[ruled,vlined,linesnumbered,resetcount,algo2e]{algorithm2e}
\usepackage{algpseudocode}
\usepackage{nicefrac}
\usepackage{xcolor}
\usepackage{xargs}      

\definecolor{OliveGreen}{rgb}{0.33, 0.42, 0.18}
\usepackage[colorinlistoftodos,prependcaption,textsize=tiny]{todonotes}
\newcommandx{\franzi}[2][1=]{\todo
  [linecolor=red,backgroundcolor=red!25,bordercolor=red,#1]{{\bf
      \underline{Franzi:}}\newline#2}}
\newcommandx{\torsten}[2][1=]{\todo
  [linecolor=blue,backgroundcolor=blue!25,bordercolor=blue,#1]{{\bf
      \underline{Torsten:}}\newline#2}}
\newcommandx{\matthias}[2][1=]{\todo
  [linecolor=orange,backgroundcolor=orange!25,bordercolor=orange,#1]{{\bf
      \underline{Matthias:}}\newline#2}}
\newcommandx{\sascha}[2][1=]{\todo
  [linecolor=OliveGreen,backgroundcolor=OliveGreen!25,bordercolor=OliveGreen,#1]{{\bf
      \underline{Sascha:}}\newline#2}}
\newcommandx{\improvement}[2][1=]{\todo[linecolor=Plum,backgroundcolor=Plum!25,bordercolor=Plum,#1]{#2}}
\newcommandx{\hidden}[2][1=]{\todo[disable,#1]{#2}}

\usepackage{thmtools,thm-restate}

\usepackage{soul}

\Crefname{equation}{Equation}{Equations}%
\Crefname{figure}{Figure}{Figures}%
\Crefname{tabular}{Table}{Tables}%
\Crefname{chapter}{Chapter}{Chapters}%
\Crefname{section}{Section}{Sections}%
\Crefname{appendix}{Appendix}{Appendices}%
\Crefname{enumi}{Item}{Items}%
\Crefname{footnote}{Footnote}{Footnotes}%
\Crefname{theorem}{Theorem}{Theorems}%
\Crefname{@theorem}{Theorem}{Theorems}%
\Crefname{lemma}{Lemma}{Lemmas}%
\Crefname{Lemma}{Lemma}{Lemmas}%
\Crefname{corollary}{Corollary}{Corollaries}%
\Crefname{observation}{Observation}{Observations}%
\Crefname{Observation}{Observation}{Observations}%
\Crefname{proposition}{Proposition}{Propositions}%
\Crefname{definition}{Definition}{Definitions}%
\Crefname{result}{Result}{Results}%
\Crefname{example}{Example}{Examples}%
\Crefname{remark}{Remark}{Remarks}%
\Crefname{note}{Note}{Notes}%
\Crefname{algorithm}{Algorithm}{Algorithms}%
\Crefname{procedure}{Procedure}{Procedures}%
\Crefname{proc}{Procedure}{Procedures}
\Crefname{line}{Line}{Lines}%
\Crefname{Line}{Line}{Lines}%
\Crefname{part}{Part}{Parts}%

\Crefname{algocf}{Algorithm}{Algorithms}
\Crefname{algocfline}{Line}{Lines}

\newcommand{\graph}{\ensuremath{G}\xspace}

\newcommand{\network}{\ensuremath{\mathcal{N}}\xspace}

\newcommand{\residualgraph}{\ensuremath{R}\xspace}
\newcommand{\linegraph}{\ensuremath{L}\xspace}
\newcommand{\vertices}{\ensuremath{V}\xspace}

\newcommand{\turbines}{\ensuremath{V_T}\xspace}
\newcommand{\substations}{\ensuremath{V_S}\xspace}
\newcommand{\substation}{\ensuremath{s}\xspace}
\newcommand{\vertex}{\ensuremath{v}\xspace}
\newcommand{\vertexa}{\ensuremath{u}\xspace}
\newcommand{\vertexb}{\ensuremath{v}\xspace}
\newcommand{\vertexc}{\ensuremath{w}\xspace}
\newcommand{\vertexd}{\ensuremath{x}\xspace}

\newcommand{\source}{\ensuremath{s}\xspace}

\newcommand{\edges}{\ensuremath{E}\xspace}

\newcommand{\edge}{\ensuremath{e}\xspace}

\newcommand{\reverse}[1]{\ensuremath{\bar{#1}}\xspace}
\newcommand{\substationCapacity}{\ensuremath{\mathrm{cap}_\mathrm{sub}}\xspace}
\newcommand{\edgelength}{\ensuremath{\mathrm{len}}\xspace}
\newcommand{\totalcost}{\ensuremath{\mathrm{cost}}\xspace}
\newcommand{\cabletypes}{\ensuremath{K}\xspace}
\newcommand{\cabletype}{\ensuremath{k}\xspace}
\newcommand{\cablecost}{\ensuremath{c}\xspace}
\newcommand{\cablecapacity}{\ensuremath{\mathrm{cap}}\xspace}
\newcommand{\decisionVar}{\ensuremath{x}\xspace}
\newcommand{\flow}{\ensuremath{f}\xspace}

\newcommand{\netflow}{\ensuremath{\flow_{\mathrm{net}}}\xspace}

\newcommand{\apath}{\ensuremath{P}\xspace}
\newcommand{\walk}{\ensuremath{W}\xspace}
\newcommand{\cycle}{\ensuremath{C}\xspace}
\newcommandx{\apathf}[1]{\apath_{#1}\xspace}

\newcommand{\changeofflow}{\ensuremath{\Delta}\xspace}
\newcommand{\residualcost}{\ensuremath{\gamma}\xspace}
\newcommand{\distance}{\ensuremath{\mathrm{\ell}}\xspace}
\newcommand{\parent}{\ensuremath{\mathrm{parent}}\xspace}
\newcommand{\strategy}[1]{\texttt{#1}\xspace}
\newcommand{\inc}{\strategy{Inc}}
\newcommand{\dec}{\strategy{Dec}}
\newcommand{\incdec}{\strategy{IncDec}}
\newcommand{\random}{\strategy{Random}}
\newcommand{\stayinc}{\strategy{StayInc}}
\newcommand{\staydec}{\strategy{StayDec}}
\newcommand{\stayincdec}{\strategy{StayIncDec}}
\newcommand{\stayrandom}{\strategy{StayRandom}}
\newcommand{\shortstayinc}{\strategy{S-Inc}}
\newcommand{\shortstaydec}{\strategy{S-Dec}}
\newcommand{\shortstayincdec}{\strategy{S-IncDec}}
\newcommand{\shortstayrandom}{\strategy{S-Random}}
\newcommand{\dijkstra}{\strategy{Dijkstra}}

\newcommand{\bfs}{\strategy{BFS}}

\newcommand{\dijkstraany}{\strategy{DijkstraAny}}

\newcommand{\collectingdijkstraany}{\strategy{CollectingDijkstraAny}}
\newcommand{\collectingdijkstralast}{\strategy{CollectingDijkstraLast}}
\newcommand{\bfsany}{\strategy{BFSAny}}

\newcommand{\collectingbfsany}{\strategy{CollectingBFSAny}}

\newcommand{\shortdijkstraany}{\strategy{Dijk-A}}
\newcommand{\shortdijkstralast}{\strategy{Dijk-L}}
\newcommand{\shortcollectingdijkstraany}{\strategy{C-Dijk-A}}
\newcommand{\shortcollectingdijkstralast}{\strategy{C-Dijk-L}}
\newcommand{\shortbfsany}{\strategy{BFS-A}}
\newcommand{\shortbfslast}{\strategy{BFS-L}}
\newcommand{\shortcollectingbfsany}{\strategy{C-BFS-A}}
\newcommand{\shortcollectingbfslast}{\strategy{C-BFS-L}}
\newcommand{\benchmarkset}{\ensuremath{\mathcal{N}}\xspace}
\newcommand{\bigO}{\ensuremath{\mathcal{O}}\xspace}

\newcommand{\NP}{\ensuremath{\mathcal{NP}}\xspace}
\newcommand{\reals}{\ensuremath{\mathbb{R}}\xspace}

\newcommand{\posreals}{\ensuremath{\reals_{\ge 0}}\xspace}

\newcommand{\naturals}{\ensuremath{\mathbb{N}}\xspace}
\newcommand{\integers}{\ensuremath{\mathbb{Z}}\xspace}

\DeclareMathOperator*{\avg}{avg}
\DeclarePairedDelimiter{\abs}{\lvert}{\rvert}
\newcommand{\increment}{\ensuremath{\hspace{-.05em}\raisebox{.4ex}{\tiny\bf ++}
}}
\newcommand{\eg}{e.\,g.\xspace}
\newcommand{\ie}{i.\,e.\xspace}

\newcommand{\percentage}[1]{\ensuremath{#1\,\%}\xspace}
\newcommand{\wcp}{\textsf{WCP}\xspace}

\newcommand{\WCP}{\textsc{Wind Farm Cabling Problem}\xspace}
\newcommand{\milp}{\textsf{MILP}\xspace}

\newcommand{\MILP}{\textsc{Mixed-integer Linear Program}\xspace}
\newcommand{\CMST}{\textsc{Capacitated Minimum Spanning Tree}\xspace}

\newcommand{\sa}{\textsf{SA}\xspace}
\newcommand{\ncc}{\textsf{NCC}\xspace}

\begin{document}

\maketitle

  \begin{abstract}
In a wind farm turbines convert wind energy into electrical energy. The 
generation of each turbine is transmitted, possibly via other 
turbines, to a substation that is connected to the power grid. 
On every possible interconnection there can be at most one of various different 
cable types. Each type comes with a cost per unit length and with a 
capacity. Designing a cost-minimal cable 
layout for a wind farm to feed all turbine production into the power grid 
is called the \mbox{\WCP~(\wcp).}

We consider a formulation of~\wcp as a flow problem on a graph where the
cost of a flow on an edge is modeled by a step function originating from 
the cable types.
Recently, we presented a proof-of-concept for a negative cycle 
canceling-based algorithm for~\wcp~\cite{Gritzbach2018}.
We extend key steps of that heuristic and build a theoretical foundation that explains how this
heuristic tackles the problems arising from the special structure of~\wcp.

A thorough experimental evaluation identifies the best setup of the algorithm
and compares it to existing methods from the literature such as 
\MILP{\textsc{ming}} (\milp) and Simulated Annealing (\sa). The heuristic runs in a range of half a 
millisecond to approximately one and a half minutes on instances with up to~500 turbines. It 
provides
solutions of similar quality compared to both competitors with running times
of one hour and one day. When comparing
the solution quality after a running time of two seconds, our algorithm 
outperforms the \milp{-} and \sa-approaches, which allows it to be applied in interactive wind 
farm planning.
\end{abstract}
  
  \newpage
  
  \section{Introduction}
  \label{sec:introduction}
  
  Wind energy becomes increasingly important to help reduce effects of 
  climate change.
  As of~2017,~\percentage{11.6} of the total electricity demand 
  in the European Union 
  is covered by wind power~\cite{online:WindEurope2017report}. 
  Across the Atlantic, the state of 
  New York aims at installing~2.4~GW of offshore wind energy capacity by~2030, 
  which could cover the demand of~1.2~million homes 
  \cite{online:NYS:OffshoreMasterPlan}.
  
  In an offshore wind farm a set of turbines generate electrical 
  energy. From offshore substations the energy is transmitted via sea cables to 
  an onshore grid point. One of the biggest wind farms currently 
  planned is Hornsea Project Three in the North Sea with up to~300~turbines 
  and twelve substations~\cite{online:4C:Hornsea3}. To transport turbine 
  production to the substations, a system of cables links turbines to 
  substations (\emph{internal cabling}) where multiple turbines may be 
  connected 
  in series. The 
  designer of a wind farm has various cable types available, 
  each 
  of which with respective costs and thermal capacities. The latter restricts 
  the amount of energy that can be transmitted through a cable.
  Planning a wind farm as a whole consists of various steps, including
  determining the locations for turbines and substations, layouting the 
  connections from substations to the grid point, and designing the internal
  cabling. The planning process comes with a high level of
  complexity, which automated approaches struggle with
  \cite{Valverde:2013:ISOPE:WindFarmLayout}. Therefore, one might opt for
  decoupling the planning steps.
  We call the task of finding a cost-minimal internal cabling of a wind 
  farm with given turbine and substation positions, as well as given turbine 
  production and substation capacities, the \WCP~(\wcp).
  This problem is strongly \NP-hard~\cite{Gritzbach23}.\footnote{The initial version of this work
  claimed that~\wcp is \NP-hard as a generalization of the~\CMST. This is incorrect
  since cable layouts in~\wcp need not be trees (cf. \cref{sec:model}). A proof
  of strong \NP-hardness can be found in the dissertation of the first author~\cite{Gritzbach23}.}

  Due to the overall cost of a wind farm, using one day of
  computation time or more arguably is a reasonable way to approach~\wcp.
  Such computation times, however, are not appropriate for an interactive 
  planning process: Imagine a wind farm planner
  uses a planning tool which allows altering turbine positions to
  explore their influence on possible cable layouts. In that case,
  computation times of at most several seconds are desirable.
  
  \subsection{Contribution and Outline}
  We extend our recent proof-of-concept, in which negative cycle canceling is 
  applied to a formulation of~\wcp as a network 
    flow problem (cf. \cref{sec:model}) with a step cost
    function representing the cable types~\cite{Gritzbach2018}.
    The idea of negative cycle canceling is to iteratively identify cycles in
    a graph in which the edges are associated with the costs of (or gains from) 
    changing the flow. Normally, a cycle of negative total cost corresponds
    to a way to decrease the cost of a previously found flow. Due to the
    step cost function, however, not every negative cycle helps improve a 
    solution to~\wcp. We 
      explore this and other issues for negative
  cycle canceling that arise from the step cost function in the flow 
  problem formulation for~\wcp.
  We present a modification of the 
  Bellman-Ford algorithm~\cite{bellman:routing,Ford:2010:FN:1942094}
  and build
  a theoretical foundation that explains how the modified algorithm
  addresses the aforementioned issues, \eg, by
  being able to identify cycles that actually improve a solution.
  This modification works on
  a subgraph of the line graph (cf. page~\pageref{footnote:linegraph}) 
  of the input graph and can be implemented in the same
  asymptotic running time as the original Bellman-Ford algorithm.
  
  We further extend that heuristic by identifying two key abstraction layers
  and applying different strategies in those layers. Using different 
  initializations is hinted at in the section on future work 
  in~\cite{Gritzbach2018}. We follow this hint and design eight concrete 
  initialization strategies.
  In another layer, we propose a total of eight so-called ``delta strategies''
  that specify the order in which different values for flow changes are 
    considered.

  In \cite{Gritzbach2018} we compared the Negative Cycle Canceling (\ncc) algorithm 
  to a~\MILP{}~(\milp) using the~\milp solver Gurobi with one-hour running 
  times on benchmark sets from the
  literature~\cite{Lehmann:2017:SAW:3077839.3077843}.
  We extend this evaluation by identifying
  the best of our variants and by comparing its results to the results 
  of \milp experiments after running times of two seconds, one hour, and 
  one day on the same benchmark sets. A running time of two seconds helps
    identify the usefulness of the \ncc algorithm to an interactive planning
      process. The other running times stand for non-time-critical planning.
  We also compare the algorithm to an 
  approach using Simulated Annealing~\cite{Lehmann:2017:SAW:3077839.3077843} 
  with different running times.
  The results show that our heuristic is very fast since it terminates on 
  instances with up to~500~turbines in under~100~seconds. At two seconds 
  our algorithm outperforms its competitors, making it feasible for 
  interactive wind farm planning. Even with longer running times for the \milp- 
  and \sa-approaches, our algorithm yields solutions to~\wcp of similar quality
  but in tens of seconds.\looseness=-1
  
  In \cref{sec:related_work} we review existing work on~\wcp and negative cycle 
  canceling. In \cref{sec:model} we define~\wcp as a flow problem.  
  We
  give theoretical insights on the difference to standard flow problems and
  present and analyze our Negative Cycle Canceling algorithm
  in \cref{sec:algorithm}.
  An extensive experimental evaluation of the algorithm is given in 
  \cref{sec:evaluation}. We conclude with a short summary of the results and 
  outline possible research directions (see \cref{sec:conclusion}).
  
  \section{Related Work}
  \label{sec:related_work}
  In one of 
  the first works on~\wcp, a hierarchical decomposition of the 
  problem was introduced~\cite{berzan}. The layers 
  relate to well-known graph problems and heuristics for various 
  settings are proposed.
  Since then, considerable effort has been put into solving variants 
  of~\wcp. Exact 
  solutions can be computed using \MILP~(\milp) formulations 
  including various degrees of technical constraints, \eg, line losses, 
  component failures, and wind stochasticity~\cite{2013Lumbreras}. However, 
  sizes of wind farms that are solved to 
  optimality in reasonable time are small. Metaheuristics such as Genetic 
  Algorithms \cite{f6079fe0004111dab4d5000ea68e967b,%
  Dahmani2015OptimizationOT} or Simulated Annealing 
  \cite{Lehmann:2017:SAW:3077839.3077843} can provide 
  good but not necessarily optimal solutions in relatively short computation times.\looseness=-1
  
  We applied negative cycle canceling
     to a suitable flow formulation for~\wcp~\cite{Gritzbach2018}, but 
  there is still an extensive agenda of open questions such as investigating 
  the effect of other cable types, a comparison to existing 
  heuristics, and using the solution as warm start for a \milp solver.
  Originally, negative cycle canceling is proposed in the context of minimum 
  cost circulations when linear cost 
  functions are considered~\cite{Klein1967}. The algorithm for the Minimum-Cost 
  Flow Problem based on cycle canceling with strongly polynomial running time 
  runs in~$\bigO(nm(\log n)\min\{\log(nC), m\log n\})$ time on a network 
  with~$n$~vertices, $m$~edges, and maximum absolute value of 
  costs~$C$~\cite{Goldberg1989}. 
  The bound for the running time of this 
  algorithm was later tightened to~$\Theta(\min\{nm\log(nC), 
  nm^2\})$~\cite{Radzik1994}. Negative cycle 
  canceling has also been used for 
  problems with non-linear cost functions. Among these are
    multicommodity flow problems with certain non-linear yet convex cost 
  functions 
  based on a queueing model~\cite{Ouorou2000} and the Capacity Expansion 
  Problem for multicommodity flow networks with certain
  non-convex 
  and non-smooth cost functions~\cite{Souza2008}. A classic algorithm for 
  finding negative 
  cycles is the Bellman-Ford 
  algorithm~\cite{bellman:routing,Ford:2010:FN:1942094} with heuristic 
  improvements~\cite{Goldfarb1991,Goldberg1993}. An experimental 
  evaluation of these heuristics and other negative cycle detection algorithms 
  is given in~\cite{Cherkassky1999}.
  
  A step cost function similar to the one in~\wcp appears in a multicommodity 
  flow problem, for which exact solutions can be obtained by a procedure based on 
  Benders Decomposition~\cite{GABREL199915}. However, this procedure is only 
  evaluated on instances with up to~20~vertices and~37~edges 
  and some running times exceed~13 hours. While our 
  approach does not guarantee to solve~\wcp to optimality, our 
  evaluation shows that the solution quality is very good compared to the~\milp 
  with running times not 
  exceeding~100~seconds on wind farms with up to~500~turbines.
  
  \section{Model}
  \label{sec:model}
  The model presented in this paper is based on an existing flow model 
  for~\wcp~\cite{Gritzbach2018}. We briefly 
  recall the model. Given a wind farm, let~\turbines 
  and \substations be the sets of turbines and substations, respectively. We define 
  a vertex set~$\vertices$ of a graph by~$\vertices = \turbines \cup 
  \substations$. For any
  two vertices \vertexa and \vertexb that can be connected by a cable 
  in the wind farm, we define exactly one directed \mbox{edge~$\edge = (\vertexa, 
    \vertexb)$,} where the direction is chosen arbitrarily. We 
  obtain a directed graph~$\graph = (\vertices, \edges)$ with
  $\vertices = \turbines \cup \substations$ and 
  \mbox{$\edges \subseteq \left( \vertices \times \vertices \right) \setminus \left( 
    \substations \times \substations \right)$}
  such that $(\vertexa, \vertexb) \in \edges$ implies 
  $(\vertexb, \vertexa) \notin \edges$. There are no edges between any two substations
    since we consider the wind farm planning step in which all positions of turbines and
    substations, as well as the cabling from substations to the
    onshore grid point have been fixed. We assume that all turbines generate 
  one unit of electricity. Note that our algorithm can be easily 
  generalized to 
  handle non-uniform integral generation. Substations have a 
  capacity~$\substationCapacity \colon \substations \to \naturals$ representing 
  the maximum amount of turbine production they can handle and each edge has 
  a length given by~$\edgelength \colon \edges \to \posreals$ 
  representing the
  geographic distance between the endpoints of the edge.\looseness=-1
  
  A \emph{flow} on~$\graph$ is a function~$\flow \colon \edges \to \reals$ and 
  for an edge~$(\vertexa, \vertexb)$ with~$\flow(\vertexa, \vertexb) > 0$ 
  (resp.~$< 0$), we say that $\flow(\vertexa, \vertexb)$ units of flow go 
  from~\vertexa to~\vertexb (resp.~$-\flow(\vertexa, \vertexb)$ units go 
  from~\vertexb to~\vertexa). For a flow~\flow and a vertex~\vertexa we define 
  the \emph{net flow in \vertexa} by $\netflow (\vertexa) = 
  \sum_{(\vertexc, \vertexa) \in \edges}
  \flow(\vertexc,\vertexa) - \sum_{(\vertexa, \vertexc) \in \edges} \flow
  (\vertexa, \vertexc)$. A flow~\flow is \emph{feasible} if the conditions on 
  flow 
  conservation for both turbines~(\cref{eq:feasibleFlow:turbines}) and 
  substations~(\cref{eq:feasibleFlow:substations}) are satisfied and if there 
  is 
  no outflow from any substation 
  ~(\cref{eq:feasibleFlow:noSubstationOutflow1,%
  eq:feasibleFlow:noSubstationOutflow2}).
  \begin{align}%
  \netflow(\vertexa) &= -1 
  &\forall \vertexa \in \turbines,
  \label{eq:feasibleFlow:turbines}\\
  \netflow(\vertexb) &\leq \substationCapacity(\vertexb) 
  &\forall \vertexb \in \substations,
  \label{eq:feasibleFlow:substations}\\
  \flow(\vertexa, \vertexb) &\geq 0
  &\forall(\vertexa, \vertexb)\in\edges : \vertexb\in\substations,
  \label{eq:feasibleFlow:noSubstationOutflow1}\\
  \flow(\vertexb, \vertexa) &\leq 0 
  &\forall(\vertexb, \vertexa)\in\edges : \vertexb\in\substations.
  \label{eq:feasibleFlow:noSubstationOutflow2}
  \end{align}%
  Let~$\cablecost \colon \posreals \to \posreals \cup \{\infty\}$ be a 
  non-decreasing, left-continuous step function 
  with~$\cablecost(0) = 0$. This function represents the cable costs
   and $\sup\{x\in\posreals : \cablecost(x) < \infty\}$ 
  is the maximum cable capacity, which we assume to be a natural number. 
  Note that such a function is neither convex nor concave in general.
  The cost of a flow on a wind farm 
  graph is then given by
  \begin{align}
  \totalcost(\flow) = \sum_{\edge \in \edges} 
  \cablecost\left(\abs{\flow(\edge)}\right) \cdot \edgelength(\edge).
  \label{eq:totalcost}
  \end{align}
  The value of $\cablecost\left(\abs{\flow(\edge)}\right)$ stands
  for the cost per unit length of the cheapest cable type with sufficient
  capacity to transmit~$\abs{\flow(\edge)}$~units of turbine production.
  With all that,~\wcp is the problem of finding a feasible flow~\flow on a given 
  wind farm graph that minimizes the cost. There is an analogon to the 
  linear-cost integer flow theorem~(\eg~\cite[Thm.~9.10]{Ahuja1993})
  that guarantees an optimal flow with integral values.
  
  \begin{restatable}{lem}{lemOptimalIntegralFlow}
    \label{lem:optimal_integral_flow}
    Suppose the cost function is discontinuous only at integers and there is a 
    feasible flow. Then, there is a cost-minimal integral flow.
  \end{restatable}

    \begin{proof}
    Suppose \flow is a (possibly non-integral) flow of minimum costs.
    We define another flow network on the same graph by setting the
    capacity~$\cablecapacity(\edge)$ of
    every edge~\edge to~$\lceil\abs{\flow(\edge)}\rceil$. Each turbine requires 
    a
    net flow of~$-1$. We model the substation capacities by adding a new
    vertex~$\substation$ and edges from all substations to~$\substation$ with
    capacities equal to the substation capacities. The net flow shall be~$0$ at 
    all substations and~$\abs{\turbines}$ at~$\substation$. We further define 
    zero
    costs for flows on all edges. By the integrality property of min-cost flow
    problems with linear cost functions (\eg, \cite[Thm.~9.10]{Ahuja1993}) there
    is a feasible integral flow~$\flow'$ in this network. Due to the 
    construction 
    of the flow network, $\flow'$ satisfies the constraints in 
    \crefrange{eq:feasibleFlow:turbines}{eq:feasibleFlow:noSubstationOutflow2}.

    Since the cost function~\cablecost is non-decreasing, it holds for all~$\edge \in \edges$
    that~$\cablecost(x) \leq \cablecost(\abs{\flow(\edge)})$ 
    for all~$x\in[0, \abs{\flow(\edge)}]$.
    Since~\cablecost is a left-continuous step function that is discontinuous
    only at integers, we also 
    have~$\cablecost(x) \leq \cablecost(\abs{\flow(\edge)})$ 
    for all~$x\in[\abs{\flow(\edge)}, \cablecapacity(\edge)]$.
    It holds in particular
    that~$\cablecost(\abs{\flow'(\edge)}) \leq \cablecost(\abs{\flow(\edge)})$.
    Thus,~$\totalcost(\flow') \leq \totalcost(\flow)$
    and~$\flow'$ is optimal in the original network.

    
  \end{proof}
  \section{Algorithm}
  \label{sec:algorithm}
  Given a wind farm graph~$\graph$ we 
  define 
  the \emph{residual 
    graph}~\residualgraph of~\graph  with vertices~$\vertices(\residualgraph)$ 
  and edges~$\edges(\residualgraph)$ by $\vertices(\residualgraph) 
  = \vertices(\graph) \cup \{\source\}$ and 
  \mbox{$\edges(\residualgraph) = \{\edge, \reverse{\edge} : \edge 
      \in\edges(\graph)\} \cup \{(\vertex, \source), (\source, \vertex) : 
      \vertex\in\substations\}$}
  \ where~\reverse{\edge}~is~the reverse of~\edge.
  The new vertex~\source, the \emph{super 
  substation}, is a virtual substation without capacity, that is connected to all 
  substations.
  The edges to and from~\source are used to model the
  substation capacity constraints and to allow the production of 
  one turbine to be reassigned to another substation.
  
\begin{figure}
    \centering
    \includegraphics{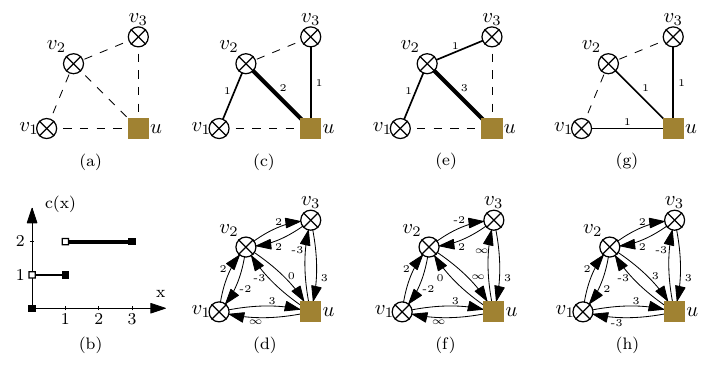}
    \caption{Examples of flows and corresponding residual graphs.
      (a)~shows a wind farm graph. Edges between turbines are of length~$2$, 
      edges between the substation~$\vertexa$ and any turbine are of length~$3$.
      (b)~depicts a cost function induced by two cable types.
      (c)~displays a feasible flow. Dashed lines do not carry any flow. The 
      thickness of solid lines represent the necessary cable type to carry the 
      respective flow.
      (d)~is the residual graph for the flow in (c) and~$\Delta = 1$. The 
      super substation is omitted for ease of presentation. There are three 
      negative cycles: $\vertexa\vertexb_2\vertexa$, 
      $\vertexa\vertexb_2\vertexb_1\vertexa$, and 
      $\vertexa\vertexb_3\vertexb_2\vertexa$.
      (e)~shows the flow obtained by sending one unit of flow along 
      $\vertexa\vertexb_3\vertexb_2\vertexa$ in (c).
      (f)~is the residual graph for (e) and~$\Delta = 1$.
      (g)~depicts the flow obtained by sending one unit of flow along 
      $\vertexa\vertexb_2\vertexb_1\vertexa$ in (c).
      (h)~displays the residual graph for (g) and~$\Delta = 1$.
    }
    \label{fig:problemsWithWCP}
  \end{figure}

  For a given feasible flow~\flow in~\graph of finite cost and 
  $\changeofflow\in\naturals$ we further 
  define \emph{residual costs}, which represent by how much the cost for the 
  edge changes if the flow on the edge is increased by~\changeofflow 
  (cf. \Cref{fig:problemsWithWCP} (a) -- (d) for an example). Note that 
  for negative quantities of flow this implies that the absolute value of the flow 
  may be reduced or even the direction of the flow on an edge may change.
  More formally, we define
  $\residualcost\colon\edges(\residualgraph) \to 
  \reals$
  \mbox{by $\residualcost(\edge) = \big(\,\cablecost(\abs{\flow(\edge) 
        + \changeofflow}) - 
      \cablecost(\abs{\flow(\edge)})\,\big)\cdot\edgelength(\edge)$}
  for all $\edge\in\edges(\residualgraph)$ that are neither incident to~\source 
  nor lead to a substation
  where we alias $\flow(\reverse{\edge}) 
  = - \flow(\edge)$ for all $\edge\in\edges(\graph)$.
  By this definition the residual costs are infinite if 
  $\cablecost(\abs{\flow(\edge)  + \changeofflow}) = \infty$, \ie, 
  if the maximal capacity on~$\edge$ is exceeded.
  For~$\vertexa \in \substations$ and~$\vertexb \in \turbines$, we 
  set~$\residualcost(\vertexa,\vertexb) = \infty$ 
  whenever~$\flow(\vertexb, \vertexa) < \changeofflow$ because 
  sending~$\flow(\vertexa, \vertexb) + \changeofflow$ units from~$\vertexa$ 
  to~$\vertexb$ would otherwise imply that flow leaves a substation. 
  On edges into~\source, we 
  set~$\residualcost(\vertexa, \source) = 0$ if and only if $\flow(\vertexa, 
  \source) + \changeofflow \leq \substationCapacity(\vertexa)$ 
  and~$\residualcost(\vertexa, \source) = \infty$ otherwise.
  On edges leaving the super substation, we 
  set~$\residualcost(\source, \vertexa) = 0$ if and only if $\flow(\vertexa, 
  \source) \geq \changeofflow$ and~$\residualcost(\source, \vertexa) = \infty$ 
  otherwise to prevent flow from leaving the substation.
  
  In a nutshell, the Negative Cycle Canceling (\ncc) 
  algorithm (\cref{alg:main}) starts
  with an initial feasible
  flow and some value 
  of~$\Delta$, computes the residual costs, and looks for a negative 
  cycle\footnote{A \emph{cycle} is a sequence of consecutive edges 
  such 
  that the first edge starts at the same vertex where the last edge ends and 
  such that no two edges start at the same vertex. That is, all cycles are 
  simple. A cycle is said to be 
  \emph{negative} if the sum of residual costs over all edges is negative.} 
  in the residual graph.
  If 
  the algorithm finds a negative cycle, it cancels the cycle, \ie, it changes 
  the flow by adding~$\Delta$ units of flow on all (residual) edges of the 
  cycle. Note that this may decrease the actual amount of flow on edges 
  of~\graph. Then this procedure is repeated with the new flow and some 
  value of~$\Delta$ which may but not need to differ from the previous one. 
  If no negative cycle is found, a new value 
  of~$\Delta$ is chosen and new residual costs are computed. This loop is 
  repeated until all sensible values of~$\Delta$ have been considered for a 
  single flow, which is then returned by the algorithm. This flow is of 
  integer value, since the initial flow is designed to only have integer values 
  and we solely consider natural values for~$\Delta$. Without loss of 
  generality we can restrict ourselves to integer flows according 
  to~\cref{lem:optimal_integral_flow}, even though our algorithm does not
  necessarily find an optimal solution of~\wcp.
  One question we answer is to what extent the algorithm 
  benefits from different initial flows and different orders in which the 
  values of~$\Delta$ are chosen. We present various initializations and
  orders for~$\Delta$ 
  in~\cref{sec:initialization_strategies,sec:delta_strategies}.

  The details of the algorithm 
  (cf.~\cref{sec:detection,sec:algorithmic_details}) address 
  problems that 
  arise from the special structure of~\wcp, namely the non-linear cost 
  function~$\cablecost$. Firstly, in classical min-cost flow problems, 
  when~$\cablecost$ is linear, the cost for changing flow by a certain amount 
  is proportional to the amount of flow change (and the length of the 
  respective edge) and does not depend on the current amount of flow on that 
  edge. Hence, there is no need for computing residual costs for different 
  values of~$\Delta$. Secondly, \emph{short} cycles, \ie, cycles of 
  two edges, may have non-zero total cost in~\wcp (cf.\ cycle 
  $\vertexa\vertexb_2\vertexa$ in \Cref{fig:problemsWithWCP} (d)). Canceling 
  such a cycle, however, does not change the 
  flow and therefore does not improve the solution. Hence, only cycles of at 
  least three edges (\emph{long} cycles) are interesting to us because they do 
  not contain both an edge and its reverse. Finding any negative cycle can be 
  done in polynomial time but finding
  long negative cycles is \NP-hard for general directed \mbox{graphs~\cite[Theorem 4 for $k = 3$]{Guo2017}.}
  Thirdly, the order of canceling cycles matters 
  (\Cref{fig:problemsWithWCP} (c) -- (g)). In (d), there are two long negative 
  cycles:~$\vertexa\vertexb_2\vertexb_1\vertexa$ and 
  $\vertexa\vertexb_3\vertexb_2\vertexa$. After canceling 
  $\vertexa\vertexb_3\vertexb_2\vertexa$ (\Cref{fig:problemsWithWCP} (e)), 
  the other cycle~$\vertexa\vertexb_2\vertexb_1\vertexa$ is not negative 
  anymore.
  Ultimately, \Cref{fig:problemsWithWCP} (e) and (f) show that the 
  non-existence 
  of negative cycles in (all) residual graphs does not imply that the
  underlying flow is
  optimal---contrary to min-cost flow problems with linear cost functions.
  In other words, there are flows that represent local but not global minima.

  
  \subsection{Detecting Long Negative Cycles}
  \label{sec:detection}
  We assume 
  that the reader is familiar with the standard Bellman-Ford 
  algorithm~\cite{bellman:routing,Ford:2010:FN:1942094}, which is
  a common approach to finding negative cycles.
  We observed in preliminary experiments that it mostly reports short cycles
  even if long cycles exist. The reason is that negative residual costs on an 
  edge are repeatedly used if the cost of the reverse edge is, say, zero. In 
  that case, the negative residual cost strongly influences the distance labels 
  on close vertices and overshadows long 
  cycles (see cycle $\vertexa\vertexb_2\vertexa$ in comparison to cycle
    $\vertexa\vertexb_2\vertexb_1\vertexa$ in \Cref{fig:problemsWithWCP} (d)).
  
  One solution is to prohibit propagating the residual cost of an edge 
  over its reverse edge. To this end, we employ the Bellman-Ford algorithm on 
  the subgraph~\linegraph of the directed line 
  graph\footnote{\label{footnote:linegraph}The \emph{line 
  graph}~$\linegraph(\graph)$ of a directed graph~\graph shows which edges are 
  incident to each other. It is defined by $\vertices(\linegraph(\graph)) = 
  \edges(\graph)$ and \mbox{$\edges(\linegraph(\graph)) = \{ 
  ((\vertexa,\vertexb),(\vertexb,\vertexc)) \colon 
  (\vertexa,\vertexb),(\vertexb,\vertexc) \in \edges(\graph)\}$}.} 
  of~\residualgraph which we obtain from the 
  line graph by removing all edges representing U-turns, \ie, edges of the form 
  $(\edge, \reverse{\edge})$ for~$\edge \in \edges(\residualgraph)$. We define 
  the cost of an edge~$(\edge_1, \edge_2)$ in~\linegraph 
  as~$\residualcost(\edge_2)$.  At every vertex~\edge of~\linegraph we 
  maintain a distance label~$\distance(\edge)$ initialized 
  as~$\residualcost(\edge)$. 
  Thus, throughout the Bellman-Ford algorithm,~$\distance(\edge)$ represents 
  the length of some walk\footnote{A \emph{walk} is a sequence of consecutive 
  edges. A walk is called \emph{closed} if the start vertex of 
  the first edge equals the target vertex of the last edge. In particular, 
  every cycle is a closed walk.} in~\linegraph starting at any vertex 
  of~\linegraph~and ending 
  at~\edge. By construction 
  of~\linegraph, the label~$\distance(\edge)$ also stands for some walk in~\residualgraph 
  which ends at the target vertex of~\edge and which does not traverse an edge 
  of~\residualgraph directly after its reverse. Consequently, a cycle~\cycle 
  in~\linegraph corresponds 
  to a closed walk~\walk without U-turns of the same cost in~\residualgraph. In 
  particular,~\walk~is not a short cycle, which is what we wanted. It may still
  occur, however, that~\walk includes an edge and its reverse. In that case,~\walk
  consists of more than one cycle that may be negative themselves. Therefore, we decompose
  the closed walk~\walk
  into cycles, which, in turn, can be canceled one after another. For 
  more details, refer to \Cref{sec:algorithmic_details}.\looseness=-1
  
  A downside of running the Bellman-Ford algorithm on the line graph 
  is that more labels have to be stored and the running time of the 
  algorithm 
  is in~$\bigO(\abs{\vertices(\linegraph)}\cdot\abs{\edges(\linegraph)})$, which 
  is worse than the running time on~\residualgraph.
  We present how to implement an algorithm that directly works 
  on~\residualgraph, that is equivalent to the Bellman-Ford algorithm 
  on~\linegraph, and that has the same asymptotic running time as 
  the original Bellman-Ford algorithm on~\residualgraph.
  To this goal, we use the special structure of~\linegraph
  to analyze what the steps of the Bellman-Ford
  algorithm on~\linegraph mean for~\residualgraph.
  When running the Bellman-Ford algorithm on~\linegraph, there is one
  label for every vertex of~\linegraph. Each of those labels gives
  rise to a label on an edge of~\residualgraph. The labels at incoming edges 
  of~$\vertex\in\vertices(\residualgraph)$ are used to compute the 
  labels at outgoing edges of~$\vertex$.
  Let~$(\vertex, \vertexc)$ and~$(\vertex, \vertexd)$ be two edges 
  leaving~$\vertex$.
  Let us assume that~$(\vertexd, \vertex)$ has the 
  smallest label of all edges entering~$\vertex$. Then,~$(\vertexd, \vertex)$ 
  is used to relax~$(\vertex, \vertexc)$. But it cannot be used to 
  relax~$(\vertex, \vertexd)$. To do so, we need the second smallest label 
  of all edges entering~\vertex. This yields the following observation.\looseness=-1
  \begin{restatable}{obs}{ObsTwoLabels}
    \label{obs:two_relevant_labels}
    For each vertex~\vertex of~\residualgraph only the two smallest 
    labels of incoming edges of~\vertex are required to correctly update the 
    labels on outgoing edges of~\vertex.
  \end{restatable}
  We call these labels \emph{relevant}.
  Consequently, throughout our modified version of the Bellman-Ford algorithm, 
  we maintain two distance labels $\distance_1(\vertexb)$ and 
  $\distance_2(\vertexb)$, and two parent pointers $\parent_1(\vertexb)$ 
  and $\parent_2(\vertexb)$ for every~$\vertexb \in \vertices(\residualgraph)$, 
  respectively. As above, $\distance_i(\vertexb)$ with~$i=1,2$ stand for the 
  length of a U-turn-free walk whose first edge is arbitrary and whose last 
  edge is $(\parent_i(\vertexb), \vertexb)$. That means that the parent 
  pointers hold the edges that have been used to build the values of the 
  distance labels. The algorithm ensures that $\parent_1(\vertexb) \neq 
  \parent_2(\vertexb)$ and $\distance_1(\vertexb) \leq \distance_2(\vertexb)$ 
  for every~$\vertexb \in \vertices(\residualgraph)$. In every iteration of the
  Bellman-Ford algorithm, each edge of~\residualgraph is considered for relaxation: For an
  edge~$\edge = (\vertexa, \vertexb)$ take~$\distance(\vertexa) = \distance_1(\vertexa)$
  if~$\parent_1(\vertexa) \neq \vertexb$ 
  and~$\distance(\vertexa) = \distance_2(\vertexa)$ otherwise. Then, check if
  $\distance(\vertexa) + \residualcost(\edge)$ yields a new relevant label at~\vertexb.
  If, during a relaxation 
  step, several incumbent labels and a newly computed candidate label have the 
  same value, we break ties in favor of the older 
  labels---as in the original algorithm.
  For each edge, checking if it yields a new relevant label at its end vertex
  can be done in constant time.
  With \Cref{obs:two_relevant_labels} we show reduced bounds for the 
  number of iterations and the overall running time compared to a 
  straightforward implementation 
  on~\linegraph.
  \begin{restatable}{theo}{ThmNumberOfIterations}
    \label{thm:number_of_iterations}
    If after~$2\cdot\abs{\vertices(\residualgraph)}$~iterations there is an 
    edge that allows reducing a label,
    then there is a negative cycle in~\linegraph.
  \end{restatable}

  \begin{proof}
    Let~$n=\abs{\vertices(\residualgraph)}$ and suppose~$\edge_{2n+1}$ is an 
    edge that allows reducing a label after 
    $2n$~iterations.
    We iteratively construct a walk backwards starting from~$\edge_{2n+1}$ by
    repeatedly applying the following procedure. At an edge~$\edge_i=(\vertexb, 
    \vertexc)$ we define~$\edge_{i-1}$ as the incoming edge of~$\vertexb$ other 
    than~$(\vertexc, \vertexb)$ with the smallest label. If there are several 
    possibilities, we pick the edge with the oldest label among them.
    The label at~$\edge_{i-1}$ is relevant by definition. We stop when an edge 
    would 
    be repeated. At this point, the walk contains a closed 
    subwalk~$\walk=(\edge_{k},\dots,\edge_{l})$ for suitable~$k,l\in\integers$ 
    with~$k<l\leq 2n+1$.
    By \cref{obs:two_relevant_labels} 
    there are at most~$2n$ edges with relevant labels.
    Hence~$k\geq1$.
    
    Since the label at~$\edge_{2n+1}$ can be updated after~$2n$~iterations, the 
    label at~$\edge_{2n}$ must have been updated in iteration~$2n$. Repeating 
    this argument inductively shows that for $i\geq k$ the label at~$\edge_{i}$ 
    was  updated in or after iteration~$i$. Therefore, 
    all 
    labels of edges in~\walk were updated after the initialization. By 
    the way the labels are computed, we therefore have
    \begin{equation}
    \distance(\edge_{i-1}) + \residualcost(\edge_i) \leq \distance(\edge_{i})
    \end{equation}
    for all edges~$\edge_{i}$ on~\walk where we alias~$\edge_{k-1} = 
    \edge_{l}$.
    
    If one of these inequalities is strict, \ie,   
    $\distance(\edge_{j-1}) + \residualcost(\edge_j) < \distance(\edge_j)$ for 
    some $j\in\{k,\dots,l\}$, then summing over the inequalities for all edges 
    in~$\walk$ will give
    \begin{equation}
    \sum_{\edge\in\walk}\big(\distance(\edge)+\residualcost(\edge)\big) <
    \sum_{\edge\in\walk} \distance(\edge),
    \end{equation}
    which can be simplified to
    \begin{equation}
    \sum_{\edge\in\walk} \residualcost(\edge) < 0.
    \end{equation}
    Hence, the total costs of~$\walk$ will be negative, which will complete the
    proof.
    
    It remains to show that there is some edge~$\edge_j=(\vertexb,\vertexc)$ 
    for 
    which the inequality is strict. To this aim let~$\edge_j$ be the edge with 
    the
    oldest label among edges in~\walk. The label~$\distance(\edge_j)$ was 
    computed from the label~$\distance'(\edge)$ of an 
    edge~$\edge=(\vertexa,\vertexb)$ with~$\vertexa \neq \vertexc$, 
    which may or may not be~$\edge_{j-1}$. That means
    \begin{equation}
    \distance'(\edge) + \residualcost(\edge_j) = \distance(\edge_j)
    \label{eq:algorithm_finds_neg_cycle:oldest_label}
    \end{equation}
    where $\distance$ denotes the labels 
    after 
    the algorithm finishes and~$\distance'$ 
    denotes the labels when~$\distance(\edge_j)$ is computed. Note 
    that the label at~$\edge$ may have been updated afterwards, 
    \ie,~$\distance(\edge) \leq \distance'(\edge)$. 

    For the sake of contradiction assume~$\distance(\edge_{j-1}) \geq \distance'(\edge)$.
    Then,
    ~$\distance'(\edge) \geq \distance(\edge) \geq \distance(\edge_{j-1}) \geq 
    \distance'(\edge)$
    where the first inequality holds since labels at the same edge do not increase during
    the algorithm and the second inequality follows from~\edge being
    an incoming edge of~\vertexb.
    Hence, all these labels are equal. The first equality implies that the label
    on~\edge was not updated after the point in time when~$\distance(\edge_j)$
    was computed and that~$\distance(\edge)$ is older than~$\distance(\edge_j)$.
    Using the second equality, we distinguish two cases: If~$\edge=\edge_{j-1}$,
    then~$\distance(\edge_{j-1})$ is older than~$\distance(\edge_j)$, which
    contradicts the choice of~$\edge_j$. If $\edge \neq \edge_{j-1}$,
    then~\edge should have been included in~\walk instead of~$\edge_{j-1}$.
    Thus, the assumption of~$\distance(\edge_{j-1}) \geq \distance'(\edge)$
    is wrong and it holds that~$\distance(\edge_{j-1}) < \distance'(\edge)$.
    Combining this inequality 
    and~\cref{eq:algorithm_finds_neg_cycle:oldest_label}
    completes the proof.\qedhere
%
    
  \end{proof}

  \begin{restatable}{coro}{CorIterationsBF}
    \label{thm:running_time_Bellman_Ford}
    A negative 
    cycle in~\linegraph can be computed
    in~$\bigO(\abs{\vertices(\residualgraph)} \cdot 
    \abs{\edges(\residualgraph)})$~time if one exists.\looseness=-1
  \end{restatable}
  
  \subsection{Algorithm in Detail}
  \label{sec:algorithmic_details}
  
  The previously described Bellman-Ford algorithm on~\linegraph is encapsulated 
  in \cref{alg:main}.
  We first compute some initial flow (line~\ref{alg:main:line:initialization}) 
  using one of eight initialization strategies presented in 
  \cref{sec:initialization_strategies}.
  In line~\ref{alg:main:line:residual_graph} we compute the residual 
  graph~\residualgraph using a given flow~\flow and a given~\changeofflow and 
  run the modified Bellman-Ford algorithm
  (line~\ref{alg:main:line:bellman_ford}). In the repeat-loop, we consider one 
  edge after another and check in 
  line \ref{alg:main:line:find_walk} if it can be relaxed (again). 
  In that case, we extract a walk~\walk 
  in~\residualgraph with 
  negative costs leading to that edge by traversing parent 
  pointers. However, canceling~\walk directly may not 
  improve the costs of the flow as~\walk may still contain an edge and 
  its reverse.
  We decompose~\walk into a set of simple
  cycles~$\mathcal{C}$ in 
  line~\ref{alg:main:line:decompose_walk} and cancel each cycle 
  independently if it is long and has negative costs 
  (lines~\ref{alg:main:line:for_each_cycle} 
  to~\ref{alg:main:line:end_cancel_cycle}). 
  Note that even though~\walk 
  has negative costs, it may happen that only short cycles in~$\mathcal{C}$ have 
  negative costs and all long cycles have non-negative costs. In this case we 
  search for another negative cycle in~\linegraph 
  (line~\ref{alg:main:line:repeat}).\looseness=-1
  
  If no negative cycle in the current graph~\linegraph is canceled, a new 
  value 
  for~\changeofflow is determined according to the \emph{delta strategy} 
  (cf. \cref{sec:delta_strategies}) in line~\ref{alg:main:line:next_delta} 
  and new 
  residual costs~\residualcost are computed.
  Line~\ref{alg:main:line:next_delta} also 
  checks if every possible value for~\changeofflow has been used after 
  the last update of~\flow without improving the solution. 
  If so,~\flow is returned.
  
  
  \SetKw{KwReturn}{return}
  \SetKw{KwBreak}{break}
  \SetKw{KwAnd}{and}
  \SetKw{KwOr}{or}
  \SetKw{KwNot}{not}
  \SetKw{KwTrue}{true}
  \SetKw{KwFalse}{false}
  \SetKwData{found}{found}
  \SetKwRepeat{innerloop}{{\bf foreach} $\edge \in 
  \edges(\residualgraph)$}{until}
  \SetKwData{initialDelta}{InitialDelta}
  \SetKwFunction{initialize}{InitializeFlow}
  \SetKwFunction{computeResidualGraph}{ComputeResidualGraph}
  \SetKwFunction{runBF}{RunBellmanFord}
  \SetKwFunction{findNegativeClosedWalk}{FindNegativeClosedWalk}
  \SetKwFunction{decomposeWalkIntoCycles}{DecomposeWalkIntoCycles}
  \SetKwFunction{addFlowOnCycle}{AddFlowOnCycle}
  \SetKwFunction{nextDelta}{NextDelta}
  \begin{algorithm2e}
    \DontPrintSemicolon
    \KwIn{Graph~\graph, costs~\cablecost, edge lengths~\edgelength}
    \KwResult{A feasible flow~\flow in~\graph}
    $\flow := \initialize(\graph, \edgelength)$, $\changeofflow := \initialDelta$
    \label{alg:main:line:initialization}\;
    \While{$\changeofflow \neq NULL$
      \label{alg:main:line:outer_while}}{
      $(\residualgraph, \residualcost) := \computeResidualGraph(\graph, 
      \cablecost, \flow, \changeofflow)$
      \label{alg:main:line:residual_graph}\;
      $\runBF(\residualgraph, 
      \residualcost)$
      \label{alg:main:line:bellman_ford}\;
      $\found := \KwFalse$\;
      \ForEach{$\edge \in \edges(\residualgraph)$}{
        $\walk := \findNegativeClosedWalk(\residualgraph, \edge)$
        \label{alg:main:line:find_walk}\;
        $\mathcal{C} := \decomposeWalkIntoCycles(W)$
        \label{alg:main:line:decompose_walk}\;
        \ForEach{$C\in\mathcal{C}$
          \label{alg:main:line:for_each_cycle}}{
          \If{$\abs{C} \geq 3$ \KwAnd $\residualcost(C) < 0$
            \label{alg:main:line:test_cycle}}{
            $\flow := \addFlowOnCycle(\flow, C, \changeofflow)$
            \label{alg:main:line:cancel_cycle}\;
            $\found := \KwTrue$
            \label{alg:main:line:end_cancel_cycle}\;
          } 
        } 
        \lIf{\found}{\label{alg:main:line:repeat}\KwBreak}
      } 
      $\changeofflow := \nextDelta(\changeofflow, \found)$
      \label{alg:main:line:next_delta}\;
    } 
    \KwReturn \flow\;
    \caption{Negative Cycle Canceling}
    \label{alg:main}
  \end{algorithm2e}

    We apply two well-known speed-up techniques to the Bellman-Ford algorithm. 
  Firstly, if one iteration does not yield any update of 
  any label, then the computation is aborted and no negative cycle can be found 
  in the current residual graph. Secondly, after sorting edges by 
  start vertices, we track whether the labels at a vertex~\vertex have been 
  updated since last considering its outgoing edges. If not, then there is no 
  need to relax the outgoing edges.
  
  \subsection{Initialization Strategies}
  \label{sec:initialization_strategies}
  Before we can start searching for and canceling negative cycles, we need some
  feasible initial flow. To obtain such a flow, we consider eight strategies,
  which all roughly work as follows. We pick a turbine~\vertexa whose production
  has not been routed to a substation yet. We then search for a shortest
  path~\apath from~\vertexa to a substation~\vertexb with free capacity using 
  Dijkstra's algorithm~\cite{Dijkstra1959}. The
  search only considers edges on which the production of the turbine can be
  routed, \ie, it ignores congested edges. We then route the production
  of~\vertexa along~\apath to~\vertexb. 
  
  We consider two metrics to compute
  shortest paths. Either we use the lengths defined 
  by~\edgelength~(cf.\ \cref{sec:model}) or we assume 
  a length of~$1$ for every edge. Turbine production can either be routed to 
  a nearest or a farthest (in the sense of the respective metric) 
  substation with free capacity. There are two ways in which the flow is 
  updated: The simpler variant routes only the
  production of~\vertexa along~\apath, \ie, the flow along~\apath is increased 
  by~$1$. The other variant greedily collects as much production from~\vertexa 
  and other
  turbines on~\apath as possible without violating any capacity constraints.
  The resulting flows are integral
  since the substation capacities and the maximum cable capacity 
  are natural numbers.
  If no feasible flow of finite cost is found during the initialization, the 
  algorithm returns without a result.
  
  This yields eight initialization strategies, which we name 
  as follows. The base part of each name is either \bfs if unit distances 
  are 
  used or \dijkstra (abbr.\ \strategy{Dijk}) if the distances given by 
  \edgelength are used.
  This part is followed by a suffix specifying the target substation: 
  \strategy{Any}~(abbr.~\strategy{A}) for the nearest and \strategy{Last} 
  (abbr.~\strategy{L}) for the farthest substation.
  An optional prefix of \strategy{Collecting} (abbr.~\strategy{C}) means that 
  the production is greedily collected along shortest paths.
  For example,
  \collectingdijkstralast (abbr.~\strategy{C-Dijk-L}) iterates over all 
  turbines 
  and for each 
  turbine~\vertexa it finds the substation~\vertexb such that the shortest path 
  given by~\edgelength from~\vertexa to~\vertexb is longest among all 
  substations. Along a shortest path from~\vertexa to~\vertexb, turbine 
  production is collected greedily.

  \subsection{Delta Strategies}
  \label{sec:delta_strategies}
  A delta strategy consists of two parts: an initial value for~\changeofflow 
  and 
  a function that returns the value of~\changeofflow for the following 
  iteration. We discuss eight delta strategies. 
  The simplest one starts 
  with $\changeofflow = 1$ and increments~\changeofflow until a negative cycle 
  is canceled. Then, $\changeofflow$ is reset to~$1$. We call this strategy 
  \inc 
  (as in increasing). Similarly, \dec (as in decreasing) starts with the 
  largest 
  possible value for \changeofflow, 
  which is twice the largest cable capacity. Then, \changeofflow is decremented 
  until a cycle is canceled and reset to the largest value.
  The third strategy \incdec behaves like \inc until a negative cycle is 
  canceled. Then, it decrements \changeofflow until~$\changeofflow=1$ and
  behaves like \inc again. To improve performance, 
  all~\changeofflow can be skipped during incrementation up to the last value 
  of~\changeofflow for which a negative cycle was canceled. The fourth strategy 
  \random returns random natural numbers between one and the maximum possible 
  value for~\changeofflow. Between any two cycle cancellations, no value is 
  repeated.
  
  For each strategy, we consider the following modification: After canceling a 
  negative cycle, we retain the current value of~\changeofflow, recompute the 
  residual costs with the new flow, and run the Bellman-Ford algorithm again. 
  We repeat this, until~\changeofflow does not yield a negative cycle. In that 
  case,~\changeofflow is changed according to the respective delta strategy. 
  We call the strategies after the modification 
  \stayinc{,} \staydec{,} \stayincdec{,} and \stayrandom 
  (or \shortstayinc, \shortstaydec, \shortstayincdec, and \shortstayrandom for short).
  
  \section{Experimental Evaluation}
  \label{sec:evaluation}
  %
  
  
  
  
  
  
  

  %
  In the previous sections, we introduced a heuristic with various strategies for
  the~\wcp. We first use statistical tests to evaluate these strategies and 
  identify the best ones (\cref{sec:best_variant}). Using the result we
  compare the best \emph{variant} (\ie, best combination of initialization and
  delta strategy) with different base line algorithms for the~\wcp namely 
  solving an exact \milp formulation (\cref{sec:milpexperiments}) 
  and a Simulated Annealing
  algorithm~\cite{Lehmann:2017:SAW:3077839.3077843} (\cref{sec:sa_experiments}).
  In preliminary experiments (\cref{sec:cplex}) we determine which of the \milp solvers Gurobi and CPLEX
  works better for~\wcp 
  to establish which solver we compare the~\ncc algorithm 
  to. 

  For our evaluation we use benchmark sets for wind farms
  from the literature~\cite{Lehmann:2017:SAW:3077839.3077843} consisting of wind farms of 
  different sizes and characteristics: small wind farms with exactly one substation 
  (\mbox{$\benchmarkset_1$: 10--79 turbines}), wind farms with multiple substations
  (\mbox{$\benchmarkset_2$: 20--79 turbines,}
  \mbox{$\benchmarkset_3$: 80--180 turbines,} \mbox{$\benchmarkset_4$: 200--499 turbines),} 
  and complete graphs \mbox{($\benchmarkset_5$: 80--180 turbines).} Our code is 
  written in C\increment{14}
  and compiled with GCC~$7.3.1$ using the~\texttt{-O3 -march=native} flags. All
  simulations are run on a 64-bit architecture with four 12-core CPUs of AMD
  clocked at~$2.1$ GHz with $256$ GB RAM running OpenSUSE Leap~$15.0$.
  All computations are run in single-threaded mode to ensure
  comparability of the different algorithms.
  
  \subsection{Comparing Variants of our Algorithm}
  \label{sec:best_variant}
  In a first step, we want to determine which delta strategy works best. To 
  this end, we randomly select 200~instances per benchmark set. 
  We run our algorithm on each instance with every pair of delta and 
  initialization strategy.
  The first eight rows of
  \cref{tbl:running_times_ncc_variants} show for every benchmark set the
  minimum, average, and maximum running times for each delta strategy across
  all initialization strategies.
  We first observe that all variants are fast, with running times between tenths
  of milliseconds to $4.5$~minutes on large instances in the worst case.
  We see that \dec is always the slowest
    strategy on average, which can be explained by the fact that \dec often tries
    large values for~$\changeofflow$, for which negative cycles are found rarely.
    The other strategies all roughly complete in the same time on average. It
    seems to be slightly faster to repeat the same~$\changeofflow$. However, for 
    our
    purpose all variants have small enough running times. We therefore
    base our decision, which variant to choose, solely on their solution 
    qualities.\looseness=-1

  \begin{table*}
      \centering
      \small
      \caption{Minimum, average and maximum of running times in milliseconds 
        of different variants. Running time 
        measurement starts before the initial flow
        is computed and ends with the termination of the algorithm prior to 
        outputting the solution. The first eight rows represent
        running times across all initialization
        strategies per delta strategies and benchmark sets. The best delta
        strategy in terms of solution quality
        is marked in green; minimal values per column are marked in 
        yellow. The last row represents the algorithm variant \incdec, 
        \collectingdijkstraany. }
        \label{tbl:running_times_ncc_variants}
      \setlength{\tabcolsep}{2pt}
      \begin{tabular}{l@{\hspace{1mm}}rrr@{\hspace{2mm}}rrr@{\hspace{2mm}}rrr@{\hspace{2mm}}rrr@{\hspace{2mm}}rrr}
      \multirow[c]{2}{*}
        {\makecell{Delta\\Strategy}} &\multicolumn{3}{c}{$\network_1$}
        &\multicolumn{3}{c}{$\network_2$}
        &\multicolumn{3}{c}{$\network_3$}
        &\multicolumn{3}{c}{$\network_4$}
        &\multicolumn{3}{c}{$\network_5$}\\
        
        \cmidrule(r){2-4}\cmidrule(r){5-7}\cmidrule(r){8-10}\cmidrule(r){11-13}\cmidrule(r){14-16}
        
        & \multicolumn{1}{c}{$\min$} & \multicolumn{1}{c}{$\avg$} & 
        \multicolumn{1}{c}{$\max$} 
        & \multicolumn{1}{c}{$\min$} & \multicolumn{1}{c}{$\avg$} & 
        \multicolumn{1}{c}{$\max$} 
        & \multicolumn{1}{c}{$\min$} & \multicolumn{1}{c}{$\avg$} & 
        \multicolumn{1}{c}{$\max$} 
        & \multicolumn{1}{c}{$\min$} & \multicolumn{1}{c}{$\avg$} & 
        \multicolumn{1}{c}{$\max$} 
        & \multicolumn{1}{c}{$\min$} & \multicolumn{1}{c}{$\avg$} & 
        \multicolumn{1}{c}{$\max$} \\
        
        \hline
        
        \dec      
        & 1.10 & 81.1  & 535 
        & 5.25 & 142.1 & 857 
        & 282  & 1.8k  & 11.3k 
        & 4.4k & 59.7k & 272k 
        & 2.4k & 30.8k & 216k
        \\
        \inc      
        & 0.45 & 46.2  & 361 
        & 2.69 & 78.0  & 531 
        & 174  & 1.2k  & 8.4k 
        & 3.0k & 49.1k & 213k 
        & 1.8k & 16.2k & 131k
        \\
        \rowcolor{green!20}
        \incdec       
        & \cellcolor{yellow!50}0.45   & 45.9  & 433 
        & \cellcolor{yellow!50}2.67   & 77.7  & 539 
        &                      174    & 1.2k  & 8.2k 
        &                      3.0k   & 48.7k & 212k 
        &                      1.9k   & 16.2k & 117k
        \\
        \random      
        &                      0.62 &                      43.9  & 
        \cellcolor{yellow!50}288 
        &                      3.50 &                      77.3  & 
                               443 
        &                      176  & \cellcolor{yellow!50}990   & 
        \cellcolor{yellow!50}5.9k 
        &                      3.2k & \cellcolor{yellow!50}32.6k & 
        \cellcolor{yellow!50}137k 
        & 1.9k &                      16.6k 
        &                      143k
        \\
        \shortstaydec    
        & 0.76 & 62.2  & 461 
        & 3.76 & 111.4 & 725 
        & 210  & 1.4k  & 9.1k 
        & 3.5k & 47.4k & 206k 
        & 1.9k & 14.7k & 133k
        \\
        \shortstayinc    
        & 0.46 & \cellcolor{yellow!50} 42.2  & 295 
        & 2.70 &                       72.8  & 438 
        & \cellcolor{yellow!50} 171  &                       1.0k  & 6.6k 
        & 2.8k &                       36.2k & 147k 
        & 1.8k &                       14.4k & \cellcolor{yellow!50} 97k
        \\
        \shortstayincdec 
        &                      0.45 &                       42.2  & 310 
        &                      2.68 & \cellcolor{yellow!50} 72.7  &  \cellcolor{yellow!50}437 
        &                      171  &                       1.0k   & 6.3k 
        & \cellcolor{yellow!50}2.8k &                       36.0k & 154k 
        &                      1.8k &                       14.4k & 120k
        \\
        \shortstayrandom 
        & 0.57 & 44.1  &                      333 
        & 3.25 & 79.1  &                      486 
        & 193  & 1.1k  &                      6.0k 
        & 3.0k & 35.1k &                      147k 
        & \cellcolor{yellow!50} 1.7k & \cellcolor{yellow!50} 14.1k & 106k
        \\

        \hline

        \strategy{BestVar}
        & 0.48 & 36.2 &  217
        & 3.51 & 52.6 &  257
        & 174  & 706  & 3.1k
        & 3.0k & 27.0k & 92.6k
        & 1.9k & 13.4k & 82.6k 
      \end{tabular}
  \bigskip
  \bigskip
      \caption{Comparison of delta strategies over all initialization 
        strategies. 
        An entry in row~$i$ and column~$j$ shows on how many instances 
        strategy~$i$ produces 
        better solutions than strategy~$j$.
        Values are marked by a star if they are significant with~$p<10^{-2}$ and
        by two stars if~$p<10^{-4}$. The best strategy is marked in green.}
      \label{tbl:statistical_tests_delta_strategies_all}
      \centering
      \small
      \setlength{\tabcolsep}{2pt}
      \begin{tabularx}{\textwidth}{l@{\hspace{1mm}}|@{\hspace{1mm}}XXXXXXXX}
        & \inc          & \dec          & \incdec           & \random 
        & \shortstayinc & \shortstaydec & \shortstayincdec  & \shortstayrandom \\
        \hline
        \inc    
        & \multicolumn{1}{c}{---} & $\percentage{60.6}^{\star\star}$ 
        & $\percentage{48.4}$     & $\percentage{60.2}^{\star\star}$  
        & $\percentage{54.2}$     & $\percentage{59.4}^{\star\star}$  
        & $\percentage{50.7}$     & $\percentage{56.5}^\star$ \\
        \dec    
        & $\percentage{39.4}$ & \multicolumn{1}{c}{---}     
        & $\percentage{38.9}$ & $\percentage{46.7}$ 
        & $\percentage{40.8}$ & $\percentage{48.4}$ 
        & $\percentage{40.6}$ & $\percentage{41.2}$ \\
        \rowcolor{green!20}\incdec    
        & $\percentage{51.6}$     & $\percentage{61.1}^{\star\star}$  
        & \multicolumn{1}{c}{---} & $\percentage{59.9}^{\star\star}$  
        & $\percentage{54.0}$     & $\percentage{60.1}^{\star\star}$  
        & $\percentage{50.8}$     & $\percentage{57.3}^\star$ \\
        \random   
        & $\percentage{39.8}$ & $\percentage{53.3}$ 
        & $\percentage{40.1}$ & \multicolumn{1}{c}{---} 
        & $\percentage{42.4}$ & $\percentage{52}$ 
        & $\percentage{42.7}$ & $\percentage{43.4}$ \\
        \shortstayinc 
        & $\percentage{45.8}$     & $\percentage{59.2}^{\star\star}$  
        & $\percentage{46.0}$     & $\percentage{57.6}^\star$ 
        & \multicolumn{1}{c}{---} & $\percentage{58.1}^{\star\star}$  
        & $\percentage{46.9}$     & $\percentage{54.7}$ \\
        \shortstaydec 
        & $\percentage{40.6}$ & $\percentage{51.6}$ 
        & $\percentage{39.9}$ & $\percentage{48.0}$ 
        & $\percentage{41.9}$ & \multicolumn{1}{c}{---} 
        & $\percentage{41.7}$ & $\percentage{41.9}$ \\
        \shortstayincdec  
        & $\percentage{49.3}$     & $\percentage{59.4}^{\star\star}$  
        & $\percentage{49.2}$     & $\percentage{57.3}^\star$ 
        & $\percentage{53.1}$     & $\percentage{58.3}^{\star\star}$ 
        & \multicolumn{1}{c}{---} & $\percentage{55.4}$ \\
        \shortstayrandom  
        & $\percentage{43.5}$ & $\percentage{58.8}^{\star\star}$  
        & $\percentage{42.7}$ & $\percentage{56.6}^\star$ 
        & $\percentage{45.3}$ & $\percentage{58.1}^{\star\star}$  
        & $\percentage{44.6}$ & \multicolumn{1}{c}{---} \\
      \end{tabularx}
\bigskip
\bigskip
    \caption{Comparison of the initialization strategies when the delta 
      strategy~\incdec is fixed. An entry in row~$i$ and column~$j$ shows on 
      how many instances strategy~$i$ produces 
      better solutions than strategy~$j$.
      Values are marked by a star if they are significant with~$p<10^{-2}$ and
      by two stars if~$p<10^{-4}$. The best strategy is marked in green.}
    \label{tbl:statistical_tests_initialization_strategies_inc_fixed}
    \centering
    \small
    \setlength{\tabcolsep}{2pt}
    \begin{tabularx}{\textwidth}{l@{\hspace{1mm}}|@{\hspace{1mm}}XXXXXXXX}
      & \shortdijkstraany & \shortbfsany  & \shortcollectingdijkstraany 
      & \shortcollectingbfsany  & \shortdijkstralast  & \shortbfslast 
      & \shortcollectingdijkstralast  & \shortcollectingbfslast \\
      \hline
      \shortdijkstraany   
      & \multicolumn{1}{c}{---}           & $\percentage{55.8}$ 
      & $\percentage{49.5}$               & $\percentage{54.9}$  
      & $\percentage{55.6}$  & $\percentage{53.7}$  
      & $\percentage{53.9}$  & $\percentage{56.5}^\star$ \\
      \shortbfsany    
      & $\percentage{44.2}$               & \multicolumn{1}{c}{---} 
      & $\percentage{42.7}$               & $\percentage{46.5}$ 
      & $\percentage{47.6}$  & $\percentage{51.1}$  
      & $\percentage{46.7}$  & $\percentage{49.3}$ \\
      \rowcolor{green!20}\shortcollectingdijkstraany    
      & $\percentage{50.5}$               & $\percentage{57.3}^\star$  
      & \multicolumn{1}{c}{---}           & $\percentage{55.3}$  
      & $\percentage{56.5}$  & $\percentage{56.5}^\star$  
      & $\percentage{54.4}$  & $\percentage{56.3}$ \\
      \shortcollectingbfsany    
      & $\percentage{45.1}$               & $\percentage{53.5}$ 
      & $\percentage{44.7}$               & \multicolumn{1}{c}{---} 
      & $\percentage{51.2}$  & $\percentage{54.5}$  
      & $\percentage{49.3}$  & $\percentage{55.4}$ \\
      \shortdijkstralast  
      & $\percentage{44.4}$     & $\percentage{52.4}$ 
      & $\percentage{43.5}$     & $\percentage{48.8}$ 
      & \multicolumn{1}{c}{---} & $\percentage{50.4}$ 
      & $\percentage{48.1}$     & $\percentage{51.7}$ \\
      \shortbfslast 
      & $\percentage{46.3}$               & $\percentage{48.9}$ 
      & $\percentage{43.5}$               & $\percentage{45.5}$ 
      & $\percentage{49.6}$  & \multicolumn{1}{c}{---} 
      & $\percentage{47.7}$         & $\percentage{53.7}$ \\
      \shortcollectingdijkstralast  
      & $\percentage{46.1}$     & $\percentage{53.3}$ 
      & $\percentage{45.6}$     & $\percentage{50.7}$ 
      & $\percentage{51.9}$     & $\percentage{52.3}$ 
      & \multicolumn{1}{c}{---} & $\percentage{53.1}$ \\
      \shortcollectingbfslast 
      & $\percentage{43.5}$         & $\percentage{50.7}$ 
      & $\percentage{43.7}$         & $\percentage{44.6}$ 
      & $\percentage{48.3}$   & $\percentage{46.3}$ 
      & $\percentage{46.9}$   & \multicolumn{1}{c}{---} \\
    \end{tabularx}
  \end{table*}

  To compare the variants in terms of solution quality,
  we compute for each delta strategy~$i$ and instance~$m$ the mean~$X^{(i)}_m$  
  of the solution values over all eight initialization strategies.
  This gives us 1000 data points per delta 
  strategy. For delta strategies~$i,j$ we perform a Binomial Sign Test 
  counting instances with~$X^{(i)}_m < X^{(j)}_m$
  and~$X^{(j)}_m < X^{(i)}_m$ (\cref{sec:signtest}), that means for this test we
    are rather interested in whether strategy~$i$ performs better than strategy~$j$
    on instance~$m$ and not by how much~$i$ is better than~$j$ on~$m$.
  \cref{tbl:statistical_tests_delta_strategies_all} summarizes the results of 
  all tests after Bonferroni-correction 
  by~112 (the number of tests from both delta and initialization strategies). 
  The percentage given in an entry in row~$i$ and 
  column~$j$ states on how many instances~$i$ performes 
  strictly better than~$j$ after averaging over all 
  initialization strategies. Note that entries $(i,j)$ and $(j,i)$ need not 
  represent 1000~instances, as two variants may return equal solution values. 
  
  In the row \incdec, all values are above~\percentage{50}, three of which are 
  significant at the $10^{-4}$ and another one at the $10^{-2}$-level. The smallest value 
  (\percentage{50.8} in column \stayincdec) stands for~460 instances on 
  which \incdec performs better than \stayincdec. To the contrary, there are~446
  instances on which \stayincdec yields better solutions (cf. entry \percentage{49.2}
  in row \stayincdec and column \incdec).
  While the differences between the four delta strategies involving \inc and \incdec 
  are not statistically significant, \incdec does seem to have a slight advantage
  over the others. Hence we consider \incdec as the best delta strategy.

\begin{figure*}
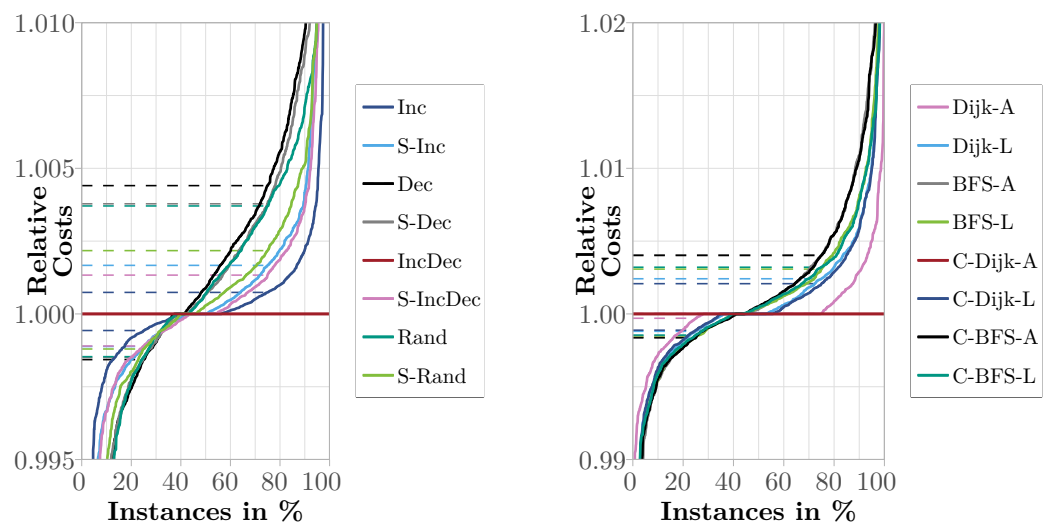

    \begin{subfigure}{.48\textwidth}%
      \centering%
      \input{06-plots/deltaStrategies_ImprovementPlot.tex}
      \label{plot:delta_strategy}%
    \end{subfigure}%
    \hfill
    \begin{subfigure}{.48\textwidth}
      \centering%
      \input{06-plots/initializationStrategies_ImprovementPlot.tex}
      \label{plot:initialization_strategy}%
    \end{subfigure}%
    \caption{Evaluation of the \ncc
      Algorithm using different strategies. For each strategy 
      and for each instance, the ratio of the best solution value 
      found 
      by that \ncc variant to the best solution value found 
      by the reference variant (marked in red) are computed. They are shown in 
      increasing order. The dashed lines represent the 25\% and
      75\% quantiles of the instances. \emph{Left:}~The delta strategies
      are presented relative to the \incdec 
      strategy. Solution values represent the average over all initialization 
      strategies. \emph{Right:}~The initialization strategies
       are presented relative to the
      \collectingdijkstraany strategy with fixed delta strategy \incdec.}%
    \label{plot:strategies_improvement}
  \end{figure*}

  In \cref{plot:strategies_improvement} (left), for the dark green curve all instances are 
  ordered by $X^{(\random)}_m/X^{(\incdec)}_m$ in ascending order. 
  For a 
  given value~$\alpha$ on the abscissa, the curve shows the relative cost 
  factor of the instance at the~$\alpha$-quantile in the computed order. The 
  other curves work accordingly.
  We see, for example, that \incdec works strictly better than \stayinc
  on~\percentage{49.6} and equally on~\percentage{8.1} of all instances 
  and on~\percentage{4.5} of all 
  instances \incdec outperforms \inc by at least~\percentage{0.5} in cost
  ratio. The minimum ratios range between~0.870 (\random) and~0.947 (\inc)
  and the maximum ratios are between~1.027 (\random) and~1.104 (\stayincdec).

  Next, we want 
  to find the best initialization strategy after fixing \incdec as the delta 
  strategy. We pair each 
  initialization strategy with \incdec on the same 1000 instances and
  summarize 
  the results of all pairwise tests after
  Bonferroni-correction with factor~112 in 
  \Cref{tbl:statistical_tests_initialization_strategies_inc_fixed}.
  We see that both initialization strategies using Euclidean distances and
  routing turbine production to the nearest free substation, \ie,
  \dijkstraany and \collectingdijkstraany, seem to work best. In particular,
  these are the only initialization strategies that show some significant
  advantage over other strategies.
  In \Cref{plot:strategies_improvement} (right) we depict ratios of solution values
    compared to \collectingdijkstraany. The minimum ratios are between~0.886 
      and~0.923 for all strategies other than \dijkstraany
        ~(0.974). The maximum ratios range between~1.054 and~1.085.
    For the main part there is hardly any
  difference between collecting strategies and their non-collecting counterparts.
  The figure shows, \eg, that on roughly \percentage{22} of all instances 
  \collectingdijkstraany is better than \bfsany and \collectingbfsany
  by~\percentage{0.5}.
  \collectingdijkstraany has a slight but not significant advantage over
  \dijkstraany.
  We therefore declare \collectingdijkstraany paired with \incdec as 
  our best variant. 

  The last row in \Cref{tbl:running_times_ncc_variants} shows the 
  running time characteristics of \collectingdijkstraany paired with \incdec.
  Running times
  range between tenths of milliseconds and~100~seconds.

\subsection{Comparing \milp solvers to establish baseline solver}
\label{sec:cplex}
  
  We conduct preliminary experiments to determine which \milp solver we use as a 
  baseline for our algorithm. To this goal, we randomly choose~35 instances each
  from benchmark sets~$\benchmarkset_1$, and~$\benchmarkset_2$ and~70 instances each 
  from benchmark sets~$\benchmarkset_3$,~$\benchmarkset_4$, and~$\benchmarkset_5$. We
  compare Gurobi~8.0.0 and IBM ILOG CPLEX Optimization Studio v12.8
  with a running time of one day per instance and solver using the~\milp
  formulation from~\cref{sec:milp_formulation}. 
  Since computing an
  optimal solution to the~\milp takes too long in almost all instances, we restrict
  the solvers to different maximum running times.
  Each 
  solver uses one thread per instance and node files are written to disk after 
  the solver uses more than 0.5 GB of memory to store node files. Other than 
  that, default values are used.
  
  During the experiments, we consider three time stamps: one hour, 
  twelve hours, and one day. For each solver, instance, 
  and time stamp we record the value of the best incumbent solution and the MIP 
  gap. If a solver terminates with a proven optimal solution after time 
  stamp~$t$, then the respective values during termination are 
  assigned to all subsequent time stamps.\looseness=-1
  
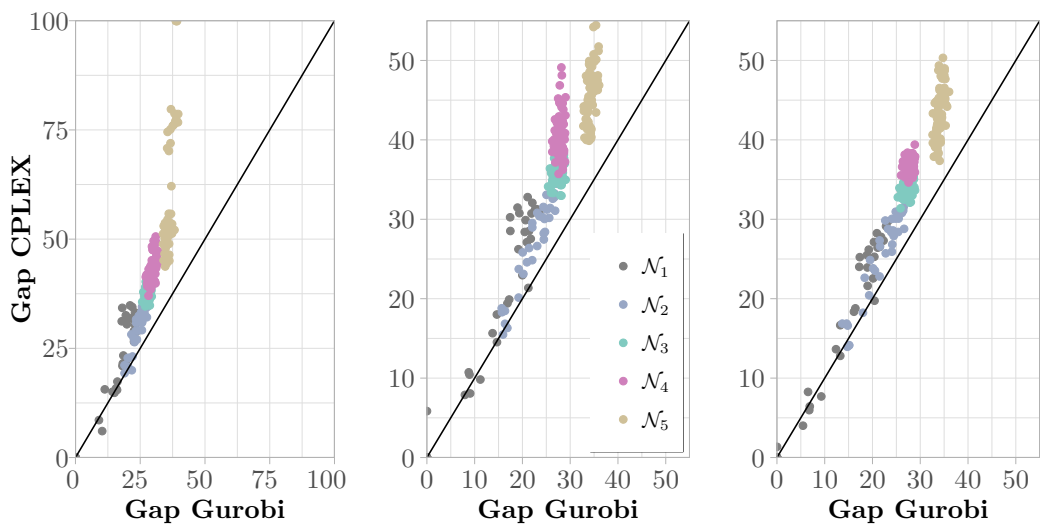
\begin{figure}
	\begin{subfigure}{.342\textwidth}%
		\centering%
  		\input{06-plots/CPLEXvsGUROBI-Gap-1Hour.tex}%
	\end{subfigure}%
	\hfill
	\begin{subfigure}{.3125\textwidth}%
  		\centering%
  		\input{06-plots/CPLEXvsGUROBI-Gap-12Hours.tex}%
	\end{subfigure}%
	\hfill
	\begin{subfigure}{.3125\textwidth}%
  		\centering%
  		\input{06-plots/CPLEXvsGUROBI-Gap-1Day.tex}%
	\end{subfigure}%
\caption{Comparison of gaps between solution values and lower bounds on the optimal value for solutions computed by CPLEX and Gurobi separated by benchmark sets 
	after different maximum running times: \emph{Left:} one hour, \emph{Middle:} twelve hours, \emph{Right:} one day.}
\label{fig:CplexVsGurobi_gaps}
\end{figure}

  The results of the experiment are depicted in 
  \Cref{fig:CplexVsGurobi_incumbent} for the quality of the 
  best solution found by the respective solver and in 
  \Cref{fig:CplexVsGurobi_gaps} for a comparison of MIP gaps.
  In \Cref{fig:CplexVsGurobi_incumbent} each data point corresponds to an
  instance and a time stamp. The value on the abscissa stands for a
  normalized difference in solution values, 
  \ie,~$\nicefrac{(\mathrm{sol}_\mathrm{Gurobi} - \mathrm{sol}_\mathrm{CPLEX})
  }{\max(\mathrm{sol}_\mathrm{Gurobi}\mkern-1.5mu,\,\mathrm{sol}_\mathrm{CPLEX})}$.
  This yields a value in~$[-1,1]$, which is negative if and only if Gurobi
  finds a better solution than CPLEX.
  \Cref{fig:CplexVsGurobi_gaps} shows the \emph{MIP gaps} computed by CPLEX and Gurobi
  for each instance and time stamps. MIP gaps (or \emph{relative gaps})
  are a standard notion from~\MILP{\textsc{ming}}.
  The best feasible solution the solver finds
  yields an upper bound ($\mathrm{ub}$) on the optimal value. The solver
  also tries to prove lower bounds ($\mathrm{lb}$). Combining the best upper and the
  best lower bound yield the MIP gap $\nicefrac{\mathrm{ub} - \mathrm{lb}}{\mathrm{ub}}$.
  This value is in the unit interval and gives
  information on how ``bad'' the solution value can be compared to the
  (unknown) optimal value. A value of zero shows that the best feasible solution
  found by the solver is optimal. Note, however, that a solution might be optimal
  even though the gap is positive.\looseness=-1

  \begin{figure}
	\centering
	\input{06-plots/CPLEXvsGUROBI-Incumbent.tex}
	\caption{Comparison solution values found by~\milp solvers CPLEX and Gurobi 
	after different running times.
	For each instance and running time the abscissa shows a normalized difference 
	in solution values, \ie,~$\nicefrac{(\mathrm{sol}_\mathrm{Gurobi} - \mathrm{sol}_\mathrm{CPLEX})
  	}{\max(\mathrm{sol}_\mathrm{Gurobi}\mkern-1.5mu,\,\mathrm{sol}_\mathrm{CPLEX})}$. There
  	are twelve instances from~$\benchmarkset_5$ with a value between~-0.29 and~-0.49 and
  	another three instances from~$\benchmarkset_5$ with a value less than~-0.99.
  	Four instances from~$\benchmarkset_5$ are infeasible.}
	\label{fig:CplexVsGurobi_incumbent}
  \end{figure}

  Evidently, Gurobi performs better across all 
  benchmark sets and time stamps. While there is evidence that the best 
  incumbent 
  solutions computed by Gurobi and CPLEX become more similar the longer the 
  experiments run, we also see that Gurobi seems to work better than CPLEX the 
  bigger the instances become.
  We therefore use Gurobi as the~\milp solver to compute the baseline to 
  which we compare the negative cycle canceling-based algorithm.  
  
  \subsection{Comparing our Best Variant with Gurobi}
  \label{sec:milpexperiments}
  We compare our algorithm in its best variant, \ie, 
  \collectingdijkstraany with \incdec, with 
  Gurobi on the~\milp formulation in \cref{sec:milp_formulation}.
  We randomly select~200 instances per benchmark set from the benchmark sets
  in \cite{Lehmann:2017:SAW:3077839.3077843}.  

    \begin{figure*}
    \begin{subfigure}[t]{.342\textwidth}%
      \centering%
      \input{06-plots/NCC2MILP-2Seconds-ImprovementPlot.tex}%
      \label{plot:improvement_plot_Algo_vs_Gurobi_two_seconds}%
    \end{subfigure}%
    \hfill
    \begin{subfigure}[t]{.3125\textwidth}%
      \centering%
      \input{06-plots/NCC2MILP-1Hour-ImprovementPlot-Shortened.tex}%
      \label{plot:improvement_plot_Algo_vs_Gurobi_one_hour}%
    \end{subfigure}%
    \hfill
    \begin{subfigure}[t]{.3125\textwidth}%
      \centering%
      \input{06-plots/NCC2MILP-1Day-ImprovementPlot-Shortened.tex}%
      \label{plot:improvement_plot_Algo_vs_Gurobi_one_day}%
    \end{subfigure}%
    \caption{Comparison of the \ncc algorithm to Gurobi 
    on~200 instances per benchmark set. The ordinate shows the ratio of 
    objective values at various maximum running times of our al\-go\-rithm to 
    objective values of Gurobi. Running times:
    \emph{Left:}~two seconds,
    \emph{Middle:}~one hour,
    \emph{Right:}~one day.}%
    \label{plot:Algo_improvement}
  \end{figure*}
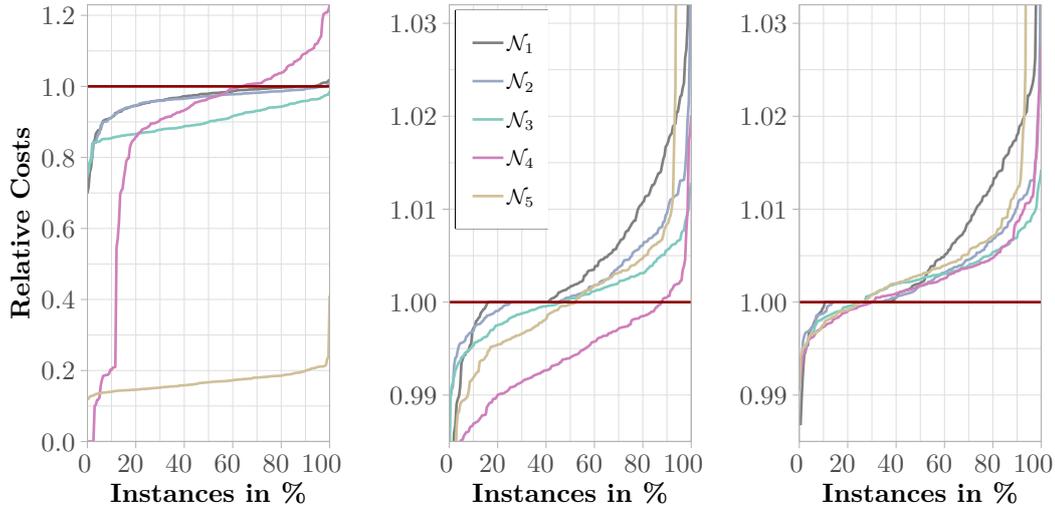
    \begin{figure}
  \centering
  \includegraphics[width=1\columnwidth]{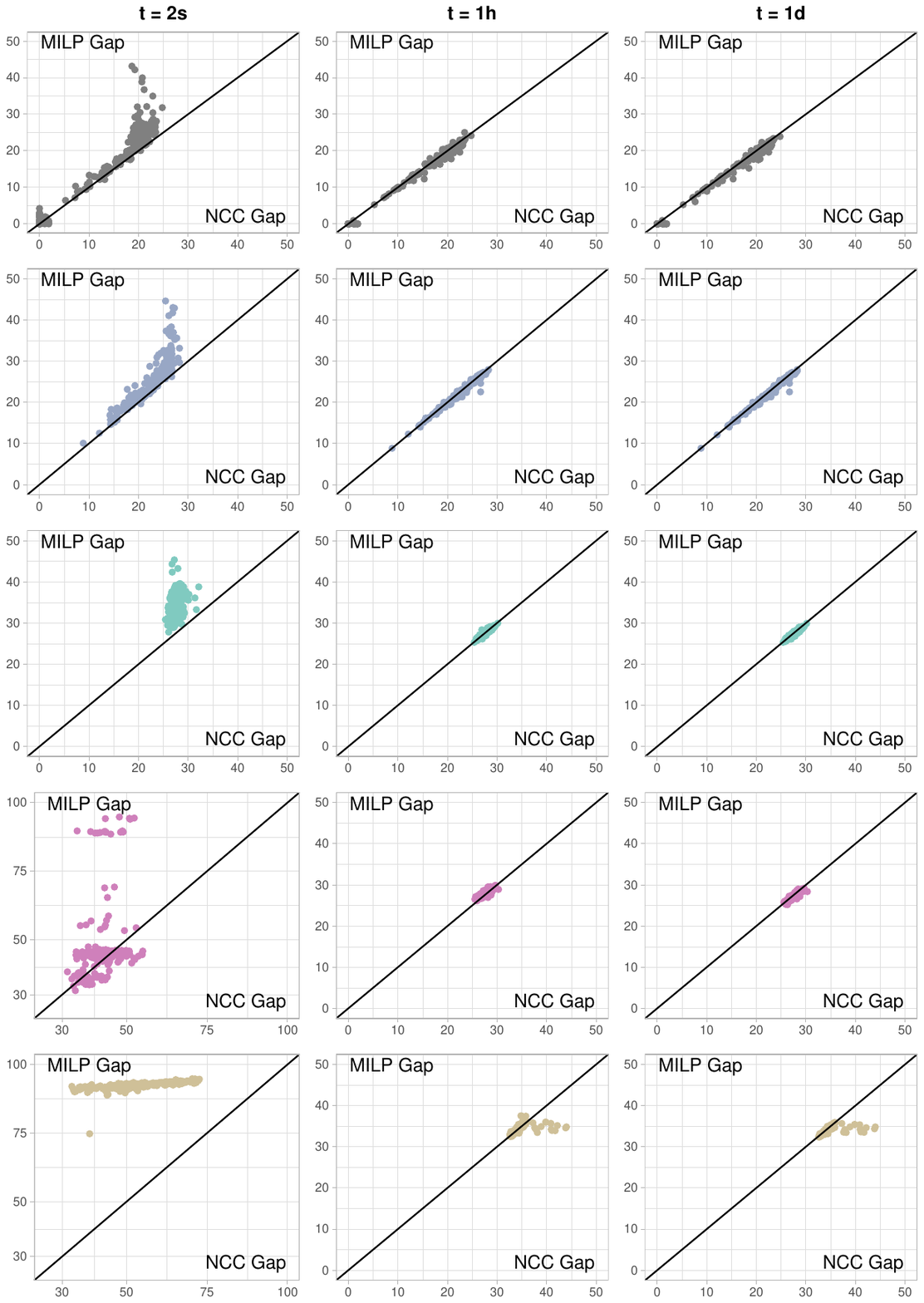}
  \caption{Comparison of gaps between solution values and lower bounds on the
  optimal value for solutions computed by Gurobi~(\milp) and the~\ncc
  algorithm separated by benchmark sets ($\benchmarkset_1$ in row~1
  through~$\benchmarkset_5$ in row~5)
  after maximum running times of two seconds, one hour, and one day.
  Lower bounds are taken from~\milp experiments with running times of one day.}
  \label{fig:milp_ncc_gaps}
  \end{figure}

  In \cref{plot:Algo_improvement} we plot the ratio of the best
  solution value found by our algorithm to Gurobi's best
  solution at running times of two seconds, one hour, and one day for each
  benchmark set separately. These running times 
  represent both interactive and non-time-critical planning. 
  Since our algorithm terminates in under~100~seconds, 
  the comparisons in \cref{plot:Algo_improvement} (middle and right)
  use the solution our algorithm 
  provides at termination. While discussing the plots, we also discuss
  an adaptation of the relative 
  gaps~$\nicefrac{\mathrm{ub} - \mathrm{lb}}{\mathrm{ub}}$ 
  we introduced in~\Cref{sec:cplex}.
  For each instance, we use the lower bounds from 
  the one-day \milp experiments.
  For each instance, each maximum running time and for both the \milp and the 
  \ncc algorithm take best solution value ($\mathrm{ub}$) found at 
  the maximum running time. 
  We refer to the relative gaps as \emph{\milp gap}
  and \emph{\ncc gap}, respectively, and show them in~\Cref{fig:milp_ncc_gaps}.
  
  After two seconds our algorithm outperforms Gurobi on all benchmark sets
  as it finds better solutions 
  on~\percentage{89} of all instances with the lowest percentage on benchmark
  set~$\benchmarkset_4$.
  On~$\benchmarkset_1$ the \ncc gaps are on average $\percentage{14.1}$ with a
  maximum of $\percentage{24.8}$ compared to \milp gaps of $\percentage{16.9}$
  on average and at most $\percentage{43.1}$. For~$\benchmarkset_3$, the \ncc
  gaps are on average $\percentage{27.6}$ with a spread of only seven
  percentage points, compared to a mean of $\percentage{34.6}$ and a maximum
  of $\percentage{45.4}$ for the \milp gap. The values for~$\benchmarkset_2$
  range between those for~$\benchmarkset_1$ and~$\benchmarkset_3$. 
  The ratios of solution values
  range between~0.699 and~1.019 for~$\benchmarkset_1$,~$\benchmarkset_2$,
  and~$\benchmarkset_3$.
  On~$\benchmarkset_4$, which contains the largest instances, our algorithm
  computes better solutions on~\percentage{62} of the
  instances. On six instances Gurobi does not find a solution. The
    instances on which Gurobi is better are on average larger than the 
  other instances
  in~$\benchmarkset_4$. There are~18~instances on which the ratio of
  solution values exceeds~1.1 with a maximum of~1.228. 
  On those very large instances, detecting negative cycles takes longer and fewer 
  iterations are performed in two seconds. The \ncc gaps 
    spread between $\percentage{31.6}$ and
    $\percentage{57.4}$ with an average of $\percentage{42.6}$. The \milp
    gaps are even worse with a mean value of $\percentage{48.3}$ and~18
    instances above $\percentage{88.5}$. 
On the complete graphs of~$\benchmarkset_5$,
  our algorithm produces solutions that are at
  least~\percentage{75}~cheaper than Gurobi's on all but one
  instance (which has a ratio of 0.411). 
  The gaps are on average at~$\percentage{53.6}$ for the
    \ncc algorithm and at $\percentage{92.3}$ for Gurobi.
  
  Within one hour (middle plot in \cref{plot:Algo_improvement})
  Gurobi finds better or equivalent solutions than our algorithm on a majority of
  the instances in benchmark sets~$\benchmarkset_1$,~$\benchmarkset_2$,
  and~$\benchmarkset_3$. 
  On~\percentage{25} of the
  instances from~$\benchmarkset_1$, on~\percentage{18.5} of the 
  instances from~$\benchmarkset_2$, and on one instance
  from~$\benchmarkset_3$ the solution values are equal.
  On~$\benchmarkset_4$ and~$\benchmarkset_5$, our
  algorithm still yields better solutions on~\percentage{87.5}
  and~\percentage{52} of the
  instances, respectively.
  Our algorithm is within~\percentage{0.5} of Gurobi's best solution 
  on~\percentage{81.4} and within~\percentage{1} 
  on~\percentage{91.3} of all instances. Only on six of~1000 instances
    (all in~$\benchmarkset_5$), the ratio exceeds~1.10 with a maximum of~1.165. 
    That means, while the \ncc algorithm is comparable
  to Gurobi in solution quality on small instances, it proves better on
  larger wind farms. Furthermore, our algorithm is much faster 
  since it terminates
  in under~100~seconds---compared to one hour of maximum running time
  for Gurobi.

  After running times of one day
  (right plot in \cref{plot:Algo_improvement}), while our
  algorithm is at least as good as Gurobi
  only on between~\percentage{25}~($\benchmarkset_5$) 
  and~\percentage{38.5}~($\benchmarkset_1$)
  of the instances, it is within~\percentage{1} of Gurobi's solution 
  on~\percentage{87.7} of all instances.
  Again, there are only six
    instances with a ratio worse than~1.10 with a maximum of~1.169.
  Our algorithm does not profit from long
  running times since it gets stuck in local minima.
  Thus, the~\milp solver is the better choice if more time is available.
  Between running times of one hour and one day, the gaps look vastly the same
  and there is hardly any difference between \ncc gaps and \milp gaps.
  They range between zero and $\percentage{25.0}$ on~$\benchmarkset_1$,
  clot around $\percentage{28}$ for~$\benchmarkset_3$ and ~$\benchmarkset_4$
  and around $\percentage{34}$ for~$\benchmarkset_5$ with seven outliers
  to the worse by the \ncc algorithm.

  In summary, these experiments show that the \ncc algorithm is a viable option
  compared to Gurobi with long running times and that it yields better
  solutions than the
  \milp solver if only a short amount of time is given.
  \subsection{Comparison to Metaheuristic Simulated Annealing}
  \label{sec:sa_experiments}
  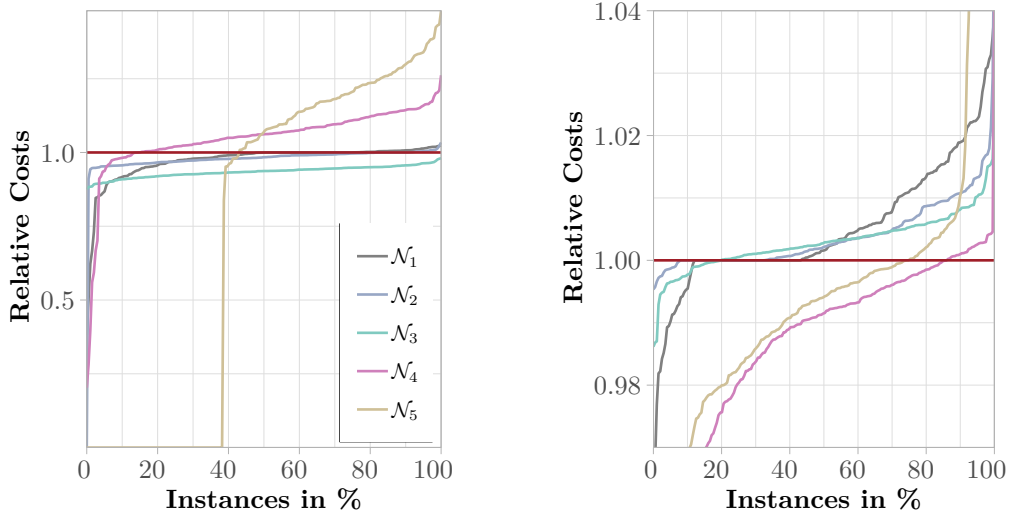
\begin{figure}
      \begin{subfigure}{.48\textwidth}%
      \centering%
      \input{06-plots/2019-SEA-SAvsAlgo-2Seconds-Shortened.tex}%
      \label{plot:improvement_plot_Algo_vs_SA_two_seconds}%
    \end{subfigure}%
    \hfill
    \begin{subfigure}{.48\textwidth}%
      \centering%
      \input{06-plots/2019-SEA-SAvsAlgo-1Hour-Shortened.tex}%
      \label{plot:improvement_plot_Algo_vs_SA__one_hour}%
    \end{subfigure}%
    \caption{Comparison of Negative Cycle Canceling algorithm to the Simulated Annealing
      algorithm on 200~instances per benchmark set. The ordinate 
      represents the ratio of objective values at different maximum running 
      times of our algorithm to 
      objective values of the Simulated Annealing algorithm.
      \emph{Left:}~Running time of two seconds.
      \emph{Right:}~Running time of one hour.}%
    \label{plot:improvement_plot_Algo_vs_SA}%
  \end{figure}
  We compare our best algorithm variant with the 
  best  
  variant of a Simulated Annealing (\sa)
  algorithm~\cite{Lehmann:2017:SAW:3077839.3077843}.
  We run the \sa algorithm on~200 randomly selected instances per benchmark set 
  (independently selected from other experiments).
  We compare the best 
  solutions found after two 
  seconds and one
  hour (\cref{plot:improvement_plot_Algo_vs_SA}).
  
  After two seconds, the \ncc algorithm performs at least as good as 
  the \sa algorithm on all 
  instances from~$\benchmarkset_3$ and on~\percentage{74.5} 
  and~\percentage{90.5} on~$\benchmarkset_1$ and~$\benchmarkset_2$, 
  respectively. The minimum ratios are~0.381 for~$\benchmarkset_1$,~0.911 
  for~$\benchmarkset_2$, and~0.875 for~$\benchmarkset_3$ with one
    instance in~$\benchmarkset_2$ where the \sa algorithm does not find
      a solution. The maximum ratio on those benchmark sets is at most~1.034.
  On the larger instances of~$\benchmarkset_4$ 
  and~$\benchmarkset_5$, our algorithm presumably cannot perform sufficient 
  iterations, as the \sa algorithm is better on~\percentage{71} of those 
  instances. Yet, the \sa algorithm does not find feasible solutions 
  on~\percentage{38.5} of instances from~$\benchmarkset_5$. The ratios
    have a wide spread: from~0.203 to~1.261 for~$\benchmarkset_4$
    and from~0.838 to~1.480 for~$\benchmarkset_5$ (save for the instances without
    a solution from the \sa algorithm).

  After one hour, the \sa algorithm provides better solutions than our algorithm
  on~\percentage{67.5} and~\percentage{80} of instances from $\benchmarkset_2$
  and $\benchmarkset_3$, respectively. Our algorithm, however,
  stays within~\percentage{1} in solution quality on~\percentage{84.2} on
  the benchmark sets~$\benchmarkset_1$--$\benchmarkset_3$. Again, our algorithm 
  seems to be stuck
  in local minima. On $\benchmarkset_4$ and $\benchmarkset_5$, our algorithm
  performs better than the \sa algorithm on~\percentage{86} 
  and~\percentage{74.5}, respectively. Apparently, the \sa algorithm needs more
  time to explore the solution space. The minimum ratios of solution values 
    are as low as~0.716 for~$\benchmarkset_1$ and between~0.905 and~0.995 for
    the other benchmark sets. The maximum ratios are at most~1.057 for all
    benchmark sets except~$\benchmarkset_5$~(1.159).
  This supports our findings from the~\milp experiments that our algorithm 
  is competitive to other approaches to solving~\wcp within very short amounts
  of time. In view of an interactive planning process, it stands out that
  the \sa algorithm struggles to find solutions quickly in dense graphs.
  \section{Conclusion}
  \label{sec:conclusion}
  Based on recently presented ideas~\cite{Gritzbach2018} we propose and compare 
  numerous 
  variants of a Negative Cycle Canceling heuristic for the~\WCP. While all 
  variants run in the order of milliseconds up to 4.5~minutes, they 
  differ significantly in quality. We identify the best variant and use it to 
  compare our heuristic to the \milp solver Gurobi and a Simulated Annealing 
  algorithm from the literature. With these comparisons we are able to solve 
  several open questions~\cite{Gritzbach2018}. 
  While the \milp solver Gurobi has the potential to find optimal solutions if 
  it 
  runs long enough, our heuristic is able to find 
  solutions of comparable quality in only a fraction of the time. Our algorithm
  beats Gurobi in finding good solutions in a matter of seconds. We make similar
  observations when we compare ourselves to a Simulated 
  Annealing approach.
%
%
  
  Moving forward, one may investigate how to improve the solution quality of 
  our heuristic. Visually comparing flows from our algorithm and other solution 
  methods may help to identify what kind of more complex circulations 
  improve the solution. It then remains to investigate how these circulations 
  can be detected. Also, methods for escaping local minima such as temporarily 
  allowing worse 
  solutions could help to improve our algorithm. It also remains open whether 
  one 
  can prove any theoretical guarantees on the solution quality or the number of 
  iterations. Along the same lines, any theoretical insights on why one delta
    or initialization strategy works better than another, or on the order in which
    cycles should be canceled could help improve the \ncc~algorithm.
  
  In a broader algorithmic view, the heuristic can be easily generalized to 
  minimum-cost flow problems 
  with other types of cost functions provided that one searches for integral 
  flows. It would be interesting to see how well the heuristic performs there. 

  \bibliography{04-bibliography/article,04-bibliography/books,%
    04-bibliography/internet}

    \newpage
  
  \appendix

  \section{Binomial Sign Test for Two Dependent Samples}
  \label{sec:signtest}
  Statistical tests help to find the best strategy variant for our algorithm 
  with
  regards to available initialization
  strategies~(\cref{sec:initialization_strategies}) and delta
  strategies~(\cref{sec:delta_strategies}). In our case, we use the 
  \emph{Binomial
    Sign Test} for two dependent samples~\cite[p.~303]{SWB-343458624}.
  
  We explain this test in a general setting here and specify how we apply the 
  test in more detail below.
  Generally speaking, we compare~$k$ variants of an algorithm. In our case 
  these 
  are the different initialization and delta strategies
  (\cref{sec:initialization_strategies,sec:delta_strategies}).
  We apply each variant to each instance. For every instance~$m$, 
  we denote the total cost of the resulting flow computed by variant~$i$ on 
  instance~$m$ by~$X^{(i)}_m$.
  
  For any ordered pair of two variants~$(i, j)$ running on a fixed 
  instance~$m$, we
  calculate its solution difference~$D = X^{(i)}_m - X^{(j)}_m$ and 
  increment---depending on the sign
  of~$D$---either~$D_{i < j}, D_{i > j}$, or $D_{i = j}$ where, for 
  example,~$D_{i < j}$ counts the instances in which~$i$ performed better 
  than~$j$. If both variants were
  equally good, then $D_{i < j} \sim
  \mathrm{Bin}(D_{i < j} + D_{i > j}, 0.5)$, \ie, $D_{i < j}$ is binomially 
  distributed on~$D_{i < j} +D_{i > j}$ trials and probability~$0.5$.
  
  We perform~$k(k-1)$ tests, one for each ordered pair of variants, and always
  test the null hypothesis~$\mathrm{H}_0 \colon \theta = 0.5$ against the
  alternative hypothesis~$\mathrm{H}_1 \colon \theta > 0.5$ where~$\theta$ is 
  the
  probability in the underlying hypothesized distribution~$D_{i < j} \sim
  \mathrm{Bin}(D_{i < j}+D_{i > j}, \theta)$. The resulting $p$-values are
  Bonferroni-corrected by the number of tests. In this setting, we interpret
  rejecting~$\mathrm{H}_0$ as algorithm variant~$i$ performing better than
  algorithm variant~$j$.

\section{\milp formulation}
\label{sec:milp_formulation}

  Recall that we introduced the notion of cable 
  types in \cref{sec:introduction}. Let~\cabletypes denote the set of cable 
  types. Each cable type~$\cabletype \in \cabletypes$ has a capacity on the 
  amount of turbine production that can be transmitted through it, which we 
  denote by~$\cablecapacity_\cabletype$, as well as a cost per unit 
  length~$\cablecost_\cabletype$ for laying a cable of type~\cabletype.
  The \milp formulation for~\wcp we used in our experiments is as follows:\looseness=-1
  \begin{align}
  \min &\sum_{\edge \in \edges} \sum_{\cabletype \in \cabletypes}
  \cablecost_\cabletype \cdot \decisionVar(\edge , \cabletype) \cdot 
  \edgelength(\edge) 
  \label{eq:milp:cost}\\
  \text{s.\ t. }\netflow(\vertexa) &= -1 
  &\forall \vertexa \in \turbines,
  \label{eq:milp:feasibleFlow:turbines}\\
  \netflow(\vertexb) &\leq \substationCapacity(\vertexb) 
  &\forall \vertexb \in \substations,
  \label{eq:milp:feasibleFlow:substations}\\
  \abs{\flow(\edge)} &\leq \sum_{\cabletype \in \cabletypes} \decisionVar(\edge 
  , \cabletype) \cdot \cablecapacity_\cabletype
  &\forall \edge \in \edges,
  \label{eq:milp:feasibleFlow:sufficentCableCapacity}\\
  \sum_{\cabletype \in \cabletypes} \decisionVar(\edge , \cabletype) &\leq 1
  &\forall \edge \in \edges,
  \label{eq:milp:feasibleFlow:oneCabletypeOnly}\\
  \flow(\vertexa, \vertexb) &\leq 0
  &\forall (\vertexa, \vertexb) \in \edges \colon \vertexa \in \substations,
  \label{eq:milp:feasibleFlow:noSubstationOutflow1}\\
  \flow(\vertexa, \vertexb) &\geq 0
  &\forall (\vertexa, \vertexb) \in \edges \colon \vertexb \in \substations,
  \label{eq:milp:feasibleFlow:noSubstationOutflow2}
  %
  %
  \end{align}
  
  where~\netflow denotes the net flow defined in~\cref{sec:model}, $\flow 
  \in 
  \reals^{\edges}$, and $\decisionVar \in \{0,1\}^{\edges \times \cabletypes}$. 
  \Cref{eq:milp:feasibleFlow:turbines,eq:milp:feasibleFlow:substations} are the 
  same as the constraints given in 
  \Cref{eq:feasibleFlow:turbines,eq:feasibleFlow:substations}. 
  \Cref{eq:milp:feasibleFlow:sufficentCableCapacity} ensures that there is 
  enough 
  cable capacity installed on every edge for the respective flow, while there 
  is 
  only one cable type on that edge due to 
  \cref{eq:milp:feasibleFlow:oneCabletypeOnly}. 
  \Cref{eq:milp:feasibleFlow:noSubstationOutflow1,eq:milp:feasibleFlow:noSubstationOutflow2}
  correspond to 
  \Cref{eq:feasibleFlow:noSubstationOutflow1,eq:feasibleFlow:noSubstationOutflow2}
  and ensure that no flow leaves any substation.

\end{document}

%% file: 06-plots/CPLEXvsGUROBI-Gap-1Hour.tex
\begin{tikzpicture}[x=1pt,y=1pt]
\definecolor{fillColor}{RGB}{255,255,255}
\path[use as bounding box,fill=fillColor,fill opacity=0.00] (0,0) rectangle (130.81,199.47);
\begin{scope}
\path[clip] (  0.00,  0.00) rectangle (130.81,199.47);
\definecolor{drawColor}{RGB}{255,255,255}
\definecolor{fillColor}{RGB}{255,255,255}

\path[draw=drawColor,line width= 0.5pt,line join=round,line cap=round,fill=fillColor] (  0.00,  0.00) rectangle (130.81,199.47);
\end{scope}
\begin{scope}
\path[clip] ( 28.84, 29.68) rectangle (125.81,194.47);
\definecolor{fillColor}{RGB}{255,255,255}

\path[fill=fillColor] ( 28.84, 29.68) rectangle (125.81,194.47);
\definecolor{drawColor}{gray}{0.87}

\path[draw=drawColor,line width= 0.1pt,line join=round] ( 28.84, 50.27) --
	(125.81, 50.27);

\path[draw=drawColor,line width= 0.1pt,line join=round] ( 28.84, 91.47) --
	(125.81, 91.47);

\path[draw=drawColor,line width= 0.1pt,line join=round] ( 28.84,132.67) --
	(125.81,132.67);

\path[draw=drawColor,line width= 0.1pt,line join=round] ( 28.84,173.87) --
	(125.81,173.87);

\path[draw=drawColor,line width= 0.1pt,line join=round] ( 40.96, 29.68) --
	( 40.96,194.47);

\path[draw=drawColor,line width= 0.1pt,line join=round] ( 65.20, 29.68) --
	( 65.20,194.47);

\path[draw=drawColor,line width= 0.1pt,line join=round] ( 89.45, 29.68) --
	( 89.45,194.47);

\path[draw=drawColor,line width= 0.1pt,line join=round] (113.69, 29.68) --
	(113.69,194.47);

\path[draw=drawColor,line width= 0.3pt,line join=round] ( 28.84, 29.68) --
	(125.81, 29.68);

\path[draw=drawColor,line width= 0.3pt,line join=round] ( 28.84, 70.87) --
	(125.81, 70.87);

\path[draw=drawColor,line width= 0.3pt,line join=round] ( 28.84,112.07) --
	(125.81,112.07);

\path[draw=drawColor,line width= 0.3pt,line join=round] ( 28.84,153.27) --
	(125.81,153.27);

\path[draw=drawColor,line width= 0.3pt,line join=round] ( 28.84,194.47) --
	(125.81,194.47);

\path[draw=drawColor,line width= 0.3pt,line join=round] ( 28.84, 29.68) --
	( 28.84,194.47);

\path[draw=drawColor,line width= 0.3pt,line join=round] ( 53.08, 29.68) --
	( 53.08,194.47);

\path[draw=drawColor,line width= 0.3pt,line join=round] ( 77.33, 29.68) --
	( 77.33,194.47);

\path[draw=drawColor,line width= 0.3pt,line join=round] (101.57, 29.68) --
	(101.57,194.47);

\path[draw=drawColor,line width= 0.3pt,line join=round] (125.81, 29.68) --
	(125.81,194.47);
\definecolor{drawColor}{RGB}{128,128,128}
\definecolor{fillColor}{RGB}{128,128,128}

\path[draw=drawColor,line width= 0.4pt,line join=round,line cap=round,fill=fillColor] ( 43.43, 54.10) circle (  1.43);

\path[draw=drawColor,line width= 0.4pt,line join=round,line cap=round,fill=fillColor] ( 46.57, 65.16) circle (  1.43);

\path[draw=drawColor,line width= 0.4pt,line join=round,line cap=round,fill=fillColor] ( 46.16, 81.07) circle (  1.43);

\path[draw=drawColor,line width= 0.4pt,line join=round,line cap=round,fill=fillColor] ( 28.85, 29.69) circle (  1.43);

\path[draw=drawColor,line width= 0.4pt,line join=round,line cap=round,fill=fillColor] ( 28.85, 29.69) circle (  1.43);

\path[draw=drawColor,line width= 0.4pt,line join=round,line cap=round,fill=fillColor] ( 38.84, 39.65) circle (  1.43);

\path[draw=drawColor,line width= 0.4pt,line join=round,line cap=round,fill=fillColor] ( 28.85, 29.69) circle (  1.43);

\path[draw=drawColor,line width= 0.4pt,line join=round,line cap=round,fill=fillColor] ( 44.54, 58.33) circle (  1.43);

\path[draw=drawColor,line width= 0.4pt,line join=round,line cap=round,fill=fillColor] ( 52.04, 83.06) circle (  1.43);

\path[draw=drawColor,line width= 0.4pt,line join=round,line cap=round,fill=fillColor] ( 50.70, 77.38) circle (  1.43);

\path[draw=drawColor,line width= 0.4pt,line join=round,line cap=round,fill=fillColor] ( 48.10, 81.85) circle (  1.43);

\path[draw=drawColor,line width= 0.4pt,line join=round,line cap=round,fill=fillColor] ( 48.09, 79.94) circle (  1.43);

\path[draw=drawColor,line width= 0.4pt,line join=round,line cap=round,fill=fillColor] ( 39.81, 55.40) circle (  1.43);

\path[draw=drawColor,line width= 0.4pt,line join=round,line cap=round,fill=fillColor] ( 47.70, 83.32) circle (  1.43);

\path[draw=drawColor,line width= 0.4pt,line join=round,line cap=round,fill=fillColor] ( 50.59, 81.90) circle (  1.43);

\path[draw=drawColor,line width= 0.4pt,line join=round,line cap=round,fill=fillColor] ( 46.77, 68.17) circle (  1.43);

\path[draw=drawColor,line width= 0.4pt,line join=round,line cap=round,fill=fillColor] ( 50.10, 86.47) circle (  1.43);

\path[draw=drawColor,line width= 0.4pt,line join=round,line cap=round,fill=fillColor] ( 47.96, 66.54) circle (  1.43);

\path[draw=drawColor,line width= 0.4pt,line join=round,line cap=round,fill=fillColor] ( 28.85, 29.69) circle (  1.43);

\path[draw=drawColor,line width= 0.4pt,line join=round,line cap=round,fill=fillColor] ( 52.81, 83.93) circle (  1.43);

\path[draw=drawColor,line width= 0.4pt,line join=round,line cap=round,fill=fillColor] ( 44.40, 55.17) circle (  1.43);

\path[draw=drawColor,line width= 0.4pt,line join=round,line cap=round,fill=fillColor] ( 42.73, 54.49) circle (  1.43);

\path[draw=drawColor,line width= 0.4pt,line join=round,line cap=round,fill=fillColor] ( 49.28, 87.07) circle (  1.43);

\path[draw=drawColor,line width= 0.4pt,line join=round,line cap=round,fill=fillColor] ( 44.06, 55.80) circle (  1.43);

\path[draw=drawColor,line width= 0.4pt,line join=round,line cap=round,fill=fillColor] ( 50.76, 73.39) circle (  1.43);

\path[draw=drawColor,line width= 0.4pt,line join=round,line cap=round,fill=fillColor] ( 51.44, 76.99) circle (  1.43);

\path[draw=drawColor,line width= 0.4pt,line join=round,line cap=round,fill=fillColor] ( 51.21, 79.61) circle (  1.43);

\path[draw=drawColor,line width= 0.4pt,line join=round,line cap=round,fill=fillColor] ( 51.72, 83.99) circle (  1.43);

\path[draw=drawColor,line width= 0.4pt,line join=round,line cap=round,fill=fillColor] ( 46.44, 64.23) circle (  1.43);

\path[draw=drawColor,line width= 0.4pt,line join=round,line cap=round,fill=fillColor] ( 46.36, 86.15) circle (  1.43);

\path[draw=drawColor,line width= 0.4pt,line join=round,line cap=round,fill=fillColor] ( 51.13, 84.43) circle (  1.43);

\path[draw=drawColor,line width= 0.4pt,line join=round,line cap=round,fill=fillColor] ( 47.79, 66.68) circle (  1.43);

\path[draw=drawColor,line width= 0.4pt,line join=round,line cap=round,fill=fillColor] ( 37.62, 43.79) circle (  1.43);

\path[draw=drawColor,line width= 0.4pt,line join=round,line cap=round,fill=fillColor] ( 49.84, 82.33) circle (  1.43);

\path[draw=drawColor,line width= 0.4pt,line join=round,line cap=round,fill=fillColor] ( 51.17, 77.53) circle (  1.43);
\definecolor{drawColor}{RGB}{152,167,197}
\definecolor{fillColor}{RGB}{152,167,197}

\path[draw=drawColor,line width= 0.4pt,line join=round,line cap=round,fill=fillColor] ( 55.34, 88.48) circle (  1.43);

\path[draw=drawColor,line width= 0.4pt,line join=round,line cap=round,fill=fillColor] ( 47.19, 64.49) circle (  1.43);

\path[draw=drawColor,line width= 0.4pt,line join=round,line cap=round,fill=fillColor] ( 51.61, 80.42) circle (  1.43);

\path[draw=drawColor,line width= 0.4pt,line join=round,line cap=round,fill=fillColor] ( 51.42, 77.42) circle (  1.43);

\path[draw=drawColor,line width= 0.4pt,line join=round,line cap=round,fill=fillColor] ( 56.06, 90.18) circle (  1.43);

\path[draw=drawColor,line width= 0.4pt,line join=round,line cap=round,fill=fillColor] ( 54.67, 84.03) circle (  1.43);

\path[draw=drawColor,line width= 0.4pt,line join=round,line cap=round,fill=fillColor] ( 54.97, 89.33) circle (  1.43);

\path[draw=drawColor,line width= 0.4pt,line join=round,line cap=round,fill=fillColor] ( 53.83, 81.04) circle (  1.43);

\path[draw=drawColor,line width= 0.4pt,line join=round,line cap=round,fill=fillColor] ( 54.55, 85.18) circle (  1.43);

\path[draw=drawColor,line width= 0.4pt,line join=round,line cap=round,fill=fillColor] ( 53.04, 83.16) circle (  1.43);

\path[draw=drawColor,line width= 0.4pt,line join=round,line cap=round,fill=fillColor] ( 50.95, 73.21) circle (  1.43);

\path[draw=drawColor,line width= 0.4pt,line join=round,line cap=round,fill=fillColor] ( 55.00, 85.90) circle (  1.43);

\path[draw=drawColor,line width= 0.4pt,line join=round,line cap=round,fill=fillColor] ( 52.36, 77.60) circle (  1.43);

\path[draw=drawColor,line width= 0.4pt,line join=round,line cap=round,fill=fillColor] ( 53.32, 83.59) circle (  1.43);

\path[draw=drawColor,line width= 0.4pt,line join=round,line cap=round,fill=fillColor] ( 55.66, 88.64) circle (  1.43);

\path[draw=drawColor,line width= 0.4pt,line join=round,line cap=round,fill=fillColor] ( 53.75, 80.55) circle (  1.43);

\path[draw=drawColor,line width= 0.4pt,line join=round,line cap=round,fill=fillColor] ( 48.52, 63.97) circle (  1.43);

\path[draw=drawColor,line width= 0.4pt,line join=round,line cap=round,fill=fillColor] ( 55.72, 86.48) circle (  1.43);

\path[draw=drawColor,line width= 0.4pt,line join=round,line cap=round,fill=fillColor] ( 48.84, 67.27) circle (  1.43);

\path[draw=drawColor,line width= 0.4pt,line join=round,line cap=round,fill=fillColor] ( 52.57, 83.15) circle (  1.43);

\path[draw=drawColor,line width= 0.4pt,line join=round,line cap=round,fill=fillColor] ( 49.78, 76.11) circle (  1.43);

\path[draw=drawColor,line width= 0.4pt,line join=round,line cap=round,fill=fillColor] ( 48.40, 67.22) circle (  1.43);

\path[draw=drawColor,line width= 0.4pt,line join=round,line cap=round,fill=fillColor] ( 50.64, 74.73) circle (  1.43);

\path[draw=drawColor,line width= 0.4pt,line join=round,line cap=round,fill=fillColor] ( 53.47, 84.58) circle (  1.43);

\path[draw=drawColor,line width= 0.4pt,line join=round,line cap=round,fill=fillColor] ( 49.93, 62.61) circle (  1.43);

\path[draw=drawColor,line width= 0.4pt,line join=round,line cap=round,fill=fillColor] ( 51.09, 74.61) circle (  1.43);

\path[draw=drawColor,line width= 0.4pt,line join=round,line cap=round,fill=fillColor] ( 51.82, 74.20) circle (  1.43);

\path[draw=drawColor,line width= 0.4pt,line join=round,line cap=round,fill=fillColor] ( 53.70, 77.73) circle (  1.43);

\path[draw=drawColor,line width= 0.4pt,line join=round,line cap=round,fill=fillColor] ( 55.57, 82.60) circle (  1.43);

\path[draw=drawColor,line width= 0.4pt,line join=round,line cap=round,fill=fillColor] ( 53.80, 81.88) circle (  1.43);

\path[draw=drawColor,line width= 0.4pt,line join=round,line cap=round,fill=fillColor] ( 51.56, 81.80) circle (  1.43);

\path[draw=drawColor,line width= 0.4pt,line join=round,line cap=round,fill=fillColor] ( 47.22, 61.49) circle (  1.43);

\path[draw=drawColor,line width= 0.4pt,line join=round,line cap=round,fill=fillColor] ( 53.77, 86.52) circle (  1.43);

\path[draw=drawColor,line width= 0.4pt,line join=round,line cap=round,fill=fillColor] ( 50.20, 67.80) circle (  1.43);

\path[draw=drawColor,line width= 0.4pt,line join=round,line cap=round,fill=fillColor] ( 54.06, 83.54) circle (  1.43);
\definecolor{drawColor}{RGB}{128,202,192}
\definecolor{fillColor}{RGB}{128,202,192}

\path[draw=drawColor,line width= 0.4pt,line join=round,line cap=round,fill=fillColor] ( 56.45, 92.65) circle (  1.43);

\path[draw=drawColor,line width= 0.4pt,line join=round,line cap=round,fill=fillColor] ( 55.30, 90.22) circle (  1.43);

\path[draw=drawColor,line width= 0.4pt,line join=round,line cap=round,fill=fillColor] ( 56.08, 94.88) circle (  1.43);

\path[draw=drawColor,line width= 0.4pt,line join=round,line cap=round,fill=fillColor] ( 56.30, 95.21) circle (  1.43);

\path[draw=drawColor,line width= 0.4pt,line join=round,line cap=round,fill=fillColor] ( 55.63, 91.07) circle (  1.43);

\path[draw=drawColor,line width= 0.4pt,line join=round,line cap=round,fill=fillColor] ( 56.38, 88.98) circle (  1.43);

\path[draw=drawColor,line width= 0.4pt,line join=round,line cap=round,fill=fillColor] ( 57.02, 91.77) circle (  1.43);

\path[draw=drawColor,line width= 0.4pt,line join=round,line cap=round,fill=fillColor] ( 56.14, 90.89) circle (  1.43);

\path[draw=drawColor,line width= 0.4pt,line join=round,line cap=round,fill=fillColor] ( 54.97, 93.97) circle (  1.43);

\path[draw=drawColor,line width= 0.4pt,line join=round,line cap=round,fill=fillColor] ( 55.68, 92.77) circle (  1.43);

\path[draw=drawColor,line width= 0.4pt,line join=round,line cap=round,fill=fillColor] ( 55.04, 92.24) circle (  1.43);

\path[draw=drawColor,line width= 0.4pt,line join=round,line cap=round,fill=fillColor] ( 56.01, 90.18) circle (  1.43);

\path[draw=drawColor,line width= 0.4pt,line join=round,line cap=round,fill=fillColor] ( 54.26, 92.11) circle (  1.43);

\path[draw=drawColor,line width= 0.4pt,line join=round,line cap=round,fill=fillColor] ( 55.96, 91.38) circle (  1.43);

\path[draw=drawColor,line width= 0.4pt,line join=round,line cap=round,fill=fillColor] ( 55.51, 92.10) circle (  1.43);

\path[draw=drawColor,line width= 0.4pt,line join=round,line cap=round,fill=fillColor] ( 56.76, 91.62) circle (  1.43);

\path[draw=drawColor,line width= 0.4pt,line join=round,line cap=round,fill=fillColor] ( 56.47, 96.07) circle (  1.43);

\path[draw=drawColor,line width= 0.4pt,line join=round,line cap=round,fill=fillColor] ( 56.10, 90.91) circle (  1.43);

\path[draw=drawColor,line width= 0.4pt,line join=round,line cap=round,fill=fillColor] ( 56.44, 92.35) circle (  1.43);

\path[draw=drawColor,line width= 0.4pt,line join=round,line cap=round,fill=fillColor] ( 56.27, 91.12) circle (  1.43);

\path[draw=drawColor,line width= 0.4pt,line join=round,line cap=round,fill=fillColor] ( 54.76, 90.11) circle (  1.43);

\path[draw=drawColor,line width= 0.4pt,line join=round,line cap=round,fill=fillColor] ( 56.30, 94.01) circle (  1.43);

\path[draw=drawColor,line width= 0.4pt,line join=round,line cap=round,fill=fillColor] ( 56.38, 88.57) circle (  1.43);

\path[draw=drawColor,line width= 0.4pt,line join=round,line cap=round,fill=fillColor] ( 56.33, 95.17) circle (  1.43);

\path[draw=drawColor,line width= 0.4pt,line join=round,line cap=round,fill=fillColor] ( 55.42, 93.32) circle (  1.43);

\path[draw=drawColor,line width= 0.4pt,line join=round,line cap=round,fill=fillColor] ( 54.84, 88.72) circle (  1.43);

\path[draw=drawColor,line width= 0.4pt,line join=round,line cap=round,fill=fillColor] ( 54.85, 90.35) circle (  1.43);

\path[draw=drawColor,line width= 0.4pt,line join=round,line cap=round,fill=fillColor] ( 55.26, 90.01) circle (  1.43);

\path[draw=drawColor,line width= 0.4pt,line join=round,line cap=round,fill=fillColor] ( 54.83, 90.45) circle (  1.43);

\path[draw=drawColor,line width= 0.4pt,line join=round,line cap=round,fill=fillColor] ( 57.26, 92.69) circle (  1.43);

\path[draw=drawColor,line width= 0.4pt,line join=round,line cap=round,fill=fillColor] ( 56.59, 96.78) circle (  1.43);

\path[draw=drawColor,line width= 0.4pt,line join=round,line cap=round,fill=fillColor] ( 54.91, 90.59) circle (  1.43);

\path[draw=drawColor,line width= 0.4pt,line join=round,line cap=round,fill=fillColor] ( 55.97, 91.92) circle (  1.43);

\path[draw=drawColor,line width= 0.4pt,line join=round,line cap=round,fill=fillColor] ( 57.38, 90.45) circle (  1.43);

\path[draw=drawColor,line width= 0.4pt,line join=round,line cap=round,fill=fillColor] ( 55.27, 90.69) circle (  1.43);

\path[draw=drawColor,line width= 0.4pt,line join=round,line cap=round,fill=fillColor] ( 55.35, 94.55) circle (  1.43);

\path[draw=drawColor,line width= 0.4pt,line join=round,line cap=round,fill=fillColor] ( 57.31, 94.20) circle (  1.43);

\path[draw=drawColor,line width= 0.4pt,line join=round,line cap=round,fill=fillColor] ( 55.46, 90.24) circle (  1.43);

\path[draw=drawColor,line width= 0.4pt,line join=round,line cap=round,fill=fillColor] ( 56.62, 92.97) circle (  1.43);

\path[draw=drawColor,line width= 0.4pt,line join=round,line cap=round,fill=fillColor] ( 56.67, 89.63) circle (  1.43);

\path[draw=drawColor,line width= 0.4pt,line join=round,line cap=round,fill=fillColor] ( 55.15, 96.02) circle (  1.43);

\path[draw=drawColor,line width= 0.4pt,line join=round,line cap=round,fill=fillColor] ( 56.67, 87.05) circle (  1.43);

\path[draw=drawColor,line width= 0.4pt,line join=round,line cap=round,fill=fillColor] ( 56.23, 89.33) circle (  1.43);

\path[draw=drawColor,line width= 0.4pt,line join=round,line cap=round,fill=fillColor] ( 53.84, 88.39) circle (  1.43);

\path[draw=drawColor,line width= 0.4pt,line join=round,line cap=round,fill=fillColor] ( 56.71, 90.64) circle (  1.43);

\path[draw=drawColor,line width= 0.4pt,line join=round,line cap=round,fill=fillColor] ( 55.88, 91.97) circle (  1.43);

\path[draw=drawColor,line width= 0.4pt,line join=round,line cap=round,fill=fillColor] ( 55.57, 98.48) circle (  1.43);

\path[draw=drawColor,line width= 0.4pt,line join=round,line cap=round,fill=fillColor] ( 55.89, 97.67) circle (  1.43);

\path[draw=drawColor,line width= 0.4pt,line join=round,line cap=round,fill=fillColor] ( 55.06, 89.25) circle (  1.43);

\path[draw=drawColor,line width= 0.4pt,line join=round,line cap=round,fill=fillColor] ( 54.81, 91.61) circle (  1.43);

\path[draw=drawColor,line width= 0.4pt,line join=round,line cap=round,fill=fillColor] ( 54.80, 91.77) circle (  1.43);

\path[draw=drawColor,line width= 0.4pt,line join=round,line cap=round,fill=fillColor] ( 57.16, 95.23) circle (  1.43);

\path[draw=drawColor,line width= 0.4pt,line join=round,line cap=round,fill=fillColor] ( 56.96, 91.92) circle (  1.43);

\path[draw=drawColor,line width= 0.4pt,line join=round,line cap=round,fill=fillColor] ( 55.17, 89.25) circle (  1.43);

\path[draw=drawColor,line width= 0.4pt,line join=round,line cap=round,fill=fillColor] ( 56.15, 94.33) circle (  1.43);

\path[draw=drawColor,line width= 0.4pt,line join=round,line cap=round,fill=fillColor] ( 55.55, 93.82) circle (  1.43);

\path[draw=drawColor,line width= 0.4pt,line join=round,line cap=round,fill=fillColor] ( 55.46, 93.88) circle (  1.43);

\path[draw=drawColor,line width= 0.4pt,line join=round,line cap=round,fill=fillColor] ( 54.99, 91.65) circle (  1.43);

\path[draw=drawColor,line width= 0.4pt,line join=round,line cap=round,fill=fillColor] ( 56.38, 89.90) circle (  1.43);

\path[draw=drawColor,line width= 0.4pt,line join=round,line cap=round,fill=fillColor] ( 55.04, 87.54) circle (  1.43);

\path[draw=drawColor,line width= 0.4pt,line join=round,line cap=round,fill=fillColor] ( 55.85, 92.99) circle (  1.43);

\path[draw=drawColor,line width= 0.4pt,line join=round,line cap=round,fill=fillColor] ( 56.33, 92.13) circle (  1.43);

\path[draw=drawColor,line width= 0.4pt,line join=round,line cap=round,fill=fillColor] ( 56.25, 96.35) circle (  1.43);

\path[draw=drawColor,line width= 0.4pt,line join=round,line cap=round,fill=fillColor] ( 55.62, 92.69) circle (  1.43);

\path[draw=drawColor,line width= 0.4pt,line join=round,line cap=round,fill=fillColor] ( 56.14, 93.01) circle (  1.43);

\path[draw=drawColor,line width= 0.4pt,line join=round,line cap=round,fill=fillColor] ( 54.74, 91.12) circle (  1.43);

\path[draw=drawColor,line width= 0.4pt,line join=round,line cap=round,fill=fillColor] ( 56.02, 90.97) circle (  1.43);

\path[draw=drawColor,line width= 0.4pt,line join=round,line cap=round,fill=fillColor] ( 54.98, 86.70) circle (  1.43);

\path[draw=drawColor,line width= 0.4pt,line join=round,line cap=round,fill=fillColor] ( 56.65, 93.33) circle (  1.43);

\path[draw=drawColor,line width= 0.4pt,line join=round,line cap=round,fill=fillColor] ( 55.57, 90.98) circle (  1.43);
\definecolor{drawColor}{RGB}{207,128,187}
\definecolor{fillColor}{RGB}{207,128,187}

\path[draw=drawColor,line width= 0.4pt,line join=round,line cap=round,fill=fillColor] ( 57.03, 95.66) circle (  1.43);

\path[draw=drawColor,line width= 0.4pt,line join=round,line cap=round,fill=fillColor] ( 56.19, 90.72) circle (  1.43);

\path[draw=drawColor,line width= 0.4pt,line join=round,line cap=round,fill=fillColor] ( 56.71, 94.19) circle (  1.43);

\path[draw=drawColor,line width= 0.4pt,line join=round,line cap=round,fill=fillColor] ( 56.93, 93.04) circle (  1.43);

\path[draw=drawColor,line width= 0.4pt,line join=round,line cap=round,fill=fillColor] ( 57.01, 93.74) circle (  1.43);

\path[draw=drawColor,line width= 0.4pt,line join=round,line cap=round,fill=fillColor] ( 59.56,102.93) circle (  1.43);

\path[draw=drawColor,line width= 0.4pt,line join=round,line cap=round,fill=fillColor] ( 59.72,106.99) circle (  1.43);

\path[draw=drawColor,line width= 0.4pt,line join=round,line cap=round,fill=fillColor] ( 58.42,104.40) circle (  1.43);

\path[draw=drawColor,line width= 0.4pt,line join=round,line cap=round,fill=fillColor] ( 55.92, 94.86) circle (  1.43);

\path[draw=drawColor,line width= 0.4pt,line join=round,line cap=round,fill=fillColor] ( 59.80,103.17) circle (  1.43);

\path[draw=drawColor,line width= 0.4pt,line join=round,line cap=round,fill=fillColor] ( 56.34, 96.04) circle (  1.43);

\path[draw=drawColor,line width= 0.4pt,line join=round,line cap=round,fill=fillColor] ( 57.85, 97.62) circle (  1.43);

\path[draw=drawColor,line width= 0.4pt,line join=round,line cap=round,fill=fillColor] ( 58.96,101.78) circle (  1.43);

\path[draw=drawColor,line width= 0.4pt,line join=round,line cap=round,fill=fillColor] ( 55.67, 95.52) circle (  1.43);

\path[draw=drawColor,line width= 0.4pt,line join=round,line cap=round,fill=fillColor] ( 56.97, 96.09) circle (  1.43);

\path[draw=drawColor,line width= 0.4pt,line join=round,line cap=round,fill=fillColor] ( 59.42,104.55) circle (  1.43);

\path[draw=drawColor,line width= 0.4pt,line join=round,line cap=round,fill=fillColor] ( 57.61, 96.66) circle (  1.43);

\path[draw=drawColor,line width= 0.4pt,line join=round,line cap=round,fill=fillColor] ( 57.79,107.96) circle (  1.43);

\path[draw=drawColor,line width= 0.4pt,line join=round,line cap=round,fill=fillColor] ( 56.58, 98.60) circle (  1.43);

\path[draw=drawColor,line width= 0.4pt,line join=round,line cap=round,fill=fillColor] ( 58.11,111.41) circle (  1.43);

\path[draw=drawColor,line width= 0.4pt,line join=round,line cap=round,fill=fillColor] ( 56.51, 94.24) circle (  1.43);

\path[draw=drawColor,line width= 0.4pt,line join=round,line cap=round,fill=fillColor] ( 59.07,107.76) circle (  1.43);

\path[draw=drawColor,line width= 0.4pt,line join=round,line cap=round,fill=fillColor] ( 59.23,102.86) circle (  1.43);

\path[draw=drawColor,line width= 0.4pt,line join=round,line cap=round,fill=fillColor] ( 58.88,113.06) circle (  1.43);

\path[draw=drawColor,line width= 0.4pt,line join=round,line cap=round,fill=fillColor] ( 58.83,111.33) circle (  1.43);

\path[draw=drawColor,line width= 0.4pt,line join=round,line cap=round,fill=fillColor] ( 57.11, 94.45) circle (  1.43);

\path[draw=drawColor,line width= 0.4pt,line join=round,line cap=round,fill=fillColor] ( 58.47,101.72) circle (  1.43);

\path[draw=drawColor,line width= 0.4pt,line join=round,line cap=round,fill=fillColor] ( 59.86,107.49) circle (  1.43);

\path[draw=drawColor,line width= 0.4pt,line join=round,line cap=round,fill=fillColor] ( 55.79, 94.26) circle (  1.43);

\path[draw=drawColor,line width= 0.4pt,line join=round,line cap=round,fill=fillColor] ( 55.46, 98.64) circle (  1.43);

\path[draw=drawColor,line width= 0.4pt,line join=round,line cap=round,fill=fillColor] ( 58.73, 98.98) circle (  1.43);

\path[draw=drawColor,line width= 0.4pt,line join=round,line cap=round,fill=fillColor] ( 56.97,100.00) circle (  1.43);

\path[draw=drawColor,line width= 0.4pt,line join=round,line cap=round,fill=fillColor] ( 57.80,101.91) circle (  1.43);

\path[draw=drawColor,line width= 0.4pt,line join=round,line cap=round,fill=fillColor] ( 56.24,100.73) circle (  1.43);

\path[draw=drawColor,line width= 0.4pt,line join=round,line cap=round,fill=fillColor] ( 57.34, 97.11) circle (  1.43);

\path[draw=drawColor,line width= 0.4pt,line join=round,line cap=round,fill=fillColor] ( 57.45,102.95) circle (  1.43);

\path[draw=drawColor,line width= 0.4pt,line join=round,line cap=round,fill=fillColor] ( 55.16, 97.80) circle (  1.43);

\path[draw=drawColor,line width= 0.4pt,line join=round,line cap=round,fill=fillColor] ( 56.18, 96.00) circle (  1.43);

\path[draw=drawColor,line width= 0.4pt,line join=round,line cap=round,fill=fillColor] ( 57.13, 96.70) circle (  1.43);

\path[draw=drawColor,line width= 0.4pt,line join=round,line cap=round,fill=fillColor] ( 58.48,108.64) circle (  1.43);

\path[draw=drawColor,line width= 0.4pt,line join=round,line cap=round,fill=fillColor] ( 58.21,101.87) circle (  1.43);

\path[draw=drawColor,line width= 0.4pt,line join=round,line cap=round,fill=fillColor] ( 59.13,106.63) circle (  1.43);

\path[draw=drawColor,line width= 0.4pt,line join=round,line cap=round,fill=fillColor] ( 57.74,108.07) circle (  1.43);

\path[draw=drawColor,line width= 0.4pt,line join=round,line cap=round,fill=fillColor] ( 59.31,104.96) circle (  1.43);

\path[draw=drawColor,line width= 0.4pt,line join=round,line cap=round,fill=fillColor] ( 56.89,101.82) circle (  1.43);

\path[draw=drawColor,line width= 0.4pt,line join=round,line cap=round,fill=fillColor] ( 57.51, 96.76) circle (  1.43);

\path[draw=drawColor,line width= 0.4pt,line join=round,line cap=round,fill=fillColor] ( 56.41, 97.93) circle (  1.43);

\path[draw=drawColor,line width= 0.4pt,line join=round,line cap=round,fill=fillColor] ( 57.19, 93.42) circle (  1.43);

\path[draw=drawColor,line width= 0.4pt,line join=round,line cap=round,fill=fillColor] ( 58.57, 96.09) circle (  1.43);

\path[draw=drawColor,line width= 0.4pt,line join=round,line cap=round,fill=fillColor] ( 57.85, 98.93) circle (  1.43);

\path[draw=drawColor,line width= 0.4pt,line join=round,line cap=round,fill=fillColor] ( 57.67, 96.28) circle (  1.43);

\path[draw=drawColor,line width= 0.4pt,line join=round,line cap=round,fill=fillColor] ( 58.47,109.99) circle (  1.43);

\path[draw=drawColor,line width= 0.4pt,line join=round,line cap=round,fill=fillColor] ( 58.68,101.99) circle (  1.43);

\path[draw=drawColor,line width= 0.4pt,line join=round,line cap=round,fill=fillColor] ( 58.88,104.90) circle (  1.43);

\path[draw=drawColor,line width= 0.4pt,line join=round,line cap=round,fill=fillColor] ( 58.56, 96.01) circle (  1.43);

\path[draw=drawColor,line width= 0.4pt,line join=round,line cap=round,fill=fillColor] ( 58.33, 99.21) circle (  1.43);

\path[draw=drawColor,line width= 0.4pt,line join=round,line cap=round,fill=fillColor] ( 57.58,100.14) circle (  1.43);

\path[draw=drawColor,line width= 0.4pt,line join=round,line cap=round,fill=fillColor] ( 57.42,104.10) circle (  1.43);

\path[draw=drawColor,line width= 0.4pt,line join=round,line cap=round,fill=fillColor] ( 59.04,101.15) circle (  1.43);

\path[draw=drawColor,line width= 0.4pt,line join=round,line cap=round,fill=fillColor] ( 57.60, 95.47) circle (  1.43);

\path[draw=drawColor,line width= 0.4pt,line join=round,line cap=round,fill=fillColor] ( 58.76,111.08) circle (  1.43);

\path[draw=drawColor,line width= 0.4pt,line join=round,line cap=round,fill=fillColor] ( 58.80, 95.38) circle (  1.43);

\path[draw=drawColor,line width= 0.4pt,line join=round,line cap=round,fill=fillColor] ( 57.08, 95.57) circle (  1.43);

\path[draw=drawColor,line width= 0.4pt,line join=round,line cap=round,fill=fillColor] ( 57.48,109.28) circle (  1.43);

\path[draw=drawColor,line width= 0.4pt,line join=round,line cap=round,fill=fillColor] ( 59.16,102.00) circle (  1.43);

\path[draw=drawColor,line width= 0.4pt,line join=round,line cap=round,fill=fillColor] ( 57.93,101.54) circle (  1.43);

\path[draw=drawColor,line width= 0.4pt,line join=round,line cap=round,fill=fillColor] ( 56.67, 99.23) circle (  1.43);

\path[draw=drawColor,line width= 0.4pt,line join=round,line cap=round,fill=fillColor] ( 56.36, 96.28) circle (  1.43);

\path[draw=drawColor,line width= 0.4pt,line join=round,line cap=round,fill=fillColor] ( 55.71, 98.64) circle (  1.43);

\path[draw=drawColor,line width= 0.4pt,line join=round,line cap=round,fill=fillColor] ( 59.07,105.00) circle (  1.43);
\definecolor{drawColor}{RGB}{207,192,152}
\definecolor{fillColor}{RGB}{207,192,152}

\path[draw=drawColor,line width= 0.4pt,line join=round,line cap=round,fill=fillColor] ( 61.86,104.04) circle (  1.43);

\path[draw=drawColor,line width= 0.4pt,line join=round,line cap=round,fill=fillColor] ( 63.60,115.71) circle (  1.43);

\path[draw=drawColor,line width= 0.4pt,line join=round,line cap=round,fill=fillColor] ( 62.58,108.67) circle (  1.43);

\path[draw=drawColor,line width= 0.4pt,line join=round,line cap=round,fill=fillColor] ( 63.82,145.31) circle (  1.43);

\path[draw=drawColor,line width= 0.4pt,line join=round,line cap=round,fill=fillColor] ( 63.92,118.20) circle (  1.43);

\path[draw=drawColor,line width= 0.4pt,line join=round,line cap=round,fill=fillColor] ( 62.98,102.88) circle (  1.43);

\path[draw=drawColor,line width= 0.4pt,line join=round,line cap=round,fill=fillColor] ( 63.97,103.91) circle (  1.43);

\path[draw=drawColor,line width= 0.4pt,line join=round,line cap=round,fill=fillColor] ( 62.42,101.74) circle (  1.43);

\path[draw=drawColor,line width= 0.4pt,line join=round,line cap=round,fill=fillColor] ( 64.37,153.46) circle (  1.43);

\path[draw=drawColor,line width= 0.4pt,line join=round,line cap=round,fill=fillColor] ( 61.63,110.22) circle (  1.43);

\path[draw=drawColor,line width= 0.4pt,line join=round,line cap=round,fill=fillColor] ( 64.74,132.04) circle (  1.43);

\path[draw=drawColor,line width= 0.4pt,line join=round,line cap=round,fill=fillColor] ( 62.38,118.18) circle (  1.43);

\path[draw=drawColor,line width= 0.4pt,line join=round,line cap=round,fill=fillColor] ( 62.61,103.95) circle (  1.43);

\path[draw=drawColor,line width= 0.4pt,line join=round,line cap=round,fill=fillColor] ( 65.91,115.53) circle (  1.43);

\path[draw=drawColor,line width= 0.4pt,line join=round,line cap=round,fill=fillColor] ( 67.30,159.27) circle (  1.43);

\path[draw=drawColor,line width= 0.4pt,line join=round,line cap=round,fill=fillColor] ( 62.51,103.53) circle (  1.43);

\path[draw=drawColor,line width= 0.4pt,line join=round,line cap=round,fill=fillColor] ( 63.18,146.37) circle (  1.43);

\path[draw=drawColor,line width= 0.4pt,line join=round,line cap=round,fill=fillColor] ( 66.84,194.28) circle (  1.43);

\path[draw=drawColor,line width= 0.4pt,line join=round,line cap=round,fill=fillColor] ( 64.52,117.98) circle (  1.43);

\path[draw=drawColor,line width= 0.4pt,line join=round,line cap=round,fill=fillColor] ( 64.61,121.60) circle (  1.43);

\path[draw=drawColor,line width= 0.4pt,line join=round,line cap=round,fill=fillColor] ( 63.90,110.28) circle (  1.43);

\path[draw=drawColor,line width= 0.4pt,line join=round,line cap=round,fill=fillColor] ( 65.01,117.78) circle (  1.43);

\path[draw=drawColor,line width= 0.4pt,line join=round,line cap=round,fill=fillColor] ( 63.86,114.16) circle (  1.43);

\path[draw=drawColor,line width= 0.4pt,line join=round,line cap=round,fill=fillColor] ( 62.69,104.34) circle (  1.43);

\path[draw=drawColor,line width= 0.4pt,line join=round,line cap=round,fill=fillColor] ( 63.08,104.04) circle (  1.43);

\path[draw=drawColor,line width= 0.4pt,line join=round,line cap=round,fill=fillColor] ( 63.21,114.76) circle (  1.43);

\path[draw=drawColor,line width= 0.4pt,line join=round,line cap=round,fill=fillColor] ( 64.83,115.84) circle (  1.43);

\path[draw=drawColor,line width= 0.4pt,line join=round,line cap=round,fill=fillColor] ( 62.39,112.38) circle (  1.43);

\path[draw=drawColor,line width= 0.4pt,line join=round,line cap=round,fill=fillColor] ( 65.94,159.41) circle (  1.43);

\path[draw=drawColor,line width= 0.4pt,line join=round,line cap=round,fill=fillColor] ( 63.32,119.50) circle (  1.43);

\path[draw=drawColor,line width= 0.4pt,line join=round,line cap=round,fill=fillColor] ( 62.75,113.67) circle (  1.43);

\path[draw=drawColor,line width= 0.4pt,line join=round,line cap=round,fill=fillColor] ( 62.83,105.70) circle (  1.43);

\path[draw=drawColor,line width= 0.4pt,line join=round,line cap=round,fill=fillColor] ( 66.52,194.38) circle (  1.43);

\path[draw=drawColor,line width= 0.4pt,line join=round,line cap=round,fill=fillColor] ( 62.99,116.14) circle (  1.43);

\path[draw=drawColor,line width= 0.4pt,line join=round,line cap=round,fill=fillColor] ( 64.97,114.15) circle (  1.43);

\path[draw=drawColor,line width= 0.4pt,line join=round,line cap=round,fill=fillColor] ( 62.71,112.37) circle (  1.43);

\path[draw=drawColor,line width= 0.4pt,line join=round,line cap=round,fill=fillColor] ( 64.59,161.08) circle (  1.43);

\path[draw=drawColor,line width= 0.4pt,line join=round,line cap=round,fill=fillColor] ( 63.32,112.65) circle (  1.43);

\path[draw=drawColor,line width= 0.4pt,line join=round,line cap=round,fill=fillColor] ( 65.70,155.28) circle (  1.43);

\path[draw=drawColor,line width= 0.4pt,line join=round,line cap=round,fill=fillColor] ( 66.41,156.75) circle (  1.43);

\path[draw=drawColor,line width= 0.4pt,line join=round,line cap=round,fill=fillColor] ( 62.29,112.54) circle (  1.43);

\path[draw=drawColor,line width= 0.4pt,line join=round,line cap=round,fill=fillColor] ( 62.82,106.37) circle (  1.43);

\path[draw=drawColor,line width= 0.4pt,line join=round,line cap=round,fill=fillColor] ( 67.21,156.13) circle (  1.43);

\path[draw=drawColor,line width= 0.4pt,line join=round,line cap=round,fill=fillColor] ( 61.63,114.08) circle (  1.43);

\path[draw=drawColor,line width= 0.4pt,line join=round,line cap=round,fill=fillColor] ( 63.25,115.16) circle (  1.43);

\path[draw=drawColor,line width= 0.4pt,line join=round,line cap=round,fill=fillColor] ( 63.98,118.22) circle (  1.43);

\path[draw=drawColor,line width= 0.4pt,line join=round,line cap=round,fill=fillColor] ( 62.52,114.55) circle (  1.43);

\path[draw=drawColor,line width= 0.4pt,line join=round,line cap=round,fill=fillColor] ( 63.77,121.66) circle (  1.43);

\path[draw=drawColor,line width= 0.4pt,line join=round,line cap=round,fill=fillColor] ( 66.55,157.26) circle (  1.43);

\path[draw=drawColor,line width= 0.4pt,line join=round,line cap=round,fill=fillColor] ( 62.67,114.71) circle (  1.43);

\path[draw=drawColor,line width= 0.4pt,line join=round,line cap=round,fill=fillColor] ( 62.67,116.11) circle (  1.43);

\path[draw=drawColor,line width= 0.4pt,line join=round,line cap=round,fill=fillColor] ( 63.30,114.92) circle (  1.43);

\path[draw=drawColor,line width= 0.4pt,line join=round,line cap=round,fill=fillColor] ( 61.87,117.02) circle (  1.43);

\path[draw=drawColor,line width= 0.4pt,line join=round,line cap=round,fill=fillColor] ( 66.43,194.39) circle (  1.43);

\path[draw=drawColor,line width= 0.4pt,line join=round,line cap=round,fill=fillColor] ( 63.33,152.52) circle (  1.43);

\path[draw=drawColor,line width= 0.4pt,line join=round,line cap=round,fill=fillColor] ( 63.00,117.32) circle (  1.43);

\path[draw=drawColor,line width= 0.4pt,line join=round,line cap=round,fill=fillColor] ( 62.20,106.51) circle (  1.43);

\path[draw=drawColor,line width= 0.4pt,line join=round,line cap=round,fill=fillColor] ( 62.54,106.40) circle (  1.43);

\path[draw=drawColor,line width= 0.4pt,line join=round,line cap=round,fill=fillColor] ( 62.37,111.98) circle (  1.43);

\path[draw=drawColor,line width= 0.4pt,line join=round,line cap=round,fill=fillColor] ( 62.44,116.02) circle (  1.43);

\path[draw=drawColor,line width= 0.4pt,line join=round,line cap=round,fill=fillColor] ( 63.08,112.20) circle (  1.43);

\path[draw=drawColor,line width= 0.4pt,line join=round,line cap=round,fill=fillColor] ( 64.42,148.26) circle (  1.43);

\path[draw=drawColor,line width= 0.4pt,line join=round,line cap=round,fill=fillColor] ( 62.63,111.71) circle (  1.43);

\path[draw=drawColor,line width= 0.4pt,line join=round,line cap=round,fill=fillColor] ( 66.23,156.40) circle (  1.43);

\path[draw=drawColor,line width= 0.4pt,line join=round,line cap=round,fill=fillColor] ( 63.93,106.18) circle (  1.43);

\path[draw=drawColor,line width= 0.4pt,line join=round,line cap=round,fill=fillColor] ( 65.26,154.80) circle (  1.43);
\definecolor{drawColor}{RGB}{0,0,0}

\path[draw=drawColor,line width= 0.6pt,line join=round] ( 28.84, 29.68) -- (125.81,194.47);
\definecolor{drawColor}{gray}{0.70}

\path[draw=drawColor,line width= 0.5pt,line join=round,line cap=round] ( 28.84, 29.68) rectangle (125.81,194.47);
\end{scope}
\begin{scope}
\path[clip] (  0.00,  0.00) rectangle (130.81,199.47);
\definecolor{drawColor}{gray}{0.30}

\node[text=drawColor,anchor=base east,inner sep=0pt, outer sep=0pt, scale=  1.00] at ( 26.34, 26.23) {0};

\node[text=drawColor,anchor=base east,inner sep=0pt, outer sep=0pt, scale=  1.00] at ( 26.34, 67.43) {25};

\node[text=drawColor,anchor=base east,inner sep=0pt, outer sep=0pt, scale=  1.00] at ( 26.34,108.63) {50};

\node[text=drawColor,anchor=base east,inner sep=0pt, outer sep=0pt, scale=  1.00] at ( 26.34,149.82) {75};

\node[text=drawColor,anchor=base east,inner sep=0pt, outer sep=0pt, scale=  1.00] at ( 26.34,191.02) {100};
\end{scope}
\begin{scope}
\path[clip] (  0.00,  0.00) rectangle (130.81,199.47);
\definecolor{drawColor}{gray}{0.70}

\path[draw=drawColor,line width= 0.3pt,line join=round] ( 26.34, 29.68) --
	( 28.84, 29.68);

\path[draw=drawColor,line width= 0.3pt,line join=round] ( 26.34, 70.87) --
	( 28.84, 70.87);

\path[draw=drawColor,line width= 0.3pt,line join=round] ( 26.34,112.07) --
	( 28.84,112.07);

\path[draw=drawColor,line width= 0.3pt,line join=round] ( 26.34,153.27) --
	( 28.84,153.27);

\path[draw=drawColor,line width= 0.3pt,line join=round] ( 26.34,194.47) --
	( 28.84,194.47);
\end{scope}
\begin{scope}
\path[clip] (  0.00,  0.00) rectangle (130.81,199.47);
\definecolor{drawColor}{gray}{0.70}

\path[draw=drawColor,line width= 0.3pt,line join=round] ( 28.84, 27.18) --
	( 28.84, 29.68);

\path[draw=drawColor,line width= 0.3pt,line join=round] ( 53.08, 27.18) --
	( 53.08, 29.68);

\path[draw=drawColor,line width= 0.3pt,line join=round] ( 77.33, 27.18) --
	( 77.33, 29.68);

\path[draw=drawColor,line width= 0.3pt,line join=round] (101.57, 27.18) --
	(101.57, 29.68);

\path[draw=drawColor,line width= 0.3pt,line join=round] (125.81, 27.18) --
	(125.81, 29.68);
\end{scope}
\begin{scope}
\path[clip] (  0.00,  0.00) rectangle (130.81,199.47);
\definecolor{drawColor}{gray}{0.30}

\node[text=drawColor,anchor=base,inner sep=0pt, outer sep=0pt, scale=  1.00] at ( 27.84, 18.29) {0};

\node[text=drawColor,anchor=base,inner sep=0pt, outer sep=0pt, scale=  1.00] at ( 51.08, 18.29) {25};

\node[text=drawColor,anchor=base,inner sep=0pt, outer sep=0pt, scale=  1.00] at ( 75.33, 18.29) {50};

\node[text=drawColor,anchor=base,inner sep=0pt, outer sep=0pt, scale=  1.00] at ( 99.57, 18.29) {75};

\node[text=drawColor,anchor=base,inner sep=0pt, outer sep=0pt, scale=  1.00] at (122.81, 18.29) {100};
\end{scope}
\begin{scope}
\path[clip] (  0.00,  0.00) rectangle (130.81,199.47);
\definecolor{drawColor}{RGB}{0,0,0}

\node[text=drawColor,anchor=base,inner sep=0pt, outer sep=0pt, scale=  1.00] at ( 77.33,  6.94) {\bfseries Gap Gurobi};
\end{scope}
\begin{scope}
\path[clip] (  0.00,  0.00) rectangle (130.81,199.47);
\definecolor{drawColor}{RGB}{0,0,0}

\node[text=drawColor,rotate= 90.00,anchor=base,inner sep=0pt, outer sep=0pt, scale=  1.00] at ( 11.90,112.07) {\bfseries Gap CPLEX};
\end{scope}
\end{tikzpicture}

%% file: 06-plots/CPLEXvsGUROBI-Gap-12Hours.tex
\begin{tikzpicture}[x=1pt,y=1pt]
\definecolor{fillColor}{RGB}{255,255,255}
\path[use as bounding box,fill=fillColor,fill opacity=0.00] (0,0) rectangle (122.86,199.47);
\begin{scope}
\path[clip] (  0.00,  0.00) rectangle (122.86,199.47);
\definecolor{drawColor}{RGB}{255,255,255}
\definecolor{fillColor}{RGB}{255,255,255}

\path[draw=drawColor,line width= 0.5pt,line join=round,line cap=round,fill=fillColor] (  0.00,  0.00) rectangle (122.86,199.47);
\end{scope}
\begin{scope}
\path[clip] ( 19.50, 29.68) rectangle (117.86,194.47);
\definecolor{fillColor}{RGB}{255,255,255}

\path[fill=fillColor] ( 19.50, 29.68) rectangle (117.86,194.47);
\definecolor{drawColor}{gray}{0.87}

\path[draw=drawColor,line width= 0.1pt,line join=round] ( 19.50, 44.66) --
	(117.86, 44.66);

\path[draw=drawColor,line width= 0.1pt,line join=round] ( 19.50, 74.62) --
	(117.86, 74.62);

\path[draw=drawColor,line width= 0.1pt,line join=round] ( 19.50,104.58) --
	(117.86,104.58);

\path[draw=drawColor,line width= 0.1pt,line join=round] ( 19.50,134.54) --
	(117.86,134.54);

\path[draw=drawColor,line width= 0.1pt,line join=round] ( 19.50,164.50) --
	(117.86,164.50);

\path[draw=drawColor,line width= 0.1pt,line join=round] ( 19.50,194.47) --
	(117.86,194.47);

\path[draw=drawColor,line width= 0.1pt,line join=round] ( 28.44, 29.68) --
	( 28.44,194.47);

\path[draw=drawColor,line width= 0.1pt,line join=round] ( 46.32, 29.68) --
	( 46.32,194.47);

\path[draw=drawColor,line width= 0.1pt,line join=round] ( 64.21, 29.68) --
	( 64.21,194.47);

\path[draw=drawColor,line width= 0.1pt,line join=round] ( 82.09, 29.68) --
	( 82.09,194.47);

\path[draw=drawColor,line width= 0.1pt,line join=round] ( 99.98, 29.68) --
	( 99.98,194.47);

\path[draw=drawColor,line width= 0.1pt,line join=round] (117.86, 29.68) --
	(117.86,194.47);

\path[draw=drawColor,line width= 0.3pt,line join=round] ( 19.50, 29.68) --
	(117.86, 29.68);

\path[draw=drawColor,line width= 0.3pt,line join=round] ( 19.50, 59.64) --
	(117.86, 59.64);

\path[draw=drawColor,line width= 0.3pt,line join=round] ( 19.50, 89.60) --
	(117.86, 89.60);

\path[draw=drawColor,line width= 0.3pt,line join=round] ( 19.50,119.56) --
	(117.86,119.56);

\path[draw=drawColor,line width= 0.3pt,line join=round] ( 19.50,149.52) --
	(117.86,149.52);

\path[draw=drawColor,line width= 0.3pt,line join=round] ( 19.50,179.48) --
	(117.86,179.48);

\path[draw=drawColor,line width= 0.3pt,line join=round] ( 19.50, 29.68) --
	( 19.50,194.47);

\path[draw=drawColor,line width= 0.3pt,line join=round] ( 37.38, 29.68) --
	( 37.38,194.47);

\path[draw=drawColor,line width= 0.3pt,line join=round] ( 55.27, 29.68) --
	( 55.27,194.47);

\path[draw=drawColor,line width= 0.3pt,line join=round] ( 73.15, 29.68) --
	( 73.15,194.47);

\path[draw=drawColor,line width= 0.3pt,line join=round] ( 91.03, 29.68) --
	( 91.03,194.47);

\path[draw=drawColor,line width= 0.3pt,line join=round] (108.92, 29.68) --
	(108.92,194.47);
\definecolor{drawColor}{RGB}{128,128,128}
\definecolor{fillColor}{RGB}{128,128,128}

\path[draw=drawColor,line width= 0.4pt,line join=round,line cap=round,fill=fillColor] ( 35.70, 53.86) circle (  1.43);

\path[draw=drawColor,line width= 0.4pt,line join=round,line cap=round,fill=fillColor] ( 44.09, 76.58) circle (  1.43);

\path[draw=drawColor,line width= 0.4pt,line join=round,line cap=round,fill=fillColor] ( 50.67,120.31) circle (  1.43);

\path[draw=drawColor,line width= 0.4pt,line join=round,line cap=round,fill=fillColor] ( 19.52, 29.71) circle (  1.43);

\path[draw=drawColor,line width= 0.4pt,line join=round,line cap=round,fill=fillColor] ( 19.52, 29.71) circle (  1.43);

\path[draw=drawColor,line width= 0.4pt,line join=round,line cap=round,fill=fillColor] ( 19.52, 29.71) circle (  1.43);

\path[draw=drawColor,line width= 0.4pt,line join=round,line cap=round,fill=fillColor] ( 19.52, 29.71) circle (  1.43);

\path[draw=drawColor,line width= 0.4pt,line join=round,line cap=round,fill=fillColor] ( 39.52, 59.10) circle (  1.43);

\path[draw=drawColor,line width= 0.4pt,line join=round,line cap=round,fill=fillColor] ( 60.67,123.35) circle (  1.43);

\path[draw=drawColor,line width= 0.4pt,line join=round,line cap=round,fill=fillColor] ( 58.38,112.19) circle (  1.43);

\path[draw=drawColor,line width= 0.4pt,line join=round,line cap=round,fill=fillColor] ( 54.12,121.86) circle (  1.43);

\path[draw=drawColor,line width= 0.4pt,line join=round,line cap=round,fill=fillColor] ( 53.78,108.26) circle (  1.43);

\path[draw=drawColor,line width= 0.4pt,line join=round,line cap=round,fill=fillColor] ( 19.52, 47.19) circle (  1.43);

\path[draw=drawColor,line width= 0.4pt,line join=round,line cap=round,fill=fillColor] ( 53.44,123.97) circle (  1.43);

\path[draw=drawColor,line width= 0.4pt,line join=round,line cap=round,fill=fillColor] ( 58.67,121.75) circle (  1.43);

\path[draw=drawColor,line width= 0.4pt,line join=round,line cap=round,fill=fillColor] ( 45.70, 83.63) circle (  1.43);

\path[draw=drawColor,line width= 0.4pt,line join=round,line cap=round,fill=fillColor] ( 57.22,127.92) circle (  1.43);

\path[draw=drawColor,line width= 0.4pt,line join=round,line cap=round,fill=fillColor] ( 49.65, 87.84) circle (  1.43);

\path[draw=drawColor,line width= 0.4pt,line join=round,line cap=round,fill=fillColor] ( 19.52, 29.71) circle (  1.43);

\path[draw=drawColor,line width= 0.4pt,line join=round,line cap=round,fill=fillColor] ( 61.67,121.61) circle (  1.43);

\path[draw=drawColor,line width= 0.4pt,line join=round,line cap=round,fill=fillColor] ( 35.45, 60.96) circle (  1.43);

\path[draw=drawColor,line width= 0.4pt,line join=round,line cap=round,fill=fillColor] ( 33.74, 53.31) circle (  1.43);

\path[draw=drawColor,line width= 0.4pt,line join=round,line cap=round,fill=fillColor] ( 56.14,114.77) circle (  1.43);

\path[draw=drawColor,line width= 0.4pt,line join=round,line cap=round,fill=fillColor] ( 35.17, 61.83) circle (  1.43);

\path[draw=drawColor,line width= 0.4pt,line join=round,line cap=round,fill=fillColor] ( 55.26, 98.46) circle (  1.43);

\path[draw=drawColor,line width= 0.4pt,line join=round,line cap=round,fill=fillColor] ( 57.48, 93.67) circle (  1.43);

\path[draw=drawColor,line width= 0.4pt,line join=round,line cap=round,fill=fillColor] ( 57.73,115.11) circle (  1.43);

\path[draw=drawColor,line width= 0.4pt,line join=round,line cap=round,fill=fillColor] ( 60.04,123.33) circle (  1.43);

\path[draw=drawColor,line width= 0.4pt,line join=round,line cap=round,fill=fillColor] ( 45.73, 73.20) circle (  1.43);

\path[draw=drawColor,line width= 0.4pt,line join=round,line cap=round,fill=fillColor] ( 50.74,115.14) circle (  1.43);

\path[draw=drawColor,line width= 0.4pt,line join=round,line cap=round,fill=fillColor] ( 58.98,125.78) circle (  1.43);

\path[draw=drawColor,line width= 0.4pt,line join=round,line cap=round,fill=fillColor] ( 50.31, 89.29) circle (  1.43);

\path[draw=drawColor,line width= 0.4pt,line join=round,line cap=round,fill=fillColor] ( 19.52, 29.71) circle (  1.43);

\path[draw=drawColor,line width= 0.4pt,line join=round,line cap=round,fill=fillColor] ( 56.44,119.24) circle (  1.43);

\path[draw=drawColor,line width= 0.4pt,line join=round,line cap=round,fill=fillColor] ( 56.93,110.91) circle (  1.43);
\definecolor{drawColor}{RGB}{152,167,197}
\definecolor{fillColor}{RGB}{152,167,197}

\path[draw=drawColor,line width= 0.4pt,line join=round,line cap=round,fill=fillColor] ( 67.26,129.30) circle (  1.43);

\path[draw=drawColor,line width= 0.4pt,line join=round,line cap=round,fill=fillColor] ( 47.60, 84.49) circle (  1.43);

\path[draw=drawColor,line width= 0.4pt,line join=round,line cap=round,fill=fillColor] ( 58.96,116.48) circle (  1.43);

\path[draw=drawColor,line width= 0.4pt,line join=round,line cap=round,fill=fillColor] ( 57.77,108.70) circle (  1.43);

\path[draw=drawColor,line width= 0.4pt,line join=round,line cap=round,fill=fillColor] ( 68.09,133.02) circle (  1.43);

\path[draw=drawColor,line width= 0.4pt,line join=round,line cap=round,fill=fillColor] ( 65.63,123.82) circle (  1.43);

\path[draw=drawColor,line width= 0.4pt,line join=round,line cap=round,fill=fillColor] ( 66.59,134.00) circle (  1.43);

\path[draw=drawColor,line width= 0.4pt,line join=round,line cap=round,fill=fillColor] ( 63.71,114.79) circle (  1.43);

\path[draw=drawColor,line width= 0.4pt,line join=round,line cap=round,fill=fillColor] ( 65.30,120.10) circle (  1.43);

\path[draw=drawColor,line width= 0.4pt,line join=round,line cap=round,fill=fillColor] ( 60.83,121.99) circle (  1.43);

\path[draw=drawColor,line width= 0.4pt,line join=round,line cap=round,fill=fillColor] ( 55.62, 98.94) circle (  1.43);

\path[draw=drawColor,line width= 0.4pt,line join=round,line cap=round,fill=fillColor] ( 66.62,127.31) circle (  1.43);

\path[draw=drawColor,line width= 0.4pt,line join=round,line cap=round,fill=fillColor] ( 60.79,109.47) circle (  1.43);

\path[draw=drawColor,line width= 0.4pt,line join=round,line cap=round,fill=fillColor] ( 62.68,120.43) circle (  1.43);

\path[draw=drawColor,line width= 0.4pt,line join=round,line cap=round,fill=fillColor] ( 67.76,133.42) circle (  1.43);

\path[draw=drawColor,line width= 0.4pt,line join=round,line cap=round,fill=fillColor] ( 63.24,114.24) circle (  1.43);

\path[draw=drawColor,line width= 0.4pt,line join=round,line cap=round,fill=fillColor] ( 48.76, 80.14) circle (  1.43);

\path[draw=drawColor,line width= 0.4pt,line join=round,line cap=round,fill=fillColor] ( 67.20,127.87) circle (  1.43);

\path[draw=drawColor,line width= 0.4pt,line join=round,line cap=round,fill=fillColor] ( 47.51, 85.99) circle (  1.43);

\path[draw=drawColor,line width= 0.4pt,line join=round,line cap=round,fill=fillColor] ( 61.63,120.72) circle (  1.43);

\path[draw=drawColor,line width= 0.4pt,line join=round,line cap=round,fill=fillColor] ( 55.36,107.06) circle (  1.43);

\path[draw=drawColor,line width= 0.4pt,line join=round,line cap=round,fill=fillColor] ( 48.73, 85.01) circle (  1.43);

\path[draw=drawColor,line width= 0.4pt,line join=round,line cap=round,fill=fillColor] ( 53.89,100.77) circle (  1.43);

\path[draw=drawColor,line width= 0.4pt,line join=round,line cap=round,fill=fillColor] ( 63.11,124.21) circle (  1.43);

\path[draw=drawColor,line width= 0.4pt,line join=round,line cap=round,fill=fillColor] ( 49.62, 78.55) circle (  1.43);

\path[draw=drawColor,line width= 0.4pt,line join=round,line cap=round,fill=fillColor] ( 56.96,103.33) circle (  1.43);

\path[draw=drawColor,line width= 0.4pt,line join=round,line cap=round,fill=fillColor] ( 58.93,104.12) circle (  1.43);

\path[draw=drawColor,line width= 0.4pt,line join=round,line cap=round,fill=fillColor] ( 63.28,111.94) circle (  1.43);

\path[draw=drawColor,line width= 0.4pt,line join=round,line cap=round,fill=fillColor] ( 67.51,122.78) circle (  1.43);

\path[draw=drawColor,line width= 0.4pt,line join=round,line cap=round,fill=fillColor] ( 63.55,119.68) circle (  1.43);

\path[draw=drawColor,line width= 0.4pt,line join=round,line cap=round,fill=fillColor] ( 58.95,118.15) circle (  1.43);

\path[draw=drawColor,line width= 0.4pt,line join=round,line cap=round,fill=fillColor] ( 48.00, 76.07) circle (  1.43);

\path[draw=drawColor,line width= 0.4pt,line join=round,line cap=round,fill=fillColor] ( 64.28,128.78) circle (  1.43);

\path[draw=drawColor,line width= 0.4pt,line join=round,line cap=round,fill=fillColor] ( 53.85, 90.02) circle (  1.43);

\path[draw=drawColor,line width= 0.4pt,line join=round,line cap=round,fill=fillColor] ( 64.43,123.25) circle (  1.43);
\definecolor{drawColor}{RGB}{128,202,192}
\definecolor{fillColor}{RGB}{128,202,192}

\path[draw=drawColor,line width= 0.4pt,line join=round,line cap=round,fill=fillColor] ( 69.20,138.79) circle (  1.43);

\path[draw=drawColor,line width= 0.4pt,line join=round,line cap=round,fill=fillColor] ( 67.15,133.99) circle (  1.43);

\path[draw=drawColor,line width= 0.4pt,line join=round,line cap=round,fill=fillColor] ( 69.08,140.98) circle (  1.43);

\path[draw=drawColor,line width= 0.4pt,line join=round,line cap=round,fill=fillColor] ( 68.97,141.24) circle (  1.43);

\path[draw=drawColor,line width= 0.4pt,line join=round,line cap=round,fill=fillColor] ( 67.92,135.58) circle (  1.43);

\path[draw=drawColor,line width= 0.4pt,line join=round,line cap=round,fill=fillColor] ( 69.50,132.45) circle (  1.43);

\path[draw=drawColor,line width= 0.4pt,line join=round,line cap=round,fill=fillColor] ( 70.38,137.14) circle (  1.43);

\path[draw=drawColor,line width= 0.4pt,line join=round,line cap=round,fill=fillColor] ( 68.70,134.69) circle (  1.43);

\path[draw=drawColor,line width= 0.4pt,line join=round,line cap=round,fill=fillColor] ( 67.15,142.60) circle (  1.43);

\path[draw=drawColor,line width= 0.4pt,line join=round,line cap=round,fill=fillColor] ( 68.50,140.00) circle (  1.43);

\path[draw=drawColor,line width= 0.4pt,line join=round,line cap=round,fill=fillColor] ( 67.31,136.17) circle (  1.43);

\path[draw=drawColor,line width= 0.4pt,line join=round,line cap=round,fill=fillColor] ( 68.91,134.95) circle (  1.43);

\path[draw=drawColor,line width= 0.4pt,line join=round,line cap=round,fill=fillColor] ( 65.58,138.71) circle (  1.43);

\path[draw=drawColor,line width= 0.4pt,line join=round,line cap=round,fill=fillColor] ( 68.64,133.60) circle (  1.43);

\path[draw=drawColor,line width= 0.4pt,line join=round,line cap=round,fill=fillColor] ( 68.04,136.85) circle (  1.43);

\path[draw=drawColor,line width= 0.4pt,line join=round,line cap=round,fill=fillColor] ( 69.91,137.13) circle (  1.43);

\path[draw=drawColor,line width= 0.4pt,line join=round,line cap=round,fill=fillColor] ( 70.00,142.42) circle (  1.43);

\path[draw=drawColor,line width= 0.4pt,line join=round,line cap=round,fill=fillColor] ( 68.92,133.62) circle (  1.43);

\path[draw=drawColor,line width= 0.4pt,line join=round,line cap=round,fill=fillColor] ( 69.57,138.67) circle (  1.43);

\path[draw=drawColor,line width= 0.4pt,line join=round,line cap=round,fill=fillColor] ( 69.50,137.54) circle (  1.43);

\path[draw=drawColor,line width= 0.4pt,line join=round,line cap=round,fill=fillColor] ( 66.51,133.64) circle (  1.43);

\path[draw=drawColor,line width= 0.4pt,line join=round,line cap=round,fill=fillColor] ( 69.51,142.59) circle (  1.43);

\path[draw=drawColor,line width= 0.4pt,line join=round,line cap=round,fill=fillColor] ( 69.45,128.59) circle (  1.43);

\path[draw=drawColor,line width= 0.4pt,line join=round,line cap=round,fill=fillColor] ( 69.38,142.62) circle (  1.43);

\path[draw=drawColor,line width= 0.4pt,line join=round,line cap=round,fill=fillColor] ( 67.81,139.11) circle (  1.43);

\path[draw=drawColor,line width= 0.4pt,line join=round,line cap=round,fill=fillColor] ( 66.87,133.28) circle (  1.43);

\path[draw=drawColor,line width= 0.4pt,line join=round,line cap=round,fill=fillColor] ( 66.67,135.62) circle (  1.43);

\path[draw=drawColor,line width= 0.4pt,line join=round,line cap=round,fill=fillColor] ( 67.09,133.80) circle (  1.43);

\path[draw=drawColor,line width= 0.4pt,line join=round,line cap=round,fill=fillColor] ( 66.51,134.39) circle (  1.43);

\path[draw=drawColor,line width= 0.4pt,line join=round,line cap=round,fill=fillColor] ( 71.11,140.53) circle (  1.43);

\path[draw=drawColor,line width= 0.4pt,line join=round,line cap=round,fill=fillColor] ( 69.79,145.39) circle (  1.43);

\path[draw=drawColor,line width= 0.4pt,line join=round,line cap=round,fill=fillColor] ( 66.90,135.18) circle (  1.43);

\path[draw=drawColor,line width= 0.4pt,line join=round,line cap=round,fill=fillColor] ( 67.88,140.72) circle (  1.43);

\path[draw=drawColor,line width= 0.4pt,line join=round,line cap=round,fill=fillColor] ( 71.43,134.46) circle (  1.43);

\path[draw=drawColor,line width= 0.4pt,line join=round,line cap=round,fill=fillColor] ( 67.48,135.11) circle (  1.43);

\path[draw=drawColor,line width= 0.4pt,line join=round,line cap=round,fill=fillColor] ( 67.44,142.47) circle (  1.43);

\path[draw=drawColor,line width= 0.4pt,line join=round,line cap=round,fill=fillColor] ( 70.87,142.55) circle (  1.43);

\path[draw=drawColor,line width= 0.4pt,line join=round,line cap=round,fill=fillColor] ( 67.71,133.35) circle (  1.43);

\path[draw=drawColor,line width= 0.4pt,line join=round,line cap=round,fill=fillColor] ( 69.89,141.60) circle (  1.43);

\path[draw=drawColor,line width= 0.4pt,line join=round,line cap=round,fill=fillColor] ( 69.92,132.46) circle (  1.43);

\path[draw=drawColor,line width= 0.4pt,line join=round,line cap=round,fill=fillColor] ( 67.63,140.11) circle (  1.43);

\path[draw=drawColor,line width= 0.4pt,line join=round,line cap=round,fill=fillColor] ( 69.87,128.45) circle (  1.43);

\path[draw=drawColor,line width= 0.4pt,line join=round,line cap=round,fill=fillColor] ( 68.26,134.76) circle (  1.43);

\path[draw=drawColor,line width= 0.4pt,line join=round,line cap=round,fill=fillColor] ( 64.89,131.91) circle (  1.43);

\path[draw=drawColor,line width= 0.4pt,line join=round,line cap=round,fill=fillColor] ( 69.89,133.07) circle (  1.43);

\path[draw=drawColor,line width= 0.4pt,line join=round,line cap=round,fill=fillColor] ( 68.18,138.11) circle (  1.43);

\path[draw=drawColor,line width= 0.4pt,line join=round,line cap=round,fill=fillColor] ( 68.10,149.78) circle (  1.43);

\path[draw=drawColor,line width= 0.4pt,line join=round,line cap=round,fill=fillColor] ( 68.61,148.36) circle (  1.43);

\path[draw=drawColor,line width= 0.4pt,line join=round,line cap=round,fill=fillColor] ( 66.27,132.27) circle (  1.43);

\path[draw=drawColor,line width= 0.4pt,line join=round,line cap=round,fill=fillColor] ( 66.88,137.15) circle (  1.43);

\path[draw=drawColor,line width= 0.4pt,line join=round,line cap=round,fill=fillColor] ( 66.93,139.13) circle (  1.43);

\path[draw=drawColor,line width= 0.4pt,line join=round,line cap=round,fill=fillColor] ( 70.80,143.84) circle (  1.43);

\path[draw=drawColor,line width= 0.4pt,line join=round,line cap=round,fill=fillColor] ( 70.43,135.33) circle (  1.43);

\path[draw=drawColor,line width= 0.4pt,line join=round,line cap=round,fill=fillColor] ( 67.21,133.15) circle (  1.43);

\path[draw=drawColor,line width= 0.4pt,line join=round,line cap=round,fill=fillColor] ( 69.49,143.76) circle (  1.43);

\path[draw=drawColor,line width= 0.4pt,line join=round,line cap=round,fill=fillColor] ( 68.13,140.47) circle (  1.43);

\path[draw=drawColor,line width= 0.4pt,line join=round,line cap=round,fill=fillColor] ( 67.88,140.22) circle (  1.43);

\path[draw=drawColor,line width= 0.4pt,line join=round,line cap=round,fill=fillColor] ( 66.86,136.50) circle (  1.43);

\path[draw=drawColor,line width= 0.4pt,line join=round,line cap=round,fill=fillColor] ( 69.24,132.83) circle (  1.43);

\path[draw=drawColor,line width= 0.4pt,line join=round,line cap=round,fill=fillColor] ( 65.99,129.54) circle (  1.43);

\path[draw=drawColor,line width= 0.4pt,line join=round,line cap=round,fill=fillColor] ( 68.59,140.58) circle (  1.43);

\path[draw=drawColor,line width= 0.4pt,line join=round,line cap=round,fill=fillColor] ( 69.37,137.02) circle (  1.43);

\path[draw=drawColor,line width= 0.4pt,line join=round,line cap=round,fill=fillColor] ( 68.83,147.39) circle (  1.43);

\path[draw=drawColor,line width= 0.4pt,line join=round,line cap=round,fill=fillColor] ( 68.15,136.91) circle (  1.43);

\path[draw=drawColor,line width= 0.4pt,line join=round,line cap=round,fill=fillColor] ( 69.46,140.08) circle (  1.43);

\path[draw=drawColor,line width= 0.4pt,line join=round,line cap=round,fill=fillColor] ( 66.03,136.97) circle (  1.43);

\path[draw=drawColor,line width= 0.4pt,line join=round,line cap=round,fill=fillColor] ( 68.71,134.86) circle (  1.43);

\path[draw=drawColor,line width= 0.4pt,line join=round,line cap=round,fill=fillColor] ( 66.78,129.46) circle (  1.43);

\path[draw=drawColor,line width= 0.4pt,line join=round,line cap=round,fill=fillColor] ( 70.10,138.96) circle (  1.43);

\path[draw=drawColor,line width= 0.4pt,line join=round,line cap=round,fill=fillColor] ( 68.26,136.39) circle (  1.43);
\definecolor{drawColor}{RGB}{207,128,187}
\definecolor{fillColor}{RGB}{207,128,187}

\path[draw=drawColor,line width= 0.4pt,line join=round,line cap=round,fill=fillColor] ( 70.22,142.98) circle (  1.43);

\path[draw=drawColor,line width= 0.4pt,line join=round,line cap=round,fill=fillColor] ( 68.82,136.68) circle (  1.43);

\path[draw=drawColor,line width= 0.4pt,line join=round,line cap=round,fill=fillColor] ( 70.45,140.94) circle (  1.43);

\path[draw=drawColor,line width= 0.4pt,line join=round,line cap=round,fill=fillColor] ( 70.39,139.77) circle (  1.43);

\path[draw=drawColor,line width= 0.4pt,line join=round,line cap=round,fill=fillColor] ( 70.37,138.27) circle (  1.43);

\path[draw=drawColor,line width= 0.4pt,line join=round,line cap=round,fill=fillColor] ( 71.21,152.07) circle (  1.43);

\path[draw=drawColor,line width= 0.4pt,line join=round,line cap=round,fill=fillColor] ( 70.33,163.33) circle (  1.43);

\path[draw=drawColor,line width= 0.4pt,line join=round,line cap=round,fill=fillColor] ( 69.30,163.91) circle (  1.43);

\path[draw=drawColor,line width= 0.4pt,line join=round,line cap=round,fill=fillColor] ( 68.10,144.50) circle (  1.43);

\path[draw=drawColor,line width= 0.4pt,line join=round,line cap=round,fill=fillColor] ( 71.05,149.65) circle (  1.43);

\path[draw=drawColor,line width= 0.4pt,line join=round,line cap=round,fill=fillColor] ( 69.69,143.75) circle (  1.43);

\path[draw=drawColor,line width= 0.4pt,line join=round,line cap=round,fill=fillColor] ( 68.25,150.49) circle (  1.43);

\path[draw=drawColor,line width= 0.4pt,line join=round,line cap=round,fill=fillColor] ( 70.24,156.45) circle (  1.43);

\path[draw=drawColor,line width= 0.4pt,line join=round,line cap=round,fill=fillColor] ( 67.84,147.53) circle (  1.43);

\path[draw=drawColor,line width= 0.4pt,line join=round,line cap=round,fill=fillColor] ( 70.20,146.84) circle (  1.43);

\path[draw=drawColor,line width= 0.4pt,line join=round,line cap=round,fill=fillColor] ( 70.56,160.98) circle (  1.43);

\path[draw=drawColor,line width= 0.4pt,line join=round,line cap=round,fill=fillColor] ( 68.79,148.59) circle (  1.43);

\path[draw=drawColor,line width= 0.4pt,line join=round,line cap=round,fill=fillColor] ( 69.05,153.81) circle (  1.43);

\path[draw=drawColor,line width= 0.4pt,line join=round,line cap=round,fill=fillColor] ( 69.91,144.78) circle (  1.43);

\path[draw=drawColor,line width= 0.4pt,line join=round,line cap=round,fill=fillColor] ( 67.94,151.97) circle (  1.43);

\path[draw=drawColor,line width= 0.4pt,line join=round,line cap=round,fill=fillColor] ( 66.46,144.98) circle (  1.43);

\path[draw=drawColor,line width= 0.4pt,line join=round,line cap=round,fill=fillColor] ( 69.35,163.28) circle (  1.43);

\path[draw=drawColor,line width= 0.4pt,line join=round,line cap=round,fill=fillColor] ( 70.86,150.88) circle (  1.43);

\path[draw=drawColor,line width= 0.4pt,line join=round,line cap=round,fill=fillColor] ( 70.02,173.88) circle (  1.43);

\path[draw=drawColor,line width= 0.4pt,line join=round,line cap=round,fill=fillColor] ( 69.85,176.85) circle (  1.43);

\path[draw=drawColor,line width= 0.4pt,line join=round,line cap=round,fill=fillColor] ( 71.21,141.23) circle (  1.43);

\path[draw=drawColor,line width= 0.4pt,line join=round,line cap=round,fill=fillColor] ( 68.05,156.57) circle (  1.43);

\path[draw=drawColor,line width= 0.4pt,line join=round,line cap=round,fill=fillColor] ( 71.40,165.57) circle (  1.43);

\path[draw=drawColor,line width= 0.4pt,line join=round,line cap=round,fill=fillColor] ( 67.60,141.00) circle (  1.43);

\path[draw=drawColor,line width= 0.4pt,line join=round,line cap=round,fill=fillColor] ( 67.65,149.27) circle (  1.43);

\path[draw=drawColor,line width= 0.4pt,line join=round,line cap=round,fill=fillColor] ( 69.69,152.69) circle (  1.43);

\path[draw=drawColor,line width= 0.4pt,line join=round,line cap=round,fill=fillColor] ( 69.91,150.36) circle (  1.43);

\path[draw=drawColor,line width= 0.4pt,line join=round,line cap=round,fill=fillColor] ( 67.46,157.24) circle (  1.43);

\path[draw=drawColor,line width= 0.4pt,line join=round,line cap=round,fill=fillColor] ( 68.70,155.03) circle (  1.43);

\path[draw=drawColor,line width= 0.4pt,line join=round,line cap=round,fill=fillColor] ( 70.36,145.46) circle (  1.43);

\path[draw=drawColor,line width= 0.4pt,line join=round,line cap=round,fill=fillColor] ( 70.72,155.42) circle (  1.43);

\path[draw=drawColor,line width= 0.4pt,line join=round,line cap=round,fill=fillColor] ( 66.43,148.96) circle (  1.43);

\path[draw=drawColor,line width= 0.4pt,line join=round,line cap=round,fill=fillColor] ( 68.80,146.35) circle (  1.43);

\path[draw=drawColor,line width= 0.4pt,line join=round,line cap=round,fill=fillColor] ( 70.85,145.56) circle (  1.43);

\path[draw=drawColor,line width= 0.4pt,line join=round,line cap=round,fill=fillColor] ( 69.25,170.08) circle (  1.43);

\path[draw=drawColor,line width= 0.4pt,line join=round,line cap=round,fill=fillColor] ( 68.20,155.90) circle (  1.43);

\path[draw=drawColor,line width= 0.4pt,line join=round,line cap=round,fill=fillColor] ( 69.13,163.47) circle (  1.43);

\path[draw=drawColor,line width= 0.4pt,line join=round,line cap=round,fill=fillColor] ( 67.16,149.67) circle (  1.43);

\path[draw=drawColor,line width= 0.4pt,line join=round,line cap=round,fill=fillColor] ( 71.36,158.76) circle (  1.43);

\path[draw=drawColor,line width= 0.4pt,line join=round,line cap=round,fill=fillColor] ( 69.90,154.37) circle (  1.43);

\path[draw=drawColor,line width= 0.4pt,line join=round,line cap=round,fill=fillColor] ( 67.21,149.20) circle (  1.43);

\path[draw=drawColor,line width= 0.4pt,line join=round,line cap=round,fill=fillColor] ( 69.51,145.22) circle (  1.43);

\path[draw=drawColor,line width= 0.4pt,line join=round,line cap=round,fill=fillColor] ( 69.80,142.32) circle (  1.43);

\path[draw=drawColor,line width= 0.4pt,line join=round,line cap=round,fill=fillColor] ( 68.36,148.25) circle (  1.43);

\path[draw=drawColor,line width= 0.4pt,line join=round,line cap=round,fill=fillColor] ( 69.52,149.58) circle (  1.43);

\path[draw=drawColor,line width= 0.4pt,line join=round,line cap=round,fill=fillColor] ( 68.33,148.63) circle (  1.43);

\path[draw=drawColor,line width= 0.4pt,line join=round,line cap=round,fill=fillColor] ( 68.38,148.95) circle (  1.43);

\path[draw=drawColor,line width= 0.4pt,line join=round,line cap=round,fill=fillColor] ( 69.63,153.75) circle (  1.43);

\path[draw=drawColor,line width= 0.4pt,line join=round,line cap=round,fill=fillColor] ( 68.65,165.11) circle (  1.43);

\path[draw=drawColor,line width= 0.4pt,line join=round,line cap=round,fill=fillColor] ( 68.74,148.62) circle (  1.43);

\path[draw=drawColor,line width= 0.4pt,line join=round,line cap=round,fill=fillColor] ( 69.39,152.36) circle (  1.43);

\path[draw=drawColor,line width= 0.4pt,line join=round,line cap=round,fill=fillColor] ( 68.55,155.69) circle (  1.43);

\path[draw=drawColor,line width= 0.4pt,line join=round,line cap=round,fill=fillColor] ( 67.05,147.20) circle (  1.43);

\path[draw=drawColor,line width= 0.4pt,line join=round,line cap=round,fill=fillColor] ( 70.21,151.30) circle (  1.43);

\path[draw=drawColor,line width= 0.4pt,line join=round,line cap=round,fill=fillColor] ( 70.81,144.72) circle (  1.43);

\path[draw=drawColor,line width= 0.4pt,line join=round,line cap=round,fill=fillColor] ( 70.36,146.60) circle (  1.43);

\path[draw=drawColor,line width= 0.4pt,line join=round,line cap=round,fill=fillColor] ( 68.71,143.77) circle (  1.43);

\path[draw=drawColor,line width= 0.4pt,line join=round,line cap=round,fill=fillColor] ( 70.44,145.77) circle (  1.43);

\path[draw=drawColor,line width= 0.4pt,line join=round,line cap=round,fill=fillColor] ( 67.36,153.07) circle (  1.43);

\path[draw=drawColor,line width= 0.4pt,line join=round,line cap=round,fill=fillColor] ( 70.28,156.61) circle (  1.43);

\path[draw=drawColor,line width= 0.4pt,line join=round,line cap=round,fill=fillColor] ( 68.94,158.36) circle (  1.43);

\path[draw=drawColor,line width= 0.4pt,line join=round,line cap=round,fill=fillColor] ( 69.48,151.87) circle (  1.43);

\path[draw=drawColor,line width= 0.4pt,line join=round,line cap=round,fill=fillColor] ( 69.57,147.67) circle (  1.43);

\path[draw=drawColor,line width= 0.4pt,line join=round,line cap=round,fill=fillColor] ( 68.64,147.82) circle (  1.43);

\path[draw=drawColor,line width= 0.4pt,line join=round,line cap=round,fill=fillColor] ( 69.92,161.84) circle (  1.43);
\definecolor{drawColor}{RGB}{207,192,152}
\definecolor{fillColor}{RGB}{207,192,152}

\path[draw=drawColor,line width= 0.4pt,line join=round,line cap=round,fill=fillColor] ( 78.51,150.32) circle (  1.43);

\path[draw=drawColor,line width= 0.4pt,line join=round,line cap=round,fill=fillColor] ( 80.37,171.62) circle (  1.43);

\path[draw=drawColor,line width= 0.4pt,line join=round,line cap=round,fill=fillColor] ( 79.56,161.57) circle (  1.43);

\path[draw=drawColor,line width= 0.4pt,line join=round,line cap=round,fill=fillColor] ( 80.30,179.26) circle (  1.43);

\path[draw=drawColor,line width= 0.4pt,line join=round,line cap=round,fill=fillColor] ( 81.91,181.76) circle (  1.43);

\path[draw=drawColor,line width= 0.4pt,line join=round,line cap=round,fill=fillColor] ( 80.88,149.93) circle (  1.43);

\path[draw=drawColor,line width= 0.4pt,line join=round,line cap=round,fill=fillColor] ( 80.83,152.01) circle (  1.43);

\path[draw=drawColor,line width= 0.4pt,line join=round,line cap=round,fill=fillColor] ( 79.24,157.40) circle (  1.43);

\path[draw=drawColor,line width= 0.4pt,line join=round,line cap=round,fill=fillColor] ( 82.19,179.02) circle (  1.43);

\path[draw=drawColor,line width= 0.4pt,line join=round,line cap=round,fill=fillColor] ( 78.09,154.74) circle (  1.43);

\path[draw=drawColor,line width= 0.4pt,line join=round,line cap=round,fill=fillColor] ( 81.86,192.08) circle (  1.43);

\path[draw=drawColor,line width= 0.4pt,line join=round,line cap=round,fill=fillColor] ( 79.09,173.62) circle (  1.43);

\path[draw=drawColor,line width= 0.4pt,line join=round,line cap=round,fill=fillColor] ( 80.23,149.18) circle (  1.43);

\path[draw=drawColor,line width= 0.4pt,line join=round,line cap=round,fill=fillColor] ( 81.24,167.61) circle (  1.43);

\path[draw=drawColor,line width= 0.4pt,line join=round,line cap=round,fill=fillColor] ( 83.98,170.12) circle (  1.43);

\path[draw=drawColor,line width= 0.4pt,line join=round,line cap=round,fill=fillColor] ( 79.23,149.48) circle (  1.43);

\path[draw=drawColor,line width= 0.4pt,line join=round,line cap=round,fill=fillColor] ( 79.96,172.18) circle (  1.43);

\path[draw=drawColor,line width= 0.4pt,line join=round,line cap=round,fill=fillColor] ( 82.83,192.78) circle (  1.43);

\path[draw=drawColor,line width= 0.4pt,line join=round,line cap=round,fill=fillColor] ( 81.99,173.17) circle (  1.43);

\path[draw=drawColor,line width= 0.4pt,line join=round,line cap=round,fill=fillColor] ( 82.45,169.87) circle (  1.43);

\path[draw=drawColor,line width= 0.4pt,line join=round,line cap=round,fill=fillColor] ( 80.36,171.19) circle (  1.43);

\path[draw=drawColor,line width= 0.4pt,line join=round,line cap=round,fill=fillColor] ( 83.46,168.33) circle (  1.43);

\path[draw=drawColor,line width= 0.4pt,line join=round,line cap=round,fill=fillColor] ( 81.57,162.67) circle (  1.43);

\path[draw=drawColor,line width= 0.4pt,line join=round,line cap=round,fill=fillColor] ( 79.73,159.88) circle (  1.43);

\path[draw=drawColor,line width= 0.4pt,line join=round,line cap=round,fill=fillColor] ( 81.06,153.41) circle (  1.43);

\path[draw=drawColor,line width= 0.4pt,line join=round,line cap=round,fill=fillColor] ( 80.94,153.92) circle (  1.43);

\path[draw=drawColor,line width= 0.4pt,line join=round,line cap=round,fill=fillColor] ( 83.17,172.33) circle (  1.43);

\path[draw=drawColor,line width= 0.4pt,line join=round,line cap=round,fill=fillColor] ( 80.00,161.87) circle (  1.43);

\path[draw=drawColor,line width= 0.4pt,line join=round,line cap=round,fill=fillColor] ( 82.19,171.86) circle (  1.43);

\path[draw=drawColor,line width= 0.4pt,line join=round,line cap=round,fill=fillColor] ( 80.62,177.37) circle (  1.43);

\path[draw=drawColor,line width= 0.4pt,line join=round,line cap=round,fill=fillColor] ( 79.52,171.80) circle (  1.43);

\path[draw=drawColor,line width= 0.4pt,line join=round,line cap=round,fill=fillColor] ( 80.37,154.75) circle (  1.43);

\path[draw=drawColor,line width= 0.4pt,line join=round,line cap=round,fill=fillColor] ( 81.67,173.59) circle (  1.43);

\path[draw=drawColor,line width= 0.4pt,line join=round,line cap=round,fill=fillColor] ( 80.62,163.80) circle (  1.43);

\path[draw=drawColor,line width= 0.4pt,line join=round,line cap=round,fill=fillColor] ( 82.84,159.85) circle (  1.43);

\path[draw=drawColor,line width= 0.4pt,line join=round,line cap=round,fill=fillColor] ( 79.83,161.82) circle (  1.43);

\path[draw=drawColor,line width= 0.4pt,line join=round,line cap=round,fill=fillColor] ( 82.90,179.80) circle (  1.43);

\path[draw=drawColor,line width= 0.4pt,line join=round,line cap=round,fill=fillColor] ( 80.47,169.13) circle (  1.43);

\path[draw=drawColor,line width= 0.4pt,line join=round,line cap=round,fill=fillColor] ( 80.24,178.06) circle (  1.43);

\path[draw=drawColor,line width= 0.4pt,line join=round,line cap=round,fill=fillColor] ( 83.51,173.82) circle (  1.43);

\path[draw=drawColor,line width= 0.4pt,line join=round,line cap=round,fill=fillColor] ( 80.20,163.79) circle (  1.43);

\path[draw=drawColor,line width= 0.4pt,line join=round,line cap=round,fill=fillColor] ( 79.57,153.46) circle (  1.43);

\path[draw=drawColor,line width= 0.4pt,line join=round,line cap=round,fill=fillColor] ( 83.75,183.05) circle (  1.43);

\path[draw=drawColor,line width= 0.4pt,line join=round,line cap=round,fill=fillColor] ( 78.28,160.42) circle (  1.43);

\path[draw=drawColor,line width= 0.4pt,line join=round,line cap=round,fill=fillColor] ( 78.97,157.38) circle (  1.43);

\path[draw=drawColor,line width= 0.4pt,line join=round,line cap=round,fill=fillColor] ( 82.27,167.89) circle (  1.43);

\path[draw=drawColor,line width= 0.4pt,line join=round,line cap=round,fill=fillColor] ( 79.54,162.30) circle (  1.43);

\path[draw=drawColor,line width= 0.4pt,line join=round,line cap=round,fill=fillColor] ( 81.93,168.85) circle (  1.43);

\path[draw=drawColor,line width= 0.4pt,line join=round,line cap=round,fill=fillColor] ( 81.21,164.52) circle (  1.43);

\path[draw=drawColor,line width= 0.4pt,line join=round,line cap=round,fill=fillColor] ( 80.03,165.91) circle (  1.43);

\path[draw=drawColor,line width= 0.4pt,line join=round,line cap=round,fill=fillColor] ( 79.67,161.10) circle (  1.43);

\path[draw=drawColor,line width= 0.4pt,line join=round,line cap=round,fill=fillColor] ( 81.51,168.85) circle (  1.43);

\path[draw=drawColor,line width= 0.4pt,line join=round,line cap=round,fill=fillColor] ( 78.24,162.14) circle (  1.43);

\path[draw=drawColor,line width= 0.4pt,line join=round,line cap=round,fill=fillColor] ( 83.78,184.72) circle (  1.43);

\path[draw=drawColor,line width= 0.4pt,line join=round,line cap=round,fill=fillColor] ( 81.26,166.53) circle (  1.43);

\path[draw=drawColor,line width= 0.4pt,line join=round,line cap=round,fill=fillColor] ( 80.58,169.22) circle (  1.43);

\path[draw=drawColor,line width= 0.4pt,line join=round,line cap=round,fill=fillColor] ( 79.06,150.32) circle (  1.43);

\path[draw=drawColor,line width= 0.4pt,line join=round,line cap=round,fill=fillColor] ( 79.95,156.32) circle (  1.43);

\path[draw=drawColor,line width= 0.4pt,line join=round,line cap=round,fill=fillColor] ( 80.40,175.09) circle (  1.43);

\path[draw=drawColor,line width= 0.4pt,line join=round,line cap=round,fill=fillColor] ( 79.82,160.42) circle (  1.43);

\path[draw=drawColor,line width= 0.4pt,line join=round,line cap=round,fill=fillColor] ( 79.71,163.31) circle (  1.43);

\path[draw=drawColor,line width= 0.4pt,line join=round,line cap=round,fill=fillColor] ( 83.36,171.85) circle (  1.43);

\path[draw=drawColor,line width= 0.4pt,line join=round,line cap=round,fill=fillColor] ( 79.73,164.61) circle (  1.43);

\path[draw=drawColor,line width= 0.4pt,line join=round,line cap=round,fill=fillColor] ( 83.17,172.56) circle (  1.43);

\path[draw=drawColor,line width= 0.4pt,line join=round,line cap=round,fill=fillColor] ( 81.06,150.58) circle (  1.43);

\path[draw=drawColor,line width= 0.4pt,line join=round,line cap=round,fill=fillColor] ( 81.34,177.13) circle (  1.43);
\definecolor{drawColor}{RGB}{0,0,0}

\path[draw=drawColor,line width= 0.6pt,line join=round] ( 19.50, 29.68) -- (117.86,194.47);
\definecolor{drawColor}{gray}{0.70}

\path[draw=drawColor,line width= 0.5pt,line join=round,line cap=round] ( 19.50, 29.68) rectangle (117.86,194.47);
\end{scope}
\begin{scope}
\path[clip] (  0.00,  0.00) rectangle (122.86,199.47);
\definecolor{drawColor}{gray}{0.30}

\node[text=drawColor,anchor=base east,inner sep=0pt, outer sep=0pt, scale=  1.00] at ( 15.00, 26.23) {0};

\node[text=drawColor,anchor=base east,inner sep=0pt, outer sep=0pt, scale=  1.00] at ( 15.00, 56.19) {10};

\node[text=drawColor,anchor=base east,inner sep=0pt, outer sep=0pt, scale=  1.00] at ( 15.00, 86.16) {20};

\node[text=drawColor,anchor=base east,inner sep=0pt, outer sep=0pt, scale=  1.00] at ( 15.00,116.12) {30};

\node[text=drawColor,anchor=base east,inner sep=0pt, outer sep=0pt, scale=  1.00] at ( 15.00,146.08) {40};

\node[text=drawColor,anchor=base east,inner sep=0pt, outer sep=0pt, scale=  1.00] at ( 15.00,176.04) {50};
\end{scope}
\begin{scope}
\path[clip] (  0.00,  0.00) rectangle (122.86,199.47);
\definecolor{drawColor}{gray}{0.70}

\path[draw=drawColor,line width= 0.3pt,line join=round] ( 17.00, 29.68) --
	( 19.50, 29.68);

\path[draw=drawColor,line width= 0.3pt,line join=round] ( 17.00, 59.64) --
	( 19.50, 59.64);

\path[draw=drawColor,line width= 0.3pt,line join=round] ( 17.00, 89.60) --
	( 19.50, 89.60);

\path[draw=drawColor,line width= 0.3pt,line join=round] ( 17.00,119.56) --
	( 19.50,119.56);

\path[draw=drawColor,line width= 0.3pt,line join=round] ( 17.00,149.52) --
	( 19.50,149.52);

\path[draw=drawColor,line width= 0.3pt,line join=round] ( 17.00,179.48) --
	( 19.50,179.48);
\end{scope}
\begin{scope}
\path[clip] (  0.00,  0.00) rectangle (122.86,199.47);
\definecolor{drawColor}{gray}{0.70}

\path[draw=drawColor,line width= 0.3pt,line join=round] ( 19.50, 27.18) --
	( 19.50, 29.68);

\path[draw=drawColor,line width= 0.3pt,line join=round] ( 37.38, 27.18) --
	( 37.38, 29.68);

\path[draw=drawColor,line width= 0.3pt,line join=round] ( 55.27, 27.18) --
	( 55.27, 29.68);

\path[draw=drawColor,line width= 0.3pt,line join=round] ( 73.15, 27.18) --
	( 73.15, 29.68);

\path[draw=drawColor,line width= 0.3pt,line join=round] ( 91.03, 27.18) --
	( 91.03, 29.68);

\path[draw=drawColor,line width= 0.3pt,line join=round] (108.92, 27.18) --
	(108.92, 29.68);
\end{scope}
\begin{scope}
\path[clip] (  0.00,  0.00) rectangle (122.86,199.47);
\definecolor{drawColor}{gray}{0.30}

\node[text=drawColor,anchor=base,inner sep=0pt, outer sep=0pt, scale=  1.00] at ( 19.50, 18.29) {0};

\node[text=drawColor,anchor=base,inner sep=0pt, outer sep=0pt, scale=  1.00] at ( 37.38, 18.29) {10};

\node[text=drawColor,anchor=base,inner sep=0pt, outer sep=0pt, scale=  1.00] at ( 55.27, 18.29) {20};

\node[text=drawColor,anchor=base,inner sep=0pt, outer sep=0pt, scale=  1.00] at ( 73.15, 18.29) {30};

\node[text=drawColor,anchor=base,inner sep=0pt, outer sep=0pt, scale=  1.00] at ( 91.03, 18.29) {40};

\node[text=drawColor,anchor=base,inner sep=0pt, outer sep=0pt, scale=  1.00] at (108.92, 18.29) {50};
\end{scope}
\begin{scope}
\path[clip] (  0.00,  0.00) rectangle (122.86,199.47);
\definecolor{drawColor}{RGB}{0,0,0}

\node[text=drawColor,anchor=base,inner sep=0pt, outer sep=0pt, scale=  1.00] at ( 68.68,  6.94) {\bfseries Gap Gurobi};
\end{scope}
\begin{scope}
\path[clip] (  0.00,  0.00) rectangle (122.86,199.47);
\definecolor{drawColor}{RGB}{0,0,0}

\path[draw=drawColor,line width= 0.4pt,line join=round,line cap=round] ( 80.79, 31.82) rectangle (115.40,114.09);
\end{scope}
\begin{scope}
\path[clip] (  0.00,  0.00) rectangle (122.86,199.47);
\definecolor{fillColor}{RGB}{255,255,255}

\path[fill=fillColor] ( 80.79, 31.82) rectangle (115.40,114.09);
\end{scope}
\begin{scope}
\path[clip] (  0.00,  0.00) rectangle (122.86,199.47);
\definecolor{fillColor}{RGB}{255,255,255}

\path[fill=fillColor] ( 85.79, 94.63) rectangle (100.25,109.09);
\end{scope}
\begin{scope}
\path[clip] (  0.00,  0.00) rectangle (122.86,199.47);
\definecolor{drawColor}{RGB}{128,128,128}
\definecolor{fillColor}{RGB}{128,128,128}

\path[draw=drawColor,line width= 0.4pt,line join=round,line cap=round,fill=fillColor] ( 93.02,101.86) circle (  1.43);
\end{scope}
\begin{scope}
\path[clip] (  0.00,  0.00) rectangle (122.86,199.47);
\definecolor{fillColor}{RGB}{255,255,255}

\path[fill=fillColor] ( 85.79, 80.18) rectangle (100.25, 94.63);
\end{scope}
\begin{scope}
\path[clip] (  0.00,  0.00) rectangle (122.86,199.47);
\definecolor{drawColor}{RGB}{152,167,197}
\definecolor{fillColor}{RGB}{152,167,197}

\path[draw=drawColor,line width= 0.4pt,line join=round,line cap=round,fill=fillColor] ( 93.02, 87.41) circle (  1.43);
\end{scope}
\begin{scope}
\path[clip] (  0.00,  0.00) rectangle (122.86,199.47);
\definecolor{fillColor}{RGB}{255,255,255}

\path[fill=fillColor] ( 85.79, 65.73) rectangle (100.25, 80.18);
\end{scope}
\begin{scope}
\path[clip] (  0.00,  0.00) rectangle (122.86,199.47);
\definecolor{drawColor}{RGB}{128,202,192}
\definecolor{fillColor}{RGB}{128,202,192}

\path[draw=drawColor,line width= 0.4pt,line join=round,line cap=round,fill=fillColor] ( 93.02, 72.95) circle (  1.43);
\end{scope}
\begin{scope}
\path[clip] (  0.00,  0.00) rectangle (122.86,199.47);
\definecolor{fillColor}{RGB}{255,255,255}

\path[fill=fillColor] ( 85.79, 51.27) rectangle (100.25, 65.73);
\end{scope}
\begin{scope}
\path[clip] (  0.00,  0.00) rectangle (122.86,199.47);
\definecolor{drawColor}{RGB}{207,128,187}
\definecolor{fillColor}{RGB}{207,128,187}

\path[draw=drawColor,line width= 0.4pt,line join=round,line cap=round,fill=fillColor] ( 93.02, 58.50) circle (  1.43);
\end{scope}
\begin{scope}
\path[clip] (  0.00,  0.00) rectangle (122.86,199.47);
\definecolor{fillColor}{RGB}{255,255,255}

\path[fill=fillColor] ( 85.79, 36.82) rectangle (100.25, 51.27);
\end{scope}
\begin{scope}
\path[clip] (  0.00,  0.00) rectangle (122.86,199.47);
\definecolor{drawColor}{RGB}{207,192,152}
\definecolor{fillColor}{RGB}{207,192,152}

\path[draw=drawColor,line width= 0.4pt,line join=round,line cap=round,fill=fillColor] ( 93.02, 44.05) circle (  1.43);
\end{scope}
\begin{scope}
\path[clip] (  0.00,  0.00) rectangle (122.86,199.47);
\definecolor{drawColor}{RGB}{0,0,0}

\node[text=drawColor,anchor=base west,inner sep=0pt, outer sep=0pt, scale=  0.80] at (100.25, 99.11) {$\ensuremath{\mathcal{N}}_1$};
\end{scope}
\begin{scope}
\path[clip] (  0.00,  0.00) rectangle (122.86,199.47);
\definecolor{drawColor}{RGB}{0,0,0}

\node[text=drawColor,anchor=base west,inner sep=0pt, outer sep=0pt, scale=  0.80] at (100.25, 84.65) {$\ensuremath{\mathcal{N}}_2$};
\end{scope}
\begin{scope}
\path[clip] (  0.00,  0.00) rectangle (122.86,199.47);
\definecolor{drawColor}{RGB}{0,0,0}

\node[text=drawColor,anchor=base west,inner sep=0pt, outer sep=0pt, scale=  0.80] at (100.25, 70.20) {$\ensuremath{\mathcal{N}}_3$};
\end{scope}
\begin{scope}
\path[clip] (  0.00,  0.00) rectangle (122.86,199.47);
\definecolor{drawColor}{RGB}{0,0,0}

\node[text=drawColor,anchor=base west,inner sep=0pt, outer sep=0pt, scale=  0.80] at (100.25, 55.74) {$\ensuremath{\mathcal{N}}_4$};
\end{scope}
\begin{scope}
\path[clip] (  0.00,  0.00) rectangle (122.86,199.47);
\definecolor{drawColor}{RGB}{0,0,0}

\node[text=drawColor,anchor=base west,inner sep=0pt, outer sep=0pt, scale=  0.80] at (100.25, 41.29) {$\ensuremath{\mathcal{N}}_5$};
\end{scope}
\end{tikzpicture}

%% file: 06-plots/CPLEXvsGUROBI-Gap-1Day.tex
\begin{tikzpicture}[x=1pt,y=1pt]
\definecolor{fillColor}{RGB}{255,255,255}
\path[use as bounding box,fill=fillColor,fill opacity=0.00] (0,0) rectangle (122.86,199.47);
\begin{scope}
\path[clip] (  0.00,  0.00) rectangle (122.86,199.47);
\definecolor{drawColor}{RGB}{255,255,255}
\definecolor{fillColor}{RGB}{255,255,255}

\path[draw=drawColor,line width= 0.5pt,line join=round,line cap=round,fill=fillColor] (  0.00,  0.00) rectangle (122.86,199.47);
\end{scope}
\begin{scope}
\path[clip] ( 19.50, 29.68) rectangle (117.86,194.47);
\definecolor{fillColor}{RGB}{255,255,255}

\path[fill=fillColor] ( 19.50, 29.68) rectangle (117.86,194.47);
\definecolor{drawColor}{gray}{0.87}

\path[draw=drawColor,line width= 0.1pt,line join=round] ( 19.50, 44.66) --
	(117.86, 44.66);

\path[draw=drawColor,line width= 0.1pt,line join=round] ( 19.50, 74.62) --
	(117.86, 74.62);

\path[draw=drawColor,line width= 0.1pt,line join=round] ( 19.50,104.58) --
	(117.86,104.58);

\path[draw=drawColor,line width= 0.1pt,line join=round] ( 19.50,134.54) --
	(117.86,134.54);

\path[draw=drawColor,line width= 0.1pt,line join=round] ( 19.50,164.50) --
	(117.86,164.50);

\path[draw=drawColor,line width= 0.1pt,line join=round] ( 19.50,194.47) --
	(117.86,194.47);

\path[draw=drawColor,line width= 0.1pt,line join=round] ( 28.44, 29.68) --
	( 28.44,194.47);

\path[draw=drawColor,line width= 0.1pt,line join=round] ( 46.32, 29.68) --
	( 46.32,194.47);

\path[draw=drawColor,line width= 0.1pt,line join=round] ( 64.21, 29.68) --
	( 64.21,194.47);

\path[draw=drawColor,line width= 0.1pt,line join=round] ( 82.09, 29.68) --
	( 82.09,194.47);

\path[draw=drawColor,line width= 0.1pt,line join=round] ( 99.98, 29.68) --
	( 99.98,194.47);

\path[draw=drawColor,line width= 0.1pt,line join=round] (117.86, 29.68) --
	(117.86,194.47);

\path[draw=drawColor,line width= 0.3pt,line join=round] ( 19.50, 29.68) --
	(117.86, 29.68);

\path[draw=drawColor,line width= 0.3pt,line join=round] ( 19.50, 59.64) --
	(117.86, 59.64);

\path[draw=drawColor,line width= 0.3pt,line join=round] ( 19.50, 89.60) --
	(117.86, 89.60);

\path[draw=drawColor,line width= 0.3pt,line join=round] ( 19.50,119.56) --
	(117.86,119.56);

\path[draw=drawColor,line width= 0.3pt,line join=round] ( 19.50,149.52) --
	(117.86,149.52);

\path[draw=drawColor,line width= 0.3pt,line join=round] ( 19.50,179.48) --
	(117.86,179.48);

\path[draw=drawColor,line width= 0.3pt,line join=round] ( 19.50, 29.68) --
	( 19.50,194.47);

\path[draw=drawColor,line width= 0.3pt,line join=round] ( 37.38, 29.68) --
	( 37.38,194.47);

\path[draw=drawColor,line width= 0.3pt,line join=round] ( 55.27, 29.68) --
	( 55.27,194.47);

\path[draw=drawColor,line width= 0.3pt,line join=round] ( 73.15, 29.68) --
	( 73.15,194.47);

\path[draw=drawColor,line width= 0.3pt,line join=round] ( 91.03, 29.68) --
	( 91.03,194.47);

\path[draw=drawColor,line width= 0.3pt,line join=round] (108.92, 29.68) --
	(108.92,194.47);
\definecolor{drawColor}{RGB}{128,128,128}
\definecolor{fillColor}{RGB}{128,128,128}

\path[draw=drawColor,line width= 0.4pt,line join=round,line cap=round,fill=fillColor] ( 31.59, 47.55) circle (  1.43);

\path[draw=drawColor,line width= 0.4pt,line join=round,line cap=round,fill=fillColor] ( 41.61, 70.54) circle (  1.43);

\path[draw=drawColor,line width= 0.4pt,line join=round,line cap=round,fill=fillColor] ( 50.33,101.61) circle (  1.43);

\path[draw=drawColor,line width= 0.4pt,line join=round,line cap=round,fill=fillColor] ( 19.52, 29.71) circle (  1.43);

\path[draw=drawColor,line width= 0.4pt,line join=round,line cap=round,fill=fillColor] ( 19.52, 29.71) circle (  1.43);

\path[draw=drawColor,line width= 0.4pt,line join=round,line cap=round,fill=fillColor] ( 19.52, 29.71) circle (  1.43);

\path[draw=drawColor,line width= 0.4pt,line join=round,line cap=round,fill=fillColor] ( 19.52, 29.71) circle (  1.43);

\path[draw=drawColor,line width= 0.4pt,line join=round,line cap=round,fill=fillColor] ( 36.10, 52.72) circle (  1.43);

\path[draw=drawColor,line width= 0.4pt,line join=round,line cap=round,fill=fillColor] ( 60.35,116.95) circle (  1.43);

\path[draw=drawColor,line width= 0.4pt,line join=round,line cap=round,fill=fillColor] ( 57.91,109.13) circle (  1.43);

\path[draw=drawColor,line width= 0.4pt,line join=round,line cap=round,fill=fillColor] ( 53.87,108.11) circle (  1.43);

\path[draw=drawColor,line width= 0.4pt,line join=round,line cap=round,fill=fillColor] ( 53.36,101.38) circle (  1.43);

\path[draw=drawColor,line width= 0.4pt,line join=round,line cap=round,fill=fillColor] ( 19.52, 33.72) circle (  1.43);

\path[draw=drawColor,line width= 0.4pt,line join=round,line cap=round,fill=fillColor] ( 53.20,106.06) circle (  1.43);

\path[draw=drawColor,line width= 0.4pt,line join=round,line cap=round,fill=fillColor] ( 58.41,111.08) circle (  1.43);

\path[draw=drawColor,line width= 0.4pt,line join=round,line cap=round,fill=fillColor] ( 43.21, 79.60) circle (  1.43);

\path[draw=drawColor,line width= 0.4pt,line join=round,line cap=round,fill=fillColor] ( 56.91,114.30) circle (  1.43);

\path[draw=drawColor,line width= 0.4pt,line join=round,line cap=round,fill=fillColor] ( 48.33, 84.68) circle (  1.43);

\path[draw=drawColor,line width= 0.4pt,line join=round,line cap=round,fill=fillColor] ( 19.52, 29.71) circle (  1.43);

\path[draw=drawColor,line width= 0.4pt,line join=round,line cap=round,fill=fillColor] ( 61.03,118.77) circle (  1.43);

\path[draw=drawColor,line width= 0.4pt,line join=round,line cap=round,fill=fillColor] ( 31.78, 48.96) circle (  1.43);

\path[draw=drawColor,line width= 0.4pt,line join=round,line cap=round,fill=fillColor] ( 29.25, 41.69) circle (  1.43);

\path[draw=drawColor,line width= 0.4pt,line join=round,line cap=round,fill=fillColor] ( 55.82,110.98) circle (  1.43);

\path[draw=drawColor,line width= 0.4pt,line join=round,line cap=round,fill=fillColor] ( 31.14, 54.44) circle (  1.43);

\path[draw=drawColor,line width= 0.4pt,line join=round,line cap=round,fill=fillColor] ( 53.46, 94.44) circle (  1.43);

\path[draw=drawColor,line width= 0.4pt,line join=round,line cap=round,fill=fillColor] ( 56.03, 88.83) circle (  1.43);

\path[draw=drawColor,line width= 0.4pt,line join=round,line cap=round,fill=fillColor] ( 57.18,108.85) circle (  1.43);

\path[draw=drawColor,line width= 0.4pt,line join=round,line cap=round,fill=fillColor] ( 59.40,111.51) circle (  1.43);

\path[draw=drawColor,line width= 0.4pt,line join=round,line cap=round,fill=fillColor] ( 43.20, 68.03) circle (  1.43);

\path[draw=drawColor,line width= 0.4pt,line join=round,line cap=round,fill=fillColor] ( 50.52,105.27) circle (  1.43);

\path[draw=drawColor,line width= 0.4pt,line join=round,line cap=round,fill=fillColor] ( 58.46,112.92) circle (  1.43);

\path[draw=drawColor,line width= 0.4pt,line join=round,line cap=round,fill=fillColor] ( 48.86, 85.96) circle (  1.43);

\path[draw=drawColor,line width= 0.4pt,line join=round,line cap=round,fill=fillColor] ( 19.52, 29.71) circle (  1.43);

\path[draw=drawColor,line width= 0.4pt,line join=round,line cap=round,fill=fillColor] ( 55.96,105.41) circle (  1.43);

\path[draw=drawColor,line width= 0.4pt,line join=round,line cap=round,fill=fillColor] ( 55.50, 97.26) circle (  1.43);
\definecolor{drawColor}{RGB}{152,167,197}
\definecolor{fillColor}{RGB}{152,167,197}

\path[draw=drawColor,line width= 0.4pt,line join=round,line cap=round,fill=fillColor] ( 66.91,124.15) circle (  1.43);

\path[draw=drawColor,line width= 0.4pt,line join=round,line cap=round,fill=fillColor] ( 45.59, 80.20) circle (  1.43);

\path[draw=drawColor,line width= 0.4pt,line join=round,line cap=round,fill=fillColor] ( 57.94,111.58) circle (  1.43);

\path[draw=drawColor,line width= 0.4pt,line join=round,line cap=round,fill=fillColor] ( 56.19,100.19) circle (  1.43);

\path[draw=drawColor,line width= 0.4pt,line join=round,line cap=round,fill=fillColor] ( 67.76,126.52) circle (  1.43);

\path[draw=drawColor,line width= 0.4pt,line join=round,line cap=round,fill=fillColor] ( 65.05,119.83) circle (  1.43);

\path[draw=drawColor,line width= 0.4pt,line join=round,line cap=round,fill=fillColor] ( 66.30,130.14) circle (  1.43);

\path[draw=drawColor,line width= 0.4pt,line join=round,line cap=round,fill=fillColor] ( 63.21,113.27) circle (  1.43);

\path[draw=drawColor,line width= 0.4pt,line join=round,line cap=round,fill=fillColor] ( 64.88,114.77) circle (  1.43);

\path[draw=drawColor,line width= 0.4pt,line join=round,line cap=round,fill=fillColor] ( 60.06,119.09) circle (  1.43);

\path[draw=drawColor,line width= 0.4pt,line join=round,line cap=round,fill=fillColor] ( 54.04, 90.87) circle (  1.43);

\path[draw=drawColor,line width= 0.4pt,line join=round,line cap=round,fill=fillColor] ( 66.27,121.95) circle (  1.43);

\path[draw=drawColor,line width= 0.4pt,line join=round,line cap=round,fill=fillColor] ( 60.22,106.71) circle (  1.43);

\path[draw=drawColor,line width= 0.4pt,line join=round,line cap=round,fill=fillColor] ( 62.32,114.67) circle (  1.43);

\path[draw=drawColor,line width= 0.4pt,line join=round,line cap=round,fill=fillColor] ( 67.41,128.83) circle (  1.43);

\path[draw=drawColor,line width= 0.4pt,line join=round,line cap=round,fill=fillColor] ( 62.53,107.33) circle (  1.43);

\path[draw=drawColor,line width= 0.4pt,line join=round,line cap=round,fill=fillColor] ( 46.48, 71.98) circle (  1.43);

\path[draw=drawColor,line width= 0.4pt,line join=round,line cap=round,fill=fillColor] ( 66.76,122.84) circle (  1.43);

\path[draw=drawColor,line width= 0.4pt,line join=round,line cap=round,fill=fillColor] ( 43.97, 80.20) circle (  1.43);

\path[draw=drawColor,line width= 0.4pt,line join=round,line cap=round,fill=fillColor] ( 61.10,115.93) circle (  1.43);

\path[draw=drawColor,line width= 0.4pt,line join=round,line cap=round,fill=fillColor] ( 54.38,104.19) circle (  1.43);

\path[draw=drawColor,line width= 0.4pt,line join=round,line cap=round,fill=fillColor] ( 45.99, 79.45) circle (  1.43);

\path[draw=drawColor,line width= 0.4pt,line join=round,line cap=round,fill=fillColor] ( 52.35, 97.50) circle (  1.43);

\path[draw=drawColor,line width= 0.4pt,line join=round,line cap=round,fill=fillColor] ( 62.58,119.72) circle (  1.43);

\path[draw=drawColor,line width= 0.4pt,line join=round,line cap=round,fill=fillColor] ( 46.54, 72.05) circle (  1.43);

\path[draw=drawColor,line width= 0.4pt,line join=round,line cap=round,fill=fillColor] ( 55.49,101.05) circle (  1.43);

\path[draw=drawColor,line width= 0.4pt,line join=round,line cap=round,fill=fillColor] ( 57.91, 98.00) circle (  1.43);

\path[draw=drawColor,line width= 0.4pt,line join=round,line cap=round,fill=fillColor] ( 62.68,110.07) circle (  1.43);

\path[draw=drawColor,line width= 0.4pt,line join=round,line cap=round,fill=fillColor] ( 67.09,115.95) circle (  1.43);

\path[draw=drawColor,line width= 0.4pt,line join=round,line cap=round,fill=fillColor] ( 62.97,115.33) circle (  1.43);

\path[draw=drawColor,line width= 0.4pt,line join=round,line cap=round,fill=fillColor] ( 58.09,109.30) circle (  1.43);

\path[draw=drawColor,line width= 0.4pt,line join=round,line cap=round,fill=fillColor] ( 45.87, 71.45) circle (  1.43);

\path[draw=drawColor,line width= 0.4pt,line join=round,line cap=round,fill=fillColor] ( 63.98,122.41) circle (  1.43);

\path[draw=drawColor,line width= 0.4pt,line join=round,line cap=round,fill=fillColor] ( 51.69, 84.30) circle (  1.43);

\path[draw=drawColor,line width= 0.4pt,line join=round,line cap=round,fill=fillColor] ( 63.90,119.82) circle (  1.43);
\definecolor{drawColor}{RGB}{128,202,192}
\definecolor{fillColor}{RGB}{128,202,192}

\path[draw=drawColor,line width= 0.4pt,line join=round,line cap=round,fill=fillColor] ( 68.95,132.11) circle (  1.43);

\path[draw=drawColor,line width= 0.4pt,line join=round,line cap=round,fill=fillColor] ( 66.94,126.98) circle (  1.43);

\path[draw=drawColor,line width= 0.4pt,line join=round,line cap=round,fill=fillColor] ( 68.92,134.91) circle (  1.43);

\path[draw=drawColor,line width= 0.4pt,line join=round,line cap=round,fill=fillColor] ( 68.70,130.88) circle (  1.43);

\path[draw=drawColor,line width= 0.4pt,line join=round,line cap=round,fill=fillColor] ( 67.60,127.13) circle (  1.43);

\path[draw=drawColor,line width= 0.4pt,line join=round,line cap=round,fill=fillColor] ( 69.29,130.32) circle (  1.43);

\path[draw=drawColor,line width= 0.4pt,line join=round,line cap=round,fill=fillColor] ( 70.16,133.16) circle (  1.43);

\path[draw=drawColor,line width= 0.4pt,line join=round,line cap=round,fill=fillColor] ( 68.50,129.20) circle (  1.43);

\path[draw=drawColor,line width= 0.4pt,line join=round,line cap=round,fill=fillColor] ( 66.93,132.93) circle (  1.43);

\path[draw=drawColor,line width= 0.4pt,line join=round,line cap=round,fill=fillColor] ( 68.37,133.18) circle (  1.43);

\path[draw=drawColor,line width= 0.4pt,line join=round,line cap=round,fill=fillColor] ( 67.18,132.56) circle (  1.43);

\path[draw=drawColor,line width= 0.4pt,line join=round,line cap=round,fill=fillColor] ( 68.74,128.60) circle (  1.43);

\path[draw=drawColor,line width= 0.4pt,line join=round,line cap=round,fill=fillColor] ( 65.36,131.38) circle (  1.43);

\path[draw=drawColor,line width= 0.4pt,line join=round,line cap=round,fill=fillColor] ( 68.35,128.61) circle (  1.43);

\path[draw=drawColor,line width= 0.4pt,line join=round,line cap=round,fill=fillColor] ( 67.85,127.27) circle (  1.43);

\path[draw=drawColor,line width= 0.4pt,line join=round,line cap=round,fill=fillColor] ( 69.57,133.38) circle (  1.43);

\path[draw=drawColor,line width= 0.4pt,line join=round,line cap=round,fill=fillColor] ( 69.92,135.47) circle (  1.43);

\path[draw=drawColor,line width= 0.4pt,line join=round,line cap=round,fill=fillColor] ( 68.69,130.56) circle (  1.43);

\path[draw=drawColor,line width= 0.4pt,line join=round,line cap=round,fill=fillColor] ( 69.48,133.45) circle (  1.43);

\path[draw=drawColor,line width= 0.4pt,line join=round,line cap=round,fill=fillColor] ( 69.36,132.70) circle (  1.43);

\path[draw=drawColor,line width= 0.4pt,line join=round,line cap=round,fill=fillColor] ( 66.38,127.67) circle (  1.43);

\path[draw=drawColor,line width= 0.4pt,line join=round,line cap=round,fill=fillColor] ( 69.40,135.81) circle (  1.43);

\path[draw=drawColor,line width= 0.4pt,line join=round,line cap=round,fill=fillColor] ( 69.24,125.83) circle (  1.43);

\path[draw=drawColor,line width= 0.4pt,line join=round,line cap=round,fill=fillColor] ( 69.28,134.76) circle (  1.43);

\path[draw=drawColor,line width= 0.4pt,line join=round,line cap=round,fill=fillColor] ( 67.68,132.51) circle (  1.43);

\path[draw=drawColor,line width= 0.4pt,line join=round,line cap=round,fill=fillColor] ( 66.71,127.92) circle (  1.43);

\path[draw=drawColor,line width= 0.4pt,line join=round,line cap=round,fill=fillColor] ( 66.43,127.38) circle (  1.43);

\path[draw=drawColor,line width= 0.4pt,line join=round,line cap=round,fill=fillColor] ( 66.86,127.92) circle (  1.43);

\path[draw=drawColor,line width= 0.4pt,line join=round,line cap=round,fill=fillColor] ( 66.25,129.65) circle (  1.43);

\path[draw=drawColor,line width= 0.4pt,line join=round,line cap=round,fill=fillColor] ( 70.94,131.67) circle (  1.43);

\path[draw=drawColor,line width= 0.4pt,line join=round,line cap=round,fill=fillColor] ( 69.66,135.69) circle (  1.43);

\path[draw=drawColor,line width= 0.4pt,line join=round,line cap=round,fill=fillColor] ( 66.75,127.92) circle (  1.43);

\path[draw=drawColor,line width= 0.4pt,line join=round,line cap=round,fill=fillColor] ( 67.77,133.18) circle (  1.43);

\path[draw=drawColor,line width= 0.4pt,line join=round,line cap=round,fill=fillColor] ( 71.19,130.49) circle (  1.43);

\path[draw=drawColor,line width= 0.4pt,line join=round,line cap=round,fill=fillColor] ( 67.34,130.49) circle (  1.43);

\path[draw=drawColor,line width= 0.4pt,line join=round,line cap=round,fill=fillColor] ( 67.33,135.88) circle (  1.43);

\path[draw=drawColor,line width= 0.4pt,line join=round,line cap=round,fill=fillColor] ( 70.77,134.71) circle (  1.43);

\path[draw=drawColor,line width= 0.4pt,line join=round,line cap=round,fill=fillColor] ( 67.54,131.33) circle (  1.43);

\path[draw=drawColor,line width= 0.4pt,line join=round,line cap=round,fill=fillColor] ( 69.50,134.45) circle (  1.43);

\path[draw=drawColor,line width= 0.4pt,line join=round,line cap=round,fill=fillColor] ( 69.73,129.73) circle (  1.43);

\path[draw=drawColor,line width= 0.4pt,line join=round,line cap=round,fill=fillColor] ( 66.93,133.07) circle (  1.43);

\path[draw=drawColor,line width= 0.4pt,line join=round,line cap=round,fill=fillColor] ( 69.61,126.05) circle (  1.43);

\path[draw=drawColor,line width= 0.4pt,line join=round,line cap=round,fill=fillColor] ( 68.03,130.32) circle (  1.43);

\path[draw=drawColor,line width= 0.4pt,line join=round,line cap=round,fill=fillColor] ( 64.75,127.86) circle (  1.43);

\path[draw=drawColor,line width= 0.4pt,line join=round,line cap=round,fill=fillColor] ( 69.61,128.20) circle (  1.43);

\path[draw=drawColor,line width= 0.4pt,line join=round,line cap=round,fill=fillColor] ( 68.07,132.59) circle (  1.43);

\path[draw=drawColor,line width= 0.4pt,line join=round,line cap=round,fill=fillColor] ( 67.96,134.56) circle (  1.43);

\path[draw=drawColor,line width= 0.4pt,line join=round,line cap=round,fill=fillColor] ( 68.27,137.07) circle (  1.43);

\path[draw=drawColor,line width= 0.4pt,line join=round,line cap=round,fill=fillColor] ( 66.07,128.85) circle (  1.43);

\path[draw=drawColor,line width= 0.4pt,line join=round,line cap=round,fill=fillColor] ( 66.77,132.31) circle (  1.43);

\path[draw=drawColor,line width= 0.4pt,line join=round,line cap=round,fill=fillColor] ( 66.83,133.44) circle (  1.43);

\path[draw=drawColor,line width= 0.4pt,line join=round,line cap=round,fill=fillColor] ( 70.68,135.65) circle (  1.43);

\path[draw=drawColor,line width= 0.4pt,line join=round,line cap=round,fill=fillColor] ( 70.21,128.57) circle (  1.43);

\path[draw=drawColor,line width= 0.4pt,line join=round,line cap=round,fill=fillColor] ( 66.94,128.47) circle (  1.43);

\path[draw=drawColor,line width= 0.4pt,line join=round,line cap=round,fill=fillColor] ( 69.41,135.95) circle (  1.43);

\path[draw=drawColor,line width= 0.4pt,line join=round,line cap=round,fill=fillColor] ( 67.95,135.03) circle (  1.43);

\path[draw=drawColor,line width= 0.4pt,line join=round,line cap=round,fill=fillColor] ( 67.67,133.16) circle (  1.43);

\path[draw=drawColor,line width= 0.4pt,line join=round,line cap=round,fill=fillColor] ( 66.69,130.03) circle (  1.43);

\path[draw=drawColor,line width= 0.4pt,line join=round,line cap=round,fill=fillColor] ( 68.90,127.96) circle (  1.43);

\path[draw=drawColor,line width= 0.4pt,line join=round,line cap=round,fill=fillColor] ( 65.77,123.70) circle (  1.43);

\path[draw=drawColor,line width= 0.4pt,line join=round,line cap=round,fill=fillColor] ( 68.48,138.13) circle (  1.43);

\path[draw=drawColor,line width= 0.4pt,line join=round,line cap=round,fill=fillColor] ( 69.22,132.45) circle (  1.43);

\path[draw=drawColor,line width= 0.4pt,line join=round,line cap=round,fill=fillColor] ( 68.74,139.80) circle (  1.43);

\path[draw=drawColor,line width= 0.4pt,line join=round,line cap=round,fill=fillColor] ( 68.02,131.52) circle (  1.43);

\path[draw=drawColor,line width= 0.4pt,line join=round,line cap=round,fill=fillColor] ( 69.34,132.46) circle (  1.43);

\path[draw=drawColor,line width= 0.4pt,line join=round,line cap=round,fill=fillColor] ( 65.87,130.66) circle (  1.43);

\path[draw=drawColor,line width= 0.4pt,line join=round,line cap=round,fill=fillColor] ( 68.48,130.25) circle (  1.43);

\path[draw=drawColor,line width= 0.4pt,line join=round,line cap=round,fill=fillColor] ( 66.54,127.44) circle (  1.43);

\path[draw=drawColor,line width= 0.4pt,line join=round,line cap=round,fill=fillColor] ( 69.96,135.12) circle (  1.43);

\path[draw=drawColor,line width= 0.4pt,line join=round,line cap=round,fill=fillColor] ( 68.07,130.20) circle (  1.43);
\definecolor{drawColor}{RGB}{207,128,187}
\definecolor{fillColor}{RGB}{207,128,187}

\path[draw=drawColor,line width= 0.4pt,line join=round,line cap=round,fill=fillColor] ( 70.17,137.77) circle (  1.43);

\path[draw=drawColor,line width= 0.4pt,line join=round,line cap=round,fill=fillColor] ( 68.78,133.45) circle (  1.43);

\path[draw=drawColor,line width= 0.4pt,line join=round,line cap=round,fill=fillColor] ( 70.38,136.29) circle (  1.43);

\path[draw=drawColor,line width= 0.4pt,line join=round,line cap=round,fill=fillColor] ( 70.30,135.22) circle (  1.43);

\path[draw=drawColor,line width= 0.4pt,line join=round,line cap=round,fill=fillColor] ( 70.28,136.46) circle (  1.43);

\path[draw=drawColor,line width= 0.4pt,line join=round,line cap=round,fill=fillColor] ( 71.07,142.53) circle (  1.43);

\path[draw=drawColor,line width= 0.4pt,line join=round,line cap=round,fill=fillColor] ( 69.86,142.95) circle (  1.43);

\path[draw=drawColor,line width= 0.4pt,line join=round,line cap=round,fill=fillColor] ( 69.14,141.58) circle (  1.43);

\path[draw=drawColor,line width= 0.4pt,line join=round,line cap=round,fill=fillColor] ( 68.03,136.02) circle (  1.43);

\path[draw=drawColor,line width= 0.4pt,line join=round,line cap=round,fill=fillColor] ( 70.92,141.17) circle (  1.43);

\path[draw=drawColor,line width= 0.4pt,line join=round,line cap=round,fill=fillColor] ( 69.32,138.04) circle (  1.43);

\path[draw=drawColor,line width= 0.4pt,line join=round,line cap=round,fill=fillColor] ( 68.19,140.25) circle (  1.43);

\path[draw=drawColor,line width= 0.4pt,line join=round,line cap=round,fill=fillColor] ( 70.08,140.86) circle (  1.43);

\path[draw=drawColor,line width= 0.4pt,line join=round,line cap=round,fill=fillColor] ( 67.77,138.22) circle (  1.43);

\path[draw=drawColor,line width= 0.4pt,line join=round,line cap=round,fill=fillColor] ( 70.14,140.75) circle (  1.43);

\path[draw=drawColor,line width= 0.4pt,line join=round,line cap=round,fill=fillColor] ( 70.28,144.47) circle (  1.43);

\path[draw=drawColor,line width= 0.4pt,line join=round,line cap=round,fill=fillColor] ( 68.72,139.42) circle (  1.43);

\path[draw=drawColor,line width= 0.4pt,line join=round,line cap=round,fill=fillColor] ( 68.48,142.60) circle (  1.43);

\path[draw=drawColor,line width= 0.4pt,line join=round,line cap=round,fill=fillColor] ( 69.87,137.40) circle (  1.43);

\path[draw=drawColor,line width= 0.4pt,line join=round,line cap=round,fill=fillColor] ( 67.56,141.76) circle (  1.43);

\path[draw=drawColor,line width= 0.4pt,line join=round,line cap=round,fill=fillColor] ( 65.95,136.13) circle (  1.43);

\path[draw=drawColor,line width= 0.4pt,line join=round,line cap=round,fill=fillColor] ( 68.97,140.31) circle (  1.43);

\path[draw=drawColor,line width= 0.4pt,line join=round,line cap=round,fill=fillColor] ( 70.76,139.49) circle (  1.43);

\path[draw=drawColor,line width= 0.4pt,line join=round,line cap=round,fill=fillColor] ( 69.53,144.34) circle (  1.43);

\path[draw=drawColor,line width= 0.4pt,line join=round,line cap=round,fill=fillColor] ( 69.43,142.94) circle (  1.43);

\path[draw=drawColor,line width= 0.4pt,line join=round,line cap=round,fill=fillColor] ( 71.14,137.26) circle (  1.43);

\path[draw=drawColor,line width= 0.4pt,line join=round,line cap=round,fill=fillColor] ( 67.48,141.28) circle (  1.43);

\path[draw=drawColor,line width= 0.4pt,line join=round,line cap=round,fill=fillColor] ( 71.10,147.72) circle (  1.43);

\path[draw=drawColor,line width= 0.4pt,line join=round,line cap=round,fill=fillColor] ( 67.55,135.31) circle (  1.43);

\path[draw=drawColor,line width= 0.4pt,line join=round,line cap=round,fill=fillColor] ( 67.60,139.53) circle (  1.43);

\path[draw=drawColor,line width= 0.4pt,line join=round,line cap=round,fill=fillColor] ( 69.16,143.36) circle (  1.43);

\path[draw=drawColor,line width= 0.4pt,line join=round,line cap=round,fill=fillColor] ( 69.85,140.28) circle (  1.43);

\path[draw=drawColor,line width= 0.4pt,line join=round,line cap=round,fill=fillColor] ( 67.34,143.92) circle (  1.43);

\path[draw=drawColor,line width= 0.4pt,line join=round,line cap=round,fill=fillColor] ( 68.42,139.84) circle (  1.43);

\path[draw=drawColor,line width= 0.4pt,line join=round,line cap=round,fill=fillColor] ( 70.11,140.95) circle (  1.43);

\path[draw=drawColor,line width= 0.4pt,line join=round,line cap=round,fill=fillColor] ( 70.68,140.14) circle (  1.43);

\path[draw=drawColor,line width= 0.4pt,line join=round,line cap=round,fill=fillColor] ( 66.38,139.35) circle (  1.43);

\path[draw=drawColor,line width= 0.4pt,line join=round,line cap=round,fill=fillColor] ( 68.71,137.20) circle (  1.43);

\path[draw=drawColor,line width= 0.4pt,line join=round,line cap=round,fill=fillColor] ( 70.79,138.24) circle (  1.43);

\path[draw=drawColor,line width= 0.4pt,line join=round,line cap=round,fill=fillColor] ( 68.94,141.61) circle (  1.43);

\path[draw=drawColor,line width= 0.4pt,line join=round,line cap=round,fill=fillColor] ( 68.04,143.72) circle (  1.43);

\path[draw=drawColor,line width= 0.4pt,line join=round,line cap=round,fill=fillColor] ( 68.79,143.97) circle (  1.43);

\path[draw=drawColor,line width= 0.4pt,line join=round,line cap=round,fill=fillColor] ( 66.91,141.28) circle (  1.43);

\path[draw=drawColor,line width= 0.4pt,line join=round,line cap=round,fill=fillColor] ( 71.22,143.24) circle (  1.43);

\path[draw=drawColor,line width= 0.4pt,line join=round,line cap=round,fill=fillColor] ( 69.80,139.60) circle (  1.43);

\path[draw=drawColor,line width= 0.4pt,line join=round,line cap=round,fill=fillColor] ( 67.02,139.88) circle (  1.43);

\path[draw=drawColor,line width= 0.4pt,line join=round,line cap=round,fill=fillColor] ( 69.46,140.69) circle (  1.43);

\path[draw=drawColor,line width= 0.4pt,line join=round,line cap=round,fill=fillColor] ( 69.74,137.51) circle (  1.43);

\path[draw=drawColor,line width= 0.4pt,line join=round,line cap=round,fill=fillColor] ( 67.85,141.43) circle (  1.43);

\path[draw=drawColor,line width= 0.4pt,line join=round,line cap=round,fill=fillColor] ( 69.44,141.22) circle (  1.43);

\path[draw=drawColor,line width= 0.4pt,line join=round,line cap=round,fill=fillColor] ( 67.98,140.28) circle (  1.43);

\path[draw=drawColor,line width= 0.4pt,line join=round,line cap=round,fill=fillColor] ( 68.10,139.80) circle (  1.43);

\path[draw=drawColor,line width= 0.4pt,line join=round,line cap=round,fill=fillColor] ( 69.35,139.84) circle (  1.43);

\path[draw=drawColor,line width= 0.4pt,line join=round,line cap=round,fill=fillColor] ( 68.39,144.69) circle (  1.43);

\path[draw=drawColor,line width= 0.4pt,line join=round,line cap=round,fill=fillColor] ( 68.35,141.71) circle (  1.43);

\path[draw=drawColor,line width= 0.4pt,line join=round,line cap=round,fill=fillColor] ( 69.32,143.12) circle (  1.43);

\path[draw=drawColor,line width= 0.4pt,line join=round,line cap=round,fill=fillColor] ( 68.32,142.01) circle (  1.43);

\path[draw=drawColor,line width= 0.4pt,line join=round,line cap=round,fill=fillColor] ( 66.68,141.40) circle (  1.43);

\path[draw=drawColor,line width= 0.4pt,line join=round,line cap=round,fill=fillColor] ( 69.96,141.27) circle (  1.43);

\path[draw=drawColor,line width= 0.4pt,line join=round,line cap=round,fill=fillColor] ( 70.41,139.15) circle (  1.43);

\path[draw=drawColor,line width= 0.4pt,line join=round,line cap=round,fill=fillColor] ( 70.19,140.67) circle (  1.43);

\path[draw=drawColor,line width= 0.4pt,line join=round,line cap=round,fill=fillColor] ( 68.59,139.07) circle (  1.43);

\path[draw=drawColor,line width= 0.4pt,line join=round,line cap=round,fill=fillColor] ( 70.37,139.06) circle (  1.43);

\path[draw=drawColor,line width= 0.4pt,line join=round,line cap=round,fill=fillColor] ( 67.30,139.60) circle (  1.43);

\path[draw=drawColor,line width= 0.4pt,line join=round,line cap=round,fill=fillColor] ( 70.23,143.90) circle (  1.43);

\path[draw=drawColor,line width= 0.4pt,line join=round,line cap=round,fill=fillColor] ( 68.51,141.26) circle (  1.43);

\path[draw=drawColor,line width= 0.4pt,line join=round,line cap=round,fill=fillColor] ( 69.38,142.74) circle (  1.43);

\path[draw=drawColor,line width= 0.4pt,line join=round,line cap=round,fill=fillColor] ( 69.50,138.36) circle (  1.43);

\path[draw=drawColor,line width= 0.4pt,line join=round,line cap=round,fill=fillColor] ( 68.57,139.25) circle (  1.43);

\path[draw=drawColor,line width= 0.4pt,line join=round,line cap=round,fill=fillColor] ( 69.56,141.29) circle (  1.43);
\definecolor{drawColor}{RGB}{207,192,152}
\definecolor{fillColor}{RGB}{207,192,152}

\path[draw=drawColor,line width= 0.4pt,line join=round,line cap=round,fill=fillColor] ( 78.24,143.16) circle (  1.43);

\path[draw=drawColor,line width= 0.4pt,line join=round,line cap=round,fill=fillColor] ( 79.90,158.70) circle (  1.43);

\path[draw=drawColor,line width= 0.4pt,line join=round,line cap=round,fill=fillColor] ( 79.16,150.25) circle (  1.43);

\path[draw=drawColor,line width= 0.4pt,line join=round,line cap=round,fill=fillColor] ( 80.14,177.56) circle (  1.43);

\path[draw=drawColor,line width= 0.4pt,line join=round,line cap=round,fill=fillColor] ( 81.38,172.66) circle (  1.43);

\path[draw=drawColor,line width= 0.4pt,line join=round,line cap=round,fill=fillColor] ( 80.42,141.62) circle (  1.43);

\path[draw=drawColor,line width= 0.4pt,line join=round,line cap=round,fill=fillColor] ( 80.66,149.98) circle (  1.43);

\path[draw=drawColor,line width= 0.4pt,line join=round,line cap=round,fill=fillColor] ( 78.73,143.34) circle (  1.43);

\path[draw=drawColor,line width= 0.4pt,line join=round,line cap=round,fill=fillColor] ( 82.15,173.88) circle (  1.43);

\path[draw=drawColor,line width= 0.4pt,line join=round,line cap=round,fill=fillColor] ( 77.85,148.35) circle (  1.43);

\path[draw=drawColor,line width= 0.4pt,line join=round,line cap=round,fill=fillColor] ( 81.28,157.96) circle (  1.43);

\path[draw=drawColor,line width= 0.4pt,line join=round,line cap=round,fill=fillColor] ( 78.59,163.38) circle (  1.43);

\path[draw=drawColor,line width= 0.4pt,line join=round,line cap=round,fill=fillColor] ( 79.96,144.69) circle (  1.43);

\path[draw=drawColor,line width= 0.4pt,line join=round,line cap=round,fill=fillColor] ( 80.72,153.97) circle (  1.43);

\path[draw=drawColor,line width= 0.4pt,line join=round,line cap=round,fill=fillColor] ( 83.93,167.65) circle (  1.43);

\path[draw=drawColor,line width= 0.4pt,line join=round,line cap=round,fill=fillColor] ( 78.83,146.63) circle (  1.43);

\path[draw=drawColor,line width= 0.4pt,line join=round,line cap=round,fill=fillColor] ( 79.65,164.44) circle (  1.43);

\path[draw=drawColor,line width= 0.4pt,line join=round,line cap=round,fill=fillColor] ( 82.54,172.49) circle (  1.43);

\path[draw=drawColor,line width= 0.4pt,line join=round,line cap=round,fill=fillColor] ( 81.53,164.77) circle (  1.43);

\path[draw=drawColor,line width= 0.4pt,line join=round,line cap=round,fill=fillColor] ( 82.41,164.79) circle (  1.43);

\path[draw=drawColor,line width= 0.4pt,line join=round,line cap=round,fill=fillColor] ( 79.84,162.17) circle (  1.43);

\path[draw=drawColor,line width= 0.4pt,line join=round,line cap=round,fill=fillColor] ( 82.86,162.71) circle (  1.43);

\path[draw=drawColor,line width= 0.4pt,line join=round,line cap=round,fill=fillColor] ( 81.23,159.18) circle (  1.43);

\path[draw=drawColor,line width= 0.4pt,line join=round,line cap=round,fill=fillColor] ( 79.35,150.73) circle (  1.43);

\path[draw=drawColor,line width= 0.4pt,line join=round,line cap=round,fill=fillColor] ( 80.76,148.46) circle (  1.43);

\path[draw=drawColor,line width= 0.4pt,line join=round,line cap=round,fill=fillColor] ( 80.47,149.78) circle (  1.43);

\path[draw=drawColor,line width= 0.4pt,line join=round,line cap=round,fill=fillColor] ( 82.47,165.21) circle (  1.43);

\path[draw=drawColor,line width= 0.4pt,line join=round,line cap=round,fill=fillColor] ( 79.78,157.12) circle (  1.43);

\path[draw=drawColor,line width= 0.4pt,line join=round,line cap=round,fill=fillColor] ( 82.19,169.12) circle (  1.43);

\path[draw=drawColor,line width= 0.4pt,line join=round,line cap=round,fill=fillColor] ( 80.36,169.05) circle (  1.43);

\path[draw=drawColor,line width= 0.4pt,line join=round,line cap=round,fill=fillColor] ( 79.02,163.90) circle (  1.43);

\path[draw=drawColor,line width= 0.4pt,line join=round,line cap=round,fill=fillColor] ( 79.89,144.18) circle (  1.43);

\path[draw=drawColor,line width= 0.4pt,line join=round,line cap=round,fill=fillColor] ( 80.39,168.23) circle (  1.43);

\path[draw=drawColor,line width= 0.4pt,line join=round,line cap=round,fill=fillColor] ( 79.93,154.73) circle (  1.43);

\path[draw=drawColor,line width= 0.4pt,line join=round,line cap=round,fill=fillColor] ( 82.64,154.91) circle (  1.43);

\path[draw=drawColor,line width= 0.4pt,line join=round,line cap=round,fill=fillColor] ( 79.48,155.07) circle (  1.43);

\path[draw=drawColor,line width= 0.4pt,line join=round,line cap=round,fill=fillColor] ( 82.26,176.47) circle (  1.43);

\path[draw=drawColor,line width= 0.4pt,line join=round,line cap=round,fill=fillColor] ( 80.24,158.34) circle (  1.43);

\path[draw=drawColor,line width= 0.4pt,line join=round,line cap=round,fill=fillColor] ( 79.93,175.77) circle (  1.43);

\path[draw=drawColor,line width= 0.4pt,line join=round,line cap=round,fill=fillColor] ( 81.98,165.04) circle (  1.43);

\path[draw=drawColor,line width= 0.4pt,line join=round,line cap=round,fill=fillColor] ( 79.72,157.34) circle (  1.43);

\path[draw=drawColor,line width= 0.4pt,line join=round,line cap=round,fill=fillColor] ( 79.23,146.77) circle (  1.43);

\path[draw=drawColor,line width= 0.4pt,line join=round,line cap=round,fill=fillColor] ( 82.26,175.57) circle (  1.43);

\path[draw=drawColor,line width= 0.4pt,line join=round,line cap=round,fill=fillColor] ( 77.77,151.43) circle (  1.43);

\path[draw=drawColor,line width= 0.4pt,line join=round,line cap=round,fill=fillColor] ( 78.57,147.21) circle (  1.43);

\path[draw=drawColor,line width= 0.4pt,line join=round,line cap=round,fill=fillColor] ( 81.96,163.87) circle (  1.43);

\path[draw=drawColor,line width= 0.4pt,line join=round,line cap=round,fill=fillColor] ( 79.23,153.34) circle (  1.43);

\path[draw=drawColor,line width= 0.4pt,line join=round,line cap=round,fill=fillColor] ( 81.63,164.07) circle (  1.43);

\path[draw=drawColor,line width= 0.4pt,line join=round,line cap=round,fill=fillColor] ( 80.74,162.77) circle (  1.43);

\path[draw=drawColor,line width= 0.4pt,line join=round,line cap=round,fill=fillColor] ( 79.64,155.79) circle (  1.43);

\path[draw=drawColor,line width= 0.4pt,line join=round,line cap=round,fill=fillColor] ( 79.22,154.01) circle (  1.43);

\path[draw=drawColor,line width= 0.4pt,line join=round,line cap=round,fill=fillColor] ( 81.09,156.69) circle (  1.43);

\path[draw=drawColor,line width= 0.4pt,line join=round,line cap=round,fill=fillColor] ( 77.89,159.50) circle (  1.43);

\path[draw=drawColor,line width= 0.4pt,line join=round,line cap=round,fill=fillColor] ( 81.65,180.45) circle (  1.43);

\path[draw=drawColor,line width= 0.4pt,line join=round,line cap=round,fill=fillColor] ( 80.94,158.74) circle (  1.43);

\path[draw=drawColor,line width= 0.4pt,line join=round,line cap=round,fill=fillColor] ( 80.12,165.34) circle (  1.43);

\path[draw=drawColor,line width= 0.4pt,line join=round,line cap=round,fill=fillColor] ( 78.74,147.87) circle (  1.43);

\path[draw=drawColor,line width= 0.4pt,line join=round,line cap=round,fill=fillColor] ( 79.60,146.34) circle (  1.43);

\path[draw=drawColor,line width= 0.4pt,line join=round,line cap=round,fill=fillColor] ( 79.91,161.30) circle (  1.43);

\path[draw=drawColor,line width= 0.4pt,line join=round,line cap=round,fill=fillColor] ( 79.42,152.67) circle (  1.43);

\path[draw=drawColor,line width= 0.4pt,line join=round,line cap=round,fill=fillColor] ( 79.43,152.42) circle (  1.43);

\path[draw=drawColor,line width= 0.4pt,line join=round,line cap=round,fill=fillColor] ( 83.21,162.05) circle (  1.43);

\path[draw=drawColor,line width= 0.4pt,line join=round,line cap=round,fill=fillColor] ( 79.43,157.80) circle (  1.43);

\path[draw=drawColor,line width= 0.4pt,line join=round,line cap=round,fill=fillColor] ( 81.26,167.16) circle (  1.43);

\path[draw=drawColor,line width= 0.4pt,line join=round,line cap=round,fill=fillColor] ( 80.72,149.16) circle (  1.43);

\path[draw=drawColor,line width= 0.4pt,line join=round,line cap=round,fill=fillColor] ( 80.83,173.57) circle (  1.43);
\definecolor{drawColor}{RGB}{0,0,0}

\path[draw=drawColor,line width= 0.6pt,line join=round] ( 19.50, 29.68) -- (117.86,194.47);
\definecolor{drawColor}{gray}{0.70}

\path[draw=drawColor,line width= 0.5pt,line join=round,line cap=round] ( 19.50, 29.68) rectangle (117.86,194.47);
\end{scope}
\begin{scope}
\path[clip] (  0.00,  0.00) rectangle (122.86,199.47);
\definecolor{drawColor}{gray}{0.30}

\node[text=drawColor,anchor=base east,inner sep=0pt, outer sep=0pt, scale=  1.00] at ( 15.00, 26.23) {0};

\node[text=drawColor,anchor=base east,inner sep=0pt, outer sep=0pt, scale=  1.00] at ( 15.00, 56.19) {10};

\node[text=drawColor,anchor=base east,inner sep=0pt, outer sep=0pt, scale=  1.00] at ( 15.00, 86.16) {20};

\node[text=drawColor,anchor=base east,inner sep=0pt, outer sep=0pt, scale=  1.00] at ( 15.00,116.12) {30};

\node[text=drawColor,anchor=base east,inner sep=0pt, outer sep=0pt, scale=  1.00] at ( 15.00,146.08) {40};

\node[text=drawColor,anchor=base east,inner sep=0pt, outer sep=0pt, scale=  1.00] at ( 15.00,176.04) {50};
\end{scope}
\begin{scope}
\path[clip] (  0.00,  0.00) rectangle (122.86,199.47);
\definecolor{drawColor}{gray}{0.70}

\path[draw=drawColor,line width= 0.3pt,line join=round] ( 17.00, 29.68) --
	( 19.50, 29.68);

\path[draw=drawColor,line width= 0.3pt,line join=round] ( 17.00, 59.64) --
	( 19.50, 59.64);

\path[draw=drawColor,line width= 0.3pt,line join=round] ( 17.00, 89.60) --
	( 19.50, 89.60);

\path[draw=drawColor,line width= 0.3pt,line join=round] ( 17.00,119.56) --
	( 19.50,119.56);

\path[draw=drawColor,line width= 0.3pt,line join=round] ( 17.00,149.52) --
	( 19.50,149.52);

\path[draw=drawColor,line width= 0.3pt,line join=round] ( 17.00,179.48) --
	( 19.50,179.48);
\end{scope}
\begin{scope}
\path[clip] (  0.00,  0.00) rectangle (122.86,199.47);
\definecolor{drawColor}{gray}{0.70}

\path[draw=drawColor,line width= 0.3pt,line join=round] ( 19.50, 27.18) --
	( 19.50, 29.68);

\path[draw=drawColor,line width= 0.3pt,line join=round] ( 37.38, 27.18) --
	( 37.38, 29.68);

\path[draw=drawColor,line width= 0.3pt,line join=round] ( 55.27, 27.18) --
	( 55.27, 29.68);

\path[draw=drawColor,line width= 0.3pt,line join=round] ( 73.15, 27.18) --
	( 73.15, 29.68);

\path[draw=drawColor,line width= 0.3pt,line join=round] ( 91.03, 27.18) --
	( 91.03, 29.68);

\path[draw=drawColor,line width= 0.3pt,line join=round] (108.92, 27.18) --
	(108.92, 29.68);
\end{scope}
\begin{scope}
\path[clip] (  0.00,  0.00) rectangle (122.86,199.47);
\definecolor{drawColor}{gray}{0.30}

\node[text=drawColor,anchor=base,inner sep=0pt, outer sep=0pt, scale=  1.00] at ( 19.50, 18.29) {0};

\node[text=drawColor,anchor=base,inner sep=0pt, outer sep=0pt, scale=  1.00] at ( 37.38, 18.29) {10};

\node[text=drawColor,anchor=base,inner sep=0pt, outer sep=0pt, scale=  1.00] at ( 55.27, 18.29) {20};

\node[text=drawColor,anchor=base,inner sep=0pt, outer sep=0pt, scale=  1.00] at ( 73.15, 18.29) {30};

\node[text=drawColor,anchor=base,inner sep=0pt, outer sep=0pt, scale=  1.00] at ( 91.03, 18.29) {40};

\node[text=drawColor,anchor=base,inner sep=0pt, outer sep=0pt, scale=  1.00] at (108.92, 18.29) {50};
\end{scope}
\begin{scope}
\path[clip] (  0.00,  0.00) rectangle (122.86,199.47);
\definecolor{drawColor}{RGB}{0,0,0}

\node[text=drawColor,anchor=base,inner sep=0pt, outer sep=0pt, scale=  1.00] at ( 68.68,  6.94) {\bfseries Gap Gurobi};
\end{scope}
\end{tikzpicture}

%% file: 06-plots/NCC2MILP-2Seconds-ImprovementPlot.tex
\begin{tikzpicture}[x=1pt,y=1pt]
\definecolor{fillColor}{RGB}{255,255,255}
\path[use as bounding box,fill=fillColor,fill opacity=0.00] (0,0) rectangle (129.36,199.47);
\begin{scope}
\path[clip] (  0.00,  0.00) rectangle (129.36,199.47);
\definecolor{drawColor}{RGB}{255,255,255}
\definecolor{fillColor}{RGB}{255,255,255}

\path[draw=drawColor,line width= 0.5pt,line join=round,line cap=round,fill=fillColor] (  0.00,  0.00) rectangle (129.36,199.47);
\end{scope}
\begin{scope}
\path[clip] ( 33.62, 29.68) rectangle (124.36,194.47);
\definecolor{fillColor}{RGB}{255,255,255}

\path[fill=fillColor] ( 33.62, 29.68) rectangle (124.36,194.47);
\definecolor{drawColor}{gray}{0.87}

\path[draw=drawColor,line width= 0.1pt,line join=round] ( 33.62, 43.07) --
	(124.36, 43.07);

\path[draw=drawColor,line width= 0.1pt,line join=round] ( 33.62, 69.87) --
	(124.36, 69.87);

\path[draw=drawColor,line width= 0.1pt,line join=round] ( 33.62, 96.66) --
	(124.36, 96.66);

\path[draw=drawColor,line width= 0.1pt,line join=round] ( 33.62,123.46) --
	(124.36,123.46);

\path[draw=drawColor,line width= 0.1pt,line join=round] ( 33.62,150.25) --
	(124.36,150.25);

\path[draw=drawColor,line width= 0.1pt,line join=round] ( 33.62,177.05) --
	(124.36,177.05);

\path[draw=drawColor,line width= 0.1pt,line join=round] ( 42.69, 29.68) --
	( 42.69,194.47);

\path[draw=drawColor,line width= 0.1pt,line join=round] ( 60.84, 29.68) --
	( 60.84,194.47);

\path[draw=drawColor,line width= 0.1pt,line join=round] ( 78.99, 29.68) --
	( 78.99,194.47);

\path[draw=drawColor,line width= 0.1pt,line join=round] ( 97.14, 29.68) --
	( 97.14,194.47);

\path[draw=drawColor,line width= 0.1pt,line join=round] (115.29, 29.68) --
	(115.29,194.47);

\path[draw=drawColor,line width= 0.3pt,line join=round] ( 33.62, 29.68) --
	(124.36, 29.68);

\path[draw=drawColor,line width= 0.3pt,line join=round] ( 33.62, 56.47) --
	(124.36, 56.47);

\path[draw=drawColor,line width= 0.3pt,line join=round] ( 33.62, 83.27) --
	(124.36, 83.27);

\path[draw=drawColor,line width= 0.3pt,line join=round] ( 33.62,110.06) --
	(124.36,110.06);

\path[draw=drawColor,line width= 0.3pt,line join=round] ( 33.62,136.86) --
	(124.36,136.86);

\path[draw=drawColor,line width= 0.3pt,line join=round] ( 33.62,163.65) --
	(124.36,163.65);

\path[draw=drawColor,line width= 0.3pt,line join=round] ( 33.62,190.45) --
	(124.36,190.45);

\path[draw=drawColor,line width= 0.3pt,line join=round] ( 33.62, 29.68) --
	( 33.62,194.47);

\path[draw=drawColor,line width= 0.3pt,line join=round] ( 51.77, 29.68) --
	( 51.77,194.47);

\path[draw=drawColor,line width= 0.3pt,line join=round] ( 69.92, 29.68) --
	( 69.92,194.47);

\path[draw=drawColor,line width= 0.3pt,line join=round] ( 88.07, 29.68) --
	( 88.07,194.47);

\path[draw=drawColor,line width= 0.3pt,line join=round] (106.21, 29.68) --
	(106.21,194.47);

\path[draw=drawColor,line width= 0.3pt,line join=round] (124.36, 29.68) --
	(124.36,194.47);
\definecolor{drawColor}{RGB}{128,128,128}

\path[draw=drawColor,line width= 1.0pt,line join=round] ( 33.62,123.32) --
	( 34.08,125.65) --
	( 34.53,131.19) --
	( 34.99,132.98) --
	( 35.44,137.17) --
	( 35.90,142.65) --
	( 36.36,143.17) --
	( 36.81,145.81) --
	( 37.27,146.52) --
	( 37.72,147.16) --
	( 38.18,147.23) --
	( 38.64,147.44) --
	( 39.09,149.67) --
	( 39.55,150.59) --
	( 40.00,150.92) --
	( 40.46,151.15) --
	( 40.92,151.19) --
	( 41.37,151.41) --
	( 41.83,151.56) --
	( 42.28,151.63) --
	( 42.74,151.99) --
	( 43.20,152.28) --
	( 43.65,152.54) --
	( 44.11,153.62) --
	( 44.56,153.66) --
	( 45.02,153.83) --
	( 45.48,154.01) --
	( 45.93,154.16) --
	( 46.39,154.20) --
	( 46.84,154.82) --
	( 47.30,155.00) --
	( 47.76,155.24) --
	( 48.21,155.33) --
	( 48.67,155.41) --
	( 49.12,155.73) --
	( 49.58,155.79) --
	( 50.04,156.08) --
	( 50.49,156.12) --
	( 50.95,156.30) --
	( 51.40,156.35) --
	( 51.86,156.39) --
	( 52.32,156.44) --
	( 52.77,156.61) --
	( 53.23,156.69) --
	( 53.68,156.85) --
	( 54.14,157.06) --
	( 54.60,157.18) --
	( 55.05,157.19) --
	( 55.51,157.41) --
	( 55.96,157.47) --
	( 56.42,157.49) --
	( 56.88,157.49) --
	( 57.33,157.92) --
	( 57.79,157.99) --
	( 58.24,158.06) --
	( 58.70,158.09) --
	( 59.16,158.09) --
	( 59.61,158.22) --
	( 60.07,158.25) --
	( 60.52,158.36) --
	( 60.98,158.36) --
	( 61.44,158.40) --
	( 61.89,158.44) --
	( 62.35,158.50) --
	( 62.80,158.76) --
	( 63.26,158.77) --
	( 63.72,158.87) --
	( 64.17,159.07) --
	( 64.63,159.12) --
	( 65.08,159.20) --
	( 65.54,159.23) --
	( 66.00,159.27) --
	( 66.45,159.27) --
	( 66.91,159.32) --
	( 67.36,159.34) --
	( 67.82,159.44) --
	( 68.28,159.50) --
	( 68.73,159.57) --
	( 69.19,159.81) --
	( 69.64,159.89) --
	( 70.10,159.90) --
	( 70.56,159.95) --
	( 71.01,160.02) --
	( 71.47,160.09) --
	( 71.92,160.12) --
	( 72.38,160.13) --
	( 72.84,160.17) --
	( 73.29,160.34) --
	( 73.75,160.34) --
	( 74.20,160.34) --
	( 74.66,160.35) --
	( 75.12,160.40) --
	( 75.57,160.64) --
	( 76.03,160.73) --
	( 76.48,160.78) --
	( 76.94,160.79) --
	( 77.40,160.83) --
	( 77.85,160.83) --
	( 78.31,160.83) --
	( 78.76,160.84) --
	( 79.22,160.95) --
	( 79.68,161.04) --
	( 80.13,161.10) --
	( 80.59,161.15) --
	( 81.04,161.16) --
	( 81.50,161.18) --
	( 81.96,161.20) --
	( 82.41,161.20) --
	( 82.87,161.23) --
	( 83.32,161.34) --
	( 83.78,161.37) --
	( 84.24,161.42) --
	( 84.69,161.42) --
	( 85.15,161.47) --
	( 85.60,161.48) --
	( 86.06,161.51) --
	( 86.52,161.57) --
	( 86.97,161.58) --
	( 87.43,161.70) --
	( 87.88,161.72) --
	( 88.34,161.91) --
	( 88.80,162.00) --
	( 89.25,162.00) --
	( 89.71,162.19) --
	( 90.16,162.19) --
	( 90.62,162.24) --
	( 91.08,162.27) --
	( 91.53,162.29) --
	( 91.99,162.29) --
	( 92.44,162.30) --
	( 92.90,162.33) --
	( 93.36,162.34) --
	( 93.81,162.35) --
	( 94.27,162.48) --
	( 94.72,162.49) --
	( 95.18,162.49) --
	( 95.64,162.50) --
	( 96.09,162.53) --
	( 96.55,162.57) --
	( 97.00,162.57) --
	( 97.46,162.57) --
	( 97.92,162.57) --
	( 98.37,162.67) --
	( 98.83,162.76) --
	( 99.28,162.79) --
	( 99.74,162.80) --
	(100.20,162.81) --
	(100.65,162.81) --
	(101.11,162.82) --
	(101.56,162.95) --
	(102.02,162.96) --
	(102.48,162.98) --
	(102.93,163.06) --
	(103.39,163.10) --
	(103.84,163.10) --
	(104.30,163.12) --
	(104.76,163.12) --
	(105.21,163.14) --
	(105.67,163.15) --
	(106.12,163.16) --
	(106.58,163.24) --
	(107.04,163.29) --
	(107.49,163.29) --
	(107.95,163.30) --
	(108.40,163.34) --
	(108.86,163.34) --
	(109.32,163.36) --
	(109.77,163.36) --
	(110.23,163.36) --
	(110.68,163.47) --
	(111.14,163.47) --
	(111.60,163.56) --
	(112.05,163.65) --
	(112.51,163.65) --
	(112.96,163.65) --
	(113.42,163.65) --
	(113.88,163.65) --
	(114.33,163.65) --
	(114.79,163.65) --
	(115.24,163.65) --
	(115.70,163.65) --
	(116.16,163.65) --
	(116.61,163.65) --
	(117.07,163.65) --
	(117.52,163.65) --
	(117.98,163.65) --
	(118.44,163.65) --
	(118.89,163.82) --
	(119.35,163.90) --
	(119.80,164.01) --
	(120.26,164.22) --
	(120.72,164.26) --
	(121.17,164.74) --
	(121.63,164.91) --
	(122.08,165.05) --
	(122.54,165.09) --
	(123.00,165.09) --
	(123.45,165.28) --
	(123.91,165.81) --
	(124.36,166.20);
\definecolor{drawColor}{RGB}{152,167,197}

\path[draw=drawColor,line width= 1.0pt,line join=round] ( 33.62,129.18) --
	( 34.08,134.14) --
	( 34.53,134.83) --
	( 34.99,136.44) --
	( 35.44,136.57) --
	( 35.90,142.26) --
	( 36.36,142.41) --
	( 36.81,142.44) --
	( 37.27,144.50) --
	( 37.72,145.42) --
	( 38.18,146.00) --
	( 38.64,149.00) --
	( 39.09,149.04) --
	( 39.55,150.17) --
	( 40.00,150.38) --
	( 40.46,150.38) --
	( 40.92,150.43) --
	( 41.37,150.96) --
	( 41.83,151.49) --
	( 42.28,151.65) --
	( 42.74,152.14) --
	( 43.20,152.48) --
	( 43.65,152.99) --
	( 44.11,153.22) --
	( 44.56,153.27) --
	( 45.02,154.01) --
	( 45.48,154.46) --
	( 45.93,154.53) --
	( 46.39,154.79) --
	( 46.84,154.86) --
	( 47.30,155.16) --
	( 47.76,155.47) --
	( 48.21,155.62) --
	( 48.67,155.70) --
	( 49.12,155.71) --
	( 49.58,155.83) --
	( 50.04,156.13) --
	( 50.49,156.19) --
	( 50.95,156.27) --
	( 51.40,156.29) --
	( 51.86,156.60) --
	( 52.32,156.62) --
	( 52.77,156.66) --
	( 53.23,157.22) --
	( 53.68,157.26) --
	( 54.14,157.36) --
	( 54.60,157.54) --
	( 55.05,157.56) --
	( 55.51,157.68) --
	( 55.96,157.70) --
	( 56.42,157.71) --
	( 56.88,157.83) --
	( 57.33,157.83) --
	( 57.79,157.94) --
	( 58.24,157.96) --
	( 58.70,158.02) --
	( 59.16,158.12) --
	( 59.61,158.17) --
	( 60.07,158.20) --
	( 60.52,158.23) --
	( 60.98,158.24) --
	( 61.44,158.25) --
	( 61.89,158.25) --
	( 62.35,158.28) --
	( 62.80,158.50) --
	( 63.26,158.52) --
	( 63.72,158.59) --
	( 64.17,158.61) --
	( 64.63,158.66) --
	( 65.08,158.73) --
	( 65.54,158.73) --
	( 66.00,158.80) --
	( 66.45,158.86) --
	( 66.91,158.86) --
	( 67.36,158.87) --
	( 67.82,158.89) --
	( 68.28,158.90) --
	( 68.73,158.94) --
	( 69.19,158.96) --
	( 69.64,159.00) --
	( 70.10,159.10) --
	( 70.56,159.11) --
	( 71.01,159.17) --
	( 71.47,159.25) --
	( 71.92,159.26) --
	( 72.38,159.35) --
	( 72.84,159.38) --
	( 73.29,159.42) --
	( 73.75,159.43) --
	( 74.20,159.51) --
	( 74.66,159.54) --
	( 75.12,159.59) --
	( 75.57,159.61) --
	( 76.03,159.73) --
	( 76.48,159.78) --
	( 76.94,159.89) --
	( 77.40,159.91) --
	( 77.85,159.91) --
	( 78.31,159.93) --
	( 78.76,159.94) --
	( 79.22,159.96) --
	( 79.68,159.98) --
	( 80.13,160.05) --
	( 80.59,160.06) --
	( 81.04,160.10) --
	( 81.50,160.13) --
	( 81.96,160.15) --
	( 82.41,160.17) --
	( 82.87,160.24) --
	( 83.32,160.35) --
	( 83.78,160.39) --
	( 84.24,160.48) --
	( 84.69,160.53) --
	( 85.15,160.55) --
	( 85.60,160.56) --
	( 86.06,160.58) --
	( 86.52,160.62) --
	( 86.97,160.64) --
	( 87.43,160.65) --
	( 87.88,160.68) --
	( 88.34,160.68) --
	( 88.80,160.68) --
	( 89.25,160.75) --
	( 89.71,160.75) --
	( 90.16,160.77) --
	( 90.62,160.78) --
	( 91.08,160.79) --
	( 91.53,160.81) --
	( 91.99,161.00) --
	( 92.44,161.01) --
	( 92.90,161.02) --
	( 93.36,161.03) --
	( 93.81,161.04) --
	( 94.27,161.09) --
	( 94.72,161.14) --
	( 95.18,161.14) --
	( 95.64,161.17) --
	( 96.09,161.22) --
	( 96.55,161.23) --
	( 97.00,161.25) --
	( 97.46,161.26) --
	( 97.92,161.26) --
	( 98.37,161.27) --
	( 98.83,161.40) --
	( 99.28,161.42) --
	( 99.74,161.46) --
	(100.20,161.47) --
	(100.65,161.49) --
	(101.11,161.50) --
	(101.56,161.61) --
	(102.02,161.68) --
	(102.48,161.68) --
	(102.93,161.71) --
	(103.39,161.74) --
	(103.84,161.75) --
	(104.30,161.75) --
	(104.76,161.76) --
	(105.21,161.77) --
	(105.67,161.77) --
	(106.12,161.83) --
	(106.58,161.84) --
	(107.04,161.87) --
	(107.49,161.99) --
	(107.95,162.04) --
	(108.40,162.07) --
	(108.86,162.09) --
	(109.32,162.11) --
	(109.77,162.19) --
	(110.23,162.24) --
	(110.68,162.25) --
	(111.14,162.26) --
	(111.60,162.30) --
	(112.05,162.33) --
	(112.51,162.35) --
	(112.96,162.36) --
	(113.42,162.37) --
	(113.88,162.42) --
	(114.33,162.42) --
	(114.79,162.49) --
	(115.24,162.71) --
	(115.70,162.83) --
	(116.16,162.83) --
	(116.61,162.87) --
	(117.07,162.96) --
	(117.52,163.00) --
	(117.98,163.00) --
	(118.44,163.04) --
	(118.89,163.11) --
	(119.35,163.27) --
	(119.80,163.33) --
	(120.26,163.42) --
	(120.72,163.49) --
	(121.17,163.58) --
	(121.63,163.59) --
	(122.08,163.65) --
	(122.54,163.69) --
	(123.00,163.70) --
	(123.45,163.94) --
	(123.91,164.49) --
	(124.36,165.05);
\definecolor{drawColor}{RGB}{128,202,192}

\path[draw=drawColor,line width= 1.0pt,line join=round] ( 33.62,130.28) --
	( 34.08,131.50) --
	( 34.53,135.15) --
	( 34.99,135.17) --
	( 35.44,142.04) --
	( 35.90,142.58) --
	( 36.36,142.66) --
	( 36.81,142.67) --
	( 37.27,142.72) --
	( 37.72,142.78) --
	( 38.18,142.82) --
	( 38.64,143.23) --
	( 39.09,143.74) --
	( 39.55,143.79) --
	( 40.00,143.83) --
	( 40.46,143.84) --
	( 40.92,143.86) --
	( 41.37,143.87) --
	( 41.83,143.90) --
	( 42.28,144.10) --
	( 42.74,144.13) --
	( 43.20,144.38) --
	( 43.65,144.40) --
	( 44.11,144.58) --
	( 44.56,144.64) --
	( 45.02,144.74) --
	( 45.48,144.78) --
	( 45.93,144.83) --
	( 46.39,145.07) --
	( 46.84,145.11) --
	( 47.30,145.13) --
	( 47.76,145.14) --
	( 48.21,145.33) --
	( 48.67,145.34) --
	( 49.12,145.38) --
	( 49.58,145.38) --
	( 50.04,145.54) --
	( 50.49,145.55) --
	( 50.95,145.59) --
	( 51.40,145.70) --
	( 51.86,145.72) --
	( 52.32,145.79) --
	( 52.77,145.83) --
	( 53.23,145.90) --
	( 53.68,146.10) --
	( 54.14,146.17) --
	( 54.60,146.17) --
	( 55.05,146.27) --
	( 55.51,146.33) --
	( 55.96,146.36) --
	( 56.42,146.39) --
	( 56.88,146.43) --
	( 57.33,146.49) --
	( 57.79,146.50) --
	( 58.24,146.59) --
	( 58.70,146.61) --
	( 59.16,146.81) --
	( 59.61,147.03) --
	( 60.07,147.21) --
	( 60.52,147.24) --
	( 60.98,147.32) --
	( 61.44,147.35) --
	( 61.89,147.53) --
	( 62.35,147.61) --
	( 62.80,147.64) --
	( 63.26,147.64) --
	( 63.72,147.68) --
	( 64.17,147.74) --
	( 64.63,147.78) --
	( 65.08,147.79) --
	( 65.54,147.81) --
	( 66.00,147.82) --
	( 66.45,147.84) --
	( 66.91,147.85) --
	( 67.36,147.99) --
	( 67.82,148.13) --
	( 68.28,148.16) --
	( 68.73,148.18) --
	( 69.19,148.34) --
	( 69.64,148.48) --
	( 70.10,148.69) --
	( 70.56,148.71) --
	( 71.01,148.74) --
	( 71.47,148.74) --
	( 71.92,148.74) --
	( 72.38,148.88) --
	( 72.84,148.93) --
	( 73.29,148.97) --
	( 73.75,149.06) --
	( 74.20,149.11) --
	( 74.66,149.20) --
	( 75.12,149.24) --
	( 75.57,149.32) --
	( 76.03,149.33) --
	( 76.48,149.46) --
	( 76.94,149.59) --
	( 77.40,149.63) --
	( 77.85,149.86) --
	( 78.31,149.97) --
	( 78.76,150.07) --
	( 79.22,150.34) --
	( 79.68,150.53) --
	( 80.13,150.60) --
	( 80.59,150.61) --
	( 81.04,150.65) --
	( 81.50,150.67) --
	( 81.96,150.90) --
	( 82.41,151.00) --
	( 82.87,151.03) --
	( 83.32,151.10) --
	( 83.78,151.16) --
	( 84.24,151.20) --
	( 84.69,151.31) --
	( 85.15,151.34) --
	( 85.60,151.39) --
	( 86.06,151.59) --
	( 86.52,151.61) --
	( 86.97,151.87) --
	( 87.43,152.15) --
	( 87.88,152.32) --
	( 88.34,152.52) --
	( 88.80,152.66) --
	( 89.25,152.70) --
	( 89.71,152.84) --
	( 90.16,153.02) --
	( 90.62,153.11) --
	( 91.08,153.14) --
	( 91.53,153.31) --
	( 91.99,153.31) --
	( 92.44,153.69) --
	( 92.90,153.82) --
	( 93.36,153.84) --
	( 93.81,153.84) --
	( 94.27,153.93) --
	( 94.72,153.93) --
	( 95.18,154.07) --
	( 95.64,154.10) --
	( 96.09,154.11) --
	( 96.55,154.38) --
	( 97.00,154.42) --
	( 97.46,154.45) --
	( 97.92,154.54) --
	( 98.37,154.57) --
	( 98.83,154.68) --
	( 99.28,154.85) --
	( 99.74,154.92) --
	(100.20,155.03) --
	(100.65,155.13) --
	(101.11,155.51) --
	(101.56,155.60) --
	(102.02,155.68) --
	(102.48,155.69) --
	(102.93,155.70) --
	(103.39,155.72) --
	(103.84,155.78) --
	(104.30,155.79) --
	(104.76,155.81) --
	(105.21,155.95) --
	(105.67,155.96) --
	(106.12,156.02) --
	(106.58,156.12) --
	(107.04,156.27) --
	(107.49,156.28) --
	(107.95,156.54) --
	(108.40,156.56) --
	(108.86,156.67) --
	(109.32,157.09) --
	(109.77,157.12) --
	(110.23,157.18) --
	(110.68,157.19) --
	(111.14,157.20) --
	(111.60,157.36) --
	(112.05,157.53) --
	(112.51,157.66) --
	(112.96,157.73) --
	(113.42,157.96) --
	(113.88,157.96) --
	(114.33,158.04) --
	(114.79,158.14) --
	(115.24,158.15) --
	(115.70,158.19) --
	(116.16,158.25) --
	(116.61,158.37) --
	(117.07,158.63) --
	(117.52,158.72) --
	(117.98,158.74) --
	(118.44,158.78) --
	(118.89,158.78) --
	(119.35,158.80) --
	(119.80,158.83) --
	(120.26,158.97) --
	(120.72,159.03) --
	(121.17,159.49) --
	(121.63,159.60) --
	(122.08,159.91) --
	(122.54,160.19) --
	(123.00,160.32) --
	(123.45,160.52) --
	(123.91,160.55) --
	(124.36,161.85);
\definecolor{drawColor}{RGB}{207,128,187}

\path[draw=drawColor,line width= 1.0pt,line join=round] ( 33.62, 29.68) --
	( 34.08, 29.68) --
	( 34.53, 29.68) --
	( 34.99, 29.68) --
	( 35.44, 29.68) --
	( 35.90, 29.68) --
	( 36.36, 43.13) --
	( 36.81, 43.53) --
	( 37.27, 45.34) --
	( 37.72, 45.52) --
	( 38.18, 46.28) --
	( 38.64, 50.85) --
	( 39.09, 52.88) --
	( 39.55, 54.44) --
	( 40.00, 54.59) --
	( 40.46, 54.72) --
	( 40.92, 54.81) --
	( 41.37, 54.86) --
	( 41.83, 55.44) --
	( 42.28, 56.79) --
	( 42.74, 56.79) --
	( 43.20, 57.54) --
	( 43.65, 57.70) --
	( 44.11, 57.79) --
	( 44.56,102.95) --
	( 45.02,106.37) --
	( 45.48,112.57) --
	( 45.93,123.04) --
	( 46.39,124.42) --
	( 46.84,125.11) --
	( 47.30,129.20) --
	( 47.76,131.94) --
	( 48.21,135.47) --
	( 48.67,136.29) --
	( 49.12,136.54) --
	( 49.58,140.66) --
	( 50.04,142.18) --
	( 50.49,142.98) --
	( 50.95,143.48) --
	( 51.40,143.99) --
	( 51.86,144.79) --
	( 52.32,144.93) --
	( 52.77,145.97) --
	( 53.23,146.37) --
	( 53.68,146.57) --
	( 54.14,147.44) --
	( 54.60,147.55) --
	( 55.05,147.78) --
	( 55.51,147.91) --
	( 55.96,147.99) --
	( 56.42,148.50) --
	( 56.88,148.68) --
	( 57.33,149.62) --
	( 57.79,149.66) --
	( 58.24,149.69) --
	( 58.70,149.77) --
	( 59.16,149.86) --
	( 59.61,150.13) --
	( 60.07,150.30) --
	( 60.52,150.54) --
	( 60.98,151.24) --
	( 61.44,151.30) --
	( 61.89,151.44) --
	( 62.35,151.63) --
	( 62.80,151.93) --
	( 63.26,151.96) --
	( 63.72,152.18) --
	( 64.17,152.24) --
	( 64.63,152.27) --
	( 65.08,152.30) --
	( 65.54,152.91) --
	( 66.00,153.09) --
	( 66.45,153.23) --
	( 66.91,153.64) --
	( 67.36,153.83) --
	( 67.82,153.84) --
	( 68.28,153.89) --
	( 68.73,154.32) --
	( 69.19,154.35) --
	( 69.64,154.50) --
	( 70.10,154.77) --
	( 70.56,154.94) --
	( 71.01,155.28) --
	( 71.47,155.48) --
	( 71.92,155.82) --
	( 72.38,156.34) --
	( 72.84,156.62) --
	( 73.29,156.81) --
	( 73.75,157.48) --
	( 74.20,157.55) --
	( 74.66,157.67) --
	( 75.12,157.81) --
	( 75.57,157.96) --
	( 76.03,158.06) --
	( 76.48,158.24) --
	( 76.94,158.31) --
	( 77.40,158.33) --
	( 77.85,158.55) --
	( 78.31,158.55) --
	( 78.76,159.05) --
	( 79.22,159.15) --
	( 79.68,159.22) --
	( 80.13,159.45) --
	( 80.59,159.54) --
	( 81.04,159.67) --
	( 81.50,159.72) --
	( 81.96,159.91) --
	( 82.41,159.98) --
	( 82.87,160.28) --
	( 83.32,160.44) --
	( 83.78,160.54) --
	( 84.24,160.71) --
	( 84.69,160.90) --
	( 85.15,161.10) --
	( 85.60,161.13) --
	( 86.06,161.82) --
	( 86.52,162.28) --
	( 86.97,162.53) --
	( 87.43,162.57) --
	( 87.88,162.70) --
	( 88.34,162.91) --
	( 88.80,163.02) --
	( 89.25,163.04) --
	( 89.71,163.06) --
	( 90.16,163.35) --
	( 90.62,163.45) --
	( 91.08,163.58) --
	( 91.53,163.68) --
	( 91.99,163.80) --
	( 92.44,163.90) --
	( 92.90,164.21) --
	( 93.36,164.26) --
	( 93.81,164.43) --
	( 94.27,164.46) --
	( 94.72,164.48) --
	( 95.18,164.57) --
	( 95.64,164.66) --
	( 96.09,164.66) --
	( 96.55,164.70) --
	( 97.00,164.76) --
	( 97.46,164.80) --
	( 97.92,164.80) --
	( 98.37,164.87) --
	( 98.83,164.94) --
	( 99.28,165.36) --
	( 99.74,165.68) --
	(100.20,166.02) --
	(100.65,166.18) --
	(101.11,166.34) --
	(101.56,166.46) --
	(102.02,166.49) --
	(102.48,166.66) --
	(102.93,167.31) --
	(103.39,167.56) --
	(103.84,167.99) --
	(104.30,168.01) --
	(104.76,168.20) --
	(105.21,168.48) --
	(105.67,168.66) --
	(106.12,168.67) --
	(106.58,168.94) --
	(107.04,169.04) --
	(107.49,169.69) --
	(107.95,169.82) --
	(108.40,170.95) --
	(108.86,171.37) --
	(109.32,171.62) --
	(109.77,171.80) --
	(110.23,171.83) --
	(110.68,172.04) --
	(111.14,173.03) --
	(111.60,173.09) --
	(112.05,173.32) --
	(112.51,173.68) --
	(112.96,173.89) --
	(113.42,174.21) --
	(113.88,175.22) --
	(114.33,175.73) --
	(114.79,175.92) --
	(115.24,176.08) --
	(115.70,176.22) --
	(116.16,176.85) --
	(116.61,177.19) --
	(117.07,177.30) --
	(117.52,177.61) --
	(117.98,177.87) --
	(118.44,178.07) --
	(118.89,179.58) --
	(119.35,179.63) --
	(119.80,179.96) --
	(120.26,181.00) --
	(120.72,181.89) --
	(121.17,182.59) --
	(121.63,188.66) --
	(122.08,190.96) --
	(122.54,191.22) --
	(123.00,191.53) --
	(123.45,191.57) --
	(123.91,192.62) --
	(124.36,194.20);
\definecolor{drawColor}{RGB}{207,192,152}

\path[draw=drawColor,line width= 1.0pt,line join=round] ( 33.62, 45.64) --
	( 34.08, 45.82) --
	( 34.53, 46.63) --
	( 34.99, 46.79) --
	( 35.44, 46.98) --
	( 35.90, 47.03) --
	( 36.36, 47.28) --
	( 36.81, 47.61) --
	( 37.27, 47.64) --
	( 37.72, 47.85) --
	( 38.18, 47.85) --
	( 38.64, 47.90) --
	( 39.09, 48.08) --
	( 39.55, 48.12) --
	( 40.00, 48.14) --
	( 40.46, 48.19) --
	( 40.92, 48.20) --
	( 41.37, 48.36) --
	( 41.83, 48.39) --
	( 42.28, 48.41) --
	( 42.74, 48.41) --
	( 43.20, 48.60) --
	( 43.65, 48.65) --
	( 44.11, 48.72) --
	( 44.56, 48.74) --
	( 45.02, 48.83) --
	( 45.48, 48.87) --
	( 45.93, 48.90) --
	( 46.39, 48.90) --
	( 46.84, 48.96) --
	( 47.30, 48.98) --
	( 47.76, 49.00) --
	( 48.21, 49.01) --
	( 48.67, 49.04) --
	( 49.12, 49.06) --
	( 49.58, 49.11) --
	( 50.04, 49.14) --
	( 50.49, 49.15) --
	( 50.95, 49.18) --
	( 51.40, 49.20) --
	( 51.86, 49.21) --
	( 52.32, 49.30) --
	( 52.77, 49.31) --
	( 53.23, 49.36) --
	( 53.68, 49.46) --
	( 54.14, 49.51) --
	( 54.60, 49.52) --
	( 55.05, 49.53) --
	( 55.51, 49.54) --
	( 55.96, 49.56) --
	( 56.42, 49.61) --
	( 56.88, 49.68) --
	( 57.33, 49.74) --
	( 57.79, 49.74) --
	( 58.24, 49.83) --
	( 58.70, 49.85) --
	( 59.16, 49.86) --
	( 59.61, 49.92) --
	( 60.07, 49.93) --
	( 60.52, 49.99) --
	( 60.98, 50.03) --
	( 61.44, 50.15) --
	( 61.89, 50.17) --
	( 62.35, 50.25) --
	( 62.80, 50.28) --
	( 63.26, 50.28) --
	( 63.72, 50.28) --
	( 64.17, 50.36) --
	( 64.63, 50.43) --
	( 65.08, 50.46) --
	( 65.54, 50.46) --
	( 66.00, 50.47) --
	( 66.45, 50.51) --
	( 66.91, 50.55) --
	( 67.36, 50.58) --
	( 67.82, 50.60) --
	( 68.28, 50.85) --
	( 68.73, 50.85) --
	( 69.19, 50.86) --
	( 69.64, 50.87) --
	( 70.10, 50.93) --
	( 70.56, 50.93) --
	( 71.01, 50.96) --
	( 71.47, 51.05) --
	( 71.92, 51.08) --
	( 72.38, 51.16) --
	( 72.84, 51.18) --
	( 73.29, 51.24) --
	( 73.75, 51.27) --
	( 74.20, 51.41) --
	( 74.66, 51.43) --
	( 75.12, 51.54) --
	( 75.57, 51.75) --
	( 76.03, 51.77) --
	( 76.48, 51.84) --
	( 76.94, 51.84) --
	( 77.40, 51.86) --
	( 77.85, 51.97) --
	( 78.31, 51.97) --
	( 78.76, 52.01) --
	( 79.22, 52.02) --
	( 79.68, 52.08) --
	( 80.13, 52.17) --
	( 80.59, 52.19) --
	( 81.04, 52.22) --
	( 81.50, 52.27) --
	( 81.96, 52.27) --
	( 82.41, 52.33) --
	( 82.87, 52.35) --
	( 83.32, 52.38) --
	( 83.78, 52.39) --
	( 84.24, 52.40) --
	( 84.69, 52.42) --
	( 85.15, 52.47) --
	( 85.60, 52.49) --
	( 86.06, 52.53) --
	( 86.52, 52.63) --
	( 86.97, 52.65) --
	( 87.43, 52.68) --
	( 87.88, 52.79) --
	( 88.34, 52.85) --
	( 88.80, 52.93) --
	( 89.25, 52.96) --
	( 89.71, 53.01) --
	( 90.16, 53.02) --
	( 90.62, 53.10) --
	( 91.08, 53.10) --
	( 91.53, 53.16) --
	( 91.99, 53.23) --
	( 92.44, 53.27) --
	( 92.90, 53.27) --
	( 93.36, 53.28) --
	( 93.81, 53.36) --
	( 94.27, 53.38) --
	( 94.72, 53.40) --
	( 95.18, 53.47) --
	( 95.64, 53.72) --
	( 96.09, 53.76) --
	( 96.55, 53.77) --
	( 97.00, 53.78) --
	( 97.46, 53.80) --
	( 97.92, 53.85) --
	( 98.37, 53.86) --
	( 98.83, 53.94) --
	( 99.28, 53.97) --
	( 99.74, 54.04) --
	(100.20, 54.05) --
	(100.65, 54.07) --
	(101.11, 54.13) --
	(101.56, 54.18) --
	(102.02, 54.28) --
	(102.48, 54.30) --
	(102.93, 54.34) --
	(103.39, 54.36) --
	(103.84, 54.38) --
	(104.30, 54.41) --
	(104.76, 54.42) --
	(105.21, 54.48) --
	(105.67, 54.48) --
	(106.12, 54.50) --
	(106.58, 54.58) --
	(107.04, 54.71) --
	(107.49, 54.86) --
	(107.95, 54.88) --
	(108.40, 54.96) --
	(108.86, 54.96) --
	(109.32, 55.03) --
	(109.77, 55.06) --
	(110.23, 55.16) --
	(110.68, 55.36) --
	(111.14, 55.41) --
	(111.60, 55.55) --
	(112.05, 55.62) --
	(112.51, 55.63) --
	(112.96, 55.64) --
	(113.42, 55.90) --
	(113.88, 55.93) --
	(114.33, 55.97) --
	(114.79, 56.15) --
	(115.24, 56.41) --
	(115.70, 56.46) --
	(116.16, 56.73) --
	(116.61, 56.73) --
	(117.07, 57.00) --
	(117.52, 57.18) --
	(117.98, 57.30) --
	(118.44, 57.35) --
	(118.89, 57.42) --
	(119.35, 57.63) --
	(119.80, 57.80) --
	(120.26, 57.85) --
	(120.72, 57.95) --
	(121.17, 57.97) --
	(121.63, 57.98) --
	(122.08, 58.07) --
	(122.54, 58.17) --
	(123.00, 58.66) --
	(123.45, 60.91) --
	(123.91, 61.55) --
	(124.36, 84.73);
\definecolor{drawColor}{RGB}{139,0,0}

\path[draw=drawColor,line width= 1.0pt,line join=round] ( 33.62,163.65) --
	( 33.62,163.65) --
	( 33.62,163.65) --
	( 33.62,163.65) --
	( 33.62,163.65) --
	( 34.08,163.65) --
	( 34.08,163.65) --
	( 34.08,163.65) --
	( 34.08,163.65) --
	( 34.08,163.65) --
	( 34.53,163.65) --
	( 34.53,163.65) --
	( 34.53,163.65) --
	( 34.53,163.65) --
	( 34.53,163.65) --
	( 34.99,163.65) --
	( 34.99,163.65) --
	( 34.99,163.65) --
	( 34.99,163.65) --
	( 34.99,163.65) --
	( 35.44,163.65) --
	( 35.44,163.65) --
	( 35.44,163.65) --
	( 35.44,163.65) --
	( 35.44,163.65) --
	( 35.90,163.65) --
	( 35.90,163.65) --
	( 35.90,163.65) --
	( 35.90,163.65) --
	( 35.90,163.65) --
	( 36.36,163.65) --
	( 36.36,163.65) --
	( 36.36,163.65) --
	( 36.36,163.65) --
	( 36.36,163.65) --
	( 36.81,163.65) --
	( 36.81,163.65) --
	( 36.81,163.65) --
	( 36.81,163.65) --
	( 36.81,163.65) --
	( 37.27,163.65) --
	( 37.27,163.65) --
	( 37.27,163.65) --
	( 37.27,163.65) --
	( 37.27,163.65) --
	( 37.72,163.65) --
	( 37.72,163.65) --
	( 37.72,163.65) --
	( 37.72,163.65) --
	( 37.72,163.65) --
	( 38.18,163.65) --
	( 38.18,163.65) --
	( 38.18,163.65) --
	( 38.18,163.65) --
	( 38.18,163.65) --
	( 38.64,163.65) --
	( 38.64,163.65) --
	( 38.64,163.65) --
	( 38.64,163.65) --
	( 38.64,163.65) --
	( 39.09,163.65) --
	( 39.09,163.65) --
	( 39.09,163.65) --
	( 39.09,163.65) --
	( 39.09,163.65) --
	( 39.55,163.65) --
	( 39.55,163.65) --
	( 39.55,163.65) --
	( 39.55,163.65) --
	( 39.55,163.65) --
	( 40.00,163.65) --
	( 40.00,163.65) --
	( 40.00,163.65) --
	( 40.00,163.65) --
	( 40.00,163.65) --
	( 40.46,163.65) --
	( 40.46,163.65) --
	( 40.46,163.65) --
	( 40.46,163.65) --
	( 40.46,163.65) --
	( 40.92,163.65) --
	( 40.92,163.65) --
	( 40.92,163.65) --
	( 40.92,163.65) --
	( 40.92,163.65) --
	( 41.37,163.65) --
	( 41.37,163.65) --
	( 41.37,163.65) --
	( 41.37,163.65) --
	( 41.37,163.65) --
	( 41.83,163.65) --
	( 41.83,163.65) --
	( 41.83,163.65) --
	( 41.83,163.65) --
	( 41.83,163.65) --
	( 42.28,163.65) --
	( 42.28,163.65) --
	( 42.28,163.65) --
	( 42.28,163.65) --
	( 42.28,163.65) --
	( 42.74,163.65) --
	( 42.74,163.65) --
	( 42.74,163.65) --
	( 42.74,163.65) --
	( 42.74,163.65) --
	( 43.20,163.65) --
	( 43.20,163.65) --
	( 43.20,163.65) --
	( 43.20,163.65) --
	( 43.20,163.65) --
	( 43.65,163.65) --
	( 43.65,163.65) --
	( 43.65,163.65) --
	( 43.65,163.65) --
	( 43.65,163.65) --
	( 44.11,163.65) --
	( 44.11,163.65) --
	( 44.11,163.65) --
	( 44.11,163.65) --
	( 44.11,163.65) --
	( 44.56,163.65) --
	( 44.56,163.65) --
	( 44.56,163.65) --
	( 44.56,163.65) --
	( 44.56,163.65) --
	( 45.02,163.65) --
	( 45.02,163.65) --
	( 45.02,163.65) --
	( 45.02,163.65) --
	( 45.02,163.65) --
	( 45.48,163.65) --
	( 45.48,163.65) --
	( 45.48,163.65) --
	( 45.48,163.65) --
	( 45.48,163.65) --
	( 45.93,163.65) --
	( 45.93,163.65) --
	( 45.93,163.65) --
	( 45.93,163.65) --
	( 45.93,163.65) --
	( 46.39,163.65) --
	( 46.39,163.65) --
	( 46.39,163.65) --
	( 46.39,163.65) --
	( 46.39,163.65) --
	( 46.84,163.65) --
	( 46.84,163.65) --
	( 46.84,163.65) --
	( 46.84,163.65) --
	( 46.84,163.65) --
	( 47.30,163.65) --
	( 47.30,163.65) --
	( 47.30,163.65) --
	( 47.30,163.65) --
	( 47.30,163.65) --
	( 47.76,163.65) --
	( 47.76,163.65) --
	( 47.76,163.65) --
	( 47.76,163.65) --
	( 47.76,163.65) --
	( 48.21,163.65) --
	( 48.21,163.65) --
	( 48.21,163.65) --
	( 48.21,163.65) --
	( 48.21,163.65) --
	( 48.67,163.65) --
	( 48.67,163.65) --
	( 48.67,163.65) --
	( 48.67,163.65) --
	( 48.67,163.65) --
	( 49.12,163.65) --
	( 49.12,163.65) --
	( 49.12,163.65) --
	( 49.12,163.65) --
	( 49.12,163.65) --
	( 49.58,163.65) --
	( 49.58,163.65) --
	( 49.58,163.65) --
	( 49.58,163.65) --
	( 49.58,163.65) --
	( 50.04,163.65) --
	( 50.04,163.65) --
	( 50.04,163.65) --
	( 50.04,163.65) --
	( 50.04,163.65) --
	( 50.49,163.65) --
	( 50.49,163.65) --
	( 50.49,163.65) --
	( 50.49,163.65) --
	( 50.49,163.65) --
	( 50.95,163.65) --
	( 50.95,163.65) --
	( 50.95,163.65) --
	( 50.95,163.65) --
	( 50.95,163.65) --
	( 51.40,163.65) --
	( 51.40,163.65) --
	( 51.40,163.65) --
	( 51.40,163.65) --
	( 51.40,163.65) --
	( 51.86,163.65) --
	( 51.86,163.65) --
	( 51.86,163.65) --
	( 51.86,163.65) --
	( 51.86,163.65) --
	( 52.32,163.65) --
	( 52.32,163.65) --
	( 52.32,163.65) --
	( 52.32,163.65) --
	( 52.32,163.65) --
	( 52.77,163.65) --
	( 52.77,163.65) --
	( 52.77,163.65) --
	( 52.77,163.65) --
	( 52.77,163.65) --
	( 53.23,163.65) --
	( 53.23,163.65) --
	( 53.23,163.65) --
	( 53.23,163.65) --
	( 53.23,163.65) --
	( 53.68,163.65) --
	( 53.68,163.65) --
	( 53.68,163.65) --
	( 53.68,163.65) --
	( 53.68,163.65) --
	( 54.14,163.65) --
	( 54.14,163.65) --
	( 54.14,163.65) --
	( 54.14,163.65) --
	( 54.14,163.65) --
	( 54.60,163.65) --
	( 54.60,163.65) --
	( 54.60,163.65) --
	( 54.60,163.65) --
	( 54.60,163.65) --
	( 55.05,163.65) --
	( 55.05,163.65) --
	( 55.05,163.65) --
	( 55.05,163.65) --
	( 55.05,163.65) --
	( 55.51,163.65) --
	( 55.51,163.65) --
	( 55.51,163.65) --
	( 55.51,163.65) --
	( 55.51,163.65) --
	( 55.96,163.65) --
	( 55.96,163.65) --
	( 55.96,163.65) --
	( 55.96,163.65) --
	( 55.96,163.65) --
	( 56.42,163.65) --
	( 56.42,163.65) --
	( 56.42,163.65) --
	( 56.42,163.65) --
	( 56.42,163.65) --
	( 56.88,163.65) --
	( 56.88,163.65) --
	( 56.88,163.65) --
	( 56.88,163.65) --
	( 56.88,163.65) --
	( 57.33,163.65) --
	( 57.33,163.65) --
	( 57.33,163.65) --
	( 57.33,163.65) --
	( 57.33,163.65) --
	( 57.79,163.65) --
	( 57.79,163.65) --
	( 57.79,163.65) --
	( 57.79,163.65) --
	( 57.79,163.65) --
	( 58.24,163.65) --
	( 58.24,163.65) --
	( 58.24,163.65) --
	( 58.24,163.65) --
	( 58.24,163.65) --
	( 58.70,163.65) --
	( 58.70,163.65) --
	( 58.70,163.65) --
	( 58.70,163.65) --
	( 58.70,163.65) --
	( 59.16,163.65) --
	( 59.16,163.65) --
	( 59.16,163.65) --
	( 59.16,163.65) --
	( 59.16,163.65) --
	( 59.61,163.65) --
	( 59.61,163.65) --
	( 59.61,163.65) --
	( 59.61,163.65) --
	( 59.61,163.65) --
	( 60.07,163.65) --
	( 60.07,163.65) --
	( 60.07,163.65) --
	( 60.07,163.65) --
	( 60.07,163.65) --
	( 60.52,163.65) --
	( 60.52,163.65) --
	( 60.52,163.65) --
	( 60.52,163.65) --
	( 60.52,163.65) --
	( 60.98,163.65) --
	( 60.98,163.65) --
	( 60.98,163.65) --
	( 60.98,163.65) --
	( 60.98,163.65) --
	( 61.44,163.65) --
	( 61.44,163.65) --
	( 61.44,163.65) --
	( 61.44,163.65) --
	( 61.44,163.65) --
	( 61.89,163.65) --
	( 61.89,163.65) --
	( 61.89,163.65) --
	( 61.89,163.65) --
	( 61.89,163.65) --
	( 62.35,163.65) --
	( 62.35,163.65) --
	( 62.35,163.65) --
	( 62.35,163.65) --
	( 62.35,163.65) --
	( 62.80,163.65) --
	( 62.80,163.65) --
	( 62.80,163.65) --
	( 62.80,163.65) --
	( 62.80,163.65) --
	( 63.26,163.65) --
	( 63.26,163.65) --
	( 63.26,163.65) --
	( 63.26,163.65) --
	( 63.26,163.65) --
	( 63.72,163.65) --
	( 63.72,163.65) --
	( 63.72,163.65) --
	( 63.72,163.65) --
	( 63.72,163.65) --
	( 64.17,163.65) --
	( 64.17,163.65) --
	( 64.17,163.65) --
	( 64.17,163.65) --
	( 64.17,163.65) --
	( 64.63,163.65) --
	( 64.63,163.65) --
	( 64.63,163.65) --
	( 64.63,163.65) --
	( 64.63,163.65) --
	( 65.08,163.65) --
	( 65.08,163.65) --
	( 65.08,163.65) --
	( 65.08,163.65) --
	( 65.08,163.65) --
	( 65.54,163.65) --
	( 65.54,163.65) --
	( 65.54,163.65) --
	( 65.54,163.65) --
	( 65.54,163.65) --
	( 66.00,163.65) --
	( 66.00,163.65) --
	( 66.00,163.65) --
	( 66.00,163.65) --
	( 66.00,163.65) --
	( 66.45,163.65) --
	( 66.45,163.65) --
	( 66.45,163.65) --
	( 66.45,163.65) --
	( 66.45,163.65) --
	( 66.91,163.65) --
	( 66.91,163.65) --
	( 66.91,163.65) --
	( 66.91,163.65) --
	( 66.91,163.65) --
	( 67.36,163.65) --
	( 67.36,163.65) --
	( 67.36,163.65) --
	( 67.36,163.65) --
	( 67.36,163.65) --
	( 67.82,163.65) --
	( 67.82,163.65) --
	( 67.82,163.65) --
	( 67.82,163.65) --
	( 67.82,163.65) --
	( 68.28,163.65) --
	( 68.28,163.65) --
	( 68.28,163.65) --
	( 68.28,163.65) --
	( 68.28,163.65) --
	( 68.73,163.65) --
	( 68.73,163.65) --
	( 68.73,163.65) --
	( 68.73,163.65) --
	( 68.73,163.65) --
	( 69.19,163.65) --
	( 69.19,163.65) --
	( 69.19,163.65) --
	( 69.19,163.65) --
	( 69.19,163.65) --
	( 69.64,163.65) --
	( 69.64,163.65) --
	( 69.64,163.65) --
	( 69.64,163.65) --
	( 69.64,163.65) --
	( 70.10,163.65) --
	( 70.10,163.65) --
	( 70.10,163.65) --
	( 70.10,163.65) --
	( 70.10,163.65) --
	( 70.56,163.65) --
	( 70.56,163.65) --
	( 70.56,163.65) --
	( 70.56,163.65) --
	( 70.56,163.65) --
	( 71.01,163.65) --
	( 71.01,163.65) --
	( 71.01,163.65) --
	( 71.01,163.65) --
	( 71.01,163.65) --
	( 71.47,163.65) --
	( 71.47,163.65) --
	( 71.47,163.65) --
	( 71.47,163.65) --
	( 71.47,163.65) --
	( 71.92,163.65) --
	( 71.92,163.65) --
	( 71.92,163.65) --
	( 71.92,163.65) --
	( 71.92,163.65) --
	( 72.38,163.65) --
	( 72.38,163.65) --
	( 72.38,163.65) --
	( 72.38,163.65) --
	( 72.38,163.65) --
	( 72.84,163.65) --
	( 72.84,163.65) --
	( 72.84,163.65) --
	( 72.84,163.65) --
	( 72.84,163.65) --
	( 73.29,163.65) --
	( 73.29,163.65) --
	( 73.29,163.65) --
	( 73.29,163.65) --
	( 73.29,163.65) --
	( 73.75,163.65) --
	( 73.75,163.65) --
	( 73.75,163.65) --
	( 73.75,163.65) --
	( 73.75,163.65) --
	( 74.20,163.65) --
	( 74.20,163.65) --
	( 74.20,163.65) --
	( 74.20,163.65) --
	( 74.20,163.65) --
	( 74.66,163.65) --
	( 74.66,163.65) --
	( 74.66,163.65) --
	( 74.66,163.65) --
	( 74.66,163.65) --
	( 75.12,163.65) --
	( 75.12,163.65) --
	( 75.12,163.65) --
	( 75.12,163.65) --
	( 75.12,163.65) --
	( 75.57,163.65) --
	( 75.57,163.65) --
	( 75.57,163.65) --
	( 75.57,163.65) --
	( 75.57,163.65) --
	( 76.03,163.65) --
	( 76.03,163.65) --
	( 76.03,163.65) --
	( 76.03,163.65) --
	( 76.03,163.65) --
	( 76.48,163.65) --
	( 76.48,163.65) --
	( 76.48,163.65) --
	( 76.48,163.65) --
	( 76.48,163.65) --
	( 76.94,163.65) --
	( 76.94,163.65) --
	( 76.94,163.65) --
	( 76.94,163.65) --
	( 76.94,163.65) --
	( 77.40,163.65) --
	( 77.40,163.65) --
	( 77.40,163.65) --
	( 77.40,163.65) --
	( 77.40,163.65) --
	( 77.85,163.65) --
	( 77.85,163.65) --
	( 77.85,163.65) --
	( 77.85,163.65) --
	( 77.85,163.65) --
	( 78.31,163.65) --
	( 78.31,163.65) --
	( 78.31,163.65) --
	( 78.31,163.65) --
	( 78.31,163.65) --
	( 78.76,163.65) --
	( 78.76,163.65) --
	( 78.76,163.65) --
	( 78.76,163.65) --
	( 78.76,163.65) --
	( 79.22,163.65) --
	( 79.22,163.65) --
	( 79.22,163.65) --
	( 79.22,163.65) --
	( 79.22,163.65) --
	( 79.68,163.65) --
	( 79.68,163.65) --
	( 79.68,163.65) --
	( 79.68,163.65) --
	( 79.68,163.65) --
	( 80.13,163.65) --
	( 80.13,163.65) --
	( 80.13,163.65) --
	( 80.13,163.65) --
	( 80.13,163.65) --
	( 80.59,163.65) --
	( 80.59,163.65) --
	( 80.59,163.65) --
	( 80.59,163.65) --
	( 80.59,163.65) --
	( 81.04,163.65) --
	( 81.04,163.65) --
	( 81.04,163.65) --
	( 81.04,163.65) --
	( 81.04,163.65) --
	( 81.50,163.65) --
	( 81.50,163.65) --
	( 81.50,163.65) --
	( 81.50,163.65) --
	( 81.50,163.65) --
	( 81.96,163.65) --
	( 81.96,163.65) --
	( 81.96,163.65) --
	( 81.96,163.65) --
	( 81.96,163.65) --
	( 82.41,163.65) --
	( 82.41,163.65) --
	( 82.41,163.65) --
	( 82.41,163.65) --
	( 82.41,163.65) --
	( 82.87,163.65) --
	( 82.87,163.65) --
	( 82.87,163.65) --
	( 82.87,163.65) --
	( 82.87,163.65) --
	( 83.32,163.65) --
	( 83.32,163.65) --
	( 83.32,163.65) --
	( 83.32,163.65) --
	( 83.32,163.65) --
	( 83.78,163.65) --
	( 83.78,163.65) --
	( 83.78,163.65) --
	( 83.78,163.65) --
	( 83.78,163.65) --
	( 84.24,163.65) --
	( 84.24,163.65) --
	( 84.24,163.65) --
	( 84.24,163.65) --
	( 84.24,163.65) --
	( 84.69,163.65) --
	( 84.69,163.65) --
	( 84.69,163.65) --
	( 84.69,163.65) --
	( 84.69,163.65) --
	( 85.15,163.65) --
	( 85.15,163.65) --
	( 85.15,163.65) --
	( 85.15,163.65) --
	( 85.15,163.65) --
	( 85.60,163.65) --
	( 85.60,163.65) --
	( 85.60,163.65) --
	( 85.60,163.65) --
	( 85.60,163.65) --
	( 86.06,163.65) --
	( 86.06,163.65) --
	( 86.06,163.65) --
	( 86.06,163.65) --
	( 86.06,163.65) --
	( 86.52,163.65) --
	( 86.52,163.65) --
	( 86.52,163.65) --
	( 86.52,163.65) --
	( 86.52,163.65) --
	( 86.97,163.65) --
	( 86.97,163.65) --
	( 86.97,163.65) --
	( 86.97,163.65) --
	( 86.97,163.65) --
	( 87.43,163.65) --
	( 87.43,163.65) --
	( 87.43,163.65) --
	( 87.43,163.65) --
	( 87.43,163.65) --
	( 87.88,163.65) --
	( 87.88,163.65) --
	( 87.88,163.65) --
	( 87.88,163.65) --
	( 87.88,163.65) --
	( 88.34,163.65) --
	( 88.34,163.65) --
	( 88.34,163.65) --
	( 88.34,163.65) --
	( 88.34,163.65) --
	( 88.80,163.65) --
	( 88.80,163.65) --
	( 88.80,163.65) --
	( 88.80,163.65) --
	( 88.80,163.65) --
	( 89.25,163.65) --
	( 89.25,163.65) --
	( 89.25,163.65) --
	( 89.25,163.65) --
	( 89.25,163.65) --
	( 89.71,163.65) --
	( 89.71,163.65) --
	( 89.71,163.65) --
	( 89.71,163.65) --
	( 89.71,163.65) --
	( 90.16,163.65) --
	( 90.16,163.65) --
	( 90.16,163.65) --
	( 90.16,163.65) --
	( 90.16,163.65) --
	( 90.62,163.65) --
	( 90.62,163.65) --
	( 90.62,163.65) --
	( 90.62,163.65) --
	( 90.62,163.65) --
	( 91.08,163.65) --
	( 91.08,163.65) --
	( 91.08,163.65) --
	( 91.08,163.65) --
	( 91.08,163.65) --
	( 91.53,163.65) --
	( 91.53,163.65) --
	( 91.53,163.65) --
	( 91.53,163.65) --
	( 91.53,163.65) --
	( 91.99,163.65) --
	( 91.99,163.65) --
	( 91.99,163.65) --
	( 91.99,163.65) --
	( 91.99,163.65) --
	( 92.44,163.65) --
	( 92.44,163.65) --
	( 92.44,163.65) --
	( 92.44,163.65) --
	( 92.44,163.65) --
	( 92.90,163.65) --
	( 92.90,163.65) --
	( 92.90,163.65) --
	( 92.90,163.65) --
	( 92.90,163.65) --
	( 93.36,163.65) --
	( 93.36,163.65) --
	( 93.36,163.65) --
	( 93.36,163.65) --
	( 93.36,163.65) --
	( 93.81,163.65) --
	( 93.81,163.65) --
	( 93.81,163.65) --
	( 93.81,163.65) --
	( 93.81,163.65) --
	( 94.27,163.65) --
	( 94.27,163.65) --
	( 94.27,163.65) --
	( 94.27,163.65) --
	( 94.27,163.65) --
	( 94.72,163.65) --
	( 94.72,163.65) --
	( 94.72,163.65) --
	( 94.72,163.65) --
	( 94.72,163.65) --
	( 95.18,163.65) --
	( 95.18,163.65) --
	( 95.18,163.65) --
	( 95.18,163.65) --
	( 95.18,163.65) --
	( 95.64,163.65) --
	( 95.64,163.65) --
	( 95.64,163.65) --
	( 95.64,163.65) --
	( 95.64,163.65) --
	( 96.09,163.65) --
	( 96.09,163.65) --
	( 96.09,163.65) --
	( 96.09,163.65) --
	( 96.09,163.65) --
	( 96.55,163.65) --
	( 96.55,163.65) --
	( 96.55,163.65) --
	( 96.55,163.65) --
	( 96.55,163.65) --
	( 97.00,163.65) --
	( 97.00,163.65) --
	( 97.00,163.65) --
	( 97.00,163.65) --
	( 97.00,163.65) --
	( 97.46,163.65) --
	( 97.46,163.65) --
	( 97.46,163.65) --
	( 97.46,163.65) --
	( 97.46,163.65) --
	( 97.92,163.65) --
	( 97.92,163.65) --
	( 97.92,163.65) --
	( 97.92,163.65) --
	( 97.92,163.65) --
	( 98.37,163.65) --
	( 98.37,163.65) --
	( 98.37,163.65) --
	( 98.37,163.65) --
	( 98.37,163.65) --
	( 98.83,163.65) --
	( 98.83,163.65) --
	( 98.83,163.65) --
	( 98.83,163.65) --
	( 98.83,163.65) --
	( 99.28,163.65) --
	( 99.28,163.65) --
	( 99.28,163.65) --
	( 99.28,163.65) --
	( 99.28,163.65) --
	( 99.74,163.65) --
	( 99.74,163.65) --
	( 99.74,163.65) --
	( 99.74,163.65) --
	( 99.74,163.65) --
	(100.20,163.65) --
	(100.20,163.65) --
	(100.20,163.65) --
	(100.20,163.65) --
	(100.20,163.65) --
	(100.65,163.65) --
	(100.65,163.65) --
	(100.65,163.65) --
	(100.65,163.65) --
	(100.65,163.65) --
	(101.11,163.65) --
	(101.11,163.65) --
	(101.11,163.65) --
	(101.11,163.65) --
	(101.11,163.65) --
	(101.56,163.65) --
	(101.56,163.65) --
	(101.56,163.65) --
	(101.56,163.65) --
	(101.56,163.65) --
	(102.02,163.65) --
	(102.02,163.65) --
	(102.02,163.65) --
	(102.02,163.65) --
	(102.02,163.65) --
	(102.48,163.65) --
	(102.48,163.65) --
	(102.48,163.65) --
	(102.48,163.65) --
	(102.48,163.65) --
	(102.93,163.65) --
	(102.93,163.65) --
	(102.93,163.65) --
	(102.93,163.65) --
	(102.93,163.65) --
	(103.39,163.65) --
	(103.39,163.65) --
	(103.39,163.65) --
	(103.39,163.65) --
	(103.39,163.65) --
	(103.84,163.65) --
	(103.84,163.65) --
	(103.84,163.65) --
	(103.84,163.65) --
	(103.84,163.65) --
	(104.30,163.65) --
	(104.30,163.65) --
	(104.30,163.65) --
	(104.30,163.65) --
	(104.30,163.65) --
	(104.76,163.65) --
	(104.76,163.65) --
	(104.76,163.65) --
	(104.76,163.65) --
	(104.76,163.65) --
	(105.21,163.65) --
	(105.21,163.65) --
	(105.21,163.65) --
	(105.21,163.65) --
	(105.21,163.65) --
	(105.67,163.65) --
	(105.67,163.65) --
	(105.67,163.65) --
	(105.67,163.65) --
	(105.67,163.65) --
	(106.12,163.65) --
	(106.12,163.65) --
	(106.12,163.65) --
	(106.12,163.65) --
	(106.12,163.65) --
	(106.58,163.65) --
	(106.58,163.65) --
	(106.58,163.65) --
	(106.58,163.65) --
	(106.58,163.65) --
	(107.04,163.65) --
	(107.04,163.65) --
	(107.04,163.65) --
	(107.04,163.65) --
	(107.04,163.65) --
	(107.49,163.65) --
	(107.49,163.65) --
	(107.49,163.65) --
	(107.49,163.65) --
	(107.49,163.65) --
	(107.95,163.65) --
	(107.95,163.65) --
	(107.95,163.65) --
	(107.95,163.65) --
	(107.95,163.65) --
	(108.40,163.65) --
	(108.40,163.65) --
	(108.40,163.65) --
	(108.40,163.65) --
	(108.40,163.65) --
	(108.86,163.65) --
	(108.86,163.65) --
	(108.86,163.65) --
	(108.86,163.65) --
	(108.86,163.65) --
	(109.32,163.65) --
	(109.32,163.65) --
	(109.32,163.65) --
	(109.32,163.65) --
	(109.32,163.65) --
	(109.77,163.65) --
	(109.77,163.65) --
	(109.77,163.65) --
	(109.77,163.65) --
	(109.77,163.65) --
	(110.23,163.65) --
	(110.23,163.65) --
	(110.23,163.65) --
	(110.23,163.65) --
	(110.23,163.65) --
	(110.68,163.65) --
	(110.68,163.65) --
	(110.68,163.65) --
	(110.68,163.65) --
	(110.68,163.65) --
	(111.14,163.65) --
	(111.14,163.65) --
	(111.14,163.65) --
	(111.14,163.65) --
	(111.14,163.65) --
	(111.60,163.65) --
	(111.60,163.65) --
	(111.60,163.65) --
	(111.60,163.65) --
	(111.60,163.65) --
	(112.05,163.65) --
	(112.05,163.65) --
	(112.05,163.65) --
	(112.05,163.65) --
	(112.05,163.65) --
	(112.51,163.65) --
	(112.51,163.65) --
	(112.51,163.65) --
	(112.51,163.65) --
	(112.51,163.65) --
	(112.96,163.65) --
	(112.96,163.65) --
	(112.96,163.65) --
	(112.96,163.65) --
	(112.96,163.65) --
	(113.42,163.65) --
	(113.42,163.65) --
	(113.42,163.65) --
	(113.42,163.65) --
	(113.42,163.65) --
	(113.88,163.65) --
	(113.88,163.65) --
	(113.88,163.65) --
	(113.88,163.65) --
	(113.88,163.65) --
	(114.33,163.65) --
	(114.33,163.65) --
	(114.33,163.65) --
	(114.33,163.65) --
	(114.33,163.65) --
	(114.79,163.65) --
	(114.79,163.65) --
	(114.79,163.65) --
	(114.79,163.65) --
	(114.79,163.65) --
	(115.24,163.65) --
	(115.24,163.65) --
	(115.24,163.65) --
	(115.24,163.65) --
	(115.24,163.65) --
	(115.70,163.65) --
	(115.70,163.65) --
	(115.70,163.65) --
	(115.70,163.65) --
	(115.70,163.65) --
	(116.16,163.65) --
	(116.16,163.65) --
	(116.16,163.65) --
	(116.16,163.65) --
	(116.16,163.65) --
	(116.61,163.65) --
	(116.61,163.65) --
	(116.61,163.65) --
	(116.61,163.65) --
	(116.61,163.65) --
	(117.07,163.65) --
	(117.07,163.65) --
	(117.07,163.65) --
	(117.07,163.65) --
	(117.07,163.65) --
	(117.52,163.65) --
	(117.52,163.65) --
	(117.52,163.65) --
	(117.52,163.65) --
	(117.52,163.65) --
	(117.98,163.65) --
	(117.98,163.65) --
	(117.98,163.65) --
	(117.98,163.65) --
	(117.98,163.65) --
	(118.44,163.65) --
	(118.44,163.65) --
	(118.44,163.65) --
	(118.44,163.65) --
	(118.44,163.65) --
	(118.89,163.65) --
	(118.89,163.65) --
	(118.89,163.65) --
	(118.89,163.65) --
	(118.89,163.65) --
	(119.35,163.65) --
	(119.35,163.65) --
	(119.35,163.65) --
	(119.35,163.65) --
	(119.35,163.65) --
	(119.80,163.65) --
	(119.80,163.65) --
	(119.80,163.65) --
	(119.80,163.65) --
	(119.80,163.65) --
	(120.26,163.65) --
	(120.26,163.65) --
	(120.26,163.65) --
	(120.26,163.65) --
	(120.26,163.65) --
	(120.72,163.65) --
	(120.72,163.65) --
	(120.72,163.65) --
	(120.72,163.65) --
	(120.72,163.65) --
	(121.17,163.65) --
	(121.17,163.65) --
	(121.17,163.65) --
	(121.17,163.65) --
	(121.17,163.65) --
	(121.63,163.65) --
	(121.63,163.65) --
	(121.63,163.65) --
	(121.63,163.65) --
	(121.63,163.65) --
	(122.08,163.65) --
	(122.08,163.65) --
	(122.08,163.65) --
	(122.08,163.65) --
	(122.08,163.65) --
	(122.54,163.65) --
	(122.54,163.65) --
	(122.54,163.65) --
	(122.54,163.65) --
	(122.54,163.65) --
	(123.00,163.65) --
	(123.00,163.65) --
	(123.00,163.65) --
	(123.00,163.65) --
	(123.00,163.65) --
	(123.45,163.65) --
	(123.45,163.65) --
	(123.45,163.65) --
	(123.45,163.65) --
	(123.45,163.65) --
	(123.91,163.65) --
	(123.91,163.65) --
	(123.91,163.65) --
	(123.91,163.65) --
	(123.91,163.65) --
	(124.36,163.65) --
	(124.36,163.65) --
	(124.36,163.65) --
	(124.36,163.65) --
	(124.36,163.65);
\definecolor{drawColor}{gray}{0.70}

\path[draw=drawColor,line width= 0.5pt,line join=round,line cap=round] ( 33.62, 29.68) rectangle (124.36,194.47);
\end{scope}
\begin{scope}
\path[clip] (  0.00,  0.00) rectangle (129.36,199.47);
\definecolor{drawColor}{gray}{0.30}

\node[text=drawColor,anchor=base east,inner sep=0pt, outer sep=0pt, scale=  1.00] at ( 29.12, 26.23) {0.0};

\node[text=drawColor,anchor=base east,inner sep=0pt, outer sep=0pt, scale=  1.00] at ( 29.12, 53.03) {0.2};

\node[text=drawColor,anchor=base east,inner sep=0pt, outer sep=0pt, scale=  1.00] at ( 29.12, 79.82) {0.4};

\node[text=drawColor,anchor=base east,inner sep=0pt, outer sep=0pt, scale=  1.00] at ( 29.12,106.62) {0.6};

\node[text=drawColor,anchor=base east,inner sep=0pt, outer sep=0pt, scale=  1.00] at ( 29.12,133.41) {0.8};

\node[text=drawColor,anchor=base east,inner sep=0pt, outer sep=0pt, scale=  1.00] at ( 29.12,160.21) {1.0};

\node[text=drawColor,anchor=base east,inner sep=0pt, outer sep=0pt, scale=  1.00] at ( 29.12,187.00) {1.2};
\end{scope}
\begin{scope}
\path[clip] (  0.00,  0.00) rectangle (129.36,199.47);
\definecolor{drawColor}{gray}{0.70}

\path[draw=drawColor,line width= 0.3pt,line join=round] ( 31.12, 29.68) --
	( 33.62, 29.68);

\path[draw=drawColor,line width= 0.3pt,line join=round] ( 31.12, 56.47) --
	( 33.62, 56.47);

\path[draw=drawColor,line width= 0.3pt,line join=round] ( 31.12, 83.27) --
	( 33.62, 83.27);

\path[draw=drawColor,line width= 0.3pt,line join=round] ( 31.12,110.06) --
	( 33.62,110.06);

\path[draw=drawColor,line width= 0.3pt,line join=round] ( 31.12,136.86) --
	( 33.62,136.86);

\path[draw=drawColor,line width= 0.3pt,line join=round] ( 31.12,163.65) --
	( 33.62,163.65);

\path[draw=drawColor,line width= 0.3pt,line join=round] ( 31.12,190.45) --
	( 33.62,190.45);
\end{scope}
\begin{scope}
\path[clip] (  0.00,  0.00) rectangle (129.36,199.47);
\definecolor{drawColor}{gray}{0.70}

\path[draw=drawColor,line width= 0.3pt,line join=round] ( 33.62, 27.18) --
	( 33.62, 29.68);

\path[draw=drawColor,line width= 0.3pt,line join=round] ( 51.77, 27.18) --
	( 51.77, 29.68);

\path[draw=drawColor,line width= 0.3pt,line join=round] ( 69.92, 27.18) --
	( 69.92, 29.68);

\path[draw=drawColor,line width= 0.3pt,line join=round] ( 88.07, 27.18) --
	( 88.07, 29.68);

\path[draw=drawColor,line width= 0.3pt,line join=round] (106.21, 27.18) --
	(106.21, 29.68);

\path[draw=drawColor,line width= 0.3pt,line join=round] (124.36, 27.18) --
	(124.36, 29.68);
\end{scope}
\begin{scope}
\path[clip] (  0.00,  0.00) rectangle (129.36,199.47);
\definecolor{drawColor}{gray}{0.30}

\node[text=drawColor,anchor=base,inner sep=0pt, outer sep=0pt, scale=  1.00] at ( 32.62, 18.29) {0};

\node[text=drawColor,anchor=base,inner sep=0pt, outer sep=0pt, scale=  1.00] at ( 49.77, 18.29) {20};

\node[text=drawColor,anchor=base,inner sep=0pt, outer sep=0pt, scale=  1.00] at ( 67.92, 18.29) {40};

\node[text=drawColor,anchor=base,inner sep=0pt, outer sep=0pt, scale=  1.00] at ( 86.07, 18.29) {60};

\node[text=drawColor,anchor=base,inner sep=0pt, outer sep=0pt, scale=  1.00] at (104.22, 18.29) {80};

\node[text=drawColor,anchor=base,inner sep=0pt, outer sep=0pt, scale=  1.00] at (121.36, 18.29) {100};
\end{scope}
\begin{scope}
\path[clip] (  0.00,  0.00) rectangle (129.36,199.47);
\definecolor{drawColor}{RGB}{0,0,0}

\node[text=drawColor,anchor=base,inner sep=0pt, outer sep=0pt, scale=  1.00] at ( 78.99,  6.94) {\bfseries Instances in \%};
\end{scope}
\begin{scope}
\path[clip] (  0.00,  0.00) rectangle (129.36,199.47);
\definecolor{drawColor}{RGB}{0,0,0}

\node[text=drawColor,rotate= 90.00,anchor=base,inner sep=0pt, outer sep=0pt, scale=  1.00] at ( 11.90,112.07) {\bfseries Relative Costs};
\end{scope}
\end{tikzpicture}

%% file: 06-plots/NCC2MILP-1Hour-ImprovementPlot-Shortened.tex
\begin{tikzpicture}[x=1pt,y=1pt]
\definecolor{fillColor}{RGB}{255,255,255}
\path[use as bounding box,fill=fillColor,fill opacity=0.00] (0,0) rectangle (122.86,199.47);
\begin{scope}
\path[clip] (  0.00,  0.00) rectangle (122.86,199.47);
\definecolor{drawColor}{RGB}{255,255,255}
\definecolor{fillColor}{RGB}{255,255,255}

\path[draw=drawColor,line width= 0.5pt,line join=round,line cap=round,fill=fillColor] (  0.00,  0.00) rectangle (122.86,199.47);
\end{scope}
\begin{scope}
\path[clip] ( 27.27, 29.68) rectangle (117.86,194.47);
\definecolor{fillColor}{RGB}{255,255,255}

\path[fill=fillColor] ( 27.27, 29.68) rectangle (117.86,194.47);
\definecolor{drawColor}{gray}{0.87}

\path[draw=drawColor,line width= 0.1pt,line join=round] ( 27.27, 29.68) --
	(117.86, 29.68);

\path[draw=drawColor,line width= 0.1pt,line join=round] ( 27.27, 64.74) --
	(117.86, 64.74);

\path[draw=drawColor,line width= 0.1pt,line join=round] ( 27.27, 99.80) --
	(117.86, 99.80);

\path[draw=drawColor,line width= 0.1pt,line join=round] ( 27.27,134.86) --
	(117.86,134.86);

\path[draw=drawColor,line width= 0.1pt,line join=round] ( 27.27,169.92) --
	(117.86,169.92);

\path[draw=drawColor,line width= 0.1pt,line join=round] ( 36.33, 29.68) --
	( 36.33,194.47);

\path[draw=drawColor,line width= 0.1pt,line join=round] ( 54.45, 29.68) --
	( 54.45,194.47);

\path[draw=drawColor,line width= 0.1pt,line join=round] ( 72.57, 29.68) --
	( 72.57,194.47);

\path[draw=drawColor,line width= 0.1pt,line join=round] ( 90.68, 29.68) --
	( 90.68,194.47);

\path[draw=drawColor,line width= 0.1pt,line join=round] (108.80, 29.68) --
	(108.80,194.47);

\path[draw=drawColor,line width= 0.3pt,line join=round] ( 27.27, 47.21) --
	(117.86, 47.21);

\path[draw=drawColor,line width= 0.3pt,line join=round] ( 27.27, 82.27) --
	(117.86, 82.27);

\path[draw=drawColor,line width= 0.3pt,line join=round] ( 27.27,117.33) --
	(117.86,117.33);

\path[draw=drawColor,line width= 0.3pt,line join=round] ( 27.27,152.39) --
	(117.86,152.39);

\path[draw=drawColor,line width= 0.3pt,line join=round] ( 27.27,187.45) --
	(117.86,187.45);

\path[draw=drawColor,line width= 0.3pt,line join=round] ( 27.27, 29.68) --
	( 27.27,194.47);

\path[draw=drawColor,line width= 0.3pt,line join=round] ( 45.39, 29.68) --
	( 45.39,194.47);

\path[draw=drawColor,line width= 0.3pt,line join=round] ( 63.51, 29.68) --
	( 63.51,194.47);

\path[draw=drawColor,line width= 0.3pt,line join=round] ( 81.62, 29.68) --
	( 81.62,194.47);

\path[draw=drawColor,line width= 0.3pt,line join=round] ( 99.74, 29.68) --
	( 99.74,194.47);

\path[draw=drawColor,line width= 0.3pt,line join=round] (117.86, 29.68) --
	(117.86,194.47);
\definecolor{drawColor}{RGB}{128,128,128}

\path[draw=drawColor,line width= 1.0pt,line join=round] ( 28.94, 29.68) --
	( 29.09, 33.06) --
	( 29.55, 35.83) --
	( 30.00, 43.84) --
	( 30.46, 45.67) --
	( 30.92, 46.83) --
	( 31.37, 50.48) --
	( 31.83, 59.25) --
	( 32.28, 60.75) --
	( 32.74, 62.20) --
	( 33.19, 62.25) --
	( 33.65, 63.46) --
	( 34.10, 63.56) --
	( 34.56, 64.04) --
	( 35.01, 65.98) --
	( 35.47, 66.45) --
	( 35.92, 71.52) --
	( 36.38, 72.61) --
	( 36.83, 73.02) --
	( 37.29, 75.05) --
	( 37.74, 75.44) --
	( 38.20, 76.24) --
	( 38.65, 76.64) --
	( 39.11, 78.27) --
	( 39.56, 78.73) --
	( 40.02, 79.52) --
	( 40.47, 79.71) --
	( 40.93, 80.62) --
	( 41.38, 81.56) --
	( 41.84, 82.27) --
	( 42.30, 82.27) --
	( 42.75, 82.27) --
	( 43.21, 82.27) --
	( 43.66, 82.27) --
	( 44.12, 82.27) --
	( 44.57, 82.27) --
	( 45.03, 82.27) --
	( 45.48, 82.27) --
	( 45.94, 82.27) --
	( 46.39, 82.27) --
	( 46.85, 82.27) --
	( 47.30, 82.27) --
	( 47.76, 82.27) --
	( 48.21, 82.27) --
	( 48.67, 82.27) --
	( 49.12, 82.27) --
	( 49.58, 82.27) --
	( 50.03, 82.27) --
	( 50.49, 82.27) --
	( 50.94, 82.27) --
	( 51.40, 82.27) --
	( 51.85, 82.27) --
	( 52.31, 82.27) --
	( 52.76, 82.27) --
	( 53.22, 82.27) --
	( 53.68, 82.27) --
	( 54.13, 82.27) --
	( 54.59, 82.27) --
	( 55.04, 82.27) --
	( 55.50, 82.27) --
	( 55.95, 82.27) --
	( 56.41, 82.27) --
	( 56.86, 82.27) --
	( 57.32, 82.27) --
	( 57.77, 82.27) --
	( 58.23, 82.27) --
	( 58.68, 82.27) --
	( 59.14, 82.27) --
	( 59.59, 82.27) --
	( 60.05, 82.27) --
	( 60.50, 82.27) --
	( 60.96, 82.27) --
	( 61.41, 82.27) --
	( 61.87, 82.27) --
	( 62.32, 82.27) --
	( 62.78, 82.27) --
	( 63.23, 82.27) --
	( 63.69, 82.27) --
	( 64.14, 82.27) --
	( 64.60, 82.74) --
	( 65.06, 83.10) --
	( 65.51, 83.57) --
	( 65.97, 83.75) --
	( 66.42, 83.82) --
	( 66.88, 84.29) --
	( 67.33, 84.49) --
	( 67.79, 85.07) --
	( 68.24, 85.90) --
	( 68.70, 86.03) --
	( 69.15, 86.26) --
	( 69.61, 86.48) --
	( 70.06, 86.65) --
	( 70.52, 86.83) --
	( 70.97, 86.84) --
	( 71.43, 87.29) --
	( 71.88, 87.62) --
	( 72.34, 87.79) --
	( 72.79, 88.32) --
	( 73.25, 88.36) --
	( 73.70, 88.54) --
	( 74.16, 88.85) --
	( 74.61, 88.96) --
	( 75.07, 89.11) --
	( 75.53, 89.28) --
	( 75.98, 89.71) --
	( 76.44, 89.91) --
	( 76.89, 90.04) --
	( 77.35, 90.46) --
	( 77.80, 92.30) --
	( 78.26, 92.34) --
	( 78.71, 92.89) --
	( 79.17, 93.88) --
	( 79.62, 94.13) --
	( 80.08, 94.20) --
	( 80.53, 94.36) --
	( 80.99, 94.42) --
	( 81.44, 94.49) --
	( 81.90, 95.86) --
	( 82.35, 95.91) --
	( 82.81, 95.91) --
	( 83.26, 97.55) --
	( 83.72, 97.89) --
	( 84.17, 98.28) --
	( 84.63, 98.36) --
	( 85.08, 98.50) --
	( 85.54, 99.19) --
	( 85.99, 99.19) --
	( 86.45, 99.41) --
	( 86.91,100.15) --
	( 87.36,100.37) --
	( 87.82,100.70) --
	( 88.27,101.06) --
	( 88.73,102.08) --
	( 89.18,102.09) --
	( 89.64,102.18) --
	( 90.09,103.36) --
	( 90.55,104.21) --
	( 91.00,104.74) --
	( 91.46,105.72) --
	( 91.91,105.92) --
	( 92.37,106.15) --
	( 92.82,107.05) --
	( 93.28,108.62) --
	( 93.73,109.12) --
	( 94.19,109.16) --
	( 94.64,110.13) --
	( 95.10,110.87) --
	( 95.55,111.76) --
	( 96.01,112.35) --
	( 96.46,112.98) --
	( 96.92,113.01) --
	( 97.37,115.05) --
	( 97.83,117.09) --
	( 98.29,118.04) --
	( 98.74,118.74) --
	( 99.20,119.08) --
	( 99.65,119.89) --
	(100.11,120.30) --
	(100.56,121.89) --
	(101.02,122.33) --
	(101.47,122.40) --
	(101.93,122.63) --
	(102.38,123.79) --
	(102.84,125.29) --
	(103.29,126.44) --
	(103.75,126.46) --
	(104.20,127.25) --
	(104.66,128.62) --
	(105.11,129.68) --
	(105.57,130.91) --
	(106.02,130.99) --
	(106.48,131.46) --
	(106.93,133.89) --
	(107.39,136.21) --
	(107.84,137.47) --
	(108.30,140.93) --
	(108.75,141.53) --
	(109.21,142.70) --
	(109.67,144.54) --
	(110.12,145.38) --
	(110.58,145.94) --
	(111.03,148.56) --
	(111.49,149.06) --
	(111.94,149.62) --
	(112.40,153.96) --
	(112.85,154.06) --
	(113.31,156.92) --
	(113.76,159.78) --
	(114.22,160.24) --
	(114.67,163.06) --
	(115.13,164.83) --
	(115.58,174.23) --
	(116.04,177.72) --
	(116.49,186.57) --
	(116.66,194.47);
\definecolor{drawColor}{RGB}{152,167,197}

\path[draw=drawColor,line width= 1.0pt,line join=round] ( 27.27, 44.04) --
	( 27.73, 46.70) --
	( 28.18, 49.11) --
	( 28.64, 50.53) --
	( 29.09, 58.27) --
	( 29.55, 61.23) --
	( 30.00, 61.63) --
	( 30.46, 64.12) --
	( 30.92, 65.90) --
	( 31.37, 66.74) --
	( 31.83, 66.87) --
	( 32.28, 67.07) --
	( 32.74, 67.31) --
	( 33.19, 67.75) --
	( 33.65, 69.30) --
	( 34.10, 69.58) --
	( 34.56, 70.24) --
	( 35.01, 70.66) --
	( 35.47, 70.94) --
	( 35.92, 70.95) --
	( 36.38, 71.53) --
	( 36.83, 73.23) --
	( 37.29, 73.49) --
	( 37.74, 73.88) --
	( 38.20, 73.98) --
	( 38.65, 74.00) --
	( 39.11, 74.31) --
	( 39.56, 75.02) --
	( 40.02, 75.25) --
	( 40.47, 76.27) --
	( 40.93, 76.36) --
	( 41.38, 77.20) --
	( 41.84, 77.25) --
	( 42.30, 77.36) --
	( 42.75, 77.62) --
	( 43.21, 77.73) --
	( 43.66, 77.94) --
	( 44.12, 78.28) --
	( 44.57, 78.84) --
	( 45.03, 78.90) --
	( 45.48, 78.96) --
	( 45.94, 79.27) --
	( 46.39, 79.83) --
	( 46.85, 79.95) --
	( 47.30, 80.46) --
	( 47.76, 80.93) --
	( 48.21, 81.20) --
	( 48.67, 81.21) --
	( 49.12, 81.28) --
	( 49.58, 81.37) --
	( 50.03, 82.12) --
	( 50.49, 82.27) --
	( 50.94, 82.27) --
	( 51.40, 82.27) --
	( 51.85, 82.27) --
	( 52.31, 82.27) --
	( 52.76, 82.27) --
	( 53.22, 82.27) --
	( 53.68, 82.27) --
	( 54.13, 82.27) --
	( 54.59, 82.27) --
	( 55.04, 82.27) --
	( 55.50, 82.27) --
	( 55.95, 82.27) --
	( 56.41, 82.27) --
	( 56.86, 82.27) --
	( 57.32, 82.27) --
	( 57.77, 82.27) --
	( 58.23, 82.27) --
	( 58.68, 82.27) --
	( 59.14, 82.27) --
	( 59.59, 82.27) --
	( 60.05, 82.27) --
	( 60.50, 82.27) --
	( 60.96, 82.27) --
	( 61.41, 82.27) --
	( 61.87, 82.27) --
	( 62.32, 82.27) --
	( 62.78, 82.27) --
	( 63.23, 82.27) --
	( 63.69, 82.27) --
	( 64.14, 82.27) --
	( 64.60, 82.27) --
	( 65.06, 82.27) --
	( 65.51, 82.27) --
	( 65.97, 82.27) --
	( 66.42, 82.27) --
	( 66.88, 82.27) --
	( 67.33, 82.27) --
	( 67.79, 82.28) --
	( 68.24, 82.41) --
	( 68.70, 82.51) --
	( 69.15, 82.96) --
	( 69.61, 83.02) --
	( 70.06, 83.04) --
	( 70.52, 83.48) --
	( 70.97, 83.85) --
	( 71.43, 83.87) --
	( 71.88, 83.90) --
	( 72.34, 84.08) --
	( 72.79, 84.35) --
	( 73.25, 84.61) --
	( 73.70, 84.74) --
	( 74.16, 84.76) --
	( 74.61, 84.77) --
	( 75.07, 84.96) --
	( 75.53, 85.29) --
	( 75.98, 85.32) --
	( 76.44, 85.54) --
	( 76.89, 85.59) --
	( 77.35, 85.91) --
	( 77.80, 85.98) --
	( 78.26, 86.11) --
	( 78.71, 86.60) --
	( 79.17, 86.83) --
	( 79.62, 87.89) --
	( 80.08, 88.10) --
	( 80.53, 88.10) --
	( 80.99, 88.19) --
	( 81.44, 88.51) --
	( 81.90, 88.67) --
	( 82.35, 88.70) --
	( 82.81, 88.91) --
	( 83.26, 88.98) --
	( 83.72, 89.92) --
	( 84.17, 90.17) --
	( 84.63, 90.33) --
	( 85.08, 90.54) --
	( 85.54, 90.58) --
	( 85.99, 90.64) --
	( 86.45, 91.01) --
	( 86.91, 91.07) --
	( 87.36, 91.63) --
	( 87.82, 92.46) --
	( 88.27, 93.18) --
	( 88.73, 93.33) --
	( 89.18, 94.22) --
	( 89.64, 94.41) --
	( 90.09, 94.64) --
	( 90.55, 95.35) --
	( 91.00, 95.54) --
	( 91.46, 95.98) --
	( 91.91, 96.12) --
	( 92.37, 97.32) --
	( 92.82, 98.10) --
	( 93.28, 98.12) --
	( 93.73, 98.22) --
	( 94.19, 98.29) --
	( 94.64, 98.35) --
	( 95.10, 99.28) --
	( 95.55, 99.61) --
	( 96.01,100.97) --
	( 96.46,101.20) --
	( 96.92,101.66) --
	( 97.37,101.86) --
	( 97.83,102.00) --
	( 98.29,102.35) --
	( 98.74,103.41) --
	( 99.20,103.50) --
	( 99.65,103.55) --
	(100.11,104.88) --
	(100.56,104.95) --
	(101.02,105.01) --
	(101.47,105.25) --
	(101.93,106.12) --
	(102.38,106.34) --
	(102.84,106.46) --
	(103.29,106.68) --
	(103.75,107.49) --
	(104.20,107.56) --
	(104.66,108.15) --
	(105.11,108.18) --
	(105.57,109.69) --
	(106.02,110.37) --
	(106.48,110.40) --
	(106.93,110.65) --
	(107.39,112.50) --
	(107.84,112.66) --
	(108.30,115.06) --
	(108.75,115.66) --
	(109.21,117.69) --
	(109.67,119.81) --
	(110.12,120.99) --
	(110.58,121.09) --
	(111.03,121.47) --
	(111.49,122.16) --
	(111.94,122.45) --
	(112.40,122.62) --
	(112.85,122.98) --
	(113.31,126.02) --
	(113.76,128.10) --
	(114.22,128.17) --
	(114.67,128.24) --
	(115.13,128.39) --
	(115.58,132.16) --
	(116.04,139.07) --
	(116.49,151.01) --
	(116.95,157.64) --
	(117.40,182.03) --
	(117.46,194.47);
\definecolor{drawColor}{RGB}{128,202,192}

\path[draw=drawColor,line width= 1.0pt,line join=round] ( 27.50, 29.68) --
	( 27.73, 45.85) --
	( 28.18, 50.59) --
	( 28.64, 52.03) --
	( 29.09, 54.16) --
	( 29.55, 56.44) --
	( 30.00, 57.51) --
	( 30.46, 58.70) --
	( 30.92, 59.69) --
	( 31.37, 60.00) --
	( 31.83, 61.46) --
	( 32.28, 61.55) --
	( 32.74, 62.22) --
	( 33.19, 63.16) --
	( 33.65, 63.21) --
	( 34.10, 63.89) --
	( 34.56, 64.15) --
	( 35.01, 64.43) --
	( 35.47, 65.05) --
	( 35.92, 65.67) --
	( 36.38, 65.88) --
	( 36.83, 67.05) --
	( 37.29, 67.17) --
	( 37.74, 67.21) --
	( 38.20, 67.75) --
	( 38.65, 67.77) --
	( 39.11, 68.03) --
	( 39.56, 68.33) --
	( 40.02, 68.79) --
	( 40.47, 69.06) --
	( 40.93, 69.11) --
	( 41.38, 69.33) --
	( 41.84, 70.14) --
	( 42.30, 70.77) --
	( 42.75, 71.69) --
	( 43.21, 71.73) --
	( 43.66, 71.76) --
	( 44.12, 72.48) --
	( 44.57, 73.30) --
	( 45.03, 73.46) --
	( 45.48, 73.62) --
	( 45.94, 73.78) --
	( 46.39, 73.84) --
	( 46.85, 74.03) --
	( 47.30, 74.12) --
	( 47.76, 74.62) --
	( 48.21, 74.64) --
	( 48.67, 75.05) --
	( 49.12, 75.11) --
	( 49.58, 75.97) --
	( 50.03, 76.21) --
	( 50.49, 76.41) --
	( 50.94, 76.70) --
	( 51.40, 76.85) --
	( 51.85, 76.95) --
	( 52.31, 77.37) --
	( 52.76, 77.50) --
	( 53.22, 77.58) --
	( 53.68, 77.72) --
	( 54.13, 77.86) --
	( 54.59, 78.02) --
	( 55.04, 78.16) --
	( 55.50, 78.29) --
	( 55.95, 78.42) --
	( 56.41, 78.50) --
	( 56.86, 78.65) --
	( 57.32, 78.99) --
	( 57.77, 79.11) --
	( 58.23, 79.31) --
	( 58.68, 79.36) --
	( 59.14, 79.41) --
	( 59.59, 79.57) --
	( 60.05, 79.84) --
	( 60.50, 80.09) --
	( 60.96, 80.24) --
	( 61.41, 80.38) --
	( 61.87, 80.43) --
	( 62.32, 80.64) --
	( 62.78, 80.68) --
	( 63.23, 80.69) --
	( 63.69, 80.69) --
	( 64.14, 80.97) --
	( 64.60, 80.98) --
	( 65.06, 81.00) --
	( 65.51, 81.09) --
	( 65.97, 81.11) --
	( 66.42, 81.18) --
	( 66.88, 81.23) --
	( 67.33, 81.54) --
	( 67.79, 81.62) --
	( 68.24, 81.79) --
	( 68.70, 82.10) --
	( 69.15, 82.11) --
	( 69.61, 82.17) --
	( 70.06, 82.27) --
	( 70.52, 82.58) --
	( 70.97, 82.65) --
	( 71.43, 82.88) --
	( 71.88, 82.93) --
	( 72.34, 83.47) --
	( 72.79, 83.63) --
	( 73.25, 83.72) --
	( 73.70, 83.74) --
	( 74.16, 83.75) --
	( 74.61, 83.79) --
	( 75.07, 83.90) --
	( 75.53, 84.01) --
	( 75.98, 84.31) --
	( 76.44, 84.59) --
	( 76.89, 84.73) --
	( 77.35, 84.91) --
	( 77.80, 85.03) --
	( 78.26, 85.18) --
	( 78.71, 85.19) --
	( 79.17, 85.25) --
	( 79.62, 85.28) --
	( 80.08, 85.61) --
	( 80.53, 86.18) --
	( 80.99, 86.26) --
	( 81.44, 86.32) --
	( 81.90, 86.50) --
	( 82.35, 86.79) --
	( 82.81, 86.80) --
	( 83.26, 86.91) --
	( 83.72, 86.91) --
	( 84.17, 87.27) --
	( 84.63, 87.29) --
	( 85.08, 87.60) --
	( 85.54, 87.87) --
	( 85.99, 87.91) --
	( 86.45, 88.09) --
	( 86.91, 88.32) --
	( 87.36, 88.41) --
	( 87.82, 88.67) --
	( 88.27, 88.67) --
	( 88.73, 88.81) --
	( 89.18, 88.94) --
	( 89.64, 89.02) --
	( 90.09, 89.14) --
	( 90.55, 89.52) --
	( 91.00, 90.15) --
	( 91.46, 90.37) --
	( 91.91, 90.38) --
	( 92.37, 90.41) --
	( 92.82, 90.57) --
	( 93.28, 90.79) --
	( 93.73, 90.98) --
	( 94.19, 91.07) --
	( 94.64, 91.30) --
	( 95.10, 91.86) --
	( 95.55, 91.88) --
	( 96.01, 91.88) --
	( 96.46, 91.88) --
	( 96.92, 92.06) --
	( 97.37, 92.40) --
	( 97.83, 92.45) --
	( 98.29, 92.80) --
	( 98.74, 93.15) --
	( 99.20, 93.16) --
	( 99.65, 93.20) --
	(100.11, 93.21) --
	(100.56, 93.67) --
	(101.02, 94.17) --
	(101.47, 94.84) --
	(101.93, 95.08) --
	(102.38, 95.80) --
	(102.84, 95.88) --
	(103.29, 96.00) --
	(103.75, 96.93) --
	(104.20, 97.46) --
	(104.66, 97.60) --
	(105.11, 98.61) --
	(105.57, 98.90) --
	(106.02, 99.14) --
	(106.48, 99.72) --
	(106.93, 99.98) --
	(107.39,100.09) --
	(107.84,100.14) --
	(108.30,101.69) --
	(108.75,101.80) --
	(109.21,102.27) --
	(109.67,102.43) --
	(110.12,103.16) --
	(110.58,103.72) --
	(111.03,103.99) --
	(111.49,104.12) --
	(111.94,104.24) --
	(112.40,104.87) --
	(112.85,105.50) --
	(113.31,105.53) --
	(113.76,105.86) --
	(114.22,108.37) --
	(114.67,109.04) --
	(115.13,109.38) --
	(115.58,112.34) --
	(116.04,114.27) --
	(116.49,115.68) --
	(116.95,118.20) --
	(117.40,124.20) --
	(117.86,127.25);
\definecolor{drawColor}{RGB}{207,128,187}

\path[draw=drawColor,line width= 1.0pt,line join=round] ( 31.18, 29.68) --
	( 31.37, 29.82) --
	( 31.83, 30.21) --
	( 32.28, 30.35) --
	( 32.74, 31.81) --
	( 33.19, 32.52) --
	( 33.65, 33.22) --
	( 34.10, 33.25) --
	( 34.56, 33.55) --
	( 35.01, 35.04) --
	( 35.47, 35.36) --
	( 35.92, 35.62) --
	( 36.38, 36.53) --
	( 36.83, 36.62) --
	( 37.29, 36.93) --
	( 37.74, 37.14) --
	( 38.20, 37.52) --
	( 38.65, 38.20) --
	( 39.11, 38.80) --
	( 39.56, 39.50) --
	( 40.02, 39.60) --
	( 40.47, 39.66) --
	( 40.93, 39.82) --
	( 41.38, 42.73) --
	( 41.84, 43.38) --
	( 42.30, 44.13) --
	( 42.75, 44.40) --
	( 43.21, 44.46) --
	( 43.66, 45.75) --
	( 44.12, 45.78) --
	( 44.57, 45.82) --
	( 45.03, 46.62) --
	( 45.48, 47.37) --
	( 45.94, 47.50) --
	( 46.39, 47.70) --
	( 46.85, 47.72) --
	( 47.30, 47.78) --
	( 47.76, 47.90) --
	( 48.21, 48.25) --
	( 48.67, 48.51) --
	( 49.12, 48.81) --
	( 49.58, 49.23) --
	( 50.03, 50.04) --
	( 50.49, 50.18) --
	( 50.94, 50.25) --
	( 51.40, 50.43) --
	( 51.85, 50.48) --
	( 52.31, 50.99) --
	( 52.76, 51.33) --
	( 53.22, 51.52) --
	( 53.68, 51.70) --
	( 54.13, 51.96) --
	( 54.59, 52.24) --
	( 55.04, 52.66) --
	( 55.50, 53.02) --
	( 55.95, 53.05) --
	( 56.41, 53.30) --
	( 56.86, 53.42) --
	( 57.32, 53.53) --
	( 57.77, 53.80) --
	( 58.23, 53.86) --
	( 58.68, 54.45) --
	( 59.14, 54.60) --
	( 59.59, 54.86) --
	( 60.05, 54.96) --
	( 60.50, 55.22) --
	( 60.96, 55.31) --
	( 61.41, 55.41) --
	( 61.87, 55.99) --
	( 62.32, 55.99) --
	( 62.78, 56.52) --
	( 63.23, 56.53) --
	( 63.69, 56.53) --
	( 64.14, 56.64) --
	( 64.60, 56.97) --
	( 65.06, 57.03) --
	( 65.51, 58.09) --
	( 65.97, 58.19) --
	( 66.42, 58.38) --
	( 66.88, 58.49) --
	( 67.33, 58.68) --
	( 67.79, 58.72) --
	( 68.24, 58.89) --
	( 68.70, 59.58) --
	( 69.15, 59.61) --
	( 69.61, 60.58) --
	( 70.06, 60.65) --
	( 70.52, 60.70) --
	( 70.97, 60.82) --
	( 71.43, 60.98) --
	( 71.88, 61.02) --
	( 72.34, 61.06) --
	( 72.79, 61.56) --
	( 73.25, 61.74) --
	( 73.70, 62.09) --
	( 74.16, 62.19) --
	( 74.61, 62.43) --
	( 75.07, 62.53) --
	( 75.53, 62.80) --
	( 75.98, 63.11) --
	( 76.44, 63.85) --
	( 76.89, 64.10) --
	( 77.35, 64.13) --
	( 77.80, 64.61) --
	( 78.26, 64.68) --
	( 78.71, 65.19) --
	( 79.17, 65.44) --
	( 79.62, 65.80) --
	( 80.08, 66.23) --
	( 80.53, 67.00) --
	( 80.99, 67.18) --
	( 81.44, 67.31) --
	( 81.90, 67.46) --
	( 82.35, 67.71) --
	( 82.81, 68.54) --
	( 83.26, 68.74) --
	( 83.72, 68.93) --
	( 84.17, 68.97) --
	( 84.63, 69.38) --
	( 85.08, 69.53) --
	( 85.54, 69.56) --
	( 85.99, 69.64) --
	( 86.45, 69.65) --
	( 86.91, 70.50) --
	( 87.36, 70.58) --
	( 87.82, 70.67) --
	( 88.27, 71.01) --
	( 88.73, 71.24) --
	( 89.18, 71.28) --
	( 89.64, 71.36) --
	( 90.09, 71.71) --
	( 90.55, 72.12) --
	( 91.00, 72.52) --
	( 91.46, 72.61) --
	( 91.91, 72.74) --
	( 92.37, 72.84) --
	( 92.82, 72.88) --
	( 93.28, 73.06) --
	( 93.73, 73.11) --
	( 94.19, 73.19) --
	( 94.64, 73.81) --
	( 95.10, 73.83) --
	( 95.55, 75.42) --
	( 96.01, 75.67) --
	( 96.46, 75.86) --
	( 96.92, 75.91) --
	( 97.37, 76.06) --
	( 97.83, 76.09) --
	( 98.29, 76.11) --
	( 98.74, 76.70) --
	( 99.20, 76.89) --
	( 99.65, 77.01) --
	(100.11, 77.02) --
	(100.56, 77.22) --
	(101.02, 77.45) --
	(101.47, 77.48) --
	(101.93, 77.91) --
	(102.38, 78.07) --
	(102.84, 78.59) --
	(103.29, 78.67) --
	(103.75, 78.70) --
	(104.20, 79.55) --
	(104.66, 80.24) --
	(105.11, 80.49) --
	(105.57, 81.09) --
	(106.02, 81.24) --
	(106.48, 81.68) --
	(106.93, 82.64) --
	(107.39, 83.58) --
	(107.84, 83.85) --
	(108.30, 83.86) --
	(108.75, 84.11) --
	(109.21, 84.55) --
	(109.67, 84.60) --
	(110.12, 86.60) --
	(110.58, 86.89) --
	(111.03, 87.70) --
	(111.49, 88.71) --
	(111.94, 88.80) --
	(112.40, 89.08) --
	(112.85, 89.18) --
	(113.31, 89.27) --
	(113.76, 90.16) --
	(114.22, 90.45) --
	(114.67, 94.35) --
	(115.13, 95.89) --
	(115.58, 98.33) --
	(116.04,109.69) --
	(116.49,116.60) --
	(116.95,143.23) --
	(117.40,145.73) --
	(117.86,152.66);
\definecolor{drawColor}{RGB}{207,192,152}

\path[draw=drawColor,line width= 1.0pt,line join=round] ( 29.86, 29.68) --
	( 30.00, 34.85) --
	( 30.46, 40.27) --
	( 30.92, 40.82) --
	( 31.37, 43.75) --
	( 31.83, 44.58) --
	( 32.28, 45.19) --
	( 32.74, 45.21) --
	( 33.19, 45.65) --
	( 33.65, 46.10) --
	( 34.10, 46.71) --
	( 34.56, 48.73) --
	( 35.01, 52.45) --
	( 35.47, 52.60) --
	( 35.92, 53.25) --
	( 36.38, 53.89) --
	( 36.83, 54.56) --
	( 37.29, 55.34) --
	( 37.74, 55.79) --
	( 38.20, 56.13) --
	( 38.65, 59.72) --
	( 39.11, 60.21) --
	( 39.56, 60.33) --
	( 40.02, 60.50) --
	( 40.47, 60.95) --
	( 40.93, 61.68) --
	( 41.38, 62.30) --
	( 41.84, 63.94) --
	( 42.30, 64.07) --
	( 42.75, 65.40) --
	( 43.21, 65.50) --
	( 43.66, 65.61) --
	( 44.12, 65.73) --
	( 44.57, 65.83) --
	( 45.03, 66.05) --
	( 45.48, 66.23) --
	( 45.94, 66.41) --
	( 46.39, 66.43) --
	( 46.85, 66.56) --
	( 47.30, 67.25) --
	( 47.76, 67.36) --
	( 48.21, 67.63) --
	( 48.67, 67.76) --
	( 49.12, 67.99) --
	( 49.58, 68.17) --
	( 50.03, 68.20) --
	( 50.49, 68.56) --
	( 50.94, 68.87) --
	( 51.40, 69.14) --
	( 51.85, 69.15) --
	( 52.31, 69.65) --
	( 52.76, 69.66) --
	( 53.22, 69.70) --
	( 53.68, 69.88) --
	( 54.13, 70.23) --
	( 54.59, 70.36) --
	( 55.04, 70.66) --
	( 55.50, 70.79) --
	( 55.95, 71.20) --
	( 56.41, 71.51) --
	( 56.86, 71.52) --
	( 57.32, 71.72) --
	( 57.77, 71.79) --
	( 58.23, 72.12) --
	( 58.68, 72.62) --
	( 59.14, 72.77) --
	( 59.59, 73.28) --
	( 60.05, 73.34) --
	( 60.50, 73.47) --
	( 60.96, 74.42) --
	( 61.41, 74.44) --
	( 61.87, 74.60) --
	( 62.32, 75.19) --
	( 62.78, 75.37) --
	( 63.23, 75.63) --
	( 63.69, 76.15) --
	( 64.14, 77.42) --
	( 64.60, 77.79) --
	( 65.06, 78.06) --
	( 65.51, 78.33) --
	( 65.97, 78.57) --
	( 66.42, 78.74) --
	( 66.88, 79.12) --
	( 67.33, 79.18) --
	( 67.79, 79.20) --
	( 68.24, 79.71) --
	( 68.70, 80.47) --
	( 69.15, 80.67) --
	( 69.61, 80.80) --
	( 70.06, 80.85) --
	( 70.52, 80.86) --
	( 70.97, 80.86) --
	( 71.43, 81.03) --
	( 71.88, 81.24) --
	( 72.34, 81.38) --
	( 72.79, 81.66) --
	( 73.25, 81.67) --
	( 73.70, 81.71) --
	( 74.16, 81.83) --
	( 74.61, 82.40) --
	( 75.07, 83.52) --
	( 75.53, 83.82) --
	( 75.98, 83.92) --
	( 76.44, 84.54) --
	( 76.89, 85.12) --
	( 77.35, 85.64) --
	( 77.80, 86.39) --
	( 78.26, 86.56) --
	( 78.71, 86.63) --
	( 79.17, 86.67) --
	( 79.62, 86.86) --
	( 80.08, 86.87) --
	( 80.53, 87.65) --
	( 80.99, 87.84) --
	( 81.44, 88.51) --
	( 81.90, 88.61) --
	( 82.35, 88.79) --
	( 82.81, 88.94) --
	( 83.26, 89.10) --
	( 83.72, 89.20) --
	( 84.17, 89.51) --
	( 84.63, 89.53) --
	( 85.08, 89.77) --
	( 85.54, 90.64) --
	( 85.99, 90.89) --
	( 86.45, 90.95) --
	( 86.91, 91.08) --
	( 87.36, 91.43) --
	( 87.82, 91.48) --
	( 88.27, 91.83) --
	( 88.73, 92.56) --
	( 89.18, 92.97) --
	( 89.64, 93.00) --
	( 90.09, 93.02) --
	( 90.55, 93.38) --
	( 91.00, 93.43) --
	( 91.46, 93.47) --
	( 91.91, 94.16) --
	( 92.37, 94.18) --
	( 92.82, 94.18) --
	( 93.28, 94.23) --
	( 93.73, 94.27) --
	( 94.19, 94.44) --
	( 94.64, 94.96) --
	( 95.10, 95.79) --
	( 95.55, 96.01) --
	( 96.01, 96.16) --
	( 96.46, 96.23) --
	( 96.92, 96.53) --
	( 97.37, 96.92) --
	( 97.83, 97.44) --
	( 98.29, 97.77) --
	( 98.74, 98.19) --
	( 99.20, 98.49) --
	( 99.65, 98.67) --
	(100.11, 99.61) --
	(100.56, 99.72) --
	(101.02,100.12) --
	(101.47,100.14) --
	(101.93,101.80) --
	(102.38,102.22) --
	(102.84,102.77) --
	(103.29,103.30) --
	(103.75,103.54) --
	(104.20,103.93) --
	(104.66,103.95) --
	(105.11,104.42) --
	(105.57,104.99) --
	(106.02,104.99) --
	(106.48,105.52) --
	(106.93,105.85) --
	(107.39,107.55) --
	(107.84,109.77) --
	(108.30,110.82) --
	(108.75,111.69) --
	(109.21,112.05) --
	(109.67,113.67) --
	(110.12,115.14) --
	(110.58,117.51) --
	(111.03,122.04) --
	(111.49,144.14) --
	(111.94,162.83) --
	(112.10,194.47);
\definecolor{drawColor}{RGB}{139,0,0}

\path[draw=drawColor,line width= 1.0pt,line join=round] ( 27.27, 82.27) --
	( 27.50, 82.27) --
	( 27.73, 82.27) --
	( 27.73, 82.27) --
	( 28.18, 82.27) --
	( 28.18, 82.27) --
	( 28.64, 82.27) --
	( 28.64, 82.27) --
	( 28.94, 82.27) --
	( 29.09, 82.27) --
	( 29.09, 82.27) --
	( 29.09, 82.27) --
	( 29.55, 82.27) --
	( 29.55, 82.27) --
	( 29.55, 82.27) --
	( 29.86, 82.27) --
	( 30.00, 82.27) --
	( 30.00, 82.27) --
	( 30.00, 82.27) --
	( 30.00, 82.27) --
	( 30.46, 82.27) --
	( 30.46, 82.27) --
	( 30.46, 82.27) --
	( 30.46, 82.27) --
	( 30.92, 82.27) --
	( 30.92, 82.27) --
	( 30.92, 82.27) --
	( 30.92, 82.27) --
	( 31.18, 82.27) --
	( 31.37, 82.27) --
	( 31.37, 82.27) --
	( 31.37, 82.27) --
	( 31.37, 82.27) --
	( 31.37, 82.27) --
	( 31.83, 82.27) --
	( 31.83, 82.27) --
	( 31.83, 82.27) --
	( 31.83, 82.27) --
	( 31.83, 82.27) --
	( 32.28, 82.27) --
	( 32.28, 82.27) --
	( 32.28, 82.27) --
	( 32.28, 82.27) --
	( 32.28, 82.27) --
	( 32.74, 82.27) --
	( 32.74, 82.27) --
	( 32.74, 82.27) --
	( 32.74, 82.27) --
	( 32.74, 82.27) --
	( 33.19, 82.27) --
	( 33.19, 82.27) --
	( 33.19, 82.27) --
	( 33.19, 82.27) --
	( 33.19, 82.27) --
	( 33.65, 82.27) --
	( 33.65, 82.27) --
	( 33.65, 82.27) --
	( 33.65, 82.27) --
	( 33.65, 82.27) --
	( 34.10, 82.27) --
	( 34.10, 82.27) --
	( 34.10, 82.27) --
	( 34.10, 82.27) --
	( 34.10, 82.27) --
	( 34.56, 82.27) --
	( 34.56, 82.27) --
	( 34.56, 82.27) --
	( 34.56, 82.27) --
	( 34.56, 82.27) --
	( 35.01, 82.27) --
	( 35.01, 82.27) --
	( 35.01, 82.27) --
	( 35.01, 82.27) --
	( 35.01, 82.27) --
	( 35.47, 82.27) --
	( 35.47, 82.27) --
	( 35.47, 82.27) --
	( 35.47, 82.27) --
	( 35.47, 82.27) --
	( 35.92, 82.27) --
	( 35.92, 82.27) --
	( 35.92, 82.27) --
	( 35.92, 82.27) --
	( 35.92, 82.27) --
	( 36.38, 82.27) --
	( 36.38, 82.27) --
	( 36.38, 82.27) --
	( 36.38, 82.27) --
	( 36.38, 82.27) --
	( 36.83, 82.27) --
	( 36.83, 82.27) --
	( 36.83, 82.27) --
	( 36.83, 82.27) --
	( 36.83, 82.27) --
	( 37.29, 82.27) --
	( 37.29, 82.27) --
	( 37.29, 82.27) --
	( 37.29, 82.27) --
	( 37.29, 82.27) --
	( 37.74, 82.27) --
	( 37.74, 82.27) --
	( 37.74, 82.27) --
	( 37.74, 82.27) --
	( 37.74, 82.27) --
	( 38.20, 82.27) --
	( 38.20, 82.27) --
	( 38.20, 82.27) --
	( 38.20, 82.27) --
	( 38.20, 82.27) --
	( 38.65, 82.27) --
	( 38.65, 82.27) --
	( 38.65, 82.27) --
	( 38.65, 82.27) --
	( 38.65, 82.27) --
	( 39.11, 82.27) --
	( 39.11, 82.27) --
	( 39.11, 82.27) --
	( 39.11, 82.27) --
	( 39.11, 82.27) --
	( 39.56, 82.27) --
	( 39.56, 82.27) --
	( 39.56, 82.27) --
	( 39.56, 82.27) --
	( 39.56, 82.27) --
	( 40.02, 82.27) --
	( 40.02, 82.27) --
	( 40.02, 82.27) --
	( 40.02, 82.27) --
	( 40.02, 82.27) --
	( 40.47, 82.27) --
	( 40.47, 82.27) --
	( 40.47, 82.27) --
	( 40.47, 82.27) --
	( 40.47, 82.27) --
	( 40.93, 82.27) --
	( 40.93, 82.27) --
	( 40.93, 82.27) --
	( 40.93, 82.27) --
	( 40.93, 82.27) --
	( 41.38, 82.27) --
	( 41.38, 82.27) --
	( 41.38, 82.27) --
	( 41.38, 82.27) --
	( 41.38, 82.27) --
	( 41.84, 82.27) --
	( 41.84, 82.27) --
	( 41.84, 82.27) --
	( 41.84, 82.27) --
	( 41.84, 82.27) --
	( 42.30, 82.27) --
	( 42.30, 82.27) --
	( 42.30, 82.27) --
	( 42.30, 82.27) --
	( 42.30, 82.27) --
	( 42.75, 82.27) --
	( 42.75, 82.27) --
	( 42.75, 82.27) --
	( 42.75, 82.27) --
	( 42.75, 82.27) --
	( 43.21, 82.27) --
	( 43.21, 82.27) --
	( 43.21, 82.27) --
	( 43.21, 82.27) --
	( 43.21, 82.27) --
	( 43.66, 82.27) --
	( 43.66, 82.27) --
	( 43.66, 82.27) --
	( 43.66, 82.27) --
	( 43.66, 82.27) --
	( 44.12, 82.27) --
	( 44.12, 82.27) --
	( 44.12, 82.27) --
	( 44.12, 82.27) --
	( 44.12, 82.27) --
	( 44.57, 82.27) --
	( 44.57, 82.27) --
	( 44.57, 82.27) --
	( 44.57, 82.27) --
	( 44.57, 82.27) --
	( 45.03, 82.27) --
	( 45.03, 82.27) --
	( 45.03, 82.27) --
	( 45.03, 82.27) --
	( 45.03, 82.27) --
	( 45.48, 82.27) --
	( 45.48, 82.27) --
	( 45.48, 82.27) --
	( 45.48, 82.27) --
	( 45.48, 82.27) --
	( 45.94, 82.27) --
	( 45.94, 82.27) --
	( 45.94, 82.27) --
	( 45.94, 82.27) --
	( 45.94, 82.27) --
	( 46.39, 82.27) --
	( 46.39, 82.27) --
	( 46.39, 82.27) --
	( 46.39, 82.27) --
	( 46.39, 82.27) --
	( 46.85, 82.27) --
	( 46.85, 82.27) --
	( 46.85, 82.27) --
	( 46.85, 82.27) --
	( 46.85, 82.27) --
	( 47.30, 82.27) --
	( 47.30, 82.27) --
	( 47.30, 82.27) --
	( 47.30, 82.27) --
	( 47.30, 82.27) --
	( 47.76, 82.27) --
	( 47.76, 82.27) --
	( 47.76, 82.27) --
	( 47.76, 82.27) --
	( 47.76, 82.27) --
	( 48.21, 82.27) --
	( 48.21, 82.27) --
	( 48.21, 82.27) --
	( 48.21, 82.27) --
	( 48.21, 82.27) --
	( 48.67, 82.27) --
	( 48.67, 82.27) --
	( 48.67, 82.27) --
	( 48.67, 82.27) --
	( 48.67, 82.27) --
	( 49.12, 82.27) --
	( 49.12, 82.27) --
	( 49.12, 82.27) --
	( 49.12, 82.27) --
	( 49.12, 82.27) --
	( 49.58, 82.27) --
	( 49.58, 82.27) --
	( 49.58, 82.27) --
	( 49.58, 82.27) --
	( 49.58, 82.27) --
	( 50.03, 82.27) --
	( 50.03, 82.27) --
	( 50.03, 82.27) --
	( 50.03, 82.27) --
	( 50.03, 82.27) --
	( 50.49, 82.27) --
	( 50.49, 82.27) --
	( 50.49, 82.27) --
	( 50.49, 82.27) --
	( 50.49, 82.27) --
	( 50.94, 82.27) --
	( 50.94, 82.27) --
	( 50.94, 82.27) --
	( 50.94, 82.27) --
	( 50.94, 82.27) --
	( 51.40, 82.27) --
	( 51.40, 82.27) --
	( 51.40, 82.27) --
	( 51.40, 82.27) --
	( 51.40, 82.27) --
	( 51.85, 82.27) --
	( 51.85, 82.27) --
	( 51.85, 82.27) --
	( 51.85, 82.27) --
	( 51.85, 82.27) --
	( 52.31, 82.27) --
	( 52.31, 82.27) --
	( 52.31, 82.27) --
	( 52.31, 82.27) --
	( 52.31, 82.27) --
	( 52.76, 82.27) --
	( 52.76, 82.27) --
	( 52.76, 82.27) --
	( 52.76, 82.27) --
	( 52.76, 82.27) --
	( 53.22, 82.27) --
	( 53.22, 82.27) --
	( 53.22, 82.27) --
	( 53.22, 82.27) --
	( 53.22, 82.27) --
	( 53.68, 82.27) --
	( 53.68, 82.27) --
	( 53.68, 82.27) --
	( 53.68, 82.27) --
	( 53.68, 82.27) --
	( 54.13, 82.27) --
	( 54.13, 82.27) --
	( 54.13, 82.27) --
	( 54.13, 82.27) --
	( 54.13, 82.27) --
	( 54.59, 82.27) --
	( 54.59, 82.27) --
	( 54.59, 82.27) --
	( 54.59, 82.27) --
	( 54.59, 82.27) --
	( 55.04, 82.27) --
	( 55.04, 82.27) --
	( 55.04, 82.27) --
	( 55.04, 82.27) --
	( 55.04, 82.27) --
	( 55.50, 82.27) --
	( 55.50, 82.27) --
	( 55.50, 82.27) --
	( 55.50, 82.27) --
	( 55.50, 82.27) --
	( 55.95, 82.27) --
	( 55.95, 82.27) --
	( 55.95, 82.27) --
	( 55.95, 82.27) --
	( 55.95, 82.27) --
	( 56.41, 82.27) --
	( 56.41, 82.27) --
	( 56.41, 82.27) --
	( 56.41, 82.27) --
	( 56.41, 82.27) --
	( 56.86, 82.27) --
	( 56.86, 82.27) --
	( 56.86, 82.27) --
	( 56.86, 82.27) --
	( 56.86, 82.27) --
	( 57.32, 82.27) --
	( 57.32, 82.27) --
	( 57.32, 82.27) --
	( 57.32, 82.27) --
	( 57.32, 82.27) --
	( 57.77, 82.27) --
	( 57.77, 82.27) --
	( 57.77, 82.27) --
	( 57.77, 82.27) --
	( 57.77, 82.27) --
	( 58.23, 82.27) --
	( 58.23, 82.27) --
	( 58.23, 82.27) --
	( 58.23, 82.27) --
	( 58.23, 82.27) --
	( 58.68, 82.27) --
	( 58.68, 82.27) --
	( 58.68, 82.27) --
	( 58.68, 82.27) --
	( 58.68, 82.27) --
	( 59.14, 82.27) --
	( 59.14, 82.27) --
	( 59.14, 82.27) --
	( 59.14, 82.27) --
	( 59.14, 82.27) --
	( 59.59, 82.27) --
	( 59.59, 82.27) --
	( 59.59, 82.27) --
	( 59.59, 82.27) --
	( 59.59, 82.27) --
	( 60.05, 82.27) --
	( 60.05, 82.27) --
	( 60.05, 82.27) --
	( 60.05, 82.27) --
	( 60.05, 82.27) --
	( 60.50, 82.27) --
	( 60.50, 82.27) --
	( 60.50, 82.27) --
	( 60.50, 82.27) --
	( 60.50, 82.27) --
	( 60.96, 82.27) --
	( 60.96, 82.27) --
	( 60.96, 82.27) --
	( 60.96, 82.27) --
	( 60.96, 82.27) --
	( 61.41, 82.27) --
	( 61.41, 82.27) --
	( 61.41, 82.27) --
	( 61.41, 82.27) --
	( 61.41, 82.27) --
	( 61.87, 82.27) --
	( 61.87, 82.27) --
	( 61.87, 82.27) --
	( 61.87, 82.27) --
	( 61.87, 82.27) --
	( 62.32, 82.27) --
	( 62.32, 82.27) --
	( 62.32, 82.27) --
	( 62.32, 82.27) --
	( 62.32, 82.27) --
	( 62.78, 82.27) --
	( 62.78, 82.27) --
	( 62.78, 82.27) --
	( 62.78, 82.27) --
	( 62.78, 82.27) --
	( 63.23, 82.27) --
	( 63.23, 82.27) --
	( 63.23, 82.27) --
	( 63.23, 82.27) --
	( 63.23, 82.27) --
	( 63.69, 82.27) --
	( 63.69, 82.27) --
	( 63.69, 82.27) --
	( 63.69, 82.27) --
	( 63.69, 82.27) --
	( 64.14, 82.27) --
	( 64.14, 82.27) --
	( 64.14, 82.27) --
	( 64.14, 82.27) --
	( 64.14, 82.27) --
	( 64.60, 82.27) --
	( 64.60, 82.27) --
	( 64.60, 82.27) --
	( 64.60, 82.27) --
	( 64.60, 82.27) --
	( 65.06, 82.27) --
	( 65.06, 82.27) --
	( 65.06, 82.27) --
	( 65.06, 82.27) --
	( 65.06, 82.27) --
	( 65.51, 82.27) --
	( 65.51, 82.27) --
	( 65.51, 82.27) --
	( 65.51, 82.27) --
	( 65.51, 82.27) --
	( 65.97, 82.27) --
	( 65.97, 82.27) --
	( 65.97, 82.27) --
	( 65.97, 82.27) --
	( 65.97, 82.27) --
	( 66.42, 82.27) --
	( 66.42, 82.27) --
	( 66.42, 82.27) --
	( 66.42, 82.27) --
	( 66.42, 82.27) --
	( 66.88, 82.27) --
	( 66.88, 82.27) --
	( 66.88, 82.27) --
	( 66.88, 82.27) --
	( 66.88, 82.27) --
	( 67.33, 82.27) --
	( 67.33, 82.27) --
	( 67.33, 82.27) --
	( 67.33, 82.27) --
	( 67.33, 82.27) --
	( 67.79, 82.27) --
	( 67.79, 82.27) --
	( 67.79, 82.27) --
	( 67.79, 82.27) --
	( 67.79, 82.27) --
	( 68.24, 82.27) --
	( 68.24, 82.27) --
	( 68.24, 82.27) --
	( 68.24, 82.27) --
	( 68.24, 82.27) --
	( 68.70, 82.27) --
	( 68.70, 82.27) --
	( 68.70, 82.27) --
	( 68.70, 82.27) --
	( 68.70, 82.27) --
	( 69.15, 82.27) --
	( 69.15, 82.27) --
	( 69.15, 82.27) --
	( 69.15, 82.27) --
	( 69.15, 82.27) --
	( 69.61, 82.27) --
	( 69.61, 82.27) --
	( 69.61, 82.27) --
	( 69.61, 82.27) --
	( 69.61, 82.27) --
	( 70.06, 82.27) --
	( 70.06, 82.27) --
	( 70.06, 82.27) --
	( 70.06, 82.27) --
	( 70.06, 82.27) --
	( 70.52, 82.27) --
	( 70.52, 82.27) --
	( 70.52, 82.27) --
	( 70.52, 82.27) --
	( 70.52, 82.27) --
	( 70.97, 82.27) --
	( 70.97, 82.27) --
	( 70.97, 82.27) --
	( 70.97, 82.27) --
	( 70.97, 82.27) --
	( 71.43, 82.27) --
	( 71.43, 82.27) --
	( 71.43, 82.27) --
	( 71.43, 82.27) --
	( 71.43, 82.27) --
	( 71.88, 82.27) --
	( 71.88, 82.27) --
	( 71.88, 82.27) --
	( 71.88, 82.27) --
	( 71.88, 82.27) --
	( 72.34, 82.27) --
	( 72.34, 82.27) --
	( 72.34, 82.27) --
	( 72.34, 82.27) --
	( 72.34, 82.27) --
	( 72.79, 82.27) --
	( 72.79, 82.27) --
	( 72.79, 82.27) --
	( 72.79, 82.27) --
	( 72.79, 82.27) --
	( 73.25, 82.27) --
	( 73.25, 82.27) --
	( 73.25, 82.27) --
	( 73.25, 82.27) --
	( 73.25, 82.27) --
	( 73.70, 82.27) --
	( 73.70, 82.27) --
	( 73.70, 82.27) --
	( 73.70, 82.27) --
	( 73.70, 82.27) --
	( 74.16, 82.27) --
	( 74.16, 82.27) --
	( 74.16, 82.27) --
	( 74.16, 82.27) --
	( 74.16, 82.27) --
	( 74.61, 82.27) --
	( 74.61, 82.27) --
	( 74.61, 82.27) --
	( 74.61, 82.27) --
	( 74.61, 82.27) --
	( 75.07, 82.27) --
	( 75.07, 82.27) --
	( 75.07, 82.27) --
	( 75.07, 82.27) --
	( 75.07, 82.27) --
	( 75.53, 82.27) --
	( 75.53, 82.27) --
	( 75.53, 82.27) --
	( 75.53, 82.27) --
	( 75.53, 82.27) --
	( 75.98, 82.27) --
	( 75.98, 82.27) --
	( 75.98, 82.27) --
	( 75.98, 82.27) --
	( 75.98, 82.27) --
	( 76.44, 82.27) --
	( 76.44, 82.27) --
	( 76.44, 82.27) --
	( 76.44, 82.27) --
	( 76.44, 82.27) --
	( 76.89, 82.27) --
	( 76.89, 82.27) --
	( 76.89, 82.27) --
	( 76.89, 82.27) --
	( 76.89, 82.27) --
	( 77.35, 82.27) --
	( 77.35, 82.27) --
	( 77.35, 82.27) --
	( 77.35, 82.27) --
	( 77.35, 82.27) --
	( 77.80, 82.27) --
	( 77.80, 82.27) --
	( 77.80, 82.27) --
	( 77.80, 82.27) --
	( 77.80, 82.27) --
	( 78.26, 82.27) --
	( 78.26, 82.27) --
	( 78.26, 82.27) --
	( 78.26, 82.27) --
	( 78.26, 82.27) --
	( 78.71, 82.27) --
	( 78.71, 82.27) --
	( 78.71, 82.27) --
	( 78.71, 82.27) --
	( 78.71, 82.27) --
	( 79.17, 82.27) --
	( 79.17, 82.27) --
	( 79.17, 82.27) --
	( 79.17, 82.27) --
	( 79.17, 82.27) --
	( 79.62, 82.27) --
	( 79.62, 82.27) --
	( 79.62, 82.27) --
	( 79.62, 82.27) --
	( 79.62, 82.27) --
	( 80.08, 82.27) --
	( 80.08, 82.27) --
	( 80.08, 82.27) --
	( 80.08, 82.27) --
	( 80.08, 82.27) --
	( 80.53, 82.27) --
	( 80.53, 82.27) --
	( 80.53, 82.27) --
	( 80.53, 82.27) --
	( 80.53, 82.27) --
	( 80.99, 82.27) --
	( 80.99, 82.27) --
	( 80.99, 82.27) --
	( 80.99, 82.27) --
	( 80.99, 82.27) --
	( 81.44, 82.27) --
	( 81.44, 82.27) --
	( 81.44, 82.27) --
	( 81.44, 82.27) --
	( 81.44, 82.27) --
	( 81.90, 82.27) --
	( 81.90, 82.27) --
	( 81.90, 82.27) --
	( 81.90, 82.27) --
	( 81.90, 82.27) --
	( 82.35, 82.27) --
	( 82.35, 82.27) --
	( 82.35, 82.27) --
	( 82.35, 82.27) --
	( 82.35, 82.27) --
	( 82.81, 82.27) --
	( 82.81, 82.27) --
	( 82.81, 82.27) --
	( 82.81, 82.27) --
	( 82.81, 82.27) --
	( 83.26, 82.27) --
	( 83.26, 82.27) --
	( 83.26, 82.27) --
	( 83.26, 82.27) --
	( 83.26, 82.27) --
	( 83.72, 82.27) --
	( 83.72, 82.27) --
	( 83.72, 82.27) --
	( 83.72, 82.27) --
	( 83.72, 82.27) --
	( 84.17, 82.27) --
	( 84.17, 82.27) --
	( 84.17, 82.27) --
	( 84.17, 82.27) --
	( 84.17, 82.27) --
	( 84.63, 82.27) --
	( 84.63, 82.27) --
	( 84.63, 82.27) --
	( 84.63, 82.27) --
	( 84.63, 82.27) --
	( 85.08, 82.27) --
	( 85.08, 82.27) --
	( 85.08, 82.27) --
	( 85.08, 82.27) --
	( 85.08, 82.27) --
	( 85.54, 82.27) --
	( 85.54, 82.27) --
	( 85.54, 82.27) --
	( 85.54, 82.27) --
	( 85.54, 82.27) --
	( 85.99, 82.27) --
	( 85.99, 82.27) --
	( 85.99, 82.27) --
	( 85.99, 82.27) --
	( 85.99, 82.27) --
	( 86.45, 82.27) --
	( 86.45, 82.27) --
	( 86.45, 82.27) --
	( 86.45, 82.27) --
	( 86.45, 82.27) --
	( 86.91, 82.27) --
	( 86.91, 82.27) --
	( 86.91, 82.27) --
	( 86.91, 82.27) --
	( 86.91, 82.27) --
	( 87.36, 82.27) --
	( 87.36, 82.27) --
	( 87.36, 82.27) --
	( 87.36, 82.27) --
	( 87.36, 82.27) --
	( 87.82, 82.27) --
	( 87.82, 82.27) --
	( 87.82, 82.27) --
	( 87.82, 82.27) --
	( 87.82, 82.27) --
	( 88.27, 82.27) --
	( 88.27, 82.27) --
	( 88.27, 82.27) --
	( 88.27, 82.27) --
	( 88.27, 82.27) --
	( 88.73, 82.27) --
	( 88.73, 82.27) --
	( 88.73, 82.27) --
	( 88.73, 82.27) --
	( 88.73, 82.27) --
	( 89.18, 82.27) --
	( 89.18, 82.27) --
	( 89.18, 82.27) --
	( 89.18, 82.27) --
	( 89.18, 82.27) --
	( 89.64, 82.27) --
	( 89.64, 82.27) --
	( 89.64, 82.27) --
	( 89.64, 82.27) --
	( 89.64, 82.27) --
	( 90.09, 82.27) --
	( 90.09, 82.27) --
	( 90.09, 82.27) --
	( 90.09, 82.27) --
	( 90.09, 82.27) --
	( 90.55, 82.27) --
	( 90.55, 82.27) --
	( 90.55, 82.27) --
	( 90.55, 82.27) --
	( 90.55, 82.27) --
	( 91.00, 82.27) --
	( 91.00, 82.27) --
	( 91.00, 82.27) --
	( 91.00, 82.27) --
	( 91.00, 82.27) --
	( 91.46, 82.27) --
	( 91.46, 82.27) --
	( 91.46, 82.27) --
	( 91.46, 82.27) --
	( 91.46, 82.27) --
	( 91.91, 82.27) --
	( 91.91, 82.27) --
	( 91.91, 82.27) --
	( 91.91, 82.27) --
	( 91.91, 82.27) --
	( 92.37, 82.27) --
	( 92.37, 82.27) --
	( 92.37, 82.27) --
	( 92.37, 82.27) --
	( 92.37, 82.27) --
	( 92.82, 82.27) --
	( 92.82, 82.27) --
	( 92.82, 82.27) --
	( 92.82, 82.27) --
	( 92.82, 82.27) --
	( 93.28, 82.27) --
	( 93.28, 82.27) --
	( 93.28, 82.27) --
	( 93.28, 82.27) --
	( 93.28, 82.27) --
	( 93.73, 82.27) --
	( 93.73, 82.27) --
	( 93.73, 82.27) --
	( 93.73, 82.27) --
	( 93.73, 82.27) --
	( 94.19, 82.27) --
	( 94.19, 82.27) --
	( 94.19, 82.27) --
	( 94.19, 82.27) --
	( 94.19, 82.27) --
	( 94.64, 82.27) --
	( 94.64, 82.27) --
	( 94.64, 82.27) --
	( 94.64, 82.27) --
	( 94.64, 82.27) --
	( 95.10, 82.27) --
	( 95.10, 82.27) --
	( 95.10, 82.27) --
	( 95.10, 82.27) --
	( 95.10, 82.27) --
	( 95.55, 82.27) --
	( 95.55, 82.27) --
	( 95.55, 82.27) --
	( 95.55, 82.27) --
	( 95.55, 82.27) --
	( 96.01, 82.27) --
	( 96.01, 82.27) --
	( 96.01, 82.27) --
	( 96.01, 82.27) --
	( 96.01, 82.27) --
	( 96.46, 82.27) --
	( 96.46, 82.27) --
	( 96.46, 82.27) --
	( 96.46, 82.27) --
	( 96.46, 82.27) --
	( 96.92, 82.27) --
	( 96.92, 82.27) --
	( 96.92, 82.27) --
	( 96.92, 82.27) --
	( 96.92, 82.27) --
	( 97.37, 82.27) --
	( 97.37, 82.27) --
	( 97.37, 82.27) --
	( 97.37, 82.27) --
	( 97.37, 82.27) --
	( 97.83, 82.27) --
	( 97.83, 82.27) --
	( 97.83, 82.27) --
	( 97.83, 82.27) --
	( 97.83, 82.27) --
	( 98.29, 82.27) --
	( 98.29, 82.27) --
	( 98.29, 82.27) --
	( 98.29, 82.27) --
	( 98.29, 82.27) --
	( 98.74, 82.27) --
	( 98.74, 82.27) --
	( 98.74, 82.27) --
	( 98.74, 82.27) --
	( 98.74, 82.27) --
	( 99.20, 82.27) --
	( 99.20, 82.27) --
	( 99.20, 82.27) --
	( 99.20, 82.27) --
	( 99.20, 82.27) --
	( 99.65, 82.27) --
	( 99.65, 82.27) --
	( 99.65, 82.27) --
	( 99.65, 82.27) --
	( 99.65, 82.27) --
	(100.11, 82.27) --
	(100.11, 82.27) --
	(100.11, 82.27) --
	(100.11, 82.27) --
	(100.11, 82.27) --
	(100.56, 82.27) --
	(100.56, 82.27) --
	(100.56, 82.27) --
	(100.56, 82.27) --
	(100.56, 82.27) --
	(101.02, 82.27) --
	(101.02, 82.27) --
	(101.02, 82.27) --
	(101.02, 82.27) --
	(101.02, 82.27) --
	(101.47, 82.27) --
	(101.47, 82.27) --
	(101.47, 82.27) --
	(101.47, 82.27) --
	(101.47, 82.27) --
	(101.93, 82.27) --
	(101.93, 82.27) --
	(101.93, 82.27) --
	(101.93, 82.27) --
	(101.93, 82.27) --
	(102.38, 82.27) --
	(102.38, 82.27) --
	(102.38, 82.27) --
	(102.38, 82.27) --
	(102.38, 82.27) --
	(102.84, 82.27) --
	(102.84, 82.27) --
	(102.84, 82.27) --
	(102.84, 82.27) --
	(102.84, 82.27) --
	(103.29, 82.27) --
	(103.29, 82.27) --
	(103.29, 82.27) --
	(103.29, 82.27) --
	(103.29, 82.27) --
	(103.75, 82.27) --
	(103.75, 82.27) --
	(103.75, 82.27) --
	(103.75, 82.27) --
	(103.75, 82.27) --
	(104.20, 82.27) --
	(104.20, 82.27) --
	(104.20, 82.27) --
	(104.20, 82.27) --
	(104.20, 82.27) --
	(104.66, 82.27) --
	(104.66, 82.27) --
	(104.66, 82.27) --
	(104.66, 82.27) --
	(104.66, 82.27) --
	(105.11, 82.27) --
	(105.11, 82.27) --
	(105.11, 82.27) --
	(105.11, 82.27) --
	(105.11, 82.27) --
	(105.57, 82.27) --
	(105.57, 82.27) --
	(105.57, 82.27) --
	(105.57, 82.27) --
	(105.57, 82.27) --
	(106.02, 82.27) --
	(106.02, 82.27) --
	(106.02, 82.27) --
	(106.02, 82.27) --
	(106.02, 82.27) --
	(106.48, 82.27) --
	(106.48, 82.27) --
	(106.48, 82.27) --
	(106.48, 82.27) --
	(106.48, 82.27) --
	(106.93, 82.27) --
	(106.93, 82.27) --
	(106.93, 82.27) --
	(106.93, 82.27) --
	(106.93, 82.27) --
	(107.39, 82.27) --
	(107.39, 82.27) --
	(107.39, 82.27) --
	(107.39, 82.27) --
	(107.39, 82.27) --
	(107.84, 82.27) --
	(107.84, 82.27) --
	(107.84, 82.27) --
	(107.84, 82.27) --
	(107.84, 82.27) --
	(108.30, 82.27) --
	(108.30, 82.27) --
	(108.30, 82.27) --
	(108.30, 82.27) --
	(108.30, 82.27) --
	(108.75, 82.27) --
	(108.75, 82.27) --
	(108.75, 82.27) --
	(108.75, 82.27) --
	(108.75, 82.27) --
	(109.21, 82.27) --
	(109.21, 82.27) --
	(109.21, 82.27) --
	(109.21, 82.27) --
	(109.21, 82.27) --
	(109.67, 82.27) --
	(109.67, 82.27) --
	(109.67, 82.27) --
	(109.67, 82.27) --
	(109.67, 82.27) --
	(110.12, 82.27) --
	(110.12, 82.27) --
	(110.12, 82.27) --
	(110.12, 82.27) --
	(110.12, 82.27) --
	(110.58, 82.27) --
	(110.58, 82.27) --
	(110.58, 82.27) --
	(110.58, 82.27) --
	(110.58, 82.27) --
	(111.03, 82.27) --
	(111.03, 82.27) --
	(111.03, 82.27) --
	(111.03, 82.27) --
	(111.03, 82.27) --
	(111.49, 82.27) --
	(111.49, 82.27) --
	(111.49, 82.27) --
	(111.49, 82.27) --
	(111.49, 82.27) --
	(111.94, 82.27) --
	(111.94, 82.27) --
	(111.94, 82.27) --
	(111.94, 82.27) --
	(111.94, 82.27) --
	(112.10, 82.27) --
	(112.40, 82.27) --
	(112.40, 82.27) --
	(112.40, 82.27) --
	(112.40, 82.27) --
	(112.85, 82.27) --
	(112.85, 82.27) --
	(112.85, 82.27) --
	(112.85, 82.27) --
	(113.31, 82.27) --
	(113.31, 82.27) --
	(113.31, 82.27) --
	(113.31, 82.27) --
	(113.76, 82.27) --
	(113.76, 82.27) --
	(113.76, 82.27) --
	(113.76, 82.27) --
	(114.22, 82.27) --
	(114.22, 82.27) --
	(114.22, 82.27) --
	(114.22, 82.27) --
	(114.67, 82.27) --
	(114.67, 82.27) --
	(114.67, 82.27) --
	(114.67, 82.27) --
	(115.13, 82.27) --
	(115.13, 82.27) --
	(115.13, 82.27) --
	(115.13, 82.27) --
	(115.58, 82.27) --
	(115.58, 82.27) --
	(115.58, 82.27) --
	(115.58, 82.27) --
	(116.04, 82.27) --
	(116.04, 82.27) --
	(116.04, 82.27) --
	(116.04, 82.27) --
	(116.49, 82.27) --
	(116.49, 82.27) --
	(116.49, 82.27) --
	(116.49, 82.27) --
	(116.66, 82.27) --
	(116.95, 82.27) --
	(116.95, 82.27) --
	(116.95, 82.27) --
	(117.40, 82.27) --
	(117.40, 82.27) --
	(117.40, 82.27) --
	(117.46, 82.27) --
	(117.86, 82.27) --
	(117.86, 82.27);
\definecolor{drawColor}{gray}{0.70}

\path[draw=drawColor,line width= 0.5pt,line join=round,line cap=round] ( 27.27, 29.68) rectangle (117.86,194.47);
\end{scope}
\begin{scope}
\path[clip] (  0.00,  0.00) rectangle (122.86,199.47);
\definecolor{drawColor}{gray}{0.30}

\node[text=drawColor,anchor=base east,inner sep=0pt, outer sep=0pt, scale=  1.00] at ( 22.77, 43.76) {0.99};

\node[text=drawColor,anchor=base east,inner sep=0pt, outer sep=0pt, scale=  1.00] at ( 22.77, 78.82) {1.00};

\node[text=drawColor,anchor=base east,inner sep=0pt, outer sep=0pt, scale=  1.00] at ( 22.77,113.89) {1.01};

\node[text=drawColor,anchor=base east,inner sep=0pt, outer sep=0pt, scale=  1.00] at ( 22.77,148.95) {1.02};

\node[text=drawColor,anchor=base east,inner sep=0pt, outer sep=0pt, scale=  1.00] at ( 22.77,184.01) {1.03};
\end{scope}
\begin{scope}
\path[clip] (  0.00,  0.00) rectangle (122.86,199.47);
\definecolor{drawColor}{gray}{0.70}

\path[draw=drawColor,line width= 0.3pt,line join=round] ( 24.77, 47.21) --
	( 27.27, 47.21);

\path[draw=drawColor,line width= 0.3pt,line join=round] ( 24.77, 82.27) --
	( 27.27, 82.27);

\path[draw=drawColor,line width= 0.3pt,line join=round] ( 24.77,117.33) --
	( 27.27,117.33);

\path[draw=drawColor,line width= 0.3pt,line join=round] ( 24.77,152.39) --
	( 27.27,152.39);

\path[draw=drawColor,line width= 0.3pt,line join=round] ( 24.77,187.45) --
	( 27.27,187.45);
\end{scope}
\begin{scope}
\path[clip] (  0.00,  0.00) rectangle (122.86,199.47);
\definecolor{drawColor}{gray}{0.70}

\path[draw=drawColor,line width= 0.3pt,line join=round] ( 27.27, 27.18) --
	( 27.27, 29.68);

\path[draw=drawColor,line width= 0.3pt,line join=round] ( 45.39, 27.18) --
	( 45.39, 29.68);

\path[draw=drawColor,line width= 0.3pt,line join=round] ( 63.51, 27.18) --
	( 63.51, 29.68);

\path[draw=drawColor,line width= 0.3pt,line join=round] ( 81.62, 27.18) --
	( 81.62, 29.68);

\path[draw=drawColor,line width= 0.3pt,line join=round] ( 99.74, 27.18) --
	( 99.74, 29.68);

\path[draw=drawColor,line width= 0.3pt,line join=round] (117.86, 27.18) --
	(117.86, 29.68);
\end{scope}
\begin{scope}
\path[clip] (  0.00,  0.00) rectangle (122.86,199.47);
\definecolor{drawColor}{gray}{0.30}

\node[text=drawColor,anchor=base,inner sep=0pt, outer sep=0pt, scale=  1.00] at ( 26.27, 18.29) {0};

\node[text=drawColor,anchor=base,inner sep=0pt, outer sep=0pt, scale=  1.00] at ( 43.39, 18.29) {20};

\node[text=drawColor,anchor=base,inner sep=0pt, outer sep=0pt, scale=  1.00] at ( 61.51, 18.29) {40};

\node[text=drawColor,anchor=base,inner sep=0pt, outer sep=0pt, scale=  1.00] at ( 79.63, 18.29) {60};

\node[text=drawColor,anchor=base,inner sep=0pt, outer sep=0pt, scale=  1.00] at ( 97.74, 18.29) {80};

\node[text=drawColor,anchor=base,inner sep=0pt, outer sep=0pt, scale=  1.00] at (114.86, 18.29) {100};
\end{scope}
\begin{scope}
\path[clip] (  0.00,  0.00) rectangle (122.86,199.47);
\definecolor{drawColor}{RGB}{0,0,0}

\node[text=drawColor,anchor=base,inner sep=0pt, outer sep=0pt, scale=  1.00] at ( 72.57,  6.94) {\bfseries Instances in \%};
\end{scope}
\begin{scope}
\path[clip] (  0.00,  0.00) rectangle (122.86,199.47);
\definecolor{drawColor}{RGB}{0,0,0}

\path[draw=drawColor,line width= 0.4pt,line join=round,line cap=round] ( 29.54,110.05) rectangle ( 64.14,192.32);
\end{scope}
\begin{scope}
\path[clip] (  0.00,  0.00) rectangle (122.86,199.47);
\definecolor{fillColor}{RGB}{255,255,255}

\path[fill=fillColor] ( 29.54,110.05) rectangle ( 64.14,192.32);
\end{scope}
\begin{scope}
\path[clip] (  0.00,  0.00) rectangle (122.86,199.47);
\definecolor{fillColor}{RGB}{255,255,255}

\path[fill=fillColor] ( 34.54,172.87) rectangle ( 48.99,187.32);
\end{scope}
\begin{scope}
\path[clip] (  0.00,  0.00) rectangle (122.86,199.47);
\definecolor{drawColor}{RGB}{128,128,128}

\path[draw=drawColor,line width= 1.0pt,line join=round] ( 35.98,180.10) -- ( 47.55,180.10);
\end{scope}
\begin{scope}
\path[clip] (  0.00,  0.00) rectangle (122.86,199.47);
\definecolor{fillColor}{RGB}{255,255,255}

\path[fill=fillColor] ( 34.54,158.41) rectangle ( 48.99,172.87);
\end{scope}
\begin{scope}
\path[clip] (  0.00,  0.00) rectangle (122.86,199.47);
\definecolor{drawColor}{RGB}{152,167,197}

\path[draw=drawColor,line width= 1.0pt,line join=round] ( 35.98,165.64) -- ( 47.55,165.64);
\end{scope}
\begin{scope}
\path[clip] (  0.00,  0.00) rectangle (122.86,199.47);
\definecolor{fillColor}{RGB}{255,255,255}

\path[fill=fillColor] ( 34.54,143.96) rectangle ( 48.99,158.41);
\end{scope}
\begin{scope}
\path[clip] (  0.00,  0.00) rectangle (122.86,199.47);
\definecolor{drawColor}{RGB}{128,202,192}

\path[draw=drawColor,line width= 1.0pt,line join=round] ( 35.98,151.19) -- ( 47.55,151.19);
\end{scope}
\begin{scope}
\path[clip] (  0.00,  0.00) rectangle (122.86,199.47);
\definecolor{fillColor}{RGB}{255,255,255}

\path[fill=fillColor] ( 34.54,129.51) rectangle ( 48.99,143.96);
\end{scope}
\begin{scope}
\path[clip] (  0.00,  0.00) rectangle (122.86,199.47);
\definecolor{drawColor}{RGB}{207,128,187}

\path[draw=drawColor,line width= 1.0pt,line join=round] ( 35.98,136.73) -- ( 47.55,136.73);
\end{scope}
\begin{scope}
\path[clip] (  0.00,  0.00) rectangle (122.86,199.47);
\definecolor{fillColor}{RGB}{255,255,255}

\path[fill=fillColor] ( 34.54,115.05) rectangle ( 48.99,129.51);
\end{scope}
\begin{scope}
\path[clip] (  0.00,  0.00) rectangle (122.86,199.47);
\definecolor{drawColor}{RGB}{207,192,152}

\path[draw=drawColor,line width= 1.0pt,line join=round] ( 35.98,122.28) -- ( 47.55,122.28);
\end{scope}
\begin{scope}
\path[clip] (  0.00,  0.00) rectangle (122.86,199.47);
\definecolor{drawColor}{RGB}{0,0,0}

\node[text=drawColor,anchor=base west,inner sep=0pt, outer sep=0pt, scale=  0.80] at ( 48.99,177.34) {$\ensuremath{\mathcal{N}}_1$};
\end{scope}
\begin{scope}
\path[clip] (  0.00,  0.00) rectangle (122.86,199.47);
\definecolor{drawColor}{RGB}{0,0,0}

\node[text=drawColor,anchor=base west,inner sep=0pt, outer sep=0pt, scale=  0.80] at ( 48.99,162.89) {$\ensuremath{\mathcal{N}}_2$};
\end{scope}
\begin{scope}
\path[clip] (  0.00,  0.00) rectangle (122.86,199.47);
\definecolor{drawColor}{RGB}{0,0,0}

\node[text=drawColor,anchor=base west,inner sep=0pt, outer sep=0pt, scale=  0.80] at ( 48.99,148.43) {$\ensuremath{\mathcal{N}}_3$};
\end{scope}
\begin{scope}
\path[clip] (  0.00,  0.00) rectangle (122.86,199.47);
\definecolor{drawColor}{RGB}{0,0,0}

\node[text=drawColor,anchor=base west,inner sep=0pt, outer sep=0pt, scale=  0.80] at ( 48.99,133.98) {$\ensuremath{\mathcal{N}}_4$};
\end{scope}
\begin{scope}
\path[clip] (  0.00,  0.00) rectangle (122.86,199.47);
\definecolor{drawColor}{RGB}{0,0,0}

\node[text=drawColor,anchor=base west,inner sep=0pt, outer sep=0pt, scale=  0.80] at ( 48.99,119.53) {$\ensuremath{\mathcal{N}}_5$};
\end{scope}
\end{tikzpicture}

%% file: 06-plots/NCC2MILP-1Day-ImprovementPlot-Shortened.tex
\begin{tikzpicture}[x=1pt,y=1pt]
\definecolor{fillColor}{RGB}{255,255,255}
\path[use as bounding box,fill=fillColor,fill opacity=0.00] (0,0) rectangle (122.86,199.47);
\begin{scope}
\path[clip] (  0.00,  0.00) rectangle (122.86,199.47);
\definecolor{drawColor}{RGB}{255,255,255}
\definecolor{fillColor}{RGB}{255,255,255}

\path[draw=drawColor,line width= 0.5pt,line join=round,line cap=round,fill=fillColor] (  0.00,  0.00) rectangle (122.86,199.47);
\end{scope}
\begin{scope}
\path[clip] ( 27.27, 29.68) rectangle (117.86,194.47);
\definecolor{fillColor}{RGB}{255,255,255}

\path[fill=fillColor] ( 27.27, 29.68) rectangle (117.86,194.47);
\definecolor{drawColor}{gray}{0.87}

\path[draw=drawColor,line width= 0.1pt,line join=round] ( 27.27, 29.68) --
	(117.86, 29.68);

\path[draw=drawColor,line width= 0.1pt,line join=round] ( 27.27, 64.74) --
	(117.86, 64.74);

\path[draw=drawColor,line width= 0.1pt,line join=round] ( 27.27, 99.80) --
	(117.86, 99.80);

\path[draw=drawColor,line width= 0.1pt,line join=round] ( 27.27,134.86) --
	(117.86,134.86);

\path[draw=drawColor,line width= 0.1pt,line join=round] ( 27.27,169.92) --
	(117.86,169.92);

\path[draw=drawColor,line width= 0.1pt,line join=round] ( 36.33, 29.68) --
	( 36.33,194.47);

\path[draw=drawColor,line width= 0.1pt,line join=round] ( 54.45, 29.68) --
	( 54.45,194.47);

\path[draw=drawColor,line width= 0.1pt,line join=round] ( 72.57, 29.68) --
	( 72.57,194.47);

\path[draw=drawColor,line width= 0.1pt,line join=round] ( 90.68, 29.68) --
	( 90.68,194.47);

\path[draw=drawColor,line width= 0.1pt,line join=round] (108.80, 29.68) --
	(108.80,194.47);

\path[draw=drawColor,line width= 0.3pt,line join=round] ( 27.27, 47.21) --
	(117.86, 47.21);

\path[draw=drawColor,line width= 0.3pt,line join=round] ( 27.27, 82.27) --
	(117.86, 82.27);

\path[draw=drawColor,line width= 0.3pt,line join=round] ( 27.27,117.33) --
	(117.86,117.33);

\path[draw=drawColor,line width= 0.3pt,line join=round] ( 27.27,152.39) --
	(117.86,152.39);

\path[draw=drawColor,line width= 0.3pt,line join=round] ( 27.27,187.45) --
	(117.86,187.45);

\path[draw=drawColor,line width= 0.3pt,line join=round] ( 27.27, 29.68) --
	( 27.27,194.47);

\path[draw=drawColor,line width= 0.3pt,line join=round] ( 45.39, 29.68) --
	( 45.39,194.47);

\path[draw=drawColor,line width= 0.3pt,line join=round] ( 63.51, 29.68) --
	( 63.51,194.47);

\path[draw=drawColor,line width= 0.3pt,line join=round] ( 81.62, 29.68) --
	( 81.62,194.47);

\path[draw=drawColor,line width= 0.3pt,line join=round] ( 99.74, 29.68) --
	( 99.74,194.47);

\path[draw=drawColor,line width= 0.3pt,line join=round] (117.86, 29.68) --
	(117.86,194.47);
\definecolor{drawColor}{RGB}{128,128,128}

\path[draw=drawColor,line width= 1.0pt,line join=round] ( 27.73, 35.83) --
	( 28.18, 46.83) --
	( 28.64, 55.77) --
	( 29.09, 62.25) --
	( 29.55, 63.56) --
	( 30.00, 65.80) --
	( 30.46, 65.92) --
	( 30.92, 70.99) --
	( 31.37, 71.52) --
	( 31.83, 72.61) --
	( 32.28, 73.02) --
	( 32.74, 74.22) --
	( 33.19, 75.44) --
	( 33.65, 76.24) --
	( 34.10, 77.17) --
	( 34.56, 77.79) --
	( 35.01, 78.27) --
	( 35.47, 78.73) --
	( 35.92, 79.71) --
	( 36.38, 80.47) --
	( 36.83, 82.27) --
	( 37.29, 82.27) --
	( 37.74, 82.27) --
	( 38.20, 82.27) --
	( 38.65, 82.27) --
	( 39.11, 82.27) --
	( 39.56, 82.27) --
	( 40.02, 82.27) --
	( 40.47, 82.27) --
	( 40.93, 82.27) --
	( 41.38, 82.27) --
	( 41.84, 82.27) --
	( 42.30, 82.27) --
	( 42.75, 82.27) --
	( 43.21, 82.27) --
	( 43.66, 82.27) --
	( 44.12, 82.27) --
	( 44.57, 82.27) --
	( 45.03, 82.27) --
	( 45.48, 82.27) --
	( 45.94, 82.27) --
	( 46.39, 82.27) --
	( 46.85, 82.27) --
	( 47.30, 82.27) --
	( 47.76, 82.27) --
	( 48.21, 82.27) --
	( 48.67, 82.27) --
	( 49.12, 82.27) --
	( 49.58, 82.27) --
	( 50.03, 82.27) --
	( 50.49, 82.27) --
	( 50.94, 82.27) --
	( 51.40, 82.27) --
	( 51.85, 82.27) --
	( 52.31, 82.27) --
	( 52.76, 82.27) --
	( 53.22, 82.27) --
	( 53.68, 82.27) --
	( 54.13, 82.27) --
	( 54.59, 82.27) --
	( 55.04, 82.27) --
	( 55.50, 82.27) --
	( 55.95, 82.27) --
	( 56.41, 82.27) --
	( 56.86, 82.27) --
	( 57.32, 82.27) --
	( 57.77, 82.27) --
	( 58.23, 82.27) --
	( 58.68, 82.27) --
	( 59.14, 82.27) --
	( 59.59, 82.27) --
	( 60.05, 82.27) --
	( 60.50, 82.27) --
	( 60.96, 82.27) --
	( 61.41, 82.27) --
	( 61.87, 82.27) --
	( 62.32, 83.08) --
	( 62.78, 83.10) --
	( 63.23, 83.89) --
	( 63.69, 84.29) --
	( 64.14, 85.07) --
	( 64.60, 85.98) --
	( 65.06, 86.03) --
	( 65.51, 86.21) --
	( 65.97, 86.26) --
	( 66.42, 86.48) --
	( 66.88, 86.65) --
	( 67.33, 86.83) --
	( 67.79, 86.84) --
	( 68.24, 87.15) --
	( 68.70, 87.29) --
	( 69.15, 87.62) --
	( 69.61, 88.36) --
	( 70.06, 88.54) --
	( 70.52, 88.85) --
	( 70.97, 88.96) --
	( 71.43, 89.11) --
	( 71.88, 89.11) --
	( 72.34, 89.28) --
	( 72.79, 89.71) --
	( 73.25, 89.91) --
	( 73.70, 90.00) --
	( 74.16, 90.65) --
	( 74.61, 92.30) --
	( 75.07, 92.89) --
	( 75.53, 94.13) --
	( 75.98, 94.49) --
	( 76.44, 95.67) --
	( 76.89, 95.86) --
	( 77.35, 95.87) --
	( 77.80, 95.91) --
	( 78.26, 97.55) --
	( 78.71, 97.90) --
	( 79.17, 98.28) --
	( 79.62, 98.36) --
	( 80.08, 98.51) --
	( 80.53, 99.15) --
	( 80.99, 99.19) --
	( 81.44, 99.41) --
	( 81.90,100.37) --
	( 82.35,100.41) --
	( 82.81,101.06) --
	( 83.26,102.08) --
	( 83.72,102.09) --
	( 84.17,103.36) --
	( 84.63,104.21) --
	( 85.08,104.64) --
	( 85.54,104.74) --
	( 85.99,105.72) --
	( 86.45,106.15) --
	( 86.91,106.64) --
	( 87.36,107.05) --
	( 87.82,107.42) --
	( 88.27,107.65) --
	( 88.73,108.62) --
	( 89.18,109.12) --
	( 89.64,110.41) --
	( 90.09,111.76) --
	( 90.55,112.05) --
	( 91.00,112.71) --
	( 91.46,113.01) --
	( 91.91,115.05) --
	( 92.37,115.14) --
	( 92.82,116.04) --
	( 93.28,117.09) --
	( 93.73,118.04) --
	( 94.19,118.74) --
	( 94.64,119.08) --
	( 95.10,119.89) --
	( 95.55,120.30) --
	( 96.01,121.89) --
	( 96.46,122.33) --
	( 96.92,122.40) --
	( 97.37,122.63) --
	( 97.83,123.79) --
	( 98.29,124.26) --
	( 98.74,125.45) --
	( 99.20,126.27) --
	( 99.65,126.46) --
	(100.11,129.00) --
	(100.56,129.09) --
	(101.02,129.68) --
	(101.47,130.99) --
	(101.93,131.46) --
	(102.38,131.78) --
	(102.84,131.78) --
	(103.29,132.29) --
	(103.75,136.21) --
	(104.20,136.73) --
	(104.66,137.47) --
	(105.11,137.74) --
	(105.57,138.43) --
	(106.02,140.51) --
	(106.48,140.93) --
	(106.93,141.53) --
	(107.39,142.70) --
	(107.84,143.21) --
	(108.30,143.32) --
	(108.75,145.25) --
	(109.21,145.94) --
	(109.67,148.56) --
	(110.12,149.06) --
	(110.58,149.62) --
	(111.03,151.26) --
	(111.49,152.30) --
	(111.94,153.44) --
	(112.40,154.06) --
	(112.85,160.24) --
	(113.31,160.29) --
	(113.76,161.34) --
	(114.22,163.06) --
	(114.67,164.83) --
	(115.13,167.85) --
	(115.58,174.23) --
	(115.85,194.47);
\definecolor{drawColor}{RGB}{152,167,197}

\path[draw=drawColor,line width= 1.0pt,line join=round] ( 27.27, 50.53) --
	( 27.73, 53.38) --
	( 28.18, 59.66) --
	( 28.64, 67.31) --
	( 29.09, 69.30) --
	( 29.55, 70.66) --
	( 30.00, 70.94) --
	( 30.46, 71.22) --
	( 30.92, 71.34) --
	( 31.37, 72.01) --
	( 31.83, 72.56) --
	( 32.28, 73.23) --
	( 32.74, 73.49) --
	( 33.19, 74.31) --
	( 33.65, 77.20) --
	( 34.10, 77.36) --
	( 34.56, 77.49) --
	( 35.01, 77.62) --
	( 35.47, 77.76) --
	( 35.92, 77.94) --
	( 36.38, 78.06) --
	( 36.83, 78.61) --
	( 37.29, 78.71) --
	( 37.74, 80.46) --
	( 38.20, 80.93) --
	( 38.65, 81.20) --
	( 39.11, 81.21) --
	( 39.56, 81.89) --
	( 40.02, 82.27) --
	( 40.47, 82.27) --
	( 40.93, 82.27) --
	( 41.38, 82.27) --
	( 41.84, 82.27) --
	( 42.30, 82.27) --
	( 42.75, 82.27) --
	( 43.21, 82.27) --
	( 43.66, 82.27) --
	( 44.12, 82.27) --
	( 44.57, 82.27) --
	( 45.03, 82.27) --
	( 45.48, 82.27) --
	( 45.94, 82.27) --
	( 46.39, 82.27) --
	( 46.85, 82.27) --
	( 47.30, 82.27) --
	( 47.76, 82.27) --
	( 48.21, 82.27) --
	( 48.67, 82.27) --
	( 49.12, 82.27) --
	( 49.58, 82.27) --
	( 50.03, 82.27) --
	( 50.49, 82.27) --
	( 50.94, 82.27) --
	( 51.40, 82.27) --
	( 51.85, 82.27) --
	( 52.31, 82.27) --
	( 52.76, 82.27) --
	( 53.22, 82.27) --
	( 53.68, 82.27) --
	( 54.13, 82.27) --
	( 54.59, 82.27) --
	( 55.04, 82.27) --
	( 55.50, 82.27) --
	( 55.95, 82.27) --
	( 56.41, 82.27) --
	( 56.86, 82.27) --
	( 57.32, 82.27) --
	( 57.77, 82.27) --
	( 58.23, 82.41) --
	( 58.68, 82.47) --
	( 59.14, 82.72) --
	( 59.59, 82.96) --
	( 60.05, 83.02) --
	( 60.50, 83.04) --
	( 60.96, 83.10) --
	( 61.41, 83.48) --
	( 61.87, 83.76) --
	( 62.32, 83.85) --
	( 62.78, 83.87) --
	( 63.23, 83.90) --
	( 63.69, 84.08) --
	( 64.14, 84.53) --
	( 64.60, 84.61) --
	( 65.06, 84.74) --
	( 65.51, 84.78) --
	( 65.97, 84.96) --
	( 66.42, 85.13) --
	( 66.88, 85.32) --
	( 67.33, 85.33) --
	( 67.79, 85.54) --
	( 68.24, 85.59) --
	( 68.70, 85.73) --
	( 69.15, 85.91) --
	( 69.61, 86.09) --
	( 70.06, 86.32) --
	( 70.52, 86.60) --
	( 70.97, 87.18) --
	( 71.43, 87.53) --
	( 71.88, 87.89) --
	( 72.34, 88.44) --
	( 72.79, 88.56) --
	( 73.25, 88.67) --
	( 73.70, 88.70) --
	( 74.16, 88.91) --
	( 74.61, 88.96) --
	( 75.07, 89.02) --
	( 75.53, 89.92) --
	( 75.98, 89.96) --
	( 76.44, 89.97) --
	( 76.89, 90.54) --
	( 77.35, 90.58) --
	( 77.80, 91.15) --
	( 78.26, 91.64) --
	( 78.71, 91.85) --
	( 79.17, 91.93) --
	( 79.62, 92.32) --
	( 80.08, 92.46) --
	( 80.53, 93.18) --
	( 80.99, 93.29) --
	( 81.44, 93.33) --
	( 81.90, 93.44) --
	( 82.35, 93.81) --
	( 82.81, 93.88) --
	( 83.26, 94.41) --
	( 83.72, 94.64) --
	( 84.17, 94.64) --
	( 84.63, 94.87) --
	( 85.08, 95.35) --
	( 85.54, 95.48) --
	( 85.99, 95.52) --
	( 86.45, 95.54) --
	( 86.91, 95.98) --
	( 87.36, 96.12) --
	( 87.82, 96.21) --
	( 88.27, 96.27) --
	( 88.73, 96.87) --
	( 89.18, 97.42) --
	( 89.64, 98.10) --
	( 90.09, 98.12) --
	( 90.55, 98.22) --
	( 91.00, 98.35) --
	( 91.46, 98.69) --
	( 91.91, 98.88) --
	( 92.37, 99.28) --
	( 92.82, 99.61) --
	( 93.28, 99.76) --
	( 93.73,100.91) --
	( 94.19,100.97) --
	( 94.64,101.00) --
	( 95.10,101.20) --
	( 95.55,101.66) --
	( 96.01,101.86) --
	( 96.46,102.00) --
	( 96.92,103.31) --
	( 97.37,103.41) --
	( 97.83,104.93) --
	( 98.29,104.95) --
	( 98.74,105.01) --
	( 99.20,105.25) --
	( 99.65,105.81) --
	(100.11,106.34) --
	(100.56,106.46) --
	(101.02,107.49) --
	(101.47,107.56) --
	(101.93,107.80) --
	(102.38,108.15) --
	(102.84,108.18) --
	(103.29,108.70) --
	(103.75,109.69) --
	(104.20,110.37) --
	(104.66,110.40) --
	(105.11,110.65) --
	(105.57,112.50) --
	(106.02,112.66) --
	(106.48,113.21) --
	(106.93,113.92) --
	(107.39,115.06) --
	(107.84,115.66) --
	(108.30,117.23) --
	(108.75,117.69) --
	(109.21,119.81) --
	(109.67,120.80) --
	(110.12,120.99) --
	(110.58,121.09) --
	(111.03,121.47) --
	(111.49,122.45) --
	(111.94,122.62) --
	(112.40,122.98) --
	(112.85,124.46) --
	(113.31,126.02) --
	(113.76,128.10) --
	(114.22,128.17) --
	(114.67,128.24) --
	(115.13,128.39) --
	(115.58,132.16) --
	(116.04,139.07) --
	(116.49,151.01) --
	(116.95,157.64) --
	(117.40,182.03) --
	(117.46,194.47);
\definecolor{drawColor}{RGB}{128,202,192}

\path[draw=drawColor,line width= 1.0pt,line join=round] ( 27.27, 58.46) --
	( 27.73, 61.11) --
	( 28.18, 62.75) --
	( 28.64, 63.16) --
	( 29.09, 63.45) --
	( 29.55, 65.71) --
	( 30.00, 66.69) --
	( 30.46, 66.92) --
	( 30.92, 69.15) --
	( 31.37, 69.84) --
	( 31.83, 70.02) --
	( 32.28, 70.79) --
	( 32.74, 71.58) --
	( 33.19, 73.45) --
	( 33.65, 74.00) --
	( 34.10, 75.11) --
	( 34.56, 75.14) --
	( 35.01, 75.47) --
	( 35.47, 75.56) --
	( 35.92, 75.99) --
	( 36.38, 76.21) --
	( 36.83, 76.61) --
	( 37.29, 76.68) --
	( 37.74, 76.97) --
	( 38.20, 77.19) --
	( 38.65, 77.26) --
	( 39.11, 77.67) --
	( 39.56, 77.82) --
	( 40.02, 77.91) --
	( 40.47, 78.39) --
	( 40.93, 78.43) --
	( 41.38, 78.73) --
	( 41.84, 78.80) --
	( 42.30, 78.82) --
	( 42.75, 78.89) --
	( 43.21, 79.30) --
	( 43.66, 80.15) --
	( 44.12, 80.33) --
	( 44.57, 80.33) --
	( 45.03, 80.38) --
	( 45.48, 80.79) --
	( 45.94, 80.80) --
	( 46.39, 80.96) --
	( 46.85, 81.00) --
	( 47.30, 81.23) --
	( 47.76, 81.30) --
	( 48.21, 81.60) --
	( 48.67, 81.61) --
	( 49.12, 81.71) --
	( 49.58, 81.95) --
	( 50.03, 82.13) --
	( 50.49, 82.14) --
	( 50.94, 82.17) --
	( 51.40, 82.27) --
	( 51.85, 82.58) --
	( 52.31, 84.38) --
	( 52.76, 84.45) --
	( 53.22, 84.49) --
	( 53.68, 84.74) --
	( 54.13, 84.95) --
	( 54.59, 85.19) --
	( 55.04, 85.38) --
	( 55.50, 85.41) --
	( 55.95, 85.51) --
	( 56.41, 85.85) --
	( 56.86, 86.15) --
	( 57.32, 86.68) --
	( 57.77, 86.74) --
	( 58.23, 86.89) --
	( 58.68, 86.91) --
	( 59.14, 87.22) --
	( 59.59, 87.29) --
	( 60.05, 87.55) --
	( 60.50, 88.09) --
	( 60.96, 88.32) --
	( 61.41, 88.62) --
	( 61.87, 88.64) --
	( 62.32, 88.68) --
	( 62.78, 89.02) --
	( 63.23, 89.11) --
	( 63.69, 89.12) --
	( 64.14, 89.43) --
	( 64.60, 89.44) --
	( 65.06, 89.45) --
	( 65.51, 89.65) --
	( 65.97, 89.69) --
	( 66.42, 89.73) --
	( 66.88, 89.79) --
	( 67.33, 89.80) --
	( 67.79, 89.80) --
	( 68.24, 89.85) --
	( 68.70, 89.91) --
	( 69.15, 89.98) --
	( 69.61, 90.16) --
	( 70.06, 90.23) --
	( 70.52, 90.38) --
	( 70.97, 90.57) --
	( 71.43, 90.59) --
	( 71.88, 90.72) --
	( 72.34, 90.86) --
	( 72.79, 91.05) --
	( 73.25, 91.25) --
	( 73.70, 91.46) --
	( 74.16, 91.51) --
	( 74.61, 91.80) --
	( 75.07, 91.86) --
	( 75.53, 91.88) --
	( 75.98, 91.88) --
	( 76.44, 91.95) --
	( 76.89, 92.21) --
	( 77.35, 92.39) --
	( 77.80, 92.45) --
	( 78.26, 92.46) --
	( 78.71, 92.53) --
	( 79.17, 92.56) --
	( 79.62, 92.63) --
	( 80.08, 92.75) --
	( 80.53, 92.80) --
	( 80.99, 92.83) --
	( 81.44, 92.90) --
	( 81.90, 92.98) --
	( 82.35, 93.09) --
	( 82.81, 93.10) --
	( 83.26, 93.41) --
	( 83.72, 93.56) --
	( 84.17, 93.68) --
	( 84.63, 93.86) --
	( 85.08, 93.91) --
	( 85.54, 93.95) --
	( 85.99, 94.01) --
	( 86.45, 94.24) --
	( 86.91, 94.41) --
	( 87.36, 94.58) --
	( 87.82, 95.19) --
	( 88.27, 95.27) --
	( 88.73, 95.33) --
	( 89.18, 95.36) --
	( 89.64, 95.44) --
	( 90.09, 95.79) --
	( 90.55, 95.90) --
	( 91.00, 96.49) --
	( 91.46, 96.66) --
	( 91.91, 96.68) --
	( 92.37, 96.72) --
	( 92.82, 97.06) --
	( 93.28, 97.10) --
	( 93.73, 97.50) --
	( 94.19, 97.65) --
	( 94.64, 97.66) --
	( 95.10, 98.46) --
	( 95.55, 98.53) --
	( 96.01, 99.11) --
	( 96.46, 99.34) --
	( 96.92, 99.52) --
	( 97.37, 99.52) --
	( 97.83, 99.62) --
	( 98.29, 99.72) --
	( 98.74,100.05) --
	( 99.20,100.14) --
	( 99.65,101.15) --
	(100.11,101.26) --
	(100.56,101.34) --
	(101.02,101.82) --
	(101.47,101.84) --
	(101.93,102.06) --
	(102.38,102.27) --
	(102.84,102.83) --
	(103.29,103.02) --
	(103.75,103.11) --
	(104.20,103.36) --
	(104.66,103.72) --
	(105.11,103.81) --
	(105.57,104.06) --
	(106.02,104.33) --
	(106.48,105.53) --
	(106.93,105.79) --
	(107.39,105.86) --
	(107.84,106.36) --
	(108.30,106.38) --
	(108.75,106.58) --
	(109.21,108.59) --
	(109.67,109.17) --
	(110.12,109.52) --
	(110.58,109.96) --
	(111.03,110.25) --
	(111.49,110.38) --
	(111.94,111.47) --
	(112.40,112.34) --
	(112.85,112.44) --
	(113.31,112.74) --
	(113.76,112.77) --
	(114.22,114.41) --
	(114.67,115.28) --
	(115.13,115.81) --
	(115.58,116.84) --
	(116.04,118.20) --
	(116.49,123.57) --
	(116.95,126.88) --
	(117.40,128.69) --
	(117.86,132.26);
\definecolor{drawColor}{RGB}{207,128,187}

\path[draw=drawColor,line width= 1.0pt,line join=round] ( 27.27, 51.40) --
	( 27.73, 55.37) --
	( 28.18, 58.33) --
	( 28.64, 62.49) --
	( 29.09, 62.91) --
	( 29.55, 64.44) --
	( 30.00, 65.84) --
	( 30.46, 67.41) --
	( 30.92, 68.05) --
	( 31.37, 68.51) --
	( 31.83, 68.94) --
	( 32.28, 68.96) --
	( 32.74, 69.43) --
	( 33.19, 70.27) --
	( 33.65, 70.97) --
	( 34.10, 71.53) --
	( 34.56, 71.76) --
	( 35.01, 71.81) --
	( 35.47, 72.53) --
	( 35.92, 72.72) --
	( 36.38, 72.96) --
	( 36.83, 73.14) --
	( 37.29, 73.69) --
	( 37.74, 74.05) --
	( 38.20, 74.17) --
	( 38.65, 74.34) --
	( 39.11, 74.46) --
	( 39.56, 74.79) --
	( 40.02, 75.80) --
	( 40.47, 75.88) --
	( 40.93, 76.15) --
	( 41.38, 76.38) --
	( 41.84, 76.40) --
	( 42.30, 76.98) --
	( 42.75, 77.22) --
	( 43.21, 77.76) --
	( 43.66, 78.14) --
	( 44.12, 78.16) --
	( 44.57, 78.32) --
	( 45.03, 78.49) --
	( 45.48, 78.94) --
	( 45.94, 79.26) --
	( 46.39, 79.43) --
	( 46.85, 79.45) --
	( 47.30, 79.63) --
	( 47.76, 79.74) --
	( 48.21, 79.87) --
	( 48.67, 79.96) --
	( 49.12, 80.36) --
	( 49.58, 80.68) --
	( 50.03, 80.74) --
	( 50.49, 80.91) --
	( 50.94, 81.03) --
	( 51.40, 81.22) --
	( 51.85, 81.26) --
	( 52.31, 81.31) --
	( 52.76, 81.50) --
	( 53.22, 81.69) --
	( 53.68, 81.73) --
	( 54.13, 82.20) --
	( 54.59, 82.31) --
	( 55.04, 82.66) --
	( 55.50, 82.78) --
	( 55.95, 83.99) --
	( 56.41, 84.02) --
	( 56.86, 84.03) --
	( 57.32, 84.03) --
	( 57.77, 84.18) --
	( 58.23, 84.47) --
	( 58.68, 84.49) --
	( 59.14, 84.56) --
	( 59.59, 84.56) --
	( 60.05, 84.68) --
	( 60.50, 84.84) --
	( 60.96, 84.86) --
	( 61.41, 84.92) --
	( 61.87, 85.01) --
	( 62.32, 85.15) --
	( 62.78, 85.18) --
	( 63.23, 85.42) --
	( 63.69, 85.65) --
	( 64.14, 85.71) --
	( 64.60, 85.75) --
	( 65.06, 85.91) --
	( 65.51, 85.94) --
	( 65.97, 86.22) --
	( 66.42, 86.42) --
	( 66.88, 86.67) --
	( 67.33, 86.69) --
	( 67.79, 86.71) --
	( 68.24, 86.86) --
	( 68.70, 86.95) --
	( 69.15, 87.11) --
	( 69.61, 87.26) --
	( 70.06, 87.31) --
	( 70.52, 87.44) --
	( 70.97, 87.86) --
	( 71.43, 88.24) --
	( 71.88, 88.24) --
	( 72.34, 88.38) --
	( 72.79, 88.43) --
	( 73.25, 88.48) --
	( 73.70, 88.67) --
	( 74.16, 88.71) --
	( 74.61, 89.18) --
	( 75.07, 89.29) --
	( 75.53, 89.36) --
	( 75.98, 89.42) --
	( 76.44, 89.43) --
	( 76.89, 89.51) --
	( 77.35, 89.55) --
	( 77.80, 89.59) --
	( 78.26, 89.78) --
	( 78.71, 89.83) --
	( 79.17, 90.07) --
	( 79.62, 90.36) --
	( 80.08, 90.74) --
	( 80.53, 90.87) --
	( 80.99, 90.93) --
	( 81.44, 91.18) --
	( 81.90, 91.34) --
	( 82.35, 91.43) --
	( 82.81, 92.07) --
	( 83.26, 92.72) --
	( 83.72, 92.75) --
	( 84.17, 92.89) --
	( 84.63, 93.12) --
	( 85.08, 93.19) --
	( 85.54, 93.51) --
	( 85.99, 93.91) --
	( 86.45, 93.96) --
	( 86.91, 94.02) --
	( 87.36, 94.27) --
	( 87.82, 94.71) --
	( 88.27, 94.83) --
	( 88.73, 94.83) --
	( 89.18, 94.88) --
	( 89.64, 95.00) --
	( 90.09, 95.01) --
	( 90.55, 95.36) --
	( 91.00, 95.37) --
	( 91.46, 95.41) --
	( 91.91, 95.68) --
	( 92.37, 95.68) --
	( 92.82, 95.94) --
	( 93.28, 96.09) --
	( 93.73, 96.27) --
	( 94.19, 96.49) --
	( 94.64, 96.55) --
	( 95.10, 97.10) --
	( 95.55, 97.13) --
	( 96.01, 97.22) --
	( 96.46, 97.24) --
	( 96.92, 97.55) --
	( 97.37, 97.65) --
	( 97.83, 97.87) --
	( 98.29, 97.92) --
	( 98.74, 98.50) --
	( 99.20, 98.63) --
	( 99.65, 98.87) --
	(100.11, 99.19) --
	(100.56, 99.99) --
	(101.02,100.02) --
	(101.47,100.83) --
	(101.93,101.05) --
	(102.38,101.07) --
	(102.84,102.41) --
	(103.29,103.14) --
	(103.75,103.30) --
	(104.20,103.45) --
	(104.66,103.71) --
	(105.11,103.99) --
	(105.57,104.79) --
	(106.02,105.38) --
	(106.48,105.51) --
	(106.93,105.71) --
	(107.39,106.32) --
	(107.84,110.86) --
	(108.30,111.97) --
	(108.75,112.63) --
	(109.21,112.65) --
	(109.67,113.40) --
	(110.12,115.02) --
	(110.58,115.31) --
	(111.03,116.66) --
	(111.49,117.50) --
	(111.94,119.42) --
	(112.40,119.58) --
	(112.85,119.68) --
	(113.31,119.68) --
	(113.76,120.41) --
	(114.22,121.88) --
	(114.67,122.05) --
	(115.13,126.00) --
	(115.58,135.14) --
	(116.04,137.52) --
	(116.49,144.59) --
	(116.95,160.65) --
	(117.40,162.63) --
	(117.86,179.64);
\definecolor{drawColor}{RGB}{207,192,152}

\path[draw=drawColor,line width= 1.0pt,line join=round] ( 27.27, 48.76) --
	( 27.73, 61.65) --
	( 28.18, 63.04) --
	( 28.64, 65.28) --
	( 29.09, 65.36) --
	( 29.55, 65.61) --
	( 30.00, 66.68) --
	( 30.46, 67.63) --
	( 30.92, 68.94) --
	( 31.37, 69.88) --
	( 31.83, 70.16) --
	( 32.28, 70.36) --
	( 32.74, 70.39) --
	( 33.19, 70.79) --
	( 33.65, 70.82) --
	( 34.10, 71.06) --
	( 34.56, 71.40) --
	( 35.01, 71.78) --
	( 35.47, 72.74) --
	( 35.92, 73.52) --
	( 36.38, 73.60) --
	( 36.83, 74.91) --
	( 37.29, 75.39) --
	( 37.74, 75.40) --
	( 38.20, 75.76) --
	( 38.65, 75.85) --
	( 39.11, 76.07) --
	( 39.56, 76.60) --
	( 40.02, 76.68) --
	( 40.47, 76.76) --
	( 40.93, 76.87) --
	( 41.38, 77.76) --
	( 41.84, 78.56) --
	( 42.30, 78.56) --
	( 42.75, 78.61) --
	( 43.21, 79.06) --
	( 43.66, 79.10) --
	( 44.12, 79.14) --
	( 44.57, 79.16) --
	( 45.03, 79.18) --
	( 45.48, 79.31) --
	( 45.94, 79.70) --
	( 46.39, 79.75) --
	( 46.85, 79.90) --
	( 47.30, 79.99) --
	( 47.76, 80.92) --
	( 48.21, 80.98) --
	( 48.67, 81.38) --
	( 49.12, 81.92) --
	( 49.58, 82.02) --
	( 50.03, 82.40) --
	( 50.49, 82.46) --
	( 50.94, 82.46) --
	( 51.40, 82.55) --
	( 51.85, 82.63) --
	( 52.31, 82.66) --
	( 52.76, 83.58) --
	( 53.22, 83.76) --
	( 53.68, 84.48) --
	( 54.13, 84.79) --
	( 54.59, 85.04) --
	( 55.04, 85.42) --
	( 55.50, 85.64) --
	( 55.95, 85.73) --
	( 56.41, 86.27) --
	( 56.86, 86.37) --
	( 57.32, 86.63) --
	( 57.77, 86.67) --
	( 58.23, 86.79) --
	( 58.68, 86.84) --
	( 59.14, 87.56) --
	( 59.59, 87.73) --
	( 60.05, 87.84) --
	( 60.50, 88.18) --
	( 60.96, 88.44) --
	( 61.41, 88.94) --
	( 61.87, 89.11) --
	( 62.32, 89.20) --
	( 62.78, 89.21) --
	( 63.23, 89.37) --
	( 63.69, 89.41) --
	( 64.14, 89.49) --
	( 64.60, 89.72) --
	( 65.06, 89.94) --
	( 65.51, 89.98) --
	( 65.97, 90.00) --
	( 66.42, 90.18) --
	( 66.88, 90.51) --
	( 67.33, 90.76) --
	( 67.79, 90.79) --
	( 68.24, 90.90) --
	( 68.70, 91.31) --
	( 69.15, 91.35) --
	( 69.61, 91.65) --
	( 70.06, 91.83) --
	( 70.52, 91.92) --
	( 70.97, 91.97) --
	( 71.43, 92.18) --
	( 71.88, 92.37) --
	( 72.34, 92.50) --
	( 72.79, 92.56) --
	( 73.25, 92.57) --
	( 73.70, 93.21) --
	( 74.16, 93.26) --
	( 74.61, 93.26) --
	( 75.07, 93.36) --
	( 75.53, 93.44) --
	( 75.98, 93.52) --
	( 76.44, 94.05) --
	( 76.89, 94.56) --
	( 77.35, 94.72) --
	( 77.80, 94.75) --
	( 78.26, 94.82) --
	( 78.71, 95.05) --
	( 79.17, 95.06) --
	( 79.62, 95.11) --
	( 80.08, 95.14) --
	( 80.53, 95.75) --
	( 80.99, 95.78) --
	( 81.44, 96.05) --
	( 81.90, 96.16) --
	( 82.35, 96.31) --
	( 82.81, 96.53) --
	( 83.26, 96.84) --
	( 83.72, 96.94) --
	( 84.17, 97.18) --
	( 84.63, 97.24) --
	( 85.08, 97.44) --
	( 85.54, 97.48) --
	( 85.99, 97.50) --
	( 86.45, 98.14) --
	( 86.91, 98.19) --
	( 87.36, 98.31) --
	( 87.82, 98.49) --
	( 88.27, 99.76) --
	( 88.73, 99.91) --
	( 89.18,100.10) --
	( 89.64,100.12) --
	( 90.09,100.15) --
	( 90.55,100.50) --
	( 91.00,101.81) --
	( 91.46,102.41) --
	( 91.91,102.62) --
	( 92.37,102.92) --
	( 92.82,102.92) --
	( 93.28,102.98) --
	( 93.73,103.32) --
	( 94.19,103.55) --
	( 94.64,104.20) --
	( 95.10,104.43) --
	( 95.55,104.60) --
	( 96.01,104.62) --
	( 96.46,104.87) --
	( 96.92,105.09) --
	( 97.37,105.11) --
	( 97.83,105.69) --
	( 98.29,106.15) --
	( 98.74,106.41) --
	( 99.20,106.89) --
	( 99.65,107.00) --
	(100.11,107.39) --
	(100.56,107.55) --
	(101.02,108.11) --
	(101.47,109.62) --
	(101.93,109.92) --
	(102.38,110.07) --
	(102.84,110.58) --
	(103.29,113.79) --
	(103.75,113.79) --
	(104.20,113.96) --
	(104.66,114.24) --
	(105.11,114.26) --
	(105.57,117.51) --
	(106.02,117.58) --
	(106.48,120.65) --
	(106.93,121.22) --
	(107.39,121.46) --
	(107.84,123.06) --
	(108.30,123.58) --
	(108.75,124.73) --
	(109.21,125.80) --
	(109.67,125.93) --
	(110.12,128.63) --
	(110.58,135.77) --
	(111.03,146.45) --
	(111.49,151.81) --
	(111.94,170.52) --
	(112.03,194.47);
\definecolor{drawColor}{RGB}{139,0,0}

\path[draw=drawColor,line width= 1.0pt,line join=round] ( 27.27, 82.27) --
	( 27.27, 82.27) --
	( 27.27, 82.27) --
	( 27.27, 82.27) --
	( 27.73, 82.27) --
	( 27.73, 82.27) --
	( 27.73, 82.27) --
	( 27.73, 82.27) --
	( 27.73, 82.27) --
	( 28.18, 82.27) --
	( 28.18, 82.27) --
	( 28.18, 82.27) --
	( 28.18, 82.27) --
	( 28.18, 82.27) --
	( 28.64, 82.27) --
	( 28.64, 82.27) --
	( 28.64, 82.27) --
	( 28.64, 82.27) --
	( 28.64, 82.27) --
	( 29.09, 82.27) --
	( 29.09, 82.27) --
	( 29.09, 82.27) --
	( 29.09, 82.27) --
	( 29.09, 82.27) --
	( 29.55, 82.27) --
	( 29.55, 82.27) --
	( 29.55, 82.27) --
	( 29.55, 82.27) --
	( 29.55, 82.27) --
	( 30.00, 82.27) --
	( 30.00, 82.27) --
	( 30.00, 82.27) --
	( 30.00, 82.27) --
	( 30.00, 82.27) --
	( 30.46, 82.27) --
	( 30.46, 82.27) --
	( 30.46, 82.27) --
	( 30.46, 82.27) --
	( 30.46, 82.27) --
	( 30.92, 82.27) --
	( 30.92, 82.27) --
	( 30.92, 82.27) --
	( 30.92, 82.27) --
	( 30.92, 82.27) --
	( 31.37, 82.27) --
	( 31.37, 82.27) --
	( 31.37, 82.27) --
	( 31.37, 82.27) --
	( 31.37, 82.27) --
	( 31.83, 82.27) --
	( 31.83, 82.27) --
	( 31.83, 82.27) --
	( 31.83, 82.27) --
	( 31.83, 82.27) --
	( 32.28, 82.27) --
	( 32.28, 82.27) --
	( 32.28, 82.27) --
	( 32.28, 82.27) --
	( 32.28, 82.27) --
	( 32.74, 82.27) --
	( 32.74, 82.27) --
	( 32.74, 82.27) --
	( 32.74, 82.27) --
	( 32.74, 82.27) --
	( 33.19, 82.27) --
	( 33.19, 82.27) --
	( 33.19, 82.27) --
	( 33.19, 82.27) --
	( 33.19, 82.27) --
	( 33.65, 82.27) --
	( 33.65, 82.27) --
	( 33.65, 82.27) --
	( 33.65, 82.27) --
	( 33.65, 82.27) --
	( 34.10, 82.27) --
	( 34.10, 82.27) --
	( 34.10, 82.27) --
	( 34.10, 82.27) --
	( 34.10, 82.27) --
	( 34.56, 82.27) --
	( 34.56, 82.27) --
	( 34.56, 82.27) --
	( 34.56, 82.27) --
	( 34.56, 82.27) --
	( 35.01, 82.27) --
	( 35.01, 82.27) --
	( 35.01, 82.27) --
	( 35.01, 82.27) --
	( 35.01, 82.27) --
	( 35.47, 82.27) --
	( 35.47, 82.27) --
	( 35.47, 82.27) --
	( 35.47, 82.27) --
	( 35.47, 82.27) --
	( 35.92, 82.27) --
	( 35.92, 82.27) --
	( 35.92, 82.27) --
	( 35.92, 82.27) --
	( 35.92, 82.27) --
	( 36.38, 82.27) --
	( 36.38, 82.27) --
	( 36.38, 82.27) --
	( 36.38, 82.27) --
	( 36.38, 82.27) --
	( 36.83, 82.27) --
	( 36.83, 82.27) --
	( 36.83, 82.27) --
	( 36.83, 82.27) --
	( 36.83, 82.27) --
	( 37.29, 82.27) --
	( 37.29, 82.27) --
	( 37.29, 82.27) --
	( 37.29, 82.27) --
	( 37.29, 82.27) --
	( 37.74, 82.27) --
	( 37.74, 82.27) --
	( 37.74, 82.27) --
	( 37.74, 82.27) --
	( 37.74, 82.27) --
	( 38.20, 82.27) --
	( 38.20, 82.27) --
	( 38.20, 82.27) --
	( 38.20, 82.27) --
	( 38.20, 82.27) --
	( 38.65, 82.27) --
	( 38.65, 82.27) --
	( 38.65, 82.27) --
	( 38.65, 82.27) --
	( 38.65, 82.27) --
	( 39.11, 82.27) --
	( 39.11, 82.27) --
	( 39.11, 82.27) --
	( 39.11, 82.27) --
	( 39.11, 82.27) --
	( 39.56, 82.27) --
	( 39.56, 82.27) --
	( 39.56, 82.27) --
	( 39.56, 82.27) --
	( 39.56, 82.27) --
	( 40.02, 82.27) --
	( 40.02, 82.27) --
	( 40.02, 82.27) --
	( 40.02, 82.27) --
	( 40.02, 82.27) --
	( 40.47, 82.27) --
	( 40.47, 82.27) --
	( 40.47, 82.27) --
	( 40.47, 82.27) --
	( 40.47, 82.27) --
	( 40.93, 82.27) --
	( 40.93, 82.27) --
	( 40.93, 82.27) --
	( 40.93, 82.27) --
	( 40.93, 82.27) --
	( 41.38, 82.27) --
	( 41.38, 82.27) --
	( 41.38, 82.27) --
	( 41.38, 82.27) --
	( 41.38, 82.27) --
	( 41.84, 82.27) --
	( 41.84, 82.27) --
	( 41.84, 82.27) --
	( 41.84, 82.27) --
	( 41.84, 82.27) --
	( 42.30, 82.27) --
	( 42.30, 82.27) --
	( 42.30, 82.27) --
	( 42.30, 82.27) --
	( 42.30, 82.27) --
	( 42.75, 82.27) --
	( 42.75, 82.27) --
	( 42.75, 82.27) --
	( 42.75, 82.27) --
	( 42.75, 82.27) --
	( 43.21, 82.27) --
	( 43.21, 82.27) --
	( 43.21, 82.27) --
	( 43.21, 82.27) --
	( 43.21, 82.27) --
	( 43.66, 82.27) --
	( 43.66, 82.27) --
	( 43.66, 82.27) --
	( 43.66, 82.27) --
	( 43.66, 82.27) --
	( 44.12, 82.27) --
	( 44.12, 82.27) --
	( 44.12, 82.27) --
	( 44.12, 82.27) --
	( 44.12, 82.27) --
	( 44.57, 82.27) --
	( 44.57, 82.27) --
	( 44.57, 82.27) --
	( 44.57, 82.27) --
	( 44.57, 82.27) --
	( 45.03, 82.27) --
	( 45.03, 82.27) --
	( 45.03, 82.27) --
	( 45.03, 82.27) --
	( 45.03, 82.27) --
	( 45.48, 82.27) --
	( 45.48, 82.27) --
	( 45.48, 82.27) --
	( 45.48, 82.27) --
	( 45.48, 82.27) --
	( 45.94, 82.27) --
	( 45.94, 82.27) --
	( 45.94, 82.27) --
	( 45.94, 82.27) --
	( 45.94, 82.27) --
	( 46.39, 82.27) --
	( 46.39, 82.27) --
	( 46.39, 82.27) --
	( 46.39, 82.27) --
	( 46.39, 82.27) --
	( 46.85, 82.27) --
	( 46.85, 82.27) --
	( 46.85, 82.27) --
	( 46.85, 82.27) --
	( 46.85, 82.27) --
	( 47.30, 82.27) --
	( 47.30, 82.27) --
	( 47.30, 82.27) --
	( 47.30, 82.27) --
	( 47.30, 82.27) --
	( 47.76, 82.27) --
	( 47.76, 82.27) --
	( 47.76, 82.27) --
	( 47.76, 82.27) --
	( 47.76, 82.27) --
	( 48.21, 82.27) --
	( 48.21, 82.27) --
	( 48.21, 82.27) --
	( 48.21, 82.27) --
	( 48.21, 82.27) --
	( 48.67, 82.27) --
	( 48.67, 82.27) --
	( 48.67, 82.27) --
	( 48.67, 82.27) --
	( 48.67, 82.27) --
	( 49.12, 82.27) --
	( 49.12, 82.27) --
	( 49.12, 82.27) --
	( 49.12, 82.27) --
	( 49.12, 82.27) --
	( 49.58, 82.27) --
	( 49.58, 82.27) --
	( 49.58, 82.27) --
	( 49.58, 82.27) --
	( 49.58, 82.27) --
	( 50.03, 82.27) --
	( 50.03, 82.27) --
	( 50.03, 82.27) --
	( 50.03, 82.27) --
	( 50.03, 82.27) --
	( 50.49, 82.27) --
	( 50.49, 82.27) --
	( 50.49, 82.27) --
	( 50.49, 82.27) --
	( 50.49, 82.27) --
	( 50.94, 82.27) --
	( 50.94, 82.27) --
	( 50.94, 82.27) --
	( 50.94, 82.27) --
	( 50.94, 82.27) --
	( 51.40, 82.27) --
	( 51.40, 82.27) --
	( 51.40, 82.27) --
	( 51.40, 82.27) --
	( 51.40, 82.27) --
	( 51.85, 82.27) --
	( 51.85, 82.27) --
	( 51.85, 82.27) --
	( 51.85, 82.27) --
	( 51.85, 82.27) --
	( 52.31, 82.27) --
	( 52.31, 82.27) --
	( 52.31, 82.27) --
	( 52.31, 82.27) --
	( 52.31, 82.27) --
	( 52.76, 82.27) --
	( 52.76, 82.27) --
	( 52.76, 82.27) --
	( 52.76, 82.27) --
	( 52.76, 82.27) --
	( 53.22, 82.27) --
	( 53.22, 82.27) --
	( 53.22, 82.27) --
	( 53.22, 82.27) --
	( 53.22, 82.27) --
	( 53.68, 82.27) --
	( 53.68, 82.27) --
	( 53.68, 82.27) --
	( 53.68, 82.27) --
	( 53.68, 82.27) --
	( 54.13, 82.27) --
	( 54.13, 82.27) --
	( 54.13, 82.27) --
	( 54.13, 82.27) --
	( 54.13, 82.27) --
	( 54.59, 82.27) --
	( 54.59, 82.27) --
	( 54.59, 82.27) --
	( 54.59, 82.27) --
	( 54.59, 82.27) --
	( 55.04, 82.27) --
	( 55.04, 82.27) --
	( 55.04, 82.27) --
	( 55.04, 82.27) --
	( 55.04, 82.27) --
	( 55.50, 82.27) --
	( 55.50, 82.27) --
	( 55.50, 82.27) --
	( 55.50, 82.27) --
	( 55.50, 82.27) --
	( 55.95, 82.27) --
	( 55.95, 82.27) --
	( 55.95, 82.27) --
	( 55.95, 82.27) --
	( 55.95, 82.27) --
	( 56.41, 82.27) --
	( 56.41, 82.27) --
	( 56.41, 82.27) --
	( 56.41, 82.27) --
	( 56.41, 82.27) --
	( 56.86, 82.27) --
	( 56.86, 82.27) --
	( 56.86, 82.27) --
	( 56.86, 82.27) --
	( 56.86, 82.27) --
	( 57.32, 82.27) --
	( 57.32, 82.27) --
	( 57.32, 82.27) --
	( 57.32, 82.27) --
	( 57.32, 82.27) --
	( 57.77, 82.27) --
	( 57.77, 82.27) --
	( 57.77, 82.27) --
	( 57.77, 82.27) --
	( 57.77, 82.27) --
	( 58.23, 82.27) --
	( 58.23, 82.27) --
	( 58.23, 82.27) --
	( 58.23, 82.27) --
	( 58.23, 82.27) --
	( 58.68, 82.27) --
	( 58.68, 82.27) --
	( 58.68, 82.27) --
	( 58.68, 82.27) --
	( 58.68, 82.27) --
	( 59.14, 82.27) --
	( 59.14, 82.27) --
	( 59.14, 82.27) --
	( 59.14, 82.27) --
	( 59.14, 82.27) --
	( 59.59, 82.27) --
	( 59.59, 82.27) --
	( 59.59, 82.27) --
	( 59.59, 82.27) --
	( 59.59, 82.27) --
	( 60.05, 82.27) --
	( 60.05, 82.27) --
	( 60.05, 82.27) --
	( 60.05, 82.27) --
	( 60.05, 82.27) --
	( 60.50, 82.27) --
	( 60.50, 82.27) --
	( 60.50, 82.27) --
	( 60.50, 82.27) --
	( 60.50, 82.27) --
	( 60.96, 82.27) --
	( 60.96, 82.27) --
	( 60.96, 82.27) --
	( 60.96, 82.27) --
	( 60.96, 82.27) --
	( 61.41, 82.27) --
	( 61.41, 82.27) --
	( 61.41, 82.27) --
	( 61.41, 82.27) --
	( 61.41, 82.27) --
	( 61.87, 82.27) --
	( 61.87, 82.27) --
	( 61.87, 82.27) --
	( 61.87, 82.27) --
	( 61.87, 82.27) --
	( 62.32, 82.27) --
	( 62.32, 82.27) --
	( 62.32, 82.27) --
	( 62.32, 82.27) --
	( 62.32, 82.27) --
	( 62.78, 82.27) --
	( 62.78, 82.27) --
	( 62.78, 82.27) --
	( 62.78, 82.27) --
	( 62.78, 82.27) --
	( 63.23, 82.27) --
	( 63.23, 82.27) --
	( 63.23, 82.27) --
	( 63.23, 82.27) --
	( 63.23, 82.27) --
	( 63.69, 82.27) --
	( 63.69, 82.27) --
	( 63.69, 82.27) --
	( 63.69, 82.27) --
	( 63.69, 82.27) --
	( 64.14, 82.27) --
	( 64.14, 82.27) --
	( 64.14, 82.27) --
	( 64.14, 82.27) --
	( 64.14, 82.27) --
	( 64.60, 82.27) --
	( 64.60, 82.27) --
	( 64.60, 82.27) --
	( 64.60, 82.27) --
	( 64.60, 82.27) --
	( 65.06, 82.27) --
	( 65.06, 82.27) --
	( 65.06, 82.27) --
	( 65.06, 82.27) --
	( 65.06, 82.27) --
	( 65.51, 82.27) --
	( 65.51, 82.27) --
	( 65.51, 82.27) --
	( 65.51, 82.27) --
	( 65.51, 82.27) --
	( 65.97, 82.27) --
	( 65.97, 82.27) --
	( 65.97, 82.27) --
	( 65.97, 82.27) --
	( 65.97, 82.27) --
	( 66.42, 82.27) --
	( 66.42, 82.27) --
	( 66.42, 82.27) --
	( 66.42, 82.27) --
	( 66.42, 82.27) --
	( 66.88, 82.27) --
	( 66.88, 82.27) --
	( 66.88, 82.27) --
	( 66.88, 82.27) --
	( 66.88, 82.27) --
	( 67.33, 82.27) --
	( 67.33, 82.27) --
	( 67.33, 82.27) --
	( 67.33, 82.27) --
	( 67.33, 82.27) --
	( 67.79, 82.27) --
	( 67.79, 82.27) --
	( 67.79, 82.27) --
	( 67.79, 82.27) --
	( 67.79, 82.27) --
	( 68.24, 82.27) --
	( 68.24, 82.27) --
	( 68.24, 82.27) --
	( 68.24, 82.27) --
	( 68.24, 82.27) --
	( 68.70, 82.27) --
	( 68.70, 82.27) --
	( 68.70, 82.27) --
	( 68.70, 82.27) --
	( 68.70, 82.27) --
	( 69.15, 82.27) --
	( 69.15, 82.27) --
	( 69.15, 82.27) --
	( 69.15, 82.27) --
	( 69.15, 82.27) --
	( 69.61, 82.27) --
	( 69.61, 82.27) --
	( 69.61, 82.27) --
	( 69.61, 82.27) --
	( 69.61, 82.27) --
	( 70.06, 82.27) --
	( 70.06, 82.27) --
	( 70.06, 82.27) --
	( 70.06, 82.27) --
	( 70.06, 82.27) --
	( 70.52, 82.27) --
	( 70.52, 82.27) --
	( 70.52, 82.27) --
	( 70.52, 82.27) --
	( 70.52, 82.27) --
	( 70.97, 82.27) --
	( 70.97, 82.27) --
	( 70.97, 82.27) --
	( 70.97, 82.27) --
	( 70.97, 82.27) --
	( 71.43, 82.27) --
	( 71.43, 82.27) --
	( 71.43, 82.27) --
	( 71.43, 82.27) --
	( 71.43, 82.27) --
	( 71.88, 82.27) --
	( 71.88, 82.27) --
	( 71.88, 82.27) --
	( 71.88, 82.27) --
	( 71.88, 82.27) --
	( 72.34, 82.27) --
	( 72.34, 82.27) --
	( 72.34, 82.27) --
	( 72.34, 82.27) --
	( 72.34, 82.27) --
	( 72.79, 82.27) --
	( 72.79, 82.27) --
	( 72.79, 82.27) --
	( 72.79, 82.27) --
	( 72.79, 82.27) --
	( 73.25, 82.27) --
	( 73.25, 82.27) --
	( 73.25, 82.27) --
	( 73.25, 82.27) --
	( 73.25, 82.27) --
	( 73.70, 82.27) --
	( 73.70, 82.27) --
	( 73.70, 82.27) --
	( 73.70, 82.27) --
	( 73.70, 82.27) --
	( 74.16, 82.27) --
	( 74.16, 82.27) --
	( 74.16, 82.27) --
	( 74.16, 82.27) --
	( 74.16, 82.27) --
	( 74.61, 82.27) --
	( 74.61, 82.27) --
	( 74.61, 82.27) --
	( 74.61, 82.27) --
	( 74.61, 82.27) --
	( 75.07, 82.27) --
	( 75.07, 82.27) --
	( 75.07, 82.27) --
	( 75.07, 82.27) --
	( 75.07, 82.27) --
	( 75.53, 82.27) --
	( 75.53, 82.27) --
	( 75.53, 82.27) --
	( 75.53, 82.27) --
	( 75.53, 82.27) --
	( 75.98, 82.27) --
	( 75.98, 82.27) --
	( 75.98, 82.27) --
	( 75.98, 82.27) --
	( 75.98, 82.27) --
	( 76.44, 82.27) --
	( 76.44, 82.27) --
	( 76.44, 82.27) --
	( 76.44, 82.27) --
	( 76.44, 82.27) --
	( 76.89, 82.27) --
	( 76.89, 82.27) --
	( 76.89, 82.27) --
	( 76.89, 82.27) --
	( 76.89, 82.27) --
	( 77.35, 82.27) --
	( 77.35, 82.27) --
	( 77.35, 82.27) --
	( 77.35, 82.27) --
	( 77.35, 82.27) --
	( 77.80, 82.27) --
	( 77.80, 82.27) --
	( 77.80, 82.27) --
	( 77.80, 82.27) --
	( 77.80, 82.27) --
	( 78.26, 82.27) --
	( 78.26, 82.27) --
	( 78.26, 82.27) --
	( 78.26, 82.27) --
	( 78.26, 82.27) --
	( 78.71, 82.27) --
	( 78.71, 82.27) --
	( 78.71, 82.27) --
	( 78.71, 82.27) --
	( 78.71, 82.27) --
	( 79.17, 82.27) --
	( 79.17, 82.27) --
	( 79.17, 82.27) --
	( 79.17, 82.27) --
	( 79.17, 82.27) --
	( 79.62, 82.27) --
	( 79.62, 82.27) --
	( 79.62, 82.27) --
	( 79.62, 82.27) --
	( 79.62, 82.27) --
	( 80.08, 82.27) --
	( 80.08, 82.27) --
	( 80.08, 82.27) --
	( 80.08, 82.27) --
	( 80.08, 82.27) --
	( 80.53, 82.27) --
	( 80.53, 82.27) --
	( 80.53, 82.27) --
	( 80.53, 82.27) --
	( 80.53, 82.27) --
	( 80.99, 82.27) --
	( 80.99, 82.27) --
	( 80.99, 82.27) --
	( 80.99, 82.27) --
	( 80.99, 82.27) --
	( 81.44, 82.27) --
	( 81.44, 82.27) --
	( 81.44, 82.27) --
	( 81.44, 82.27) --
	( 81.44, 82.27) --
	( 81.90, 82.27) --
	( 81.90, 82.27) --
	( 81.90, 82.27) --
	( 81.90, 82.27) --
	( 81.90, 82.27) --
	( 82.35, 82.27) --
	( 82.35, 82.27) --
	( 82.35, 82.27) --
	( 82.35, 82.27) --
	( 82.35, 82.27) --
	( 82.81, 82.27) --
	( 82.81, 82.27) --
	( 82.81, 82.27) --
	( 82.81, 82.27) --
	( 82.81, 82.27) --
	( 83.26, 82.27) --
	( 83.26, 82.27) --
	( 83.26, 82.27) --
	( 83.26, 82.27) --
	( 83.26, 82.27) --
	( 83.72, 82.27) --
	( 83.72, 82.27) --
	( 83.72, 82.27) --
	( 83.72, 82.27) --
	( 83.72, 82.27) --
	( 84.17, 82.27) --
	( 84.17, 82.27) --
	( 84.17, 82.27) --
	( 84.17, 82.27) --
	( 84.17, 82.27) --
	( 84.63, 82.27) --
	( 84.63, 82.27) --
	( 84.63, 82.27) --
	( 84.63, 82.27) --
	( 84.63, 82.27) --
	( 85.08, 82.27) --
	( 85.08, 82.27) --
	( 85.08, 82.27) --
	( 85.08, 82.27) --
	( 85.08, 82.27) --
	( 85.54, 82.27) --
	( 85.54, 82.27) --
	( 85.54, 82.27) --
	( 85.54, 82.27) --
	( 85.54, 82.27) --
	( 85.99, 82.27) --
	( 85.99, 82.27) --
	( 85.99, 82.27) --
	( 85.99, 82.27) --
	( 85.99, 82.27) --
	( 86.45, 82.27) --
	( 86.45, 82.27) --
	( 86.45, 82.27) --
	( 86.45, 82.27) --
	( 86.45, 82.27) --
	( 86.91, 82.27) --
	( 86.91, 82.27) --
	( 86.91, 82.27) --
	( 86.91, 82.27) --
	( 86.91, 82.27) --
	( 87.36, 82.27) --
	( 87.36, 82.27) --
	( 87.36, 82.27) --
	( 87.36, 82.27) --
	( 87.36, 82.27) --
	( 87.82, 82.27) --
	( 87.82, 82.27) --
	( 87.82, 82.27) --
	( 87.82, 82.27) --
	( 87.82, 82.27) --
	( 88.27, 82.27) --
	( 88.27, 82.27) --
	( 88.27, 82.27) --
	( 88.27, 82.27) --
	( 88.27, 82.27) --
	( 88.73, 82.27) --
	( 88.73, 82.27) --
	( 88.73, 82.27) --
	( 88.73, 82.27) --
	( 88.73, 82.27) --
	( 89.18, 82.27) --
	( 89.18, 82.27) --
	( 89.18, 82.27) --
	( 89.18, 82.27) --
	( 89.18, 82.27) --
	( 89.64, 82.27) --
	( 89.64, 82.27) --
	( 89.64, 82.27) --
	( 89.64, 82.27) --
	( 89.64, 82.27) --
	( 90.09, 82.27) --
	( 90.09, 82.27) --
	( 90.09, 82.27) --
	( 90.09, 82.27) --
	( 90.09, 82.27) --
	( 90.55, 82.27) --
	( 90.55, 82.27) --
	( 90.55, 82.27) --
	( 90.55, 82.27) --
	( 90.55, 82.27) --
	( 91.00, 82.27) --
	( 91.00, 82.27) --
	( 91.00, 82.27) --
	( 91.00, 82.27) --
	( 91.00, 82.27) --
	( 91.46, 82.27) --
	( 91.46, 82.27) --
	( 91.46, 82.27) --
	( 91.46, 82.27) --
	( 91.46, 82.27) --
	( 91.91, 82.27) --
	( 91.91, 82.27) --
	( 91.91, 82.27) --
	( 91.91, 82.27) --
	( 91.91, 82.27) --
	( 92.37, 82.27) --
	( 92.37, 82.27) --
	( 92.37, 82.27) --
	( 92.37, 82.27) --
	( 92.37, 82.27) --
	( 92.82, 82.27) --
	( 92.82, 82.27) --
	( 92.82, 82.27) --
	( 92.82, 82.27) --
	( 92.82, 82.27) --
	( 93.28, 82.27) --
	( 93.28, 82.27) --
	( 93.28, 82.27) --
	( 93.28, 82.27) --
	( 93.28, 82.27) --
	( 93.73, 82.27) --
	( 93.73, 82.27) --
	( 93.73, 82.27) --
	( 93.73, 82.27) --
	( 93.73, 82.27) --
	( 94.19, 82.27) --
	( 94.19, 82.27) --
	( 94.19, 82.27) --
	( 94.19, 82.27) --
	( 94.19, 82.27) --
	( 94.64, 82.27) --
	( 94.64, 82.27) --
	( 94.64, 82.27) --
	( 94.64, 82.27) --
	( 94.64, 82.27) --
	( 95.10, 82.27) --
	( 95.10, 82.27) --
	( 95.10, 82.27) --
	( 95.10, 82.27) --
	( 95.10, 82.27) --
	( 95.55, 82.27) --
	( 95.55, 82.27) --
	( 95.55, 82.27) --
	( 95.55, 82.27) --
	( 95.55, 82.27) --
	( 96.01, 82.27) --
	( 96.01, 82.27) --
	( 96.01, 82.27) --
	( 96.01, 82.27) --
	( 96.01, 82.27) --
	( 96.46, 82.27) --
	( 96.46, 82.27) --
	( 96.46, 82.27) --
	( 96.46, 82.27) --
	( 96.46, 82.27) --
	( 96.92, 82.27) --
	( 96.92, 82.27) --
	( 96.92, 82.27) --
	( 96.92, 82.27) --
	( 96.92, 82.27) --
	( 97.37, 82.27) --
	( 97.37, 82.27) --
	( 97.37, 82.27) --
	( 97.37, 82.27) --
	( 97.37, 82.27) --
	( 97.83, 82.27) --
	( 97.83, 82.27) --
	( 97.83, 82.27) --
	( 97.83, 82.27) --
	( 97.83, 82.27) --
	( 98.29, 82.27) --
	( 98.29, 82.27) --
	( 98.29, 82.27) --
	( 98.29, 82.27) --
	( 98.29, 82.27) --
	( 98.74, 82.27) --
	( 98.74, 82.27) --
	( 98.74, 82.27) --
	( 98.74, 82.27) --
	( 98.74, 82.27) --
	( 99.20, 82.27) --
	( 99.20, 82.27) --
	( 99.20, 82.27) --
	( 99.20, 82.27) --
	( 99.20, 82.27) --
	( 99.65, 82.27) --
	( 99.65, 82.27) --
	( 99.65, 82.27) --
	( 99.65, 82.27) --
	( 99.65, 82.27) --
	(100.11, 82.27) --
	(100.11, 82.27) --
	(100.11, 82.27) --
	(100.11, 82.27) --
	(100.11, 82.27) --
	(100.56, 82.27) --
	(100.56, 82.27) --
	(100.56, 82.27) --
	(100.56, 82.27) --
	(100.56, 82.27) --
	(101.02, 82.27) --
	(101.02, 82.27) --
	(101.02, 82.27) --
	(101.02, 82.27) --
	(101.02, 82.27) --
	(101.47, 82.27) --
	(101.47, 82.27) --
	(101.47, 82.27) --
	(101.47, 82.27) --
	(101.47, 82.27) --
	(101.93, 82.27) --
	(101.93, 82.27) --
	(101.93, 82.27) --
	(101.93, 82.27) --
	(101.93, 82.27) --
	(102.38, 82.27) --
	(102.38, 82.27) --
	(102.38, 82.27) --
	(102.38, 82.27) --
	(102.38, 82.27) --
	(102.84, 82.27) --
	(102.84, 82.27) --
	(102.84, 82.27) --
	(102.84, 82.27) --
	(102.84, 82.27) --
	(103.29, 82.27) --
	(103.29, 82.27) --
	(103.29, 82.27) --
	(103.29, 82.27) --
	(103.29, 82.27) --
	(103.75, 82.27) --
	(103.75, 82.27) --
	(103.75, 82.27) --
	(103.75, 82.27) --
	(103.75, 82.27) --
	(104.20, 82.27) --
	(104.20, 82.27) --
	(104.20, 82.27) --
	(104.20, 82.27) --
	(104.20, 82.27) --
	(104.66, 82.27) --
	(104.66, 82.27) --
	(104.66, 82.27) --
	(104.66, 82.27) --
	(104.66, 82.27) --
	(105.11, 82.27) --
	(105.11, 82.27) --
	(105.11, 82.27) --
	(105.11, 82.27) --
	(105.11, 82.27) --
	(105.57, 82.27) --
	(105.57, 82.27) --
	(105.57, 82.27) --
	(105.57, 82.27) --
	(105.57, 82.27) --
	(106.02, 82.27) --
	(106.02, 82.27) --
	(106.02, 82.27) --
	(106.02, 82.27) --
	(106.02, 82.27) --
	(106.48, 82.27) --
	(106.48, 82.27) --
	(106.48, 82.27) --
	(106.48, 82.27) --
	(106.48, 82.27) --
	(106.93, 82.27) --
	(106.93, 82.27) --
	(106.93, 82.27) --
	(106.93, 82.27) --
	(106.93, 82.27) --
	(107.39, 82.27) --
	(107.39, 82.27) --
	(107.39, 82.27) --
	(107.39, 82.27) --
	(107.39, 82.27) --
	(107.84, 82.27) --
	(107.84, 82.27) --
	(107.84, 82.27) --
	(107.84, 82.27) --
	(107.84, 82.27) --
	(108.30, 82.27) --
	(108.30, 82.27) --
	(108.30, 82.27) --
	(108.30, 82.27) --
	(108.30, 82.27) --
	(108.75, 82.27) --
	(108.75, 82.27) --
	(108.75, 82.27) --
	(108.75, 82.27) --
	(108.75, 82.27) --
	(109.21, 82.27) --
	(109.21, 82.27) --
	(109.21, 82.27) --
	(109.21, 82.27) --
	(109.21, 82.27) --
	(109.67, 82.27) --
	(109.67, 82.27) --
	(109.67, 82.27) --
	(109.67, 82.27) --
	(109.67, 82.27) --
	(110.12, 82.27) --
	(110.12, 82.27) --
	(110.12, 82.27) --
	(110.12, 82.27) --
	(110.12, 82.27) --
	(110.58, 82.27) --
	(110.58, 82.27) --
	(110.58, 82.27) --
	(110.58, 82.27) --
	(110.58, 82.27) --
	(111.03, 82.27) --
	(111.03, 82.27) --
	(111.03, 82.27) --
	(111.03, 82.27) --
	(111.03, 82.27) --
	(111.49, 82.27) --
	(111.49, 82.27) --
	(111.49, 82.27) --
	(111.49, 82.27) --
	(111.49, 82.27) --
	(111.94, 82.27) --
	(111.94, 82.27) --
	(111.94, 82.27) --
	(111.94, 82.27) --
	(111.94, 82.27) --
	(112.03, 82.27) --
	(112.40, 82.27) --
	(112.40, 82.27) --
	(112.40, 82.27) --
	(112.40, 82.27) --
	(112.85, 82.27) --
	(112.85, 82.27) --
	(112.85, 82.27) --
	(112.85, 82.27) --
	(113.31, 82.27) --
	(113.31, 82.27) --
	(113.31, 82.27) --
	(113.31, 82.27) --
	(113.76, 82.27) --
	(113.76, 82.27) --
	(113.76, 82.27) --
	(113.76, 82.27) --
	(114.22, 82.27) --
	(114.22, 82.27) --
	(114.22, 82.27) --
	(114.22, 82.27) --
	(114.67, 82.27) --
	(114.67, 82.27) --
	(114.67, 82.27) --
	(114.67, 82.27) --
	(115.13, 82.27) --
	(115.13, 82.27) --
	(115.13, 82.27) --
	(115.13, 82.27) --
	(115.58, 82.27) --
	(115.58, 82.27) --
	(115.58, 82.27) --
	(115.58, 82.27) --
	(115.85, 82.27) --
	(116.04, 82.27) --
	(116.04, 82.27) --
	(116.04, 82.27) --
	(116.49, 82.27) --
	(116.49, 82.27) --
	(116.49, 82.27) --
	(116.95, 82.27) --
	(116.95, 82.27) --
	(116.95, 82.27) --
	(117.40, 82.27) --
	(117.40, 82.27) --
	(117.40, 82.27) --
	(117.46, 82.27) --
	(117.86, 82.27) --
	(117.86, 82.27);
\definecolor{drawColor}{gray}{0.70}

\path[draw=drawColor,line width= 0.5pt,line join=round,line cap=round] ( 27.27, 29.68) rectangle (117.86,194.47);
\end{scope}
\begin{scope}
\path[clip] (  0.00,  0.00) rectangle (122.86,199.47);
\definecolor{drawColor}{gray}{0.30}

\node[text=drawColor,anchor=base east,inner sep=0pt, outer sep=0pt, scale=  1.00] at ( 22.77, 43.76) {0.99};

\node[text=drawColor,anchor=base east,inner sep=0pt, outer sep=0pt, scale=  1.00] at ( 22.77, 78.82) {1.00};

\node[text=drawColor,anchor=base east,inner sep=0pt, outer sep=0pt, scale=  1.00] at ( 22.77,113.89) {1.01};

\node[text=drawColor,anchor=base east,inner sep=0pt, outer sep=0pt, scale=  1.00] at ( 22.77,148.95) {1.02};

\node[text=drawColor,anchor=base east,inner sep=0pt, outer sep=0pt, scale=  1.00] at ( 22.77,184.01) {1.03};
\end{scope}
\begin{scope}
\path[clip] (  0.00,  0.00) rectangle (122.86,199.47);
\definecolor{drawColor}{gray}{0.70}

\path[draw=drawColor,line width= 0.3pt,line join=round] ( 24.77, 47.21) --
	( 27.27, 47.21);

\path[draw=drawColor,line width= 0.3pt,line join=round] ( 24.77, 82.27) --
	( 27.27, 82.27);

\path[draw=drawColor,line width= 0.3pt,line join=round] ( 24.77,117.33) --
	( 27.27,117.33);

\path[draw=drawColor,line width= 0.3pt,line join=round] ( 24.77,152.39) --
	( 27.27,152.39);

\path[draw=drawColor,line width= 0.3pt,line join=round] ( 24.77,187.45) --
	( 27.27,187.45);
\end{scope}
\begin{scope}
\path[clip] (  0.00,  0.00) rectangle (122.86,199.47);
\definecolor{drawColor}{gray}{0.70}

\path[draw=drawColor,line width= 0.3pt,line join=round] ( 27.27, 27.18) --
	( 27.27, 29.68);

\path[draw=drawColor,line width= 0.3pt,line join=round] ( 45.39, 27.18) --
	( 45.39, 29.68);

\path[draw=drawColor,line width= 0.3pt,line join=round] ( 63.51, 27.18) --
	( 63.51, 29.68);

\path[draw=drawColor,line width= 0.3pt,line join=round] ( 81.62, 27.18) --
	( 81.62, 29.68);

\path[draw=drawColor,line width= 0.3pt,line join=round] ( 99.74, 27.18) --
	( 99.74, 29.68);

\path[draw=drawColor,line width= 0.3pt,line join=round] (117.86, 27.18) --
	(117.86, 29.68);
\end{scope}
\begin{scope}
\path[clip] (  0.00,  0.00) rectangle (122.86,199.47);
\definecolor{drawColor}{gray}{0.30}

\node[text=drawColor,anchor=base,inner sep=0pt, outer sep=0pt, scale=  1.00] at ( 26.27, 18.29) {0};

\node[text=drawColor,anchor=base,inner sep=0pt, outer sep=0pt, scale=  1.00] at ( 43.39, 18.29) {20};

\node[text=drawColor,anchor=base,inner sep=0pt, outer sep=0pt, scale=  1.00] at ( 61.51, 18.29) {40};

\node[text=drawColor,anchor=base,inner sep=0pt, outer sep=0pt, scale=  1.00] at ( 79.63, 18.29) {60};

\node[text=drawColor,anchor=base,inner sep=0pt, outer sep=0pt, scale=  1.00] at ( 97.74, 18.29) {80};

\node[text=drawColor,anchor=base,inner sep=0pt, outer sep=0pt, scale=  1.00] at (114.86, 18.29) {100};
\end{scope}
\begin{scope}
\path[clip] (  0.00,  0.00) rectangle (122.86,199.47);
\definecolor{drawColor}{RGB}{0,0,0}

\node[text=drawColor,anchor=base,inner sep=0pt, outer sep=0pt, scale=  1.00] at ( 72.57,  6.94) {\bfseries Instances in \%};
\end{scope}
\end{tikzpicture}

%% file: 06-plots/2019-SEA-SAvsAlgo-2Seconds-Shortened.tex
\begin{tikzpicture}[x=1pt,y=1pt]
\definecolor{fillColor}{RGB}{255,255,255}
\path[use as bounding box,fill=fillColor,fill opacity=0.00] (0,0) rectangle (171.28,199.47);
\begin{scope}
\path[clip] (  0.00,  0.00) rectangle (171.28,199.47);
\definecolor{drawColor}{RGB}{255,255,255}
\definecolor{fillColor}{RGB}{255,255,255}

\path[draw=drawColor,line width= 0.5pt,line join=round,line cap=round,fill=fillColor] (  0.00,  0.00) rectangle (171.28,199.47);
\end{scope}
\begin{scope}
\path[clip] ( 33.62, 29.68) rectangle (166.28,194.47);
\definecolor{fillColor}{RGB}{255,255,255}

\path[fill=fillColor] ( 33.62, 29.68) rectangle (166.28,194.47);
\definecolor{drawColor}{gray}{0.87}

\path[draw=drawColor,line width= 0.1pt,line join=round] ( 33.62, 57.50) --
	(166.28, 57.50);

\path[draw=drawColor,line width= 0.1pt,line join=round] ( 33.62,113.16) --
	(166.28,113.16);

\path[draw=drawColor,line width= 0.1pt,line join=round] ( 33.62,168.81) --
	(166.28,168.81);

\path[draw=drawColor,line width= 0.1pt,line join=round] ( 46.89, 29.68) --
	( 46.89,194.47);

\path[draw=drawColor,line width= 0.1pt,line join=round] ( 73.42, 29.68) --
	( 73.42,194.47);

\path[draw=drawColor,line width= 0.1pt,line join=round] ( 99.95, 29.68) --
	( 99.95,194.47);

\path[draw=drawColor,line width= 0.1pt,line join=round] (126.48, 29.68) --
	(126.48,194.47);

\path[draw=drawColor,line width= 0.1pt,line join=round] (153.01, 29.68) --
	(153.01,194.47);

\path[draw=drawColor,line width= 0.3pt,line join=round] ( 33.62, 85.33) --
	(166.28, 85.33);

\path[draw=drawColor,line width= 0.3pt,line join=round] ( 33.62,140.99) --
	(166.28,140.99);

\path[draw=drawColor,line width= 0.3pt,line join=round] ( 33.62, 29.68) --
	( 33.62,194.47);

\path[draw=drawColor,line width= 0.3pt,line join=round] ( 60.15, 29.68) --
	( 60.15,194.47);

\path[draw=drawColor,line width= 0.3pt,line join=round] ( 86.68, 29.68) --
	( 86.68,194.47);

\path[draw=drawColor,line width= 0.3pt,line join=round] (113.22, 29.68) --
	(113.22,194.47);

\path[draw=drawColor,line width= 0.3pt,line join=round] (139.75, 29.68) --
	(139.75,194.47);

\path[draw=drawColor,line width= 0.3pt,line join=round] (166.28, 29.68) --
	(166.28,194.47);
\definecolor{drawColor}{RGB}{128,128,128}

\path[draw=drawColor,line width= 1.0pt,line join=round] ( 33.62, 72.13) --
	( 34.29, 85.66) --
	( 34.95, 98.97) --
	( 35.62,103.82) --
	( 36.29,111.05) --
	( 36.95,123.84) --
	( 37.62,124.11) --
	( 38.29,124.67) --
	( 38.95,124.94) --
	( 39.62,125.68) --
	( 40.29,126.30) --
	( 40.95,128.59) --
	( 41.62,129.74) --
	( 42.29,130.02) --
	( 42.95,130.11) --
	( 43.62,130.38) --
	( 44.29,130.75) --
	( 44.95,130.77) --
	( 45.62,131.36) --
	( 46.29,131.52) --
	( 46.95,131.78) --
	( 47.62,132.02) --
	( 48.29,132.07) --
	( 48.95,132.19) --
	( 49.62,132.81) --
	( 50.29,132.83) --
	( 50.95,133.41) --
	( 51.62,133.82) --
	( 52.29,134.15) --
	( 52.95,134.27) --
	( 53.62,134.65) --
	( 54.29,135.07) --
	( 54.95,135.11) --
	( 55.62,135.48) --
	( 56.29,135.49) --
	( 56.95,135.55) --
	( 57.62,135.66) --
	( 58.29,135.71) --
	( 58.95,135.88) --
	( 59.62,135.95) --
	( 60.29,136.26) --
	( 60.95,136.67) --
	( 61.62,136.69) --
	( 62.28,136.73) --
	( 62.95,136.76) --
	( 63.62,137.10) --
	( 64.28,137.41) --
	( 64.95,137.44) --
	( 65.62,137.47) --
	( 66.28,137.64) --
	( 66.95,137.66) --
	( 67.62,138.02) --
	( 68.28,138.09) --
	( 68.95,138.14) --
	( 69.62,138.17) --
	( 70.28,138.27) --
	( 70.95,138.27) --
	( 71.62,138.44) --
	( 72.28,138.47) --
	( 72.95,138.59) --
	( 73.62,138.68) --
	( 74.28,138.70) --
	( 74.95,138.73) --
	( 75.62,138.78) --
	( 76.28,138.79) --
	( 76.95,138.81) --
	( 77.62,138.86) --
	( 78.28,138.87) --
	( 78.95,138.88) --
	( 79.62,138.93) --
	( 80.28,139.00) --
	( 80.95,139.07) --
	( 81.62,139.28) --
	( 82.28,139.33) --
	( 82.95,139.56) --
	( 83.62,139.63) --
	( 84.28,139.70) --
	( 84.95,139.79) --
	( 85.62,139.81) --
	( 86.28,139.91) --
	( 86.95,139.94) --
	( 87.62,139.97) --
	( 88.28,139.98) --
	( 88.95,140.00) --
	( 89.62,140.02) --
	( 90.28,140.03) --
	( 90.95,140.17) --
	( 91.62,140.19) --
	( 92.28,140.32) --
	( 92.95,140.44) --
	( 93.62,140.46) --
	( 94.28,140.53) --
	( 94.95,140.54) --
	( 95.62,140.58) --
	( 96.28,140.67) --
	( 96.95,140.85) --
	( 97.62,140.88) --
	( 98.28,140.90) --
	( 98.95,140.91) --
	( 99.62,140.96) --
	(100.28,140.98) --
	(100.95,140.99) --
	(101.62,140.99) --
	(102.28,140.99) --
	(102.95,140.99) --
	(103.62,140.99) --
	(104.28,140.99) --
	(104.95,140.99) --
	(105.62,140.99) --
	(106.28,140.99) --
	(106.95,140.99) --
	(107.62,140.99) --
	(108.28,140.99) --
	(108.95,140.99) --
	(109.62,140.99) --
	(110.28,140.99) --
	(110.95,140.99) --
	(111.62,140.99) --
	(112.28,140.99) --
	(112.95,140.99) --
	(113.62,140.99) --
	(114.28,140.99) --
	(114.95,140.99) --
	(115.62,140.99) --
	(116.28,140.99) --
	(116.95,140.99) --
	(117.62,140.99) --
	(118.28,140.99) --
	(118.95,140.99) --
	(119.62,140.99) --
	(120.28,140.99) --
	(120.95,140.99) --
	(121.62,140.99) --
	(122.28,140.99) --
	(122.95,140.99) --
	(123.62,140.99) --
	(124.28,140.99) --
	(124.95,140.99) --
	(125.62,140.99) --
	(126.28,140.99) --
	(126.95,140.99) --
	(127.62,140.99) --
	(128.28,140.99) --
	(128.95,140.99) --
	(129.62,140.99) --
	(130.28,140.99) --
	(130.95,140.99) --
	(131.61,140.99) --
	(132.28,140.99) --
	(132.95,141.00) --
	(133.61,141.04) --
	(134.28,141.05) --
	(134.95,141.06) --
	(135.61,141.07) --
	(136.28,141.11) --
	(136.95,141.12) --
	(137.61,141.16) --
	(138.28,141.21) --
	(138.95,141.21) --
	(139.61,141.22) --
	(140.28,141.23) --
	(140.95,141.24) --
	(141.61,141.25) --
	(142.28,141.37) --
	(142.95,141.46) --
	(143.61,141.47) --
	(144.28,141.48) --
	(144.95,141.50) --
	(145.61,141.50) --
	(146.28,141.51) --
	(146.95,141.56) --
	(147.61,141.57) --
	(148.28,141.59) --
	(148.95,141.62) --
	(149.61,141.62) --
	(150.28,141.64) --
	(150.95,141.64) --
	(151.61,141.72) --
	(152.28,141.74) --
	(152.95,141.80) --
	(153.61,141.82) --
	(154.28,141.83) --
	(154.95,141.84) --
	(155.61,141.91) --
	(156.28,142.08) --
	(156.95,142.18) --
	(157.61,142.20) --
	(158.28,142.25) --
	(158.95,142.46) --
	(159.61,142.47) --
	(160.28,142.58) --
	(160.95,142.67) --
	(161.61,142.92) --
	(162.28,142.97) --
	(162.95,143.09) --
	(163.61,143.11) --
	(164.28,143.12) --
	(164.95,143.35) --
	(165.61,143.47) --
	(166.28,143.57);
\definecolor{drawColor}{RGB}{152,167,197}

\path[draw=drawColor,line width= 1.0pt,line join=round] ( 33.62, 29.68) --
	( 34.29,131.10) --
	( 34.95,134.67) --
	( 35.62,135.11) --
	( 36.29,135.17) --
	( 36.95,135.18) --
	( 37.62,135.30) --
	( 38.29,135.39) --
	( 38.95,135.71) --
	( 39.62,135.74) --
	( 40.29,135.75) --
	( 40.95,135.76) --
	( 41.62,135.88) --
	( 42.29,135.91) --
	( 42.95,135.96) --
	( 43.62,136.00) --
	( 44.29,136.02) --
	( 44.95,136.04) --
	( 45.62,136.07) --
	( 46.29,136.11) --
	( 46.95,136.13) --
	( 47.62,136.23) --
	( 48.29,136.32) --
	( 48.95,136.33) --
	( 49.62,136.50) --
	( 50.29,136.54) --
	( 50.95,136.57) --
	( 51.62,136.66) --
	( 52.29,136.69) --
	( 52.95,136.69) --
	( 53.62,136.72) --
	( 54.29,136.73) --
	( 54.95,136.82) --
	( 55.62,136.86) --
	( 56.29,136.87) --
	( 56.95,136.88) --
	( 57.62,136.94) --
	( 58.29,137.00) --
	( 58.95,137.07) --
	( 59.62,137.17) --
	( 60.29,137.25) --
	( 60.95,137.31) --
	( 61.62,137.38) --
	( 62.28,137.38) --
	( 62.95,137.39) --
	( 63.62,137.40) --
	( 64.28,137.43) --
	( 64.95,137.45) --
	( 65.62,137.47) --
	( 66.28,137.48) --
	( 66.95,137.53) --
	( 67.62,137.58) --
	( 68.28,137.63) --
	( 68.95,137.67) --
	( 69.62,137.72) --
	( 70.28,137.77) --
	( 70.95,137.86) --
	( 71.62,137.90) --
	( 72.28,137.92) --
	( 72.95,137.94) --
	( 73.62,137.95) --
	( 74.28,137.98) --
	( 74.95,138.04) --
	( 75.62,138.12) --
	( 76.28,138.17) --
	( 76.95,138.19) --
	( 77.62,138.21) --
	( 78.28,138.22) --
	( 78.95,138.29) --
	( 79.62,138.33) --
	( 80.28,138.34) --
	( 80.95,138.38) --
	( 81.62,138.45) --
	( 82.28,138.45) --
	( 82.95,138.47) --
	( 83.62,138.51) --
	( 84.28,138.51) --
	( 84.95,138.54) --
	( 85.62,138.54) --
	( 86.28,138.55) --
	( 86.95,138.60) --
	( 87.62,138.60) --
	( 88.28,138.64) --
	( 88.95,138.66) --
	( 89.62,138.69) --
	( 90.28,138.70) --
	( 90.95,138.71) --
	( 91.62,138.79) --
	( 92.28,138.80) --
	( 92.95,138.81) --
	( 93.62,138.82) --
	( 94.28,138.85) --
	( 94.95,138.86) --
	( 95.62,138.92) --
	( 96.28,138.94) --
	( 96.95,138.94) --
	( 97.62,139.00) --
	( 98.28,139.05) --
	( 98.95,139.05) --
	( 99.62,139.18) --
	(100.28,139.20) --
	(100.95,139.23) --
	(101.62,139.30) --
	(102.28,139.30) --
	(102.95,139.35) --
	(103.62,139.43) --
	(104.28,139.46) --
	(104.95,139.53) --
	(105.62,139.62) --
	(106.28,139.72) --
	(106.95,139.72) --
	(107.62,139.74) --
	(108.28,139.75) --
	(108.95,139.76) --
	(109.62,139.77) --
	(110.28,139.90) --
	(110.95,139.90) --
	(111.62,139.91) --
	(112.28,139.92) --
	(112.95,139.92) --
	(113.62,139.96) --
	(114.28,139.96) --
	(114.95,139.96) --
	(115.62,139.97) --
	(116.28,139.97) --
	(116.95,140.03) --
	(117.62,140.04) --
	(118.28,140.07) --
	(118.95,140.07) --
	(119.62,140.08) --
	(120.28,140.13) --
	(120.95,140.13) --
	(121.62,140.15) --
	(122.28,140.15) --
	(122.95,140.16) --
	(123.62,140.20) --
	(124.28,140.20) --
	(124.95,140.21) --
	(125.62,140.21) --
	(126.28,140.27) --
	(126.95,140.28) --
	(127.62,140.38) --
	(128.28,140.41) --
	(128.95,140.42) --
	(129.62,140.43) --
	(130.28,140.44) --
	(130.95,140.49) --
	(131.61,140.53) --
	(132.28,140.60) --
	(132.95,140.62) --
	(133.61,140.63) --
	(134.28,140.65) --
	(134.95,140.67) --
	(135.61,140.68) --
	(136.28,140.70) --
	(136.95,140.70) --
	(137.61,140.73) --
	(138.28,140.73) --
	(138.95,140.83) --
	(139.61,140.86) --
	(140.28,140.87) --
	(140.95,140.87) --
	(141.61,140.91) --
	(142.28,140.94) --
	(142.95,140.95) --
	(143.61,140.97) --
	(144.28,140.99) --
	(144.95,140.99) --
	(145.61,140.99) --
	(146.28,140.99) --
	(146.95,140.99) --
	(147.61,140.99) --
	(148.28,140.99) --
	(148.95,140.99) --
	(149.61,140.99) --
	(150.28,140.99) --
	(150.95,140.99) --
	(151.61,140.99) --
	(152.28,140.99) --
	(152.95,140.99) --
	(153.61,140.99) --
	(154.28,141.02) --
	(154.95,141.07) --
	(155.61,141.11) --
	(156.28,141.22) --
	(156.95,141.28) --
	(157.61,141.33) --
	(158.28,141.35) --
	(158.95,141.40) --
	(159.61,141.44) --
	(160.28,141.47) --
	(160.95,141.57) --
	(161.61,141.63) --
	(162.28,141.82) --
	(162.95,141.83) --
	(163.61,141.97) --
	(164.28,142.03) --
	(164.95,142.52) --
	(165.61,143.52) --
	(166.28,144.77);
\definecolor{drawColor}{RGB}{128,202,192}

\path[draw=drawColor,line width= 1.0pt,line join=round] ( 33.62,127.06) --
	( 34.29,128.00) --
	( 34.95,128.08) --
	( 35.62,128.16) --
	( 36.29,128.99) --
	( 36.95,129.14) --
	( 37.62,129.20) --
	( 38.29,129.24) --
	( 38.95,129.30) --
	( 39.62,129.55) --
	( 40.29,129.55) --
	( 40.95,129.58) --
	( 41.62,129.68) --
	( 42.29,130.05) --
	( 42.95,130.05) --
	( 43.62,130.06) --
	( 44.29,130.51) --
	( 44.95,130.53) --
	( 45.62,130.76) --
	( 46.29,130.78) --
	( 46.95,130.90) --
	( 47.62,130.90) --
	( 48.29,130.97) --
	( 48.95,131.08) --
	( 49.62,131.10) --
	( 50.29,131.11) --
	( 50.95,131.30) --
	( 51.62,131.32) --
	( 52.29,131.33) --
	( 52.95,131.46) --
	( 53.62,131.49) --
	( 54.29,131.51) --
	( 54.95,131.55) --
	( 55.62,131.69) --
	( 56.29,131.69) --
	( 56.95,131.74) --
	( 57.62,131.80) --
	( 58.29,131.85) --
	( 58.95,131.88) --
	( 59.62,132.01) --
	( 60.29,132.02) --
	( 60.95,132.12) --
	( 61.62,132.16) --
	( 62.28,132.24) --
	( 62.95,132.25) --
	( 63.62,132.25) --
	( 64.28,132.28) --
	( 64.95,132.37) --
	( 65.62,132.46) --
	( 66.28,132.53) --
	( 66.95,132.54) --
	( 67.62,132.58) --
	( 68.28,132.60) --
	( 68.95,132.63) --
	( 69.62,132.68) --
	( 70.28,132.75) --
	( 70.95,132.75) --
	( 71.62,132.76) --
	( 72.28,132.76) --
	( 72.95,132.77) --
	( 73.62,132.77) --
	( 74.28,132.81) --
	( 74.95,132.83) --
	( 75.62,132.86) --
	( 76.28,132.94) --
	( 76.95,132.94) --
	( 77.62,133.03) --
	( 78.28,133.09) --
	( 78.95,133.10) --
	( 79.62,133.10) --
	( 80.28,133.13) --
	( 80.95,133.15) --
	( 81.62,133.17) --
	( 82.28,133.17) --
	( 82.95,133.18) --
	( 83.62,133.19) --
	( 84.28,133.28) --
	( 84.95,133.30) --
	( 85.62,133.32) --
	( 86.28,133.40) --
	( 86.95,133.45) --
	( 87.62,133.45) --
	( 88.28,133.48) --
	( 88.95,133.51) --
	( 89.62,133.52) --
	( 90.28,133.52) --
	( 90.95,133.55) --
	( 91.62,133.60) --
	( 92.28,133.63) --
	( 92.95,133.67) --
	( 93.62,133.67) --
	( 94.28,133.68) --
	( 94.95,133.68) --
	( 95.62,133.73) --
	( 96.28,133.74) --
	( 96.95,133.79) --
	( 97.62,133.94) --
	( 98.28,133.94) --
	( 98.95,133.95) --
	( 99.62,133.98) --
	(100.28,134.00) --
	(100.95,134.00) --
	(101.62,134.00) --
	(102.28,134.04) --
	(102.95,134.04) --
	(103.62,134.05) --
	(104.28,134.06) --
	(104.95,134.09) --
	(105.62,134.10) --
	(106.28,134.15) --
	(106.95,134.16) --
	(107.62,134.16) --
	(108.28,134.17) --
	(108.95,134.22) --
	(109.62,134.24) --
	(110.28,134.25) --
	(110.95,134.26) --
	(111.62,134.34) --
	(112.28,134.42) --
	(112.95,134.51) --
	(113.62,134.51) --
	(114.28,134.51) --
	(114.95,134.56) --
	(115.62,134.57) --
	(116.28,134.59) --
	(116.95,134.61) --
	(117.62,134.62) --
	(118.28,134.63) --
	(118.95,134.65) --
	(119.62,134.66) --
	(120.28,134.66) --
	(120.95,134.73) --
	(121.62,134.75) --
	(122.28,134.79) --
	(122.95,134.84) --
	(123.62,134.86) --
	(124.28,134.89) --
	(124.95,134.90) --
	(125.62,134.93) --
	(126.28,134.95) --
	(126.95,134.96) --
	(127.62,134.98) --
	(128.28,135.01) --
	(128.95,135.03) --
	(129.62,135.12) --
	(130.28,135.16) --
	(130.95,135.18) --
	(131.61,135.23) --
	(132.28,135.23) --
	(132.95,135.23) --
	(133.61,135.26) --
	(134.28,135.27) --
	(134.95,135.28) --
	(135.61,135.29) --
	(136.28,135.30) --
	(136.95,135.34) --
	(137.61,135.38) --
	(138.28,135.40) --
	(138.95,135.42) --
	(139.61,135.43) --
	(140.28,135.44) --
	(140.95,135.47) --
	(141.61,135.48) --
	(142.28,135.48) --
	(142.95,135.49) --
	(143.61,135.50) --
	(144.28,135.51) --
	(144.95,135.57) --
	(145.61,135.60) --
	(146.28,135.62) --
	(146.95,135.63) --
	(147.61,135.68) --
	(148.28,135.80) --
	(148.95,135.84) --
	(149.61,135.93) --
	(150.28,135.97) --
	(150.95,135.97) --
	(151.61,136.06) --
	(152.28,136.07) --
	(152.95,136.14) --
	(153.61,136.14) --
	(154.28,136.17) --
	(154.95,136.21) --
	(155.61,136.29) --
	(156.28,136.31) --
	(156.95,136.36) --
	(157.61,136.50) --
	(158.28,136.62) --
	(158.95,136.86) --
	(159.61,136.92) --
	(160.28,136.94) --
	(160.95,136.98) --
	(161.61,137.00) --
	(162.28,137.08) --
	(162.95,137.22) --
	(163.61,137.36) --
	(164.28,137.78) --
	(164.95,138.37) --
	(165.61,138.63) --
	(166.28,138.81);
\definecolor{drawColor}{RGB}{207,128,187}

\path[draw=drawColor,line width= 1.0pt,line join=round] ( 33.62, 52.25) --
	( 34.29, 60.90) --
	( 34.95, 73.87) --
	( 35.62, 91.99) --
	( 36.29, 95.66) --
	( 36.95,103.96) --
	( 37.62,107.73) --
	( 38.29,131.11) --
	( 38.95,132.01) --
	( 39.62,133.78) --
	( 40.29,133.97) --
	( 40.95,135.52) --
	( 41.62,136.30) --
	( 42.29,136.82) --
	( 42.95,138.04) --
	( 43.62,138.22) --
	( 44.29,138.43) --
	( 44.95,138.53) --
	( 45.62,138.71) --
	( 46.29,138.85) --
	( 46.95,138.96) --
	( 47.62,139.11) --
	( 48.29,139.13) --
	( 48.95,139.60) --
	( 49.62,139.66) --
	( 50.29,139.94) --
	( 50.95,140.72) --
	( 51.62,140.90) --
	( 52.29,140.91) --
	( 52.95,140.97) --
	( 53.62,141.03) --
	( 54.29,141.23) --
	( 54.95,141.26) --
	( 55.62,141.43) --
	( 56.29,141.46) --
	( 56.95,141.48) --
	( 57.62,141.49) --
	( 58.29,141.61) --
	( 58.95,141.64) --
	( 59.62,141.84) --
	( 60.29,142.04) --
	( 60.95,142.24) --
	( 61.62,142.25) --
	( 62.28,142.35) --
	( 62.95,142.35) --
	( 63.62,142.56) --
	( 64.28,142.75) --
	( 64.95,142.84) --
	( 65.62,142.92) --
	( 66.28,143.20) --
	( 66.95,143.25) --
	( 67.62,143.31) --
	( 68.28,143.37) --
	( 68.95,143.39) --
	( 69.62,143.60) --
	( 70.28,143.62) --
	( 70.95,143.68) --
	( 71.62,143.72) --
	( 72.28,143.82) --
	( 72.95,143.83) --
	( 73.62,144.18) --
	( 74.28,144.19) --
	( 74.95,144.31) --
	( 75.62,144.32) --
	( 76.28,144.33) --
	( 76.95,144.64) --
	( 77.62,144.84) --
	( 78.28,144.85) --
	( 78.95,144.92) --
	( 79.62,145.03) --
	( 80.28,145.18) --
	( 80.95,145.25) --
	( 81.62,145.45) --
	( 82.28,145.54) --
	( 82.95,145.60) --
	( 83.62,145.90) --
	( 84.28,146.03) --
	( 84.95,146.15) --
	( 85.62,146.29) --
	( 86.28,146.52) --
	( 86.95,146.52) --
	( 87.62,146.52) --
	( 88.28,146.52) --
	( 88.95,146.65) --
	( 89.62,146.77) --
	( 90.28,146.90) --
	( 90.95,146.95) --
	( 91.62,146.98) --
	( 92.28,146.98) --
	( 92.95,146.99) --
	( 93.62,147.06) --
	( 94.28,147.15) --
	( 94.95,147.34) --
	( 95.62,147.51) --
	( 96.28,147.53) --
	( 96.95,147.56) --
	( 97.62,147.77) --
	( 98.28,147.78) --
	( 98.95,147.81) --
	( 99.62,147.83) --
	(100.28,147.90) --
	(100.95,147.93) --
	(101.62,147.95) --
	(102.28,148.03) --
	(102.95,148.05) --
	(103.62,148.19) --
	(104.28,148.26) --
	(104.95,148.32) --
	(105.62,148.38) --
	(106.28,148.39) --
	(106.95,148.62) --
	(107.62,148.71) --
	(108.28,148.75) --
	(108.95,148.85) --
	(109.62,148.91) --
	(110.28,149.02) --
	(110.95,149.03) --
	(111.62,149.17) --
	(112.28,149.25) --
	(112.95,149.27) --
	(113.62,149.53) --
	(114.28,149.58) --
	(114.95,149.60) --
	(115.62,150.15) --
	(116.28,150.16) --
	(116.95,150.18) --
	(117.62,150.38) --
	(118.28,150.48) --
	(118.95,150.50) --
	(119.62,150.56) --
	(120.28,150.59) --
	(120.95,150.62) --
	(121.62,150.62) --
	(122.28,150.63) --
	(122.95,150.75) --
	(123.62,151.22) --
	(124.28,151.27) --
	(124.95,151.40) --
	(125.62,151.55) --
	(126.28,151.62) --
	(126.95,151.64) --
	(127.62,151.74) --
	(128.28,151.75) --
	(128.95,151.80) --
	(129.62,152.10) --
	(130.28,152.56) --
	(130.95,152.62) --
	(131.61,152.85) --
	(132.28,152.97) --
	(132.95,153.07) --
	(133.61,153.23) --
	(134.28,153.28) --
	(134.95,153.31) --
	(135.61,153.42) --
	(136.28,153.77) --
	(136.95,153.90) --
	(137.61,154.08) --
	(138.28,154.26) --
	(138.95,154.29) --
	(139.61,154.52) --
	(140.28,154.75) --
	(140.95,154.78) --
	(141.61,155.00) --
	(142.28,155.24) --
	(142.95,155.34) --
	(143.61,155.36) --
	(144.28,155.45) --
	(144.95,155.53) --
	(145.61,155.53) --
	(146.28,155.76) --
	(146.95,156.01) --
	(147.61,156.07) --
	(148.28,156.18) --
	(148.95,156.23) --
	(149.61,156.29) --
	(150.28,156.46) --
	(150.95,156.69) --
	(151.61,156.75) --
	(152.28,156.81) --
	(152.95,156.95) --
	(153.61,157.20) --
	(154.28,157.30) --
	(154.95,157.30) --
	(155.61,157.31) --
	(156.28,157.33) --
	(156.95,157.38) --
	(157.61,157.56) --
	(158.28,157.77) --
	(158.95,158.29) --
	(159.61,158.66) --
	(160.28,159.11) --
	(160.95,159.24) --
	(161.61,159.26) --
	(162.28,159.99) --
	(162.95,160.13) --
	(163.61,162.09) --
	(164.28,163.36) --
	(164.95,163.60) --
	(165.61,164.41) --
	(166.28,170.13);
\definecolor{drawColor}{RGB}{207,192,152}

\path[draw=drawColor,line width= 1.0pt,line join=round] ( 33.62, 29.68) --
	( 34.29, 29.68) --
	( 34.95, 29.68) --
	( 35.62, 29.68) --
	( 36.29, 29.68) --
	( 36.95, 29.68) --
	( 37.62, 29.68) --
	( 38.29, 29.68) --
	( 38.95, 29.68) --
	( 39.62, 29.68) --
	( 40.29, 29.68) --
	( 40.95, 29.68) --
	( 41.62, 29.68) --
	( 42.29, 29.68) --
	( 42.95, 29.68) --
	( 43.62, 29.68) --
	( 44.29, 29.68) --
	( 44.95, 29.68) --
	( 45.62, 29.68) --
	( 46.29, 29.68) --
	( 46.95, 29.68) --
	( 47.62, 29.68) --
	( 48.29, 29.68) --
	( 48.95, 29.68) --
	( 49.62, 29.68) --
	( 50.29, 29.68) --
	( 50.95, 29.68) --
	( 51.62, 29.68) --
	( 52.29, 29.68) --
	( 52.95, 29.68) --
	( 53.62, 29.68) --
	( 54.29, 29.68) --
	( 54.95, 29.68) --
	( 55.62, 29.68) --
	( 56.29, 29.68) --
	( 56.95, 29.68) --
	( 57.62, 29.68) --
	( 58.29, 29.68) --
	( 58.95, 29.68) --
	( 59.62, 29.68) --
	( 60.29, 29.68) --
	( 60.95, 29.68) --
	( 61.62, 29.68) --
	( 62.28, 29.68) --
	( 62.95, 29.68) --
	( 63.62, 29.68) --
	( 64.28, 29.68) --
	( 64.95, 29.68) --
	( 65.62, 29.68) --
	( 66.28, 29.68) --
	( 66.95, 29.68) --
	( 67.62, 29.68) --
	( 68.28, 29.68) --
	( 68.95, 29.68) --
	( 69.62, 29.68) --
	( 70.28, 29.68) --
	( 70.95, 29.68) --
	( 71.62, 29.68) --
	( 72.28, 29.68) --
	( 72.95, 29.68) --
	( 73.62, 29.68) --
	( 74.28, 29.68) --
	( 74.95, 29.68) --
	( 75.62, 29.68) --
	( 76.28, 29.68) --
	( 76.95, 29.68) --
	( 77.62, 29.68) --
	( 78.28, 29.68) --
	( 78.95, 29.68) --
	( 79.62, 29.68) --
	( 80.28, 29.68) --
	( 80.95, 29.68) --
	( 81.62, 29.68) --
	( 82.28, 29.68) --
	( 82.95, 29.68) --
	( 83.62, 29.68) --
	( 84.28, 29.68) --
	( 84.95,122.97) --
	( 85.62,135.66) --
	( 86.28,135.69) --
	( 86.95,136.75) --
	( 87.62,136.97) --
	( 88.28,138.59) --
	( 88.95,138.76) --
	( 89.62,138.77) --
	( 90.28,139.74) --
	( 90.95,141.99) --
	( 91.62,142.03) --
	( 92.28,142.48) --
	( 92.95,142.62) --
	( 93.62,144.83) --
	( 94.28,144.86) --
	( 94.95,144.86) --
	( 95.62,145.09) --
	( 96.28,145.37) --
	( 96.95,145.65) --
	( 97.62,145.88) --
	( 98.28,147.23) --
	( 98.95,147.77) --
	( 99.62,148.28) --
	(100.28,149.21) --
	(100.95,149.74) --
	(101.62,149.94) --
	(102.28,150.08) --
	(102.95,150.32) --
	(103.62,150.53) --
	(104.28,150.69) --
	(104.95,150.87) --
	(105.62,151.55) --
	(106.28,151.57) --
	(106.95,151.60) --
	(107.62,151.75) --
	(108.28,151.75) --
	(108.95,153.35) --
	(109.62,153.59) --
	(110.28,153.72) --
	(110.95,154.43) --
	(111.62,154.94) --
	(112.28,155.90) --
	(112.95,156.31) --
	(113.62,156.36) --
	(114.28,156.48) --
	(114.95,156.73) --
	(115.62,157.47) --
	(116.28,157.49) --
	(116.95,157.66) --
	(117.62,157.84) --
	(118.28,158.01) --
	(118.95,158.40) --
	(119.62,159.47) --
	(120.28,159.61) --
	(120.95,160.14) --
	(121.62,160.18) --
	(122.28,160.22) --
	(122.95,160.28) --
	(123.62,160.41) --
	(124.28,160.44) --
	(124.95,160.52) --
	(125.62,160.95) --
	(126.28,161.15) --
	(126.95,161.21) --
	(127.62,161.56) --
	(128.28,161.63) --
	(128.95,162.48) --
	(129.62,162.64) --
	(130.28,162.95) --
	(130.95,163.00) --
	(131.61,163.10) --
	(132.28,163.50) --
	(132.95,163.83) --
	(133.61,164.11) --
	(134.28,165.42) --
	(134.95,165.47) --
	(135.61,165.76) --
	(136.28,166.14) --
	(136.95,166.52) --
	(137.61,166.53) --
	(138.28,166.64) --
	(138.95,167.11) --
	(139.61,167.18) --
	(140.28,167.30) --
	(140.95,167.32) --
	(141.61,167.44) --
	(142.28,168.31) --
	(142.95,168.40) --
	(143.61,168.93) --
	(144.28,169.24) --
	(144.95,169.62) --
	(145.61,169.84) --
	(146.28,169.88) --
	(146.95,170.34) --
	(147.61,170.42) --
	(148.28,171.23) --
	(148.95,171.38) --
	(149.61,171.98) --
	(150.28,172.67) --
	(150.95,172.84) --
	(151.61,172.94) --
	(152.28,173.72) --
	(152.95,174.21) --
	(153.61,174.78) --
	(154.28,175.24) --
	(154.95,175.37) --
	(155.61,176.42) --
	(156.28,177.48) --
	(156.95,177.84) --
	(157.61,178.12) --
	(158.28,178.25) --
	(158.95,178.48) --
	(159.61,178.77) --
	(160.28,179.88) --
	(160.95,181.95) --
	(161.61,182.65) --
	(162.28,183.63) --
	(162.95,183.75) --
	(163.61,184.15) --
	(164.28,188.24) --
	(164.95,188.93) --
	(165.61,189.28) --
	(166.28,194.47);
\definecolor{drawColor}{RGB}{160,30,40}

\path[draw=drawColor,line width= 1.0pt,line join=round] ( 33.62,140.99) --
	( 33.62,140.99) --
	( 33.62,140.99) --
	( 33.62,140.99) --
	( 33.62,140.99) --
	( 34.29,140.99) --
	( 34.29,140.99) --
	( 34.29,140.99) --
	( 34.29,140.99) --
	( 34.29,140.99) --
	( 34.95,140.99) --
	( 34.95,140.99) --
	( 34.95,140.99) --
	( 34.95,140.99) --
	( 34.95,140.99) --
	( 35.62,140.99) --
	( 35.62,140.99) --
	( 35.62,140.99) --
	( 35.62,140.99) --
	( 35.62,140.99) --
	( 36.29,140.99) --
	( 36.29,140.99) --
	( 36.29,140.99) --
	( 36.29,140.99) --
	( 36.29,140.99) --
	( 36.95,140.99) --
	( 36.95,140.99) --
	( 36.95,140.99) --
	( 36.95,140.99) --
	( 36.95,140.99) --
	( 37.62,140.99) --
	( 37.62,140.99) --
	( 37.62,140.99) --
	( 37.62,140.99) --
	( 37.62,140.99) --
	( 38.29,140.99) --
	( 38.29,140.99) --
	( 38.29,140.99) --
	( 38.29,140.99) --
	( 38.29,140.99) --
	( 38.95,140.99) --
	( 38.95,140.99) --
	( 38.95,140.99) --
	( 38.95,140.99) --
	( 38.95,140.99) --
	( 39.62,140.99) --
	( 39.62,140.99) --
	( 39.62,140.99) --
	( 39.62,140.99) --
	( 39.62,140.99) --
	( 40.29,140.99) --
	( 40.29,140.99) --
	( 40.29,140.99) --
	( 40.29,140.99) --
	( 40.29,140.99) --
	( 40.95,140.99) --
	( 40.95,140.99) --
	( 40.95,140.99) --
	( 40.95,140.99) --
	( 40.95,140.99) --
	( 41.62,140.99) --
	( 41.62,140.99) --
	( 41.62,140.99) --
	( 41.62,140.99) --
	( 41.62,140.99) --
	( 42.29,140.99) --
	( 42.29,140.99) --
	( 42.29,140.99) --
	( 42.29,140.99) --
	( 42.29,140.99) --
	( 42.95,140.99) --
	( 42.95,140.99) --
	( 42.95,140.99) --
	( 42.95,140.99) --
	( 42.95,140.99) --
	( 43.62,140.99) --
	( 43.62,140.99) --
	( 43.62,140.99) --
	( 43.62,140.99) --
	( 43.62,140.99) --
	( 44.29,140.99) --
	( 44.29,140.99) --
	( 44.29,140.99) --
	( 44.29,140.99) --
	( 44.29,140.99) --
	( 44.95,140.99) --
	( 44.95,140.99) --
	( 44.95,140.99) --
	( 44.95,140.99) --
	( 44.95,140.99) --
	( 45.62,140.99) --
	( 45.62,140.99) --
	( 45.62,140.99) --
	( 45.62,140.99) --
	( 45.62,140.99) --
	( 46.29,140.99) --
	( 46.29,140.99) --
	( 46.29,140.99) --
	( 46.29,140.99) --
	( 46.29,140.99) --
	( 46.95,140.99) --
	( 46.95,140.99) --
	( 46.95,140.99) --
	( 46.95,140.99) --
	( 46.95,140.99) --
	( 47.62,140.99) --
	( 47.62,140.99) --
	( 47.62,140.99) --
	( 47.62,140.99) --
	( 47.62,140.99) --
	( 48.29,140.99) --
	( 48.29,140.99) --
	( 48.29,140.99) --
	( 48.29,140.99) --
	( 48.29,140.99) --
	( 48.95,140.99) --
	( 48.95,140.99) --
	( 48.95,140.99) --
	( 48.95,140.99) --
	( 48.95,140.99) --
	( 49.62,140.99) --
	( 49.62,140.99) --
	( 49.62,140.99) --
	( 49.62,140.99) --
	( 49.62,140.99) --
	( 50.29,140.99) --
	( 50.29,140.99) --
	( 50.29,140.99) --
	( 50.29,140.99) --
	( 50.29,140.99) --
	( 50.95,140.99) --
	( 50.95,140.99) --
	( 50.95,140.99) --
	( 50.95,140.99) --
	( 50.95,140.99) --
	( 51.62,140.99) --
	( 51.62,140.99) --
	( 51.62,140.99) --
	( 51.62,140.99) --
	( 51.62,140.99) --
	( 52.29,140.99) --
	( 52.29,140.99) --
	( 52.29,140.99) --
	( 52.29,140.99) --
	( 52.29,140.99) --
	( 52.95,140.99) --
	( 52.95,140.99) --
	( 52.95,140.99) --
	( 52.95,140.99) --
	( 52.95,140.99) --
	( 53.62,140.99) --
	( 53.62,140.99) --
	( 53.62,140.99) --
	( 53.62,140.99) --
	( 53.62,140.99) --
	( 54.29,140.99) --
	( 54.29,140.99) --
	( 54.29,140.99) --
	( 54.29,140.99) --
	( 54.29,140.99) --
	( 54.95,140.99) --
	( 54.95,140.99) --
	( 54.95,140.99) --
	( 54.95,140.99) --
	( 54.95,140.99) --
	( 55.62,140.99) --
	( 55.62,140.99) --
	( 55.62,140.99) --
	( 55.62,140.99) --
	( 55.62,140.99) --
	( 56.29,140.99) --
	( 56.29,140.99) --
	( 56.29,140.99) --
	( 56.29,140.99) --
	( 56.29,140.99) --
	( 56.95,140.99) --
	( 56.95,140.99) --
	( 56.95,140.99) --
	( 56.95,140.99) --
	( 56.95,140.99) --
	( 57.62,140.99) --
	( 57.62,140.99) --
	( 57.62,140.99) --
	( 57.62,140.99) --
	( 57.62,140.99) --
	( 58.29,140.99) --
	( 58.29,140.99) --
	( 58.29,140.99) --
	( 58.29,140.99) --
	( 58.29,140.99) --
	( 58.95,140.99) --
	( 58.95,140.99) --
	( 58.95,140.99) --
	( 58.95,140.99) --
	( 58.95,140.99) --
	( 59.62,140.99) --
	( 59.62,140.99) --
	( 59.62,140.99) --
	( 59.62,140.99) --
	( 59.62,140.99) --
	( 60.29,140.99) --
	( 60.29,140.99) --
	( 60.29,140.99) --
	( 60.29,140.99) --
	( 60.29,140.99) --
	( 60.95,140.99) --
	( 60.95,140.99) --
	( 60.95,140.99) --
	( 60.95,140.99) --
	( 60.95,140.99) --
	( 61.62,140.99) --
	( 61.62,140.99) --
	( 61.62,140.99) --
	( 61.62,140.99) --
	( 61.62,140.99) --
	( 62.28,140.99) --
	( 62.28,140.99) --
	( 62.28,140.99) --
	( 62.28,140.99) --
	( 62.28,140.99) --
	( 62.95,140.99) --
	( 62.95,140.99) --
	( 62.95,140.99) --
	( 62.95,140.99) --
	( 62.95,140.99) --
	( 63.62,140.99) --
	( 63.62,140.99) --
	( 63.62,140.99) --
	( 63.62,140.99) --
	( 63.62,140.99) --
	( 64.28,140.99) --
	( 64.28,140.99) --
	( 64.28,140.99) --
	( 64.28,140.99) --
	( 64.28,140.99) --
	( 64.95,140.99) --
	( 64.95,140.99) --
	( 64.95,140.99) --
	( 64.95,140.99) --
	( 64.95,140.99) --
	( 65.62,140.99) --
	( 65.62,140.99) --
	( 65.62,140.99) --
	( 65.62,140.99) --
	( 65.62,140.99) --
	( 66.28,140.99) --
	( 66.28,140.99) --
	( 66.28,140.99) --
	( 66.28,140.99) --
	( 66.28,140.99) --
	( 66.95,140.99) --
	( 66.95,140.99) --
	( 66.95,140.99) --
	( 66.95,140.99) --
	( 66.95,140.99) --
	( 67.62,140.99) --
	( 67.62,140.99) --
	( 67.62,140.99) --
	( 67.62,140.99) --
	( 67.62,140.99) --
	( 68.28,140.99) --
	( 68.28,140.99) --
	( 68.28,140.99) --
	( 68.28,140.99) --
	( 68.28,140.99) --
	( 68.95,140.99) --
	( 68.95,140.99) --
	( 68.95,140.99) --
	( 68.95,140.99) --
	( 68.95,140.99) --
	( 69.62,140.99) --
	( 69.62,140.99) --
	( 69.62,140.99) --
	( 69.62,140.99) --
	( 69.62,140.99) --
	( 70.28,140.99) --
	( 70.28,140.99) --
	( 70.28,140.99) --
	( 70.28,140.99) --
	( 70.28,140.99) --
	( 70.95,140.99) --
	( 70.95,140.99) --
	( 70.95,140.99) --
	( 70.95,140.99) --
	( 70.95,140.99) --
	( 71.62,140.99) --
	( 71.62,140.99) --
	( 71.62,140.99) --
	( 71.62,140.99) --
	( 71.62,140.99) --
	( 72.28,140.99) --
	( 72.28,140.99) --
	( 72.28,140.99) --
	( 72.28,140.99) --
	( 72.28,140.99) --
	( 72.95,140.99) --
	( 72.95,140.99) --
	( 72.95,140.99) --
	( 72.95,140.99) --
	( 72.95,140.99) --
	( 73.62,140.99) --
	( 73.62,140.99) --
	( 73.62,140.99) --
	( 73.62,140.99) --
	( 73.62,140.99) --
	( 74.28,140.99) --
	( 74.28,140.99) --
	( 74.28,140.99) --
	( 74.28,140.99) --
	( 74.28,140.99) --
	( 74.95,140.99) --
	( 74.95,140.99) --
	( 74.95,140.99) --
	( 74.95,140.99) --
	( 74.95,140.99) --
	( 75.62,140.99) --
	( 75.62,140.99) --
	( 75.62,140.99) --
	( 75.62,140.99) --
	( 75.62,140.99) --
	( 76.28,140.99) --
	( 76.28,140.99) --
	( 76.28,140.99) --
	( 76.28,140.99) --
	( 76.28,140.99) --
	( 76.95,140.99) --
	( 76.95,140.99) --
	( 76.95,140.99) --
	( 76.95,140.99) --
	( 76.95,140.99) --
	( 77.62,140.99) --
	( 77.62,140.99) --
	( 77.62,140.99) --
	( 77.62,140.99) --
	( 77.62,140.99) --
	( 78.28,140.99) --
	( 78.28,140.99) --
	( 78.28,140.99) --
	( 78.28,140.99) --
	( 78.28,140.99) --
	( 78.95,140.99) --
	( 78.95,140.99) --
	( 78.95,140.99) --
	( 78.95,140.99) --
	( 78.95,140.99) --
	( 79.62,140.99) --
	( 79.62,140.99) --
	( 79.62,140.99) --
	( 79.62,140.99) --
	( 79.62,140.99) --
	( 80.28,140.99) --
	( 80.28,140.99) --
	( 80.28,140.99) --
	( 80.28,140.99) --
	( 80.28,140.99) --
	( 80.95,140.99) --
	( 80.95,140.99) --
	( 80.95,140.99) --
	( 80.95,140.99) --
	( 80.95,140.99) --
	( 81.62,140.99) --
	( 81.62,140.99) --
	( 81.62,140.99) --
	( 81.62,140.99) --
	( 81.62,140.99) --
	( 82.28,140.99) --
	( 82.28,140.99) --
	( 82.28,140.99) --
	( 82.28,140.99) --
	( 82.28,140.99) --
	( 82.95,140.99) --
	( 82.95,140.99) --
	( 82.95,140.99) --
	( 82.95,140.99) --
	( 82.95,140.99) --
	( 83.62,140.99) --
	( 83.62,140.99) --
	( 83.62,140.99) --
	( 83.62,140.99) --
	( 83.62,140.99) --
	( 84.28,140.99) --
	( 84.28,140.99) --
	( 84.28,140.99) --
	( 84.28,140.99) --
	( 84.28,140.99) --
	( 84.95,140.99) --
	( 84.95,140.99) --
	( 84.95,140.99) --
	( 84.95,140.99) --
	( 84.95,140.99) --
	( 85.62,140.99) --
	( 85.62,140.99) --
	( 85.62,140.99) --
	( 85.62,140.99) --
	( 85.62,140.99) --
	( 86.28,140.99) --
	( 86.28,140.99) --
	( 86.28,140.99) --
	( 86.28,140.99) --
	( 86.28,140.99) --
	( 86.95,140.99) --
	( 86.95,140.99) --
	( 86.95,140.99) --
	( 86.95,140.99) --
	( 86.95,140.99) --
	( 87.62,140.99) --
	( 87.62,140.99) --
	( 87.62,140.99) --
	( 87.62,140.99) --
	( 87.62,140.99) --
	( 88.28,140.99) --
	( 88.28,140.99) --
	( 88.28,140.99) --
	( 88.28,140.99) --
	( 88.28,140.99) --
	( 88.95,140.99) --
	( 88.95,140.99) --
	( 88.95,140.99) --
	( 88.95,140.99) --
	( 88.95,140.99) --
	( 89.62,140.99) --
	( 89.62,140.99) --
	( 89.62,140.99) --
	( 89.62,140.99) --
	( 89.62,140.99) --
	( 90.28,140.99) --
	( 90.28,140.99) --
	( 90.28,140.99) --
	( 90.28,140.99) --
	( 90.28,140.99) --
	( 90.95,140.99) --
	( 90.95,140.99) --
	( 90.95,140.99) --
	( 90.95,140.99) --
	( 90.95,140.99) --
	( 91.62,140.99) --
	( 91.62,140.99) --
	( 91.62,140.99) --
	( 91.62,140.99) --
	( 91.62,140.99) --
	( 92.28,140.99) --
	( 92.28,140.99) --
	( 92.28,140.99) --
	( 92.28,140.99) --
	( 92.28,140.99) --
	( 92.95,140.99) --
	( 92.95,140.99) --
	( 92.95,140.99) --
	( 92.95,140.99) --
	( 92.95,140.99) --
	( 93.62,140.99) --
	( 93.62,140.99) --
	( 93.62,140.99) --
	( 93.62,140.99) --
	( 93.62,140.99) --
	( 94.28,140.99) --
	( 94.28,140.99) --
	( 94.28,140.99) --
	( 94.28,140.99) --
	( 94.28,140.99) --
	( 94.95,140.99) --
	( 94.95,140.99) --
	( 94.95,140.99) --
	( 94.95,140.99) --
	( 94.95,140.99) --
	( 95.62,140.99) --
	( 95.62,140.99) --
	( 95.62,140.99) --
	( 95.62,140.99) --
	( 95.62,140.99) --
	( 96.28,140.99) --
	( 96.28,140.99) --
	( 96.28,140.99) --
	( 96.28,140.99) --
	( 96.28,140.99) --
	( 96.95,140.99) --
	( 96.95,140.99) --
	( 96.95,140.99) --
	( 96.95,140.99) --
	( 96.95,140.99) --
	( 97.62,140.99) --
	( 97.62,140.99) --
	( 97.62,140.99) --
	( 97.62,140.99) --
	( 97.62,140.99) --
	( 98.28,140.99) --
	( 98.28,140.99) --
	( 98.28,140.99) --
	( 98.28,140.99) --
	( 98.28,140.99) --
	( 98.95,140.99) --
	( 98.95,140.99) --
	( 98.95,140.99) --
	( 98.95,140.99) --
	( 98.95,140.99) --
	( 99.62,140.99) --
	( 99.62,140.99) --
	( 99.62,140.99) --
	( 99.62,140.99) --
	( 99.62,140.99) --
	(100.28,140.99) --
	(100.28,140.99) --
	(100.28,140.99) --
	(100.28,140.99) --
	(100.28,140.99) --
	(100.95,140.99) --
	(100.95,140.99) --
	(100.95,140.99) --
	(100.95,140.99) --
	(100.95,140.99) --
	(101.62,140.99) --
	(101.62,140.99) --
	(101.62,140.99) --
	(101.62,140.99) --
	(101.62,140.99) --
	(102.28,140.99) --
	(102.28,140.99) --
	(102.28,140.99) --
	(102.28,140.99) --
	(102.28,140.99) --
	(102.95,140.99) --
	(102.95,140.99) --
	(102.95,140.99) --
	(102.95,140.99) --
	(102.95,140.99) --
	(103.62,140.99) --
	(103.62,140.99) --
	(103.62,140.99) --
	(103.62,140.99) --
	(103.62,140.99) --
	(104.28,140.99) --
	(104.28,140.99) --
	(104.28,140.99) --
	(104.28,140.99) --
	(104.28,140.99) --
	(104.95,140.99) --
	(104.95,140.99) --
	(104.95,140.99) --
	(104.95,140.99) --
	(104.95,140.99) --
	(105.62,140.99) --
	(105.62,140.99) --
	(105.62,140.99) --
	(105.62,140.99) --
	(105.62,140.99) --
	(106.28,140.99) --
	(106.28,140.99) --
	(106.28,140.99) --
	(106.28,140.99) --
	(106.28,140.99) --
	(106.95,140.99) --
	(106.95,140.99) --
	(106.95,140.99) --
	(106.95,140.99) --
	(106.95,140.99) --
	(107.62,140.99) --
	(107.62,140.99) --
	(107.62,140.99) --
	(107.62,140.99) --
	(107.62,140.99) --
	(108.28,140.99) --
	(108.28,140.99) --
	(108.28,140.99) --
	(108.28,140.99) --
	(108.28,140.99) --
	(108.95,140.99) --
	(108.95,140.99) --
	(108.95,140.99) --
	(108.95,140.99) --
	(108.95,140.99) --
	(109.62,140.99) --
	(109.62,140.99) --
	(109.62,140.99) --
	(109.62,140.99) --
	(109.62,140.99) --
	(110.28,140.99) --
	(110.28,140.99) --
	(110.28,140.99) --
	(110.28,140.99) --
	(110.28,140.99) --
	(110.95,140.99) --
	(110.95,140.99) --
	(110.95,140.99) --
	(110.95,140.99) --
	(110.95,140.99) --
	(111.62,140.99) --
	(111.62,140.99) --
	(111.62,140.99) --
	(111.62,140.99) --
	(111.62,140.99) --
	(112.28,140.99) --
	(112.28,140.99) --
	(112.28,140.99) --
	(112.28,140.99) --
	(112.28,140.99) --
	(112.95,140.99) --
	(112.95,140.99) --
	(112.95,140.99) --
	(112.95,140.99) --
	(112.95,140.99) --
	(113.62,140.99) --
	(113.62,140.99) --
	(113.62,140.99) --
	(113.62,140.99) --
	(113.62,140.99) --
	(114.28,140.99) --
	(114.28,140.99) --
	(114.28,140.99) --
	(114.28,140.99) --
	(114.28,140.99) --
	(114.95,140.99) --
	(114.95,140.99) --
	(114.95,140.99) --
	(114.95,140.99) --
	(114.95,140.99) --
	(115.62,140.99) --
	(115.62,140.99) --
	(115.62,140.99) --
	(115.62,140.99) --
	(115.62,140.99) --
	(116.28,140.99) --
	(116.28,140.99) --
	(116.28,140.99) --
	(116.28,140.99) --
	(116.28,140.99) --
	(116.95,140.99) --
	(116.95,140.99) --
	(116.95,140.99) --
	(116.95,140.99) --
	(116.95,140.99) --
	(117.62,140.99) --
	(117.62,140.99) --
	(117.62,140.99) --
	(117.62,140.99) --
	(117.62,140.99) --
	(118.28,140.99) --
	(118.28,140.99) --
	(118.28,140.99) --
	(118.28,140.99) --
	(118.28,140.99) --
	(118.95,140.99) --
	(118.95,140.99) --
	(118.95,140.99) --
	(118.95,140.99) --
	(118.95,140.99) --
	(119.62,140.99) --
	(119.62,140.99) --
	(119.62,140.99) --
	(119.62,140.99) --
	(119.62,140.99) --
	(120.28,140.99) --
	(120.28,140.99) --
	(120.28,140.99) --
	(120.28,140.99) --
	(120.28,140.99) --
	(120.95,140.99) --
	(120.95,140.99) --
	(120.95,140.99) --
	(120.95,140.99) --
	(120.95,140.99) --
	(121.62,140.99) --
	(121.62,140.99) --
	(121.62,140.99) --
	(121.62,140.99) --
	(121.62,140.99) --
	(122.28,140.99) --
	(122.28,140.99) --
	(122.28,140.99) --
	(122.28,140.99) --
	(122.28,140.99) --
	(122.95,140.99) --
	(122.95,140.99) --
	(122.95,140.99) --
	(122.95,140.99) --
	(122.95,140.99) --
	(123.62,140.99) --
	(123.62,140.99) --
	(123.62,140.99) --
	(123.62,140.99) --
	(123.62,140.99) --
	(124.28,140.99) --
	(124.28,140.99) --
	(124.28,140.99) --
	(124.28,140.99) --
	(124.28,140.99) --
	(124.95,140.99) --
	(124.95,140.99) --
	(124.95,140.99) --
	(124.95,140.99) --
	(124.95,140.99) --
	(125.62,140.99) --
	(125.62,140.99) --
	(125.62,140.99) --
	(125.62,140.99) --
	(125.62,140.99) --
	(126.28,140.99) --
	(126.28,140.99) --
	(126.28,140.99) --
	(126.28,140.99) --
	(126.28,140.99) --
	(126.95,140.99) --
	(126.95,140.99) --
	(126.95,140.99) --
	(126.95,140.99) --
	(126.95,140.99) --
	(127.62,140.99) --
	(127.62,140.99) --
	(127.62,140.99) --
	(127.62,140.99) --
	(127.62,140.99) --
	(128.28,140.99) --
	(128.28,140.99) --
	(128.28,140.99) --
	(128.28,140.99) --
	(128.28,140.99) --
	(128.95,140.99) --
	(128.95,140.99) --
	(128.95,140.99) --
	(128.95,140.99) --
	(128.95,140.99) --
	(129.62,140.99) --
	(129.62,140.99) --
	(129.62,140.99) --
	(129.62,140.99) --
	(129.62,140.99) --
	(130.28,140.99) --
	(130.28,140.99) --
	(130.28,140.99) --
	(130.28,140.99) --
	(130.28,140.99) --
	(130.95,140.99) --
	(130.95,140.99) --
	(130.95,140.99) --
	(130.95,140.99) --
	(130.95,140.99) --
	(131.61,140.99) --
	(131.61,140.99) --
	(131.61,140.99) --
	(131.61,140.99) --
	(131.61,140.99) --
	(132.28,140.99) --
	(132.28,140.99) --
	(132.28,140.99) --
	(132.28,140.99) --
	(132.28,140.99) --
	(132.95,140.99) --
	(132.95,140.99) --
	(132.95,140.99) --
	(132.95,140.99) --
	(132.95,140.99) --
	(133.61,140.99) --
	(133.61,140.99) --
	(133.61,140.99) --
	(133.61,140.99) --
	(133.61,140.99) --
	(134.28,140.99) --
	(134.28,140.99) --
	(134.28,140.99) --
	(134.28,140.99) --
	(134.28,140.99) --
	(134.95,140.99) --
	(134.95,140.99) --
	(134.95,140.99) --
	(134.95,140.99) --
	(134.95,140.99) --
	(135.61,140.99) --
	(135.61,140.99) --
	(135.61,140.99) --
	(135.61,140.99) --
	(135.61,140.99) --
	(136.28,140.99) --
	(136.28,140.99) --
	(136.28,140.99) --
	(136.28,140.99) --
	(136.28,140.99) --
	(136.95,140.99) --
	(136.95,140.99) --
	(136.95,140.99) --
	(136.95,140.99) --
	(136.95,140.99) --
	(137.61,140.99) --
	(137.61,140.99) --
	(137.61,140.99) --
	(137.61,140.99) --
	(137.61,140.99) --
	(138.28,140.99) --
	(138.28,140.99) --
	(138.28,140.99) --
	(138.28,140.99) --
	(138.28,140.99) --
	(138.95,140.99) --
	(138.95,140.99) --
	(138.95,140.99) --
	(138.95,140.99) --
	(138.95,140.99) --
	(139.61,140.99) --
	(139.61,140.99) --
	(139.61,140.99) --
	(139.61,140.99) --
	(139.61,140.99) --
	(140.28,140.99) --
	(140.28,140.99) --
	(140.28,140.99) --
	(140.28,140.99) --
	(140.28,140.99) --
	(140.95,140.99) --
	(140.95,140.99) --
	(140.95,140.99) --
	(140.95,140.99) --
	(140.95,140.99) --
	(141.61,140.99) --
	(141.61,140.99) --
	(141.61,140.99) --
	(141.61,140.99) --
	(141.61,140.99) --
	(142.28,140.99) --
	(142.28,140.99) --
	(142.28,140.99) --
	(142.28,140.99) --
	(142.28,140.99) --
	(142.95,140.99) --
	(142.95,140.99) --
	(142.95,140.99) --
	(142.95,140.99) --
	(142.95,140.99) --
	(143.61,140.99) --
	(143.61,140.99) --
	(143.61,140.99) --
	(143.61,140.99) --
	(143.61,140.99) --
	(144.28,140.99) --
	(144.28,140.99) --
	(144.28,140.99) --
	(144.28,140.99) --
	(144.28,140.99) --
	(144.95,140.99) --
	(144.95,140.99) --
	(144.95,140.99) --
	(144.95,140.99) --
	(144.95,140.99) --
	(145.61,140.99) --
	(145.61,140.99) --
	(145.61,140.99) --
	(145.61,140.99) --
	(145.61,140.99) --
	(146.28,140.99) --
	(146.28,140.99) --
	(146.28,140.99) --
	(146.28,140.99) --
	(146.28,140.99) --
	(146.95,140.99) --
	(146.95,140.99) --
	(146.95,140.99) --
	(146.95,140.99) --
	(146.95,140.99) --
	(147.61,140.99) --
	(147.61,140.99) --
	(147.61,140.99) --
	(147.61,140.99) --
	(147.61,140.99) --
	(148.28,140.99) --
	(148.28,140.99) --
	(148.28,140.99) --
	(148.28,140.99) --
	(148.28,140.99) --
	(148.95,140.99) --
	(148.95,140.99) --
	(148.95,140.99) --
	(148.95,140.99) --
	(148.95,140.99) --
	(149.61,140.99) --
	(149.61,140.99) --
	(149.61,140.99) --
	(149.61,140.99) --
	(149.61,140.99) --
	(150.28,140.99) --
	(150.28,140.99) --
	(150.28,140.99) --
	(150.28,140.99) --
	(150.28,140.99) --
	(150.95,140.99) --
	(150.95,140.99) --
	(150.95,140.99) --
	(150.95,140.99) --
	(150.95,140.99) --
	(151.61,140.99) --
	(151.61,140.99) --
	(151.61,140.99) --
	(151.61,140.99) --
	(151.61,140.99) --
	(152.28,140.99) --
	(152.28,140.99) --
	(152.28,140.99) --
	(152.28,140.99) --
	(152.28,140.99) --
	(152.95,140.99) --
	(152.95,140.99) --
	(152.95,140.99) --
	(152.95,140.99) --
	(152.95,140.99) --
	(153.61,140.99) --
	(153.61,140.99) --
	(153.61,140.99) --
	(153.61,140.99) --
	(153.61,140.99) --
	(154.28,140.99) --
	(154.28,140.99) --
	(154.28,140.99) --
	(154.28,140.99) --
	(154.28,140.99) --
	(154.95,140.99) --
	(154.95,140.99) --
	(154.95,140.99) --
	(154.95,140.99) --
	(154.95,140.99) --
	(155.61,140.99) --
	(155.61,140.99) --
	(155.61,140.99) --
	(155.61,140.99) --
	(155.61,140.99) --
	(156.28,140.99) --
	(156.28,140.99) --
	(156.28,140.99) --
	(156.28,140.99) --
	(156.28,140.99) --
	(156.95,140.99) --
	(156.95,140.99) --
	(156.95,140.99) --
	(156.95,140.99) --
	(156.95,140.99) --
	(157.61,140.99) --
	(157.61,140.99) --
	(157.61,140.99) --
	(157.61,140.99) --
	(157.61,140.99) --
	(158.28,140.99) --
	(158.28,140.99) --
	(158.28,140.99) --
	(158.28,140.99) --
	(158.28,140.99) --
	(158.95,140.99) --
	(158.95,140.99) --
	(158.95,140.99) --
	(158.95,140.99) --
	(158.95,140.99) --
	(159.61,140.99) --
	(159.61,140.99) --
	(159.61,140.99) --
	(159.61,140.99) --
	(159.61,140.99) --
	(160.28,140.99) --
	(160.28,140.99) --
	(160.28,140.99) --
	(160.28,140.99) --
	(160.28,140.99) --
	(160.95,140.99) --
	(160.95,140.99) --
	(160.95,140.99) --
	(160.95,140.99) --
	(160.95,140.99) --
	(161.61,140.99) --
	(161.61,140.99) --
	(161.61,140.99) --
	(161.61,140.99) --
	(161.61,140.99) --
	(162.28,140.99) --
	(162.28,140.99) --
	(162.28,140.99) --
	(162.28,140.99) --
	(162.28,140.99) --
	(162.95,140.99) --
	(162.95,140.99) --
	(162.95,140.99) --
	(162.95,140.99) --
	(162.95,140.99) --
	(163.61,140.99) --
	(163.61,140.99) --
	(163.61,140.99) --
	(163.61,140.99) --
	(163.61,140.99) --
	(164.28,140.99) --
	(164.28,140.99) --
	(164.28,140.99) --
	(164.28,140.99) --
	(164.28,140.99) --
	(164.95,140.99) --
	(164.95,140.99) --
	(164.95,140.99) --
	(164.95,140.99) --
	(164.95,140.99) --
	(165.61,140.99) --
	(165.61,140.99) --
	(165.61,140.99) --
	(165.61,140.99) --
	(165.61,140.99) --
	(166.28,140.99) --
	(166.28,140.99) --
	(166.28,140.99) --
	(166.28,140.99) --
	(166.28,140.99);
\definecolor{drawColor}{gray}{0.70}

\path[draw=drawColor,line width= 0.5pt,line join=round,line cap=round] ( 33.62, 29.68) rectangle (166.28,194.47);
\end{scope}
\begin{scope}
\path[clip] (  0.00,  0.00) rectangle (171.28,199.47);
\definecolor{drawColor}{gray}{0.30}

\node[text=drawColor,anchor=base east,inner sep=0pt, outer sep=0pt, scale=  1.00] at ( 29.12, 81.89) {0.5};

\node[text=drawColor,anchor=base east,inner sep=0pt, outer sep=0pt, scale=  1.00] at ( 29.12,137.54) {1.0};
\end{scope}
\begin{scope}
\path[clip] (  0.00,  0.00) rectangle (171.28,199.47);
\definecolor{drawColor}{gray}{0.70}

\path[draw=drawColor,line width= 0.3pt,line join=round] ( 31.12, 85.33) --
	( 33.62, 85.33);

\path[draw=drawColor,line width= 0.3pt,line join=round] ( 31.12,140.99) --
	( 33.62,140.99);
\end{scope}
\begin{scope}
\path[clip] (  0.00,  0.00) rectangle (171.28,199.47);
\definecolor{drawColor}{gray}{0.70}

\path[draw=drawColor,line width= 0.3pt,line join=round] ( 33.62, 27.18) --
	( 33.62, 29.68);

\path[draw=drawColor,line width= 0.3pt,line join=round] ( 60.15, 27.18) --
	( 60.15, 29.68);

\path[draw=drawColor,line width= 0.3pt,line join=round] ( 86.68, 27.18) --
	( 86.68, 29.68);

\path[draw=drawColor,line width= 0.3pt,line join=round] (113.22, 27.18) --
	(113.22, 29.68);

\path[draw=drawColor,line width= 0.3pt,line join=round] (139.75, 27.18) --
	(139.75, 29.68);

\path[draw=drawColor,line width= 0.3pt,line join=round] (166.28, 27.18) --
	(166.28, 29.68);
\end{scope}
\begin{scope}
\path[clip] (  0.00,  0.00) rectangle (171.28,199.47);
\definecolor{drawColor}{gray}{0.30}

\node[text=drawColor,anchor=base,inner sep=0pt, outer sep=0pt, scale=  1.00] at ( 32.62, 18.29) {0};

\node[text=drawColor,anchor=base,inner sep=0pt, outer sep=0pt, scale=  1.00] at ( 58.15, 18.29) {20};

\node[text=drawColor,anchor=base,inner sep=0pt, outer sep=0pt, scale=  1.00] at ( 84.68, 18.29) {40};

\node[text=drawColor,anchor=base,inner sep=0pt, outer sep=0pt, scale=  1.00] at (111.22, 18.29) {60};

\node[text=drawColor,anchor=base,inner sep=0pt, outer sep=0pt, scale=  1.00] at (137.75, 18.29) {80};

\node[text=drawColor,anchor=base,inner sep=0pt, outer sep=0pt, scale=  1.00] at (163.28, 18.29) {100};
\end{scope}
\begin{scope}
\path[clip] (  0.00,  0.00) rectangle (171.28,199.47);
\definecolor{drawColor}{RGB}{0,0,0}

\node[text=drawColor,anchor=base,inner sep=0pt, outer sep=0pt, scale=  1.00] at ( 99.95,  6.94) {\bfseries Instances in \%};
\end{scope}
\begin{scope}
\path[clip] (  0.00,  0.00) rectangle (171.28,199.47);
\definecolor{drawColor}{RGB}{0,0,0}

\node[text=drawColor,rotate= 90.00,anchor=base,inner sep=0pt, outer sep=0pt, scale=  1.00] at ( 11.90,112.99) {\bfseries Relative Costs};
\end{scope}
\begin{scope}
\path[clip] (  0.00,  0.00) rectangle (171.28,199.47);
\definecolor{drawColor}{RGB}{0,0,0}

\path[draw=drawColor,line width= 0.4pt,line join=round,line cap=round] (128.36, 31.82) rectangle (162.96,114.09);
\end{scope}
\begin{scope}
\path[clip] (  0.00,  0.00) rectangle (171.28,199.47);
\definecolor{fillColor}{RGB}{255,255,255}

\path[fill=fillColor] (128.36, 31.82) rectangle (162.96,114.09);
\end{scope}
\begin{scope}
\path[clip] (  0.00,  0.00) rectangle (171.28,199.47);
\definecolor{fillColor}{RGB}{255,255,255}

\path[fill=fillColor] (133.36, 94.63) rectangle (147.81,109.09);
\end{scope}
\begin{scope}
\path[clip] (  0.00,  0.00) rectangle (171.28,199.47);
\definecolor{drawColor}{RGB}{128,128,128}

\path[draw=drawColor,line width= 1.0pt,line join=round] (134.80,101.86) -- (146.37,101.86);
\end{scope}
\begin{scope}
\path[clip] (  0.00,  0.00) rectangle (171.28,199.47);
\definecolor{fillColor}{RGB}{255,255,255}

\path[fill=fillColor] (133.36, 80.18) rectangle (147.81, 94.63);
\end{scope}
\begin{scope}
\path[clip] (  0.00,  0.00) rectangle (171.28,199.47);
\definecolor{drawColor}{RGB}{152,167,197}

\path[draw=drawColor,line width= 1.0pt,line join=round] (134.80, 87.41) -- (146.37, 87.41);
\end{scope}
\begin{scope}
\path[clip] (  0.00,  0.00) rectangle (171.28,199.47);
\definecolor{fillColor}{RGB}{255,255,255}

\path[fill=fillColor] (133.36, 65.73) rectangle (147.81, 80.18);
\end{scope}
\begin{scope}
\path[clip] (  0.00,  0.00) rectangle (171.28,199.47);
\definecolor{drawColor}{RGB}{128,202,192}

\path[draw=drawColor,line width= 1.0pt,line join=round] (134.80, 72.95) -- (146.37, 72.95);
\end{scope}
\begin{scope}
\path[clip] (  0.00,  0.00) rectangle (171.28,199.47);
\definecolor{fillColor}{RGB}{255,255,255}

\path[fill=fillColor] (133.36, 51.27) rectangle (147.81, 65.73);
\end{scope}
\begin{scope}
\path[clip] (  0.00,  0.00) rectangle (171.28,199.47);
\definecolor{drawColor}{RGB}{207,128,187}

\path[draw=drawColor,line width= 1.0pt,line join=round] (134.80, 58.50) -- (146.37, 58.50);
\end{scope}
\begin{scope}
\path[clip] (  0.00,  0.00) rectangle (171.28,199.47);
\definecolor{fillColor}{RGB}{255,255,255}

\path[fill=fillColor] (133.36, 36.82) rectangle (147.81, 51.27);
\end{scope}
\begin{scope}
\path[clip] (  0.00,  0.00) rectangle (171.28,199.47);
\definecolor{drawColor}{RGB}{207,192,152}

\path[draw=drawColor,line width= 1.0pt,line join=round] (134.80, 44.05) -- (146.37, 44.05);
\end{scope}
\begin{scope}
\path[clip] (  0.00,  0.00) rectangle (171.28,199.47);
\definecolor{drawColor}{RGB}{0,0,0}

\node[text=drawColor,anchor=base west,inner sep=0pt, outer sep=0pt, scale=  0.80] at (147.81, 99.11) {$\ensuremath{\mathcal{N}}_1$};
\end{scope}
\begin{scope}
\path[clip] (  0.00,  0.00) rectangle (171.28,199.47);
\definecolor{drawColor}{RGB}{0,0,0}

\node[text=drawColor,anchor=base west,inner sep=0pt, outer sep=0pt, scale=  0.80] at (147.81, 84.65) {$\ensuremath{\mathcal{N}}_2$};
\end{scope}
\begin{scope}
\path[clip] (  0.00,  0.00) rectangle (171.28,199.47);
\definecolor{drawColor}{RGB}{0,0,0}

\node[text=drawColor,anchor=base west,inner sep=0pt, outer sep=0pt, scale=  0.80] at (147.81, 70.20) {$\ensuremath{\mathcal{N}}_3$};
\end{scope}
\begin{scope}
\path[clip] (  0.00,  0.00) rectangle (171.28,199.47);
\definecolor{drawColor}{RGB}{0,0,0}

\node[text=drawColor,anchor=base west,inner sep=0pt, outer sep=0pt, scale=  0.80] at (147.81, 55.74) {$\ensuremath{\mathcal{N}}_4$};
\end{scope}
\begin{scope}
\path[clip] (  0.00,  0.00) rectangle (171.28,199.47);
\definecolor{drawColor}{RGB}{0,0,0}

\node[text=drawColor,anchor=base west,inner sep=0pt, outer sep=0pt, scale=  0.80] at (147.81, 41.29) {$\ensuremath{\mathcal{N}}_5$};
\end{scope}
\end{tikzpicture}

%% file: 06-plots/2019-SEA-SAvsAlgo-1Hour-Shortened.tex
\begin{tikzpicture}[x=1pt,y=1pt]
\definecolor{fillColor}{RGB}{255,255,255}
\path[use as bounding box,fill=fillColor,fill opacity=0.00] (0,0) rectangle (171.28,199.47);
\begin{scope}
\path[clip] (  0.00,  0.00) rectangle (171.28,199.47);
\definecolor{drawColor}{RGB}{255,255,255}
\definecolor{fillColor}{RGB}{255,255,255}

\path[draw=drawColor,line width= 0.5pt,line join=round,line cap=round,fill=fillColor] (  0.00,  0.00) rectangle (171.28,199.47);
\end{scope}
\begin{scope}
\path[clip] ( 38.62, 29.68) rectangle (166.28,194.47);
\definecolor{fillColor}{RGB}{255,255,255}

\path[fill=fillColor] ( 38.62, 29.68) rectangle (166.28,194.47);
\definecolor{drawColor}{gray}{0.87}

\path[draw=drawColor,line width= 0.1pt,line join=round] ( 38.62, 29.68) --
	(166.28, 29.68);

\path[draw=drawColor,line width= 0.1pt,line join=round] ( 38.62, 76.76) --
	(166.28, 76.76);

\path[draw=drawColor,line width= 0.1pt,line join=round] ( 38.62,123.84) --
	(166.28,123.84);

\path[draw=drawColor,line width= 0.1pt,line join=round] ( 38.62,170.92) --
	(166.28,170.92);

\path[draw=drawColor,line width= 0.1pt,line join=round] ( 51.38, 29.68) --
	( 51.38,194.47);

\path[draw=drawColor,line width= 0.1pt,line join=round] ( 76.92, 29.68) --
	( 76.92,194.47);

\path[draw=drawColor,line width= 0.1pt,line join=round] (102.45, 29.68) --
	(102.45,194.47);

\path[draw=drawColor,line width= 0.1pt,line join=round] (127.98, 29.68) --
	(127.98,194.47);

\path[draw=drawColor,line width= 0.1pt,line join=round] (153.51, 29.68) --
	(153.51,194.47);

\path[draw=drawColor,line width= 0.3pt,line join=round] ( 38.62, 53.22) --
	(166.28, 53.22);

\path[draw=drawColor,line width= 0.3pt,line join=round] ( 38.62,100.30) --
	(166.28,100.30);

\path[draw=drawColor,line width= 0.3pt,line join=round] ( 38.62,147.38) --
	(166.28,147.38);

\path[draw=drawColor,line width= 0.3pt,line join=round] ( 38.62,194.47) --
	(166.28,194.47);

\path[draw=drawColor,line width= 0.3pt,line join=round] ( 38.62, 29.68) --
	( 38.62,194.47);

\path[draw=drawColor,line width= 0.3pt,line join=round] ( 64.15, 29.68) --
	( 64.15,194.47);

\path[draw=drawColor,line width= 0.3pt,line join=round] ( 89.68, 29.68) --
	( 89.68,194.47);

\path[draw=drawColor,line width= 0.3pt,line join=round] (115.22, 29.68) --
	(115.22,194.47);

\path[draw=drawColor,line width= 0.3pt,line join=round] (140.75, 29.68) --
	(140.75,194.47);

\path[draw=drawColor,line width= 0.3pt,line join=round] (166.28, 29.68) --
	(166.28,194.47);
\definecolor{drawColor}{RGB}{128,128,128}

\path[draw=drawColor,line width= 1.0pt,line join=round] ( 39.42, 29.68) --
	( 39.90, 45.95) --
	( 40.54, 57.65) --
	( 41.18, 58.45) --
	( 41.83, 63.15) --
	( 42.47, 65.37) --
	( 43.11, 68.77) --
	( 43.75, 75.09) --
	( 44.39, 75.51) --
	( 45.03, 77.26) --
	( 45.68, 79.86) --
	( 46.32, 80.08) --
	( 46.96, 81.33) --
	( 47.60, 83.00) --
	( 48.24, 83.78) --
	( 48.88, 86.38) --
	( 49.52, 86.70) --
	( 50.17, 88.24) --
	( 50.81, 89.15) --
	( 51.45, 89.60) --
	( 52.09, 92.73) --
	( 52.73, 95.73) --
	( 53.37, 98.72) --
	( 54.01,100.30) --
	( 54.66,100.30) --
	( 55.30,100.30) --
	( 55.94,100.30) --
	( 56.58,100.30) --
	( 57.22,100.30) --
	( 57.86,100.30) --
	( 58.51,100.30) --
	( 59.15,100.30) --
	( 59.79,100.30) --
	( 60.43,100.30) --
	( 61.07,100.30) --
	( 61.71,100.30) --
	( 62.35,100.30) --
	( 63.00,100.30) --
	( 63.64,100.30) --
	( 64.28,100.30) --
	( 64.92,100.30) --
	( 65.56,100.30) --
	( 66.20,100.30) --
	( 66.85,100.30) --
	( 67.49,100.30) --
	( 68.13,100.30) --
	( 68.77,100.30) --
	( 69.41,100.30) --
	( 70.05,100.30) --
	( 70.69,100.30) --
	( 71.34,100.30) --
	( 71.98,100.30) --
	( 72.62,100.30) --
	( 73.26,100.30) --
	( 73.90,100.30) --
	( 74.54,100.30) --
	( 75.18,100.30) --
	( 75.83,100.30) --
	( 76.47,100.30) --
	( 77.11,100.30) --
	( 77.75,100.30) --
	( 78.39,100.30) --
	( 79.03,100.30) --
	( 79.68,100.30) --
	( 80.32,100.30) --
	( 80.96,100.30) --
	( 81.60,100.30) --
	( 82.24,100.30) --
	( 82.88,100.30) --
	( 83.52,100.30) --
	( 84.17,100.30) --
	( 84.81,100.30) --
	( 85.45,100.30) --
	( 86.09,100.30) --
	( 86.73,100.30) --
	( 87.37,100.30) --
	( 88.02,100.30) --
	( 88.66,100.30) --
	( 89.30,100.30) --
	( 89.94,100.30) --
	( 90.58,100.30) --
	( 91.22,100.30) --
	( 91.86,100.30) --
	( 92.51,100.30) --
	( 93.15,100.30) --
	( 93.79,100.30) --
	( 94.43,100.84) --
	( 95.07,100.94) --
	( 95.71,101.35) --
	( 96.35,101.39) --
	( 97.00,101.75) --
	( 97.64,101.93) --
	( 98.28,102.18) --
	( 98.92,102.60) --
	( 99.56,102.98) --
	(100.20,103.05) --
	(100.85,103.13) --
	(101.49,103.91) --
	(102.13,103.95) --
	(102.77,104.12) --
	(103.41,104.45) --
	(104.05,105.43) --
	(104.69,105.86) --
	(105.34,106.00) --
	(105.98,106.29) --
	(106.62,106.41) --
	(107.26,106.65) --
	(107.90,108.18) --
	(108.54,108.31) --
	(109.19,108.41) --
	(109.83,108.43) --
	(110.47,108.81) --
	(111.11,109.76) --
	(111.75,110.36) --
	(112.39,110.54) --
	(113.03,110.56) --
	(113.68,110.73) --
	(114.32,111.10) --
	(114.96,111.42) --
	(115.60,112.16) --
	(116.24,112.34) --
	(116.88,112.37) --
	(117.52,112.62) --
	(118.17,112.92) --
	(118.81,113.07) --
	(119.45,113.61) --
	(120.09,113.62) --
	(120.73,114.10) --
	(121.37,114.14) --
	(122.02,114.39) --
	(122.66,114.41) --
	(123.30,114.45) --
	(123.94,114.56) --
	(124.58,115.68) --
	(125.22,117.43) --
	(125.86,117.99) --
	(126.51,118.03) --
	(127.15,118.16) --
	(127.79,118.33) --
	(128.43,120.13) --
	(129.07,122.25) --
	(129.71,122.37) --
	(130.36,124.12) --
	(131.00,124.40) --
	(131.64,124.68) --
	(132.28,125.12) --
	(132.92,125.56) --
	(133.56,125.75) --
	(134.20,125.98) --
	(134.85,126.79) --
	(135.49,127.25) --
	(136.13,127.30) --
	(136.77,127.48) --
	(137.41,128.12) --
	(138.05,128.86) --
	(138.69,129.32) --
	(139.34,130.74) --
	(139.98,130.85) --
	(140.62,131.57) --
	(141.26,132.95) --
	(141.90,133.01) --
	(142.54,133.33) --
	(143.19,133.42) --
	(143.83,134.02) --
	(144.47,134.03) --
	(145.11,135.32) --
	(145.75,135.86) --
	(146.39,136.46) --
	(147.03,136.52) --
	(147.68,136.70) --
	(148.32,137.49) --
	(148.96,137.55) --
	(149.60,140.87) --
	(150.24,141.30) --
	(150.88,142.18) --
	(151.53,142.81) --
	(152.17,143.12) --
	(152.81,144.72) --
	(153.45,144.81) --
	(154.09,145.14) --
	(154.73,145.51) --
	(155.37,146.86) --
	(156.02,150.31) --
	(156.66,151.21) --
	(157.30,152.32) --
	(157.94,152.38) --
	(158.58,152.80) --
	(159.22,152.84) --
	(159.86,153.54) --
	(160.51,154.53) --
	(161.15,158.22) --
	(161.79,163.83) --
	(162.43,167.29) --
	(163.07,172.55) --
	(163.71,173.17) --
	(164.36,176.18) --
	(165.00,177.83) --
	(165.64,184.82) --
	(166.28,193.82);
\definecolor{drawColor}{RGB}{152,167,197}

\path[draw=drawColor,line width= 1.0pt,line join=round] ( 38.62, 89.33) --
	( 39.26, 90.21) --
	( 39.90, 92.31) --
	( 40.54, 93.69) --
	( 41.18, 94.23) --
	( 41.83, 94.41) --
	( 42.47, 95.74) --
	( 43.11, 96.61) --
	( 43.75, 96.83) --
	( 44.39, 96.98) --
	( 45.03, 97.18) --
	( 45.68, 97.56) --
	( 46.32, 97.68) --
	( 46.96, 98.08) --
	( 47.60, 99.34) --
	( 48.24, 99.93) --
	( 48.88,100.30) --
	( 49.52,100.30) --
	( 50.17,100.30) --
	( 50.81,100.30) --
	( 51.45,100.30) --
	( 52.09,100.30) --
	( 52.73,100.30) --
	( 53.37,100.30) --
	( 54.01,100.30) --
	( 54.66,100.30) --
	( 55.30,100.30) --
	( 55.94,100.30) --
	( 56.58,100.30) --
	( 57.22,100.30) --
	( 57.86,100.30) --
	( 58.51,100.30) --
	( 59.15,100.30) --
	( 59.79,100.30) --
	( 60.43,100.30) --
	( 61.07,100.30) --
	( 61.71,100.30) --
	( 62.35,100.30) --
	( 63.00,100.30) --
	( 63.64,100.30) --
	( 64.28,100.30) --
	( 64.92,100.30) --
	( 65.56,100.30) --
	( 66.20,100.30) --
	( 66.85,100.30) --
	( 67.49,100.30) --
	( 68.13,100.30) --
	( 68.77,100.30) --
	( 69.41,100.30) --
	( 70.05,100.30) --
	( 70.69,100.30) --
	( 71.34,100.30) --
	( 71.98,100.30) --
	( 72.62,100.30) --
	( 73.26,100.30) --
	( 73.90,100.30) --
	( 74.54,100.30) --
	( 75.18,100.30) --
	( 75.83,100.30) --
	( 76.47,100.30) --
	( 77.11,100.30) --
	( 77.75,100.30) --
	( 78.39,100.30) --
	( 79.03,100.30) --
	( 79.68,100.30) --
	( 80.32,100.34) --
	( 80.96,100.44) --
	( 81.60,100.58) --
	( 82.24,100.81) --
	( 82.88,100.81) --
	( 83.52,100.82) --
	( 84.17,100.84) --
	( 84.81,101.34) --
	( 85.45,101.40) --
	( 86.09,101.44) --
	( 86.73,101.52) --
	( 87.37,101.65) --
	( 88.02,101.73) --
	( 88.66,101.84) --
	( 89.30,101.93) --
	( 89.94,101.93) --
	( 90.58,102.08) --
	( 91.22,102.35) --
	( 91.86,102.61) --
	( 92.51,102.62) --
	( 93.15,102.71) --
	( 93.79,102.93) --
	( 94.43,103.05) --
	( 95.07,103.48) --
	( 95.71,103.66) --
	( 96.35,103.76) --
	( 97.00,104.25) --
	( 97.64,104.32) --
	( 98.28,104.33) --
	( 98.92,104.59) --
	( 99.56,104.62) --
	(100.20,104.65) --
	(100.85,104.79) --
	(101.49,104.83) --
	(102.13,105.20) --
	(102.77,105.46) --
	(103.41,105.60) --
	(104.05,105.93) --
	(104.69,106.20) --
	(105.34,106.24) --
	(105.98,106.57) --
	(106.62,106.59) --
	(107.26,106.66) --
	(107.90,107.09) --
	(108.54,107.14) --
	(109.19,107.34) --
	(109.83,107.35) --
	(110.47,107.54) --
	(111.11,107.97) --
	(111.75,108.08) --
	(112.39,108.08) --
	(113.03,108.35) --
	(113.68,108.37) --
	(114.32,108.45) --
	(114.96,108.81) --
	(115.60,108.84) --
	(116.24,108.86) --
	(116.88,109.06) --
	(117.52,109.09) --
	(118.17,109.17) --
	(118.81,109.21) --
	(119.45,109.31) --
	(120.09,109.50) --
	(120.73,109.65) --
	(121.37,109.83) --
	(122.02,110.00) --
	(122.66,110.10) --
	(123.30,110.31) --
	(123.94,110.39) --
	(124.58,110.49) --
	(125.22,110.50) --
	(125.86,110.59) --
	(126.51,110.70) --
	(127.15,111.20) --
	(127.79,111.58) --
	(128.43,112.31) --
	(129.07,112.38) --
	(129.71,112.71) --
	(130.36,112.85) --
	(131.00,113.32) --
	(131.64,113.55) --
	(132.28,114.75) --
	(132.92,115.21) --
	(133.56,115.23) --
	(134.20,115.29) --
	(134.85,115.45) --
	(135.49,115.53) --
	(136.13,116.01) --
	(136.77,116.54) --
	(137.41,117.09) --
	(138.05,117.28) --
	(138.69,118.08) --
	(139.34,120.20) --
	(139.98,120.67) --
	(140.62,120.71) --
	(141.26,120.88) --
	(141.90,120.93) --
	(142.54,121.08) --
	(143.19,121.14) --
	(143.83,121.16) --
	(144.47,121.67) --
	(145.11,122.05) --
	(145.75,122.31) --
	(146.39,122.57) --
	(147.03,122.67) --
	(147.68,122.72) --
	(148.32,122.74) --
	(148.96,122.89) --
	(149.60,123.77) --
	(150.24,124.19) --
	(150.88,124.56) --
	(151.53,124.99) --
	(152.17,125.29) --
	(152.81,125.48) --
	(153.45,125.57) --
	(154.09,125.85) --
	(154.73,126.22) --
	(155.37,126.39) --
	(156.02,127.30) --
	(156.66,127.78) --
	(157.30,128.66) --
	(157.94,128.71) --
	(158.58,130.99) --
	(159.22,131.26) --
	(159.86,131.59) --
	(160.51,131.76) --
	(161.15,132.67) --
	(161.79,137.82) --
	(162.43,138.99) --
	(163.07,139.92) --
	(163.71,140.27) --
	(164.36,142.46) --
	(165.00,150.54) --
	(165.64,169.23) --
	(165.89,194.47);
\definecolor{drawColor}{RGB}{128,202,192}

\path[draw=drawColor,line width= 1.0pt,line join=round] ( 38.62, 67.72) --
	( 39.26, 69.27) --
	( 39.90, 69.74) --
	( 40.54, 83.65) --
	( 41.18, 87.41) --
	( 41.83, 88.14) --
	( 42.47, 88.40) --
	( 43.11, 90.20) --
	( 43.75, 91.61) --
	( 44.39, 91.70) --
	( 45.03, 92.20) --
	( 45.68, 92.24) --
	( 46.32, 92.46) --
	( 46.96, 92.80) --
	( 47.60, 92.88) --
	( 48.24, 93.10) --
	( 48.88, 93.61) --
	( 49.52, 94.19) --
	( 50.17, 94.25) --
	( 50.81, 94.41) --
	( 51.45, 94.96) --
	( 52.09, 95.06) --
	( 52.73, 96.43) --
	( 53.37, 97.75) --
	( 54.01, 97.85) --
	( 54.66, 97.95) --
	( 55.30, 98.00) --
	( 55.94, 98.15) --
	( 56.58, 98.48) --
	( 57.22, 98.67) --
	( 57.86, 99.03) --
	( 58.51, 99.19) --
	( 59.15, 99.39) --
	( 59.79, 99.49) --
	( 60.43, 99.61) --
	( 61.07, 99.63) --
	( 61.71, 99.79) --
	( 62.35, 99.92) --
	( 63.00,100.04) --
	( 63.64,100.18) --
	( 64.28,100.55) --
	( 64.92,100.60) --
	( 65.56,100.75) --
	( 66.20,100.75) --
	( 66.85,100.81) --
	( 67.49,100.82) --
	( 68.13,101.07) --
	( 68.77,101.17) --
	( 69.41,101.27) --
	( 70.05,101.51) --
	( 70.69,101.55) --
	( 71.34,101.76) --
	( 71.98,101.95) --
	( 72.62,102.31) --
	( 73.26,102.32) --
	( 73.90,102.47) --
	( 74.54,102.57) --
	( 75.18,102.58) --
	( 75.83,102.70) --
	( 76.47,102.74) --
	( 77.11,102.75) --
	( 77.75,102.92) --
	( 78.39,102.94) --
	( 79.03,103.00) --
	( 79.68,103.20) --
	( 80.32,103.22) --
	( 80.96,103.34) --
	( 81.60,103.35) --
	( 82.24,103.47) --
	( 82.88,103.64) --
	( 83.52,103.82) --
	( 84.17,103.87) --
	( 84.81,103.98) --
	( 85.45,103.99) --
	( 86.09,104.06) --
	( 86.73,104.08) --
	( 87.37,104.28) --
	( 88.02,104.35) --
	( 88.66,104.46) --
	( 89.30,104.51) --
	( 89.94,104.74) --
	( 90.58,104.74) --
	( 91.22,104.80) --
	( 91.86,104.92) --
	( 92.51,104.97) --
	( 93.15,105.09) --
	( 93.79,105.25) --
	( 94.43,105.28) --
	( 95.07,105.57) --
	( 95.71,105.58) --
	( 96.35,105.79) --
	( 97.00,105.79) --
	( 97.64,105.81) --
	( 98.28,105.84) --
	( 98.92,105.89) --
	( 99.56,105.96) --
	(100.20,106.13) --
	(100.85,106.23) --
	(101.49,106.62) --
	(102.13,106.71) --
	(102.77,107.12) --
	(103.41,107.12) --
	(104.05,107.22) --
	(104.69,107.35) --
	(105.34,107.49) --
	(105.98,107.62) --
	(106.62,107.64) --
	(107.26,107.78) --
	(107.90,107.80) --
	(108.54,107.85) --
	(109.19,107.97) --
	(109.83,108.12) --
	(110.47,108.15) --
	(111.11,108.18) --
	(111.75,108.21) --
	(112.39,108.35) --
	(113.03,108.49) --
	(113.68,108.52) --
	(114.32,108.58) --
	(114.96,108.59) --
	(115.60,108.60) --
	(116.24,108.88) --
	(116.88,109.14) --
	(117.52,109.32) --
	(118.17,109.46) --
	(118.81,109.53) --
	(119.45,109.59) --
	(120.09,109.70) --
	(120.73,109.70) --
	(121.37,109.89) --
	(122.02,110.17) --
	(122.66,110.32) --
	(123.30,110.35) --
	(123.94,110.42) --
	(124.58,110.72) --
	(125.22,110.80) --
	(125.86,110.84) --
	(126.51,111.12) --
	(127.15,111.18) --
	(127.79,111.37) --
	(128.43,111.54) --
	(129.07,111.60) --
	(129.71,111.62) --
	(130.36,111.73) --
	(131.00,111.83) --
	(131.64,111.94) --
	(132.28,112.05) --
	(132.92,112.06) --
	(133.56,112.56) --
	(134.20,112.69) --
	(134.85,112.90) --
	(135.49,113.10) --
	(136.13,113.20) --
	(136.77,113.53) --
	(137.41,113.56) --
	(138.05,113.64) --
	(138.69,113.64) --
	(139.34,113.70) --
	(139.98,113.71) --
	(140.62,113.94) --
	(141.26,114.35) --
	(141.90,114.49) --
	(142.54,114.69) --
	(143.19,114.79) --
	(143.83,114.82) --
	(144.47,114.84) --
	(145.11,115.13) --
	(145.75,115.48) --
	(146.39,115.63) --
	(147.03,115.69) --
	(147.68,115.80) --
	(148.32,115.96) --
	(148.96,115.98) --
	(149.60,117.09) --
	(150.24,117.76) --
	(150.88,117.80) --
	(151.53,117.84) --
	(152.17,118.30) --
	(152.81,119.12) --
	(153.45,119.43) --
	(154.09,120.11) --
	(154.73,120.18) --
	(155.37,120.23) --
	(156.02,120.30) --
	(156.66,120.36) --
	(157.30,120.83) --
	(157.94,121.13) --
	(158.58,121.19) --
	(159.22,124.09) --
	(159.86,124.33) --
	(160.51,125.12) --
	(161.15,125.16) --
	(161.79,125.69) --
	(162.43,127.64) --
	(163.07,130.95) --
	(163.71,135.69) --
	(164.36,135.99) --
	(165.00,136.67) --
	(165.64,138.46) --
	(166.28,158.12);
\definecolor{drawColor}{RGB}{207,128,187}

\path[draw=drawColor,line width= 1.0pt,line join=round] ( 58.20, 29.68) --
	( 58.51, 29.91) --
	( 59.15, 31.29) --
	( 59.79, 32.18) --
	( 60.43, 34.32) --
	( 61.07, 35.04) --
	( 61.71, 37.11) --
	( 62.35, 39.02) --
	( 63.00, 41.87) --
	( 63.64, 42.78) --
	( 64.28, 42.98) --
	( 64.92, 46.28) --
	( 65.56, 46.95) --
	( 66.20, 47.17) --
	( 66.85, 47.80) --
	( 67.49, 48.12) --
	( 68.13, 49.18) --
	( 68.77, 50.38) --
	( 69.41, 52.42) --
	( 70.05, 53.53) --
	( 70.69, 54.25) --
	( 71.34, 55.31) --
	( 71.98, 55.46) --
	( 72.62, 56.75) --
	( 73.26, 58.56) --
	( 73.90, 58.59) --
	( 74.54, 59.17) --
	( 75.18, 59.31) --
	( 75.83, 60.98) --
	( 76.47, 61.52) --
	( 77.11, 62.28) --
	( 77.75, 63.59) --
	( 78.39, 63.60) --
	( 79.03, 64.62) --
	( 79.68, 65.38) --
	( 80.32, 65.46) --
	( 80.96, 66.77) --
	( 81.60, 67.61) --
	( 82.24, 68.76) --
	( 82.88, 69.76) --
	( 83.52, 70.39) --
	( 84.17, 70.46) --
	( 84.81, 70.97) --
	( 85.45, 71.39) --
	( 86.09, 71.69) --
	( 86.73, 71.74) --
	( 87.37, 73.13) --
	( 88.02, 73.19) --
	( 88.66, 74.00) --
	( 89.30, 74.24) --
	( 89.94, 75.08) --
	( 90.58, 75.12) --
	( 91.22, 75.20) --
	( 91.86, 75.68) --
	( 92.51, 76.07) --
	( 93.15, 76.14) --
	( 93.79, 76.24) --
	( 94.43, 77.44) --
	( 95.07, 77.66) --
	( 95.71, 77.78) --
	( 96.35, 77.91) --
	( 97.00, 78.07) --
	( 97.64, 78.41) --
	( 98.28, 78.64) --
	( 98.92, 78.75) --
	( 99.56, 78.85) --
	(100.20, 79.49) --
	(100.85, 79.70) --
	(101.49, 79.76) --
	(102.13, 80.21) --
	(102.77, 80.42) --
	(103.41, 81.28) --
	(104.05, 81.40) --
	(104.69, 81.97) --
	(105.34, 82.01) --
	(105.98, 82.11) --
	(106.62, 82.27) --
	(107.26, 82.34) --
	(107.90, 82.54) --
	(108.54, 82.69) --
	(109.19, 82.72) --
	(109.83, 82.95) --
	(110.47, 82.99) --
	(111.11, 83.50) --
	(111.75, 83.70) --
	(112.39, 83.98) --
	(113.03, 84.02) --
	(113.68, 84.06) --
	(114.32, 84.13) --
	(114.96, 84.38) --
	(115.60, 84.44) --
	(116.24, 85.68) --
	(116.88, 86.16) --
	(117.52, 86.48) --
	(118.17, 86.55) --
	(118.81, 86.74) --
	(119.45, 86.84) --
	(120.09, 87.01) --
	(120.73, 87.55) --
	(121.37, 87.73) --
	(122.02, 88.40) --
	(122.66, 88.80) --
	(123.30, 89.10) --
	(123.94, 89.59) --
	(124.58, 89.68) --
	(125.22, 90.00) --
	(125.86, 90.46) --
	(126.51, 90.50) --
	(127.15, 90.54) --
	(127.79, 90.98) --
	(128.43, 91.14) --
	(129.07, 91.64) --
	(129.71, 91.97) --
	(130.36, 92.14) --
	(131.00, 92.19) --
	(131.64, 92.56) --
	(132.28, 92.58) --
	(132.92, 92.84) --
	(133.56, 93.40) --
	(134.20, 93.84) --
	(134.85, 94.19) --
	(135.49, 94.60) --
	(136.13, 94.81) --
	(136.77, 94.86) --
	(137.41, 94.92) --
	(138.05, 95.52) --
	(138.69, 95.72) --
	(139.34, 96.12) --
	(139.98, 96.30) --
	(140.62, 96.76) --
	(141.26, 96.85) --
	(141.90, 96.95) --
	(142.54, 97.50) --
	(143.19, 97.71) --
	(143.83, 98.04) --
	(144.47, 98.29) --
	(145.11, 98.43) --
	(145.75, 99.13) --
	(146.39, 99.44) --
	(147.03,100.05) --
	(147.68,100.16) --
	(148.32,100.23) --
	(148.96,100.84) --
	(149.60,101.03) --
	(150.24,101.65) --
	(150.88,101.85) --
	(151.53,101.90) --
	(152.17,102.26) --
	(152.81,102.28) --
	(153.45,102.61) --
	(154.09,102.99) --
	(154.73,103.01) --
	(155.37,103.26) --
	(156.02,103.41) --
	(156.66,104.55) --
	(157.30,104.77) --
	(157.94,105.27) --
	(158.58,105.47) --
	(159.22,105.92) --
	(159.86,106.26) --
	(160.51,106.69) --
	(161.15,106.88) --
	(161.79,106.89) --
	(162.43,107.04) --
	(163.07,107.20) --
	(163.71,109.76) --
	(164.36,110.13) --
	(165.00,110.30) --
	(165.64,111.03) --
	(166.23,194.47);
\definecolor{drawColor}{RGB}{207,192,152}

\path[draw=drawColor,line width= 1.0pt,line join=round] ( 52.24, 29.68) --
	( 52.73, 30.84) --
	( 53.37, 34.31) --
	( 54.01, 37.59) --
	( 54.66, 40.64) --
	( 55.30, 40.92) --
	( 55.94, 42.17) --
	( 56.58, 42.84) --
	( 57.22, 46.90) --
	( 57.86, 47.53) --
	( 58.51, 48.54) --
	( 59.15, 49.36) --
	( 59.79, 49.73) --
	( 60.43, 49.73) --
	( 61.07, 50.52) --
	( 61.71, 51.15) --
	( 62.35, 51.63) --
	( 63.00, 52.31) --
	( 63.64, 52.57) --
	( 64.28, 53.17) --
	( 64.92, 53.20) --
	( 65.56, 53.60) --
	( 66.20, 54.13) --
	( 66.85, 56.93) --
	( 67.49, 57.04) --
	( 68.13, 57.65) --
	( 68.77, 57.98) --
	( 69.41, 58.12) --
	( 70.05, 59.00) --
	( 70.69, 59.78) --
	( 71.34, 61.05) --
	( 71.98, 61.62) --
	( 72.62, 62.03) --
	( 73.26, 63.49) --
	( 73.90, 63.58) --
	( 74.54, 64.07) --
	( 75.18, 64.50) --
	( 75.83, 65.47) --
	( 76.47, 66.54) --
	( 77.11, 67.22) --
	( 77.75, 67.96) --
	( 78.39, 68.95) --
	( 79.03, 70.35) --
	( 79.68, 70.43) --
	( 80.32, 70.98) --
	( 80.96, 71.53) --
	( 81.60, 71.62) --
	( 82.24, 72.20) --
	( 82.88, 73.10) --
	( 83.52, 73.35) --
	( 84.17, 73.53) --
	( 84.81, 74.07) --
	( 85.45, 74.24) --
	( 86.09, 74.96) --
	( 86.73, 75.80) --
	( 87.37, 77.02) --
	( 88.02, 77.60) --
	( 88.66, 78.22) --
	( 89.30, 78.54) --
	( 89.94, 78.60) --
	( 90.58, 79.64) --
	( 91.22, 79.76) --
	( 91.86, 80.01) --
	( 92.51, 80.42) --
	( 93.15, 81.93) --
	( 93.79, 82.22) --
	( 94.43, 83.08) --
	( 95.07, 83.24) --
	( 95.71, 83.41) --
	( 96.35, 83.67) --
	( 97.00, 83.82) --
	( 97.64, 84.78) --
	( 98.28, 84.95) --
	( 98.92, 84.96) --
	( 99.56, 85.17) --
	(100.20, 85.50) --
	(100.85, 85.93) --
	(101.49, 86.05) --
	(102.13, 86.28) --
	(102.77, 86.72) --
	(103.41, 86.97) --
	(104.05, 87.43) --
	(104.69, 88.04) --
	(105.34, 88.68) --
	(105.98, 88.69) --
	(106.62, 88.84) --
	(107.26, 89.62) --
	(107.90, 89.79) --
	(108.54, 89.86) --
	(109.19, 90.22) --
	(109.83, 90.29) --
	(110.47, 90.51) --
	(111.11, 90.59) --
	(111.75, 90.84) --
	(112.39, 91.59) --
	(113.03, 91.63) --
	(113.68, 91.85) --
	(114.32, 91.93) --
	(114.96, 91.97) --
	(115.60, 92.14) --
	(116.24, 92.84) --
	(116.88, 93.35) --
	(117.52, 93.79) --
	(118.17, 93.94) --
	(118.81, 94.36) --
	(119.45, 95.19) --
	(120.09, 95.54) --
	(120.73, 95.57) --
	(121.37, 96.29) --
	(122.02, 96.43) --
	(122.66, 96.60) --
	(123.30, 96.79) --
	(123.94, 96.93) --
	(124.58, 96.98) --
	(125.22, 97.07) --
	(125.86, 97.13) --
	(126.51, 97.22) --
	(127.15, 97.47) --
	(127.79, 97.64) --
	(128.43, 98.07) --
	(129.07, 98.27) --
	(129.71, 99.02) --
	(130.36, 99.10) --
	(131.00, 99.24) --
	(131.64, 99.49) --
	(132.28,100.05) --
	(132.92,100.16) --
	(133.56,100.28) --
	(134.20,100.30) --
	(134.85,100.91) --
	(135.49,100.95) --
	(136.13,101.15) --
	(136.77,101.88) --
	(137.41,102.04) --
	(138.05,103.00) --
	(138.69,103.50) --
	(139.34,103.99) --
	(139.98,104.48) --
	(140.62,105.13) --
	(141.26,105.54) --
	(141.90,106.45) --
	(142.54,106.89) --
	(143.19,108.06) --
	(143.83,108.53) --
	(144.47,108.62) --
	(145.11,108.79) --
	(145.75,109.10) --
	(146.39,110.69) --
	(147.03,110.74) --
	(147.68,111.85) --
	(148.32,112.41) --
	(148.96,113.59) --
	(149.60,114.24) --
	(150.24,114.73) --
	(150.88,115.30) --
	(151.53,116.23) --
	(152.17,118.31) --
	(152.81,120.58) --
	(153.45,124.53) --
	(154.09,127.34) --
	(154.73,131.26) --
	(155.37,143.42) --
	(156.02,174.02) --
	(156.66,191.41) --
	(156.73,194.47);
\definecolor{drawColor}{RGB}{160,30,40}

\path[draw=drawColor,line width= 1.0pt,line join=round] ( 38.62,100.30) --
	( 38.62,100.30) --
	( 38.62,100.30) --
	( 38.62,100.30) --
	( 38.62,100.30) --
	( 39.26,100.30) --
	( 39.26,100.30) --
	( 39.26,100.30) --
	( 39.26,100.30) --
	( 39.26,100.30) --
	( 39.90,100.30) --
	( 39.90,100.30) --
	( 39.90,100.30) --
	( 39.90,100.30) --
	( 39.90,100.30) --
	( 40.54,100.30) --
	( 40.54,100.30) --
	( 40.54,100.30) --
	( 40.54,100.30) --
	( 40.54,100.30) --
	( 41.18,100.30) --
	( 41.18,100.30) --
	( 41.18,100.30) --
	( 41.18,100.30) --
	( 41.18,100.30) --
	( 41.83,100.30) --
	( 41.83,100.30) --
	( 41.83,100.30) --
	( 41.83,100.30) --
	( 41.83,100.30) --
	( 42.47,100.30) --
	( 42.47,100.30) --
	( 42.47,100.30) --
	( 42.47,100.30) --
	( 42.47,100.30) --
	( 43.11,100.30) --
	( 43.11,100.30) --
	( 43.11,100.30) --
	( 43.11,100.30) --
	( 43.11,100.30) --
	( 43.75,100.30) --
	( 43.75,100.30) --
	( 43.75,100.30) --
	( 43.75,100.30) --
	( 43.75,100.30) --
	( 44.39,100.30) --
	( 44.39,100.30) --
	( 44.39,100.30) --
	( 44.39,100.30) --
	( 44.39,100.30) --
	( 45.03,100.30) --
	( 45.03,100.30) --
	( 45.03,100.30) --
	( 45.03,100.30) --
	( 45.03,100.30) --
	( 45.68,100.30) --
	( 45.68,100.30) --
	( 45.68,100.30) --
	( 45.68,100.30) --
	( 45.68,100.30) --
	( 46.32,100.30) --
	( 46.32,100.30) --
	( 46.32,100.30) --
	( 46.32,100.30) --
	( 46.32,100.30) --
	( 46.96,100.30) --
	( 46.96,100.30) --
	( 46.96,100.30) --
	( 46.96,100.30) --
	( 46.96,100.30) --
	( 47.60,100.30) --
	( 47.60,100.30) --
	( 47.60,100.30) --
	( 47.60,100.30) --
	( 47.60,100.30) --
	( 48.24,100.30) --
	( 48.24,100.30) --
	( 48.24,100.30) --
	( 48.24,100.30) --
	( 48.24,100.30) --
	( 48.88,100.30) --
	( 48.88,100.30) --
	( 48.88,100.30) --
	( 48.88,100.30) --
	( 48.88,100.30) --
	( 49.52,100.30) --
	( 49.52,100.30) --
	( 49.52,100.30) --
	( 49.52,100.30) --
	( 49.52,100.30) --
	( 50.17,100.30) --
	( 50.17,100.30) --
	( 50.17,100.30) --
	( 50.17,100.30) --
	( 50.17,100.30) --
	( 50.81,100.30) --
	( 50.81,100.30) --
	( 50.81,100.30) --
	( 50.81,100.30) --
	( 50.81,100.30) --
	( 51.45,100.30) --
	( 51.45,100.30) --
	( 51.45,100.30) --
	( 51.45,100.30) --
	( 51.45,100.30) --
	( 52.09,100.30) --
	( 52.09,100.30) --
	( 52.09,100.30) --
	( 52.09,100.30) --
	( 52.09,100.30) --
	( 52.73,100.30) --
	( 52.73,100.30) --
	( 52.73,100.30) --
	( 52.73,100.30) --
	( 52.73,100.30) --
	( 53.37,100.30) --
	( 53.37,100.30) --
	( 53.37,100.30) --
	( 53.37,100.30) --
	( 53.37,100.30) --
	( 54.01,100.30) --
	( 54.01,100.30) --
	( 54.01,100.30) --
	( 54.01,100.30) --
	( 54.01,100.30) --
	( 54.66,100.30) --
	( 54.66,100.30) --
	( 54.66,100.30) --
	( 54.66,100.30) --
	( 54.66,100.30) --
	( 55.30,100.30) --
	( 55.30,100.30) --
	( 55.30,100.30) --
	( 55.30,100.30) --
	( 55.30,100.30) --
	( 55.94,100.30) --
	( 55.94,100.30) --
	( 55.94,100.30) --
	( 55.94,100.30) --
	( 55.94,100.30) --
	( 56.58,100.30) --
	( 56.58,100.30) --
	( 56.58,100.30) --
	( 56.58,100.30) --
	( 56.58,100.30) --
	( 57.22,100.30) --
	( 57.22,100.30) --
	( 57.22,100.30) --
	( 57.22,100.30) --
	( 57.22,100.30) --
	( 57.86,100.30) --
	( 57.86,100.30) --
	( 57.86,100.30) --
	( 57.86,100.30) --
	( 57.86,100.30) --
	( 58.51,100.30) --
	( 58.51,100.30) --
	( 58.51,100.30) --
	( 58.51,100.30) --
	( 58.51,100.30) --
	( 59.15,100.30) --
	( 59.15,100.30) --
	( 59.15,100.30) --
	( 59.15,100.30) --
	( 59.15,100.30) --
	( 59.79,100.30) --
	( 59.79,100.30) --
	( 59.79,100.30) --
	( 59.79,100.30) --
	( 59.79,100.30) --
	( 60.43,100.30) --
	( 60.43,100.30) --
	( 60.43,100.30) --
	( 60.43,100.30) --
	( 60.43,100.30) --
	( 61.07,100.30) --
	( 61.07,100.30) --
	( 61.07,100.30) --
	( 61.07,100.30) --
	( 61.07,100.30) --
	( 61.71,100.30) --
	( 61.71,100.30) --
	( 61.71,100.30) --
	( 61.71,100.30) --
	( 61.71,100.30) --
	( 62.35,100.30) --
	( 62.35,100.30) --
	( 62.35,100.30) --
	( 62.35,100.30) --
	( 62.35,100.30) --
	( 63.00,100.30) --
	( 63.00,100.30) --
	( 63.00,100.30) --
	( 63.00,100.30) --
	( 63.00,100.30) --
	( 63.64,100.30) --
	( 63.64,100.30) --
	( 63.64,100.30) --
	( 63.64,100.30) --
	( 63.64,100.30) --
	( 64.28,100.30) --
	( 64.28,100.30) --
	( 64.28,100.30) --
	( 64.28,100.30) --
	( 64.28,100.30) --
	( 64.92,100.30) --
	( 64.92,100.30) --
	( 64.92,100.30) --
	( 64.92,100.30) --
	( 64.92,100.30) --
	( 65.56,100.30) --
	( 65.56,100.30) --
	( 65.56,100.30) --
	( 65.56,100.30) --
	( 65.56,100.30) --
	( 66.20,100.30) --
	( 66.20,100.30) --
	( 66.20,100.30) --
	( 66.20,100.30) --
	( 66.20,100.30) --
	( 66.85,100.30) --
	( 66.85,100.30) --
	( 66.85,100.30) --
	( 66.85,100.30) --
	( 66.85,100.30) --
	( 67.49,100.30) --
	( 67.49,100.30) --
	( 67.49,100.30) --
	( 67.49,100.30) --
	( 67.49,100.30) --
	( 68.13,100.30) --
	( 68.13,100.30) --
	( 68.13,100.30) --
	( 68.13,100.30) --
	( 68.13,100.30) --
	( 68.77,100.30) --
	( 68.77,100.30) --
	( 68.77,100.30) --
	( 68.77,100.30) --
	( 68.77,100.30) --
	( 69.41,100.30) --
	( 69.41,100.30) --
	( 69.41,100.30) --
	( 69.41,100.30) --
	( 69.41,100.30) --
	( 70.05,100.30) --
	( 70.05,100.30) --
	( 70.05,100.30) --
	( 70.05,100.30) --
	( 70.05,100.30) --
	( 70.69,100.30) --
	( 70.69,100.30) --
	( 70.69,100.30) --
	( 70.69,100.30) --
	( 70.69,100.30) --
	( 71.34,100.30) --
	( 71.34,100.30) --
	( 71.34,100.30) --
	( 71.34,100.30) --
	( 71.34,100.30) --
	( 71.98,100.30) --
	( 71.98,100.30) --
	( 71.98,100.30) --
	( 71.98,100.30) --
	( 71.98,100.30) --
	( 72.62,100.30) --
	( 72.62,100.30) --
	( 72.62,100.30) --
	( 72.62,100.30) --
	( 72.62,100.30) --
	( 73.26,100.30) --
	( 73.26,100.30) --
	( 73.26,100.30) --
	( 73.26,100.30) --
	( 73.26,100.30) --
	( 73.90,100.30) --
	( 73.90,100.30) --
	( 73.90,100.30) --
	( 73.90,100.30) --
	( 73.90,100.30) --
	( 74.54,100.30) --
	( 74.54,100.30) --
	( 74.54,100.30) --
	( 74.54,100.30) --
	( 74.54,100.30) --
	( 75.18,100.30) --
	( 75.18,100.30) --
	( 75.18,100.30) --
	( 75.18,100.30) --
	( 75.18,100.30) --
	( 75.83,100.30) --
	( 75.83,100.30) --
	( 75.83,100.30) --
	( 75.83,100.30) --
	( 75.83,100.30) --
	( 76.47,100.30) --
	( 76.47,100.30) --
	( 76.47,100.30) --
	( 76.47,100.30) --
	( 76.47,100.30) --
	( 77.11,100.30) --
	( 77.11,100.30) --
	( 77.11,100.30) --
	( 77.11,100.30) --
	( 77.11,100.30) --
	( 77.75,100.30) --
	( 77.75,100.30) --
	( 77.75,100.30) --
	( 77.75,100.30) --
	( 77.75,100.30) --
	( 78.39,100.30) --
	( 78.39,100.30) --
	( 78.39,100.30) --
	( 78.39,100.30) --
	( 78.39,100.30) --
	( 79.03,100.30) --
	( 79.03,100.30) --
	( 79.03,100.30) --
	( 79.03,100.30) --
	( 79.03,100.30) --
	( 79.68,100.30) --
	( 79.68,100.30) --
	( 79.68,100.30) --
	( 79.68,100.30) --
	( 79.68,100.30) --
	( 80.32,100.30) --
	( 80.32,100.30) --
	( 80.32,100.30) --
	( 80.32,100.30) --
	( 80.32,100.30) --
	( 80.96,100.30) --
	( 80.96,100.30) --
	( 80.96,100.30) --
	( 80.96,100.30) --
	( 80.96,100.30) --
	( 81.60,100.30) --
	( 81.60,100.30) --
	( 81.60,100.30) --
	( 81.60,100.30) --
	( 81.60,100.30) --
	( 82.24,100.30) --
	( 82.24,100.30) --
	( 82.24,100.30) --
	( 82.24,100.30) --
	( 82.24,100.30) --
	( 82.88,100.30) --
	( 82.88,100.30) --
	( 82.88,100.30) --
	( 82.88,100.30) --
	( 82.88,100.30) --
	( 83.52,100.30) --
	( 83.52,100.30) --
	( 83.52,100.30) --
	( 83.52,100.30) --
	( 83.52,100.30) --
	( 84.17,100.30) --
	( 84.17,100.30) --
	( 84.17,100.30) --
	( 84.17,100.30) --
	( 84.17,100.30) --
	( 84.81,100.30) --
	( 84.81,100.30) --
	( 84.81,100.30) --
	( 84.81,100.30) --
	( 84.81,100.30) --
	( 85.45,100.30) --
	( 85.45,100.30) --
	( 85.45,100.30) --
	( 85.45,100.30) --
	( 85.45,100.30) --
	( 86.09,100.30) --
	( 86.09,100.30) --
	( 86.09,100.30) --
	( 86.09,100.30) --
	( 86.09,100.30) --
	( 86.73,100.30) --
	( 86.73,100.30) --
	( 86.73,100.30) --
	( 86.73,100.30) --
	( 86.73,100.30) --
	( 87.37,100.30) --
	( 87.37,100.30) --
	( 87.37,100.30) --
	( 87.37,100.30) --
	( 87.37,100.30) --
	( 88.02,100.30) --
	( 88.02,100.30) --
	( 88.02,100.30) --
	( 88.02,100.30) --
	( 88.02,100.30) --
	( 88.66,100.30) --
	( 88.66,100.30) --
	( 88.66,100.30) --
	( 88.66,100.30) --
	( 88.66,100.30) --
	( 89.30,100.30) --
	( 89.30,100.30) --
	( 89.30,100.30) --
	( 89.30,100.30) --
	( 89.30,100.30) --
	( 89.94,100.30) --
	( 89.94,100.30) --
	( 89.94,100.30) --
	( 89.94,100.30) --
	( 89.94,100.30) --
	( 90.58,100.30) --
	( 90.58,100.30) --
	( 90.58,100.30) --
	( 90.58,100.30) --
	( 90.58,100.30) --
	( 91.22,100.30) --
	( 91.22,100.30) --
	( 91.22,100.30) --
	( 91.22,100.30) --
	( 91.22,100.30) --
	( 91.86,100.30) --
	( 91.86,100.30) --
	( 91.86,100.30) --
	( 91.86,100.30) --
	( 91.86,100.30) --
	( 92.51,100.30) --
	( 92.51,100.30) --
	( 92.51,100.30) --
	( 92.51,100.30) --
	( 92.51,100.30) --
	( 93.15,100.30) --
	( 93.15,100.30) --
	( 93.15,100.30) --
	( 93.15,100.30) --
	( 93.15,100.30) --
	( 93.79,100.30) --
	( 93.79,100.30) --
	( 93.79,100.30) --
	( 93.79,100.30) --
	( 93.79,100.30) --
	( 94.43,100.30) --
	( 94.43,100.30) --
	( 94.43,100.30) --
	( 94.43,100.30) --
	( 94.43,100.30) --
	( 95.07,100.30) --
	( 95.07,100.30) --
	( 95.07,100.30) --
	( 95.07,100.30) --
	( 95.07,100.30) --
	( 95.71,100.30) --
	( 95.71,100.30) --
	( 95.71,100.30) --
	( 95.71,100.30) --
	( 95.71,100.30) --
	( 96.35,100.30) --
	( 96.35,100.30) --
	( 96.35,100.30) --
	( 96.35,100.30) --
	( 96.35,100.30) --
	( 97.00,100.30) --
	( 97.00,100.30) --
	( 97.00,100.30) --
	( 97.00,100.30) --
	( 97.00,100.30) --
	( 97.64,100.30) --
	( 97.64,100.30) --
	( 97.64,100.30) --
	( 97.64,100.30) --
	( 97.64,100.30) --
	( 98.28,100.30) --
	( 98.28,100.30) --
	( 98.28,100.30) --
	( 98.28,100.30) --
	( 98.28,100.30) --
	( 98.92,100.30) --
	( 98.92,100.30) --
	( 98.92,100.30) --
	( 98.92,100.30) --
	( 98.92,100.30) --
	( 99.56,100.30) --
	( 99.56,100.30) --
	( 99.56,100.30) --
	( 99.56,100.30) --
	( 99.56,100.30) --
	(100.20,100.30) --
	(100.20,100.30) --
	(100.20,100.30) --
	(100.20,100.30) --
	(100.20,100.30) --
	(100.85,100.30) --
	(100.85,100.30) --
	(100.85,100.30) --
	(100.85,100.30) --
	(100.85,100.30) --
	(101.49,100.30) --
	(101.49,100.30) --
	(101.49,100.30) --
	(101.49,100.30) --
	(101.49,100.30) --
	(102.13,100.30) --
	(102.13,100.30) --
	(102.13,100.30) --
	(102.13,100.30) --
	(102.13,100.30) --
	(102.77,100.30) --
	(102.77,100.30) --
	(102.77,100.30) --
	(102.77,100.30) --
	(102.77,100.30) --
	(103.41,100.30) --
	(103.41,100.30) --
	(103.41,100.30) --
	(103.41,100.30) --
	(103.41,100.30) --
	(104.05,100.30) --
	(104.05,100.30) --
	(104.05,100.30) --
	(104.05,100.30) --
	(104.05,100.30) --
	(104.69,100.30) --
	(104.69,100.30) --
	(104.69,100.30) --
	(104.69,100.30) --
	(104.69,100.30) --
	(105.34,100.30) --
	(105.34,100.30) --
	(105.34,100.30) --
	(105.34,100.30) --
	(105.34,100.30) --
	(105.98,100.30) --
	(105.98,100.30) --
	(105.98,100.30) --
	(105.98,100.30) --
	(105.98,100.30) --
	(106.62,100.30) --
	(106.62,100.30) --
	(106.62,100.30) --
	(106.62,100.30) --
	(106.62,100.30) --
	(107.26,100.30) --
	(107.26,100.30) --
	(107.26,100.30) --
	(107.26,100.30) --
	(107.26,100.30) --
	(107.90,100.30) --
	(107.90,100.30) --
	(107.90,100.30) --
	(107.90,100.30) --
	(107.90,100.30) --
	(108.54,100.30) --
	(108.54,100.30) --
	(108.54,100.30) --
	(108.54,100.30) --
	(108.54,100.30) --
	(109.19,100.30) --
	(109.19,100.30) --
	(109.19,100.30) --
	(109.19,100.30) --
	(109.19,100.30) --
	(109.83,100.30) --
	(109.83,100.30) --
	(109.83,100.30) --
	(109.83,100.30) --
	(109.83,100.30) --
	(110.47,100.30) --
	(110.47,100.30) --
	(110.47,100.30) --
	(110.47,100.30) --
	(110.47,100.30) --
	(111.11,100.30) --
	(111.11,100.30) --
	(111.11,100.30) --
	(111.11,100.30) --
	(111.11,100.30) --
	(111.75,100.30) --
	(111.75,100.30) --
	(111.75,100.30) --
	(111.75,100.30) --
	(111.75,100.30) --
	(112.39,100.30) --
	(112.39,100.30) --
	(112.39,100.30) --
	(112.39,100.30) --
	(112.39,100.30) --
	(113.03,100.30) --
	(113.03,100.30) --
	(113.03,100.30) --
	(113.03,100.30) --
	(113.03,100.30) --
	(113.68,100.30) --
	(113.68,100.30) --
	(113.68,100.30) --
	(113.68,100.30) --
	(113.68,100.30) --
	(114.32,100.30) --
	(114.32,100.30) --
	(114.32,100.30) --
	(114.32,100.30) --
	(114.32,100.30) --
	(114.96,100.30) --
	(114.96,100.30) --
	(114.96,100.30) --
	(114.96,100.30) --
	(114.96,100.30) --
	(115.60,100.30) --
	(115.60,100.30) --
	(115.60,100.30) --
	(115.60,100.30) --
	(115.60,100.30) --
	(116.24,100.30) --
	(116.24,100.30) --
	(116.24,100.30) --
	(116.24,100.30) --
	(116.24,100.30) --
	(116.88,100.30) --
	(116.88,100.30) --
	(116.88,100.30) --
	(116.88,100.30) --
	(116.88,100.30) --
	(117.52,100.30) --
	(117.52,100.30) --
	(117.52,100.30) --
	(117.52,100.30) --
	(117.52,100.30) --
	(118.17,100.30) --
	(118.17,100.30) --
	(118.17,100.30) --
	(118.17,100.30) --
	(118.17,100.30) --
	(118.81,100.30) --
	(118.81,100.30) --
	(118.81,100.30) --
	(118.81,100.30) --
	(118.81,100.30) --
	(119.45,100.30) --
	(119.45,100.30) --
	(119.45,100.30) --
	(119.45,100.30) --
	(119.45,100.30) --
	(120.09,100.30) --
	(120.09,100.30) --
	(120.09,100.30) --
	(120.09,100.30) --
	(120.09,100.30) --
	(120.73,100.30) --
	(120.73,100.30) --
	(120.73,100.30) --
	(120.73,100.30) --
	(120.73,100.30) --
	(121.37,100.30) --
	(121.37,100.30) --
	(121.37,100.30) --
	(121.37,100.30) --
	(121.37,100.30) --
	(122.02,100.30) --
	(122.02,100.30) --
	(122.02,100.30) --
	(122.02,100.30) --
	(122.02,100.30) --
	(122.66,100.30) --
	(122.66,100.30) --
	(122.66,100.30) --
	(122.66,100.30) --
	(122.66,100.30) --
	(123.30,100.30) --
	(123.30,100.30) --
	(123.30,100.30) --
	(123.30,100.30) --
	(123.30,100.30) --
	(123.94,100.30) --
	(123.94,100.30) --
	(123.94,100.30) --
	(123.94,100.30) --
	(123.94,100.30) --
	(124.58,100.30) --
	(124.58,100.30) --
	(124.58,100.30) --
	(124.58,100.30) --
	(124.58,100.30) --
	(125.22,100.30) --
	(125.22,100.30) --
	(125.22,100.30) --
	(125.22,100.30) --
	(125.22,100.30) --
	(125.86,100.30) --
	(125.86,100.30) --
	(125.86,100.30) --
	(125.86,100.30) --
	(125.86,100.30) --
	(126.51,100.30) --
	(126.51,100.30) --
	(126.51,100.30) --
	(126.51,100.30) --
	(126.51,100.30) --
	(127.15,100.30) --
	(127.15,100.30) --
	(127.15,100.30) --
	(127.15,100.30) --
	(127.15,100.30) --
	(127.79,100.30) --
	(127.79,100.30) --
	(127.79,100.30) --
	(127.79,100.30) --
	(127.79,100.30) --
	(128.43,100.30) --
	(128.43,100.30) --
	(128.43,100.30) --
	(128.43,100.30) --
	(128.43,100.30) --
	(129.07,100.30) --
	(129.07,100.30) --
	(129.07,100.30) --
	(129.07,100.30) --
	(129.07,100.30) --
	(129.71,100.30) --
	(129.71,100.30) --
	(129.71,100.30) --
	(129.71,100.30) --
	(129.71,100.30) --
	(130.36,100.30) --
	(130.36,100.30) --
	(130.36,100.30) --
	(130.36,100.30) --
	(130.36,100.30) --
	(131.00,100.30) --
	(131.00,100.30) --
	(131.00,100.30) --
	(131.00,100.30) --
	(131.00,100.30) --
	(131.64,100.30) --
	(131.64,100.30) --
	(131.64,100.30) --
	(131.64,100.30) --
	(131.64,100.30) --
	(132.28,100.30) --
	(132.28,100.30) --
	(132.28,100.30) --
	(132.28,100.30) --
	(132.28,100.30) --
	(132.92,100.30) --
	(132.92,100.30) --
	(132.92,100.30) --
	(132.92,100.30) --
	(132.92,100.30) --
	(133.56,100.30) --
	(133.56,100.30) --
	(133.56,100.30) --
	(133.56,100.30) --
	(133.56,100.30) --
	(134.20,100.30) --
	(134.20,100.30) --
	(134.20,100.30) --
	(134.20,100.30) --
	(134.20,100.30) --
	(134.85,100.30) --
	(134.85,100.30) --
	(134.85,100.30) --
	(134.85,100.30) --
	(134.85,100.30) --
	(135.49,100.30) --
	(135.49,100.30) --
	(135.49,100.30) --
	(135.49,100.30) --
	(135.49,100.30) --
	(136.13,100.30) --
	(136.13,100.30) --
	(136.13,100.30) --
	(136.13,100.30) --
	(136.13,100.30) --
	(136.77,100.30) --
	(136.77,100.30) --
	(136.77,100.30) --
	(136.77,100.30) --
	(136.77,100.30) --
	(137.41,100.30) --
	(137.41,100.30) --
	(137.41,100.30) --
	(137.41,100.30) --
	(137.41,100.30) --
	(138.05,100.30) --
	(138.05,100.30) --
	(138.05,100.30) --
	(138.05,100.30) --
	(138.05,100.30) --
	(138.69,100.30) --
	(138.69,100.30) --
	(138.69,100.30) --
	(138.69,100.30) --
	(138.69,100.30) --
	(139.34,100.30) --
	(139.34,100.30) --
	(139.34,100.30) --
	(139.34,100.30) --
	(139.34,100.30) --
	(139.98,100.30) --
	(139.98,100.30) --
	(139.98,100.30) --
	(139.98,100.30) --
	(139.98,100.30) --
	(140.62,100.30) --
	(140.62,100.30) --
	(140.62,100.30) --
	(140.62,100.30) --
	(140.62,100.30) --
	(141.26,100.30) --
	(141.26,100.30) --
	(141.26,100.30) --
	(141.26,100.30) --
	(141.26,100.30) --
	(141.90,100.30) --
	(141.90,100.30) --
	(141.90,100.30) --
	(141.90,100.30) --
	(141.90,100.30) --
	(142.54,100.30) --
	(142.54,100.30) --
	(142.54,100.30) --
	(142.54,100.30) --
	(142.54,100.30) --
	(143.19,100.30) --
	(143.19,100.30) --
	(143.19,100.30) --
	(143.19,100.30) --
	(143.19,100.30) --
	(143.83,100.30) --
	(143.83,100.30) --
	(143.83,100.30) --
	(143.83,100.30) --
	(143.83,100.30) --
	(144.47,100.30) --
	(144.47,100.30) --
	(144.47,100.30) --
	(144.47,100.30) --
	(144.47,100.30) --
	(145.11,100.30) --
	(145.11,100.30) --
	(145.11,100.30) --
	(145.11,100.30) --
	(145.11,100.30) --
	(145.75,100.30) --
	(145.75,100.30) --
	(145.75,100.30) --
	(145.75,100.30) --
	(145.75,100.30) --
	(146.39,100.30) --
	(146.39,100.30) --
	(146.39,100.30) --
	(146.39,100.30) --
	(146.39,100.30) --
	(147.03,100.30) --
	(147.03,100.30) --
	(147.03,100.30) --
	(147.03,100.30) --
	(147.03,100.30) --
	(147.68,100.30) --
	(147.68,100.30) --
	(147.68,100.30) --
	(147.68,100.30) --
	(147.68,100.30) --
	(148.32,100.30) --
	(148.32,100.30) --
	(148.32,100.30) --
	(148.32,100.30) --
	(148.32,100.30) --
	(148.96,100.30) --
	(148.96,100.30) --
	(148.96,100.30) --
	(148.96,100.30) --
	(148.96,100.30) --
	(149.60,100.30) --
	(149.60,100.30) --
	(149.60,100.30) --
	(149.60,100.30) --
	(149.60,100.30) --
	(150.24,100.30) --
	(150.24,100.30) --
	(150.24,100.30) --
	(150.24,100.30) --
	(150.24,100.30) --
	(150.88,100.30) --
	(150.88,100.30) --
	(150.88,100.30) --
	(150.88,100.30) --
	(150.88,100.30) --
	(151.53,100.30) --
	(151.53,100.30) --
	(151.53,100.30) --
	(151.53,100.30) --
	(151.53,100.30) --
	(152.17,100.30) --
	(152.17,100.30) --
	(152.17,100.30) --
	(152.17,100.30) --
	(152.17,100.30) --
	(152.81,100.30) --
	(152.81,100.30) --
	(152.81,100.30) --
	(152.81,100.30) --
	(152.81,100.30) --
	(153.45,100.30) --
	(153.45,100.30) --
	(153.45,100.30) --
	(153.45,100.30) --
	(153.45,100.30) --
	(154.09,100.30) --
	(154.09,100.30) --
	(154.09,100.30) --
	(154.09,100.30) --
	(154.09,100.30) --
	(154.73,100.30) --
	(154.73,100.30) --
	(154.73,100.30) --
	(154.73,100.30) --
	(154.73,100.30) --
	(155.37,100.30) --
	(155.37,100.30) --
	(155.37,100.30) --
	(155.37,100.30) --
	(155.37,100.30) --
	(156.02,100.30) --
	(156.02,100.30) --
	(156.02,100.30) --
	(156.02,100.30) --
	(156.02,100.30) --
	(156.66,100.30) --
	(156.66,100.30) --
	(156.66,100.30) --
	(156.66,100.30) --
	(156.66,100.30) --
	(157.30,100.30) --
	(157.30,100.30) --
	(157.30,100.30) --
	(157.30,100.30) --
	(157.30,100.30) --
	(157.94,100.30) --
	(157.94,100.30) --
	(157.94,100.30) --
	(157.94,100.30) --
	(157.94,100.30) --
	(158.58,100.30) --
	(158.58,100.30) --
	(158.58,100.30) --
	(158.58,100.30) --
	(158.58,100.30) --
	(159.22,100.30) --
	(159.22,100.30) --
	(159.22,100.30) --
	(159.22,100.30) --
	(159.22,100.30) --
	(159.86,100.30) --
	(159.86,100.30) --
	(159.86,100.30) --
	(159.86,100.30) --
	(159.86,100.30) --
	(160.51,100.30) --
	(160.51,100.30) --
	(160.51,100.30) --
	(160.51,100.30) --
	(160.51,100.30) --
	(161.15,100.30) --
	(161.15,100.30) --
	(161.15,100.30) --
	(161.15,100.30) --
	(161.15,100.30) --
	(161.79,100.30) --
	(161.79,100.30) --
	(161.79,100.30) --
	(161.79,100.30) --
	(161.79,100.30) --
	(162.43,100.30) --
	(162.43,100.30) --
	(162.43,100.30) --
	(162.43,100.30) --
	(162.43,100.30) --
	(163.07,100.30) --
	(163.07,100.30) --
	(163.07,100.30) --
	(163.07,100.30) --
	(163.07,100.30) --
	(163.71,100.30) --
	(163.71,100.30) --
	(163.71,100.30) --
	(163.71,100.30) --
	(163.71,100.30) --
	(164.36,100.30) --
	(164.36,100.30) --
	(164.36,100.30) --
	(164.36,100.30) --
	(164.36,100.30) --
	(165.00,100.30) --
	(165.00,100.30) --
	(165.00,100.30) --
	(165.00,100.30) --
	(165.00,100.30) --
	(165.64,100.30) --
	(165.64,100.30) --
	(165.64,100.30) --
	(165.64,100.30) --
	(165.64,100.30) --
	(166.28,100.30) --
	(166.28,100.30) --
	(166.28,100.30) --
	(166.28,100.30) --
	(166.28,100.30);
\definecolor{drawColor}{gray}{0.70}

\path[draw=drawColor,line width= 0.5pt,line join=round,line cap=round] ( 38.62, 29.68) rectangle (166.28,194.47);
\end{scope}
\begin{scope}
\path[clip] (  0.00,  0.00) rectangle (171.28,199.47);
\definecolor{drawColor}{gray}{0.30}

\node[text=drawColor,anchor=base east,inner sep=0pt, outer sep=0pt, scale=  1.00] at ( 34.12, 49.77) {0.98};

\node[text=drawColor,anchor=base east,inner sep=0pt, outer sep=0pt, scale=  1.00] at ( 34.12, 96.86) {1.00};

\node[text=drawColor,anchor=base east,inner sep=0pt, outer sep=0pt, scale=  1.00] at ( 34.12,143.94) {1.02};

\node[text=drawColor,anchor=base east,inner sep=0pt, outer sep=0pt, scale=  1.00] at ( 34.12,191.02) {1.04};
\end{scope}
\begin{scope}
\path[clip] (  0.00,  0.00) rectangle (171.28,199.47);
\definecolor{drawColor}{gray}{0.70}

\path[draw=drawColor,line width= 0.3pt,line join=round] ( 36.12, 53.22) --
	( 38.62, 53.22);

\path[draw=drawColor,line width= 0.3pt,line join=round] ( 36.12,100.30) --
	( 38.62,100.30);

\path[draw=drawColor,line width= 0.3pt,line join=round] ( 36.12,147.38) --
	( 38.62,147.38);

\path[draw=drawColor,line width= 0.3pt,line join=round] ( 36.12,194.47) --
	( 38.62,194.47);
\end{scope}
\begin{scope}
\path[clip] (  0.00,  0.00) rectangle (171.28,199.47);
\definecolor{drawColor}{gray}{0.70}

\path[draw=drawColor,line width= 0.3pt,line join=round] ( 38.62, 27.18) --
	( 38.62, 29.68);

\path[draw=drawColor,line width= 0.3pt,line join=round] ( 64.15, 27.18) --
	( 64.15, 29.68);

\path[draw=drawColor,line width= 0.3pt,line join=round] ( 89.68, 27.18) --
	( 89.68, 29.68);

\path[draw=drawColor,line width= 0.3pt,line join=round] (115.22, 27.18) --
	(115.22, 29.68);

\path[draw=drawColor,line width= 0.3pt,line join=round] (140.75, 27.18) --
	(140.75, 29.68);

\path[draw=drawColor,line width= 0.3pt,line join=round] (166.28, 27.18) --
	(166.28, 29.68);
\end{scope}
\begin{scope}
\path[clip] (  0.00,  0.00) rectangle (171.28,199.47);
\definecolor{drawColor}{gray}{0.30}

\node[text=drawColor,anchor=base,inner sep=0pt, outer sep=0pt, scale=  1.00] at ( 37.62, 18.29) {0};

\node[text=drawColor,anchor=base,inner sep=0pt, outer sep=0pt, scale=  1.00] at ( 62.15, 18.29) {20};

\node[text=drawColor,anchor=base,inner sep=0pt, outer sep=0pt, scale=  1.00] at ( 87.68, 18.29) {40};

\node[text=drawColor,anchor=base,inner sep=0pt, outer sep=0pt, scale=  1.00] at (113.22, 18.29) {60};

\node[text=drawColor,anchor=base,inner sep=0pt, outer sep=0pt, scale=  1.00] at (138.75, 18.29) {80};

\node[text=drawColor,anchor=base,inner sep=0pt, outer sep=0pt, scale=  1.00] at (163.28, 18.29) {100};
\end{scope}
\begin{scope}
\path[clip] (  0.00,  0.00) rectangle (171.28,199.47);
\definecolor{drawColor}{RGB}{0,0,0}

\node[text=drawColor,anchor=base,inner sep=0pt, outer sep=0pt, scale=  1.00] at (102.45,  6.94) {\bfseries Instances in \%};
\end{scope}
\begin{scope}
\path[clip] (  0.00,  0.00) rectangle (171.28,199.47);
\definecolor{drawColor}{RGB}{0,0,0}

\node[text=drawColor,rotate= 90.00,anchor=base,inner sep=0pt, outer sep=0pt, scale=  1.00] at ( 11.90,120.35) {\bfseries Relative Costs};
\end{scope}
\end{tikzpicture}

%% file: arxiv_version.bbl
\begin{thebibliography}{10}

\bibitem{online:4C:Hornsea3}
4C Offshore Ltd.
\newblock {\em Hornsea Project Three Offshore Wind Farm}, 2018.
\newblock
  \url{www.4coffshore.com/windfarms/hornsea-project-three-united-kingdom-uk1k.html},
  \mbox{Accessed: 2018-08-15}.

\bibitem{Ahuja1993}
Ravindra~K. Ahuja, Thomas~L. Magnanti, and James~B. Orlin.
\newblock {\em Network flows: theory, algorithms, and applications}.
\newblock Prentice Hall, Upper Saddle River, NJ [u.a.], 1993.

\bibitem{bellman:routing}
Richard Bellman.
\newblock On a routing problem.
\newblock {\em Quarterly of Applied Mathematics}, 16:87--90, 1958.
\newblock \href {http://dx.doi.org/10.1090/qam/102435}
  {\path{doi:10.1090/qam/102435}}.

\bibitem{berzan}
Constantin Berzan, Kalyan Veeramachaneni, James McDermott, and Una-May
  O'Reilly.
\newblock Algorithms for cable network design on large-scale wind farms.
\newblock Technical report, Massachusetts Institute of Technology, 2011.

\bibitem{Cherkassky1999}
Boris~V. Cherkassky and Andrew~V. Goldberg.
\newblock Negative-cycle detection algorithms.
\newblock {\em Mathematical Programming}, 85(2):277--311, Jun 1999.
\newblock \href {http://dx.doi.org/10.1007/s101070050058}
  {\path{doi:10.1007/s101070050058}}.

\bibitem{Dahmani2015OptimizationOT}
Ouahid Dahmani, Salvy Bourguet, Mohamed Machmoum, Patrick Guerin, Pauline
  Rhein, and Lionel Josse.
\newblock Optimization of the connection topology of an offshore wind farm
  network.
\newblock {\em {IEEE} Systems Journal}, 9(4):1519--1528, 2015.
\newblock \href {http://dx.doi.org/10.1109/JSYST.2014.2330064}
  {\path{doi:10.1109/JSYST.2014.2330064}}.

\bibitem{Souza2008}
Mauricio~C. de~Souza, Philippe Mahey, and Bernard Gendron.
\newblock Cycle‐based algorithms for multicommodity network flow problems
  with separable piecewise convex costs.
\newblock {\em Networks}, 51(2):133--141, 2008.
\newblock \href {http://dx.doi.org/10.1002/net.20208}
  {\path{doi:10.1002/net.20208}}.

\bibitem{Dijkstra1959}
Edsger~W. Dijkstra.
\newblock A note on two problems in connexion with graphs.
\newblock {\em Numerische Mathematik}, 1(1):269--271, Dec 1959.
\newblock \href {http://dx.doi.org/10.1007/BF01386390}
  {\path{doi:10.1007/BF01386390}}.

\bibitem{Ford:2010:FN:1942094}
Lester~R. Ford, Jr. and Delbert~R. Fulkerson.
\newblock {\em Flows in Networks}.
\newblock Princeton University Press, Princeton, NJ, USA, 2010.

\bibitem{GABREL199915}
Virginie Gabrel, Arnaud Knippel, and Michel Minoux.
\newblock Exact solution of multicommodity network optimization problems with
  general step cost functions.
\newblock {\em Operations Research Letters}, 25(1):15 -- 23, 1999.
\newblock \href {http://dx.doi.org/10.1016/S0167-6377(99)00020-6}
  {\path{doi:10.1016/S0167-6377(99)00020-6}}.

\bibitem{Goldberg1993}
Andrew~V. Goldberg and Tomasz Radzik.
\newblock A heuristic improvement of the {B}ellman-{F}ord algorithm.
\newblock {\em Applied Mathematics Letters}, 6(3):3 -- 6, 1993.
\newblock \href {http://dx.doi.org/10.1016/0893-9659(93)90022-F}
  {\path{doi:10.1016/0893-9659(93)90022-F}}.

\bibitem{Goldberg1989}
Andrew~V. Goldberg and Robert~E. Tarjan.
\newblock Finding minimum-cost circulations by canceling negative cycles.
\newblock {\em Journal of the ACM}, 36(4):873--886, October 1989.
\newblock \href {http://dx.doi.org/10.1145/76359.76368}
  {\path{doi:10.1145/76359.76368}}.

\bibitem{Goldfarb1991}
Donald Goldfarb, Jianxiu Hao, and Sheng-Roan Kai.
\newblock Shortest path algorithms using dynamic breadth-first search.
\newblock {\em Networks}, 21(1):29--50, 1991.
\newblock \href {http://dx.doi.org/10.1002/net.3230210105}
  {\path{doi:10.1002/net.3230210105}}.

\bibitem{Gritzbach23}
Sascha Gritzbach.
\newblock Cable layout optimization problems in the context of renewable energy
  sources, 2023.
\newblock \href {http://dx.doi.org/10.5445/IR/1000158746}
  {\path{doi:10.5445/IR/1000158746}}.

\bibitem{Gritzbach2018}
Sascha Gritzbach, Torsten Ueckerdt, Dorothea Wagner, Franziska Wegner, and
  Matthias Wolf.
\newblock {Towards negative cycle canceling in wind farm cable layout
  optimization}.
\newblock In {\em Proceedings of the 7th DACH+ Conference on Energy
  Informatics}, volume 1 (Suppl 1). Springer, 2018.
\newblock \href {http://dx.doi.org/10.1186/s42162-018-0030-6}
  {\path{doi:10.1186/s42162-018-0030-6}}.

\bibitem{Guo2017}
Longkun Guo and Peng Li.
\newblock On the complexity of detecting $k$-length negative cost cycles.
\newblock In {\em Combinatorial Optimization and Applications, COCOA 2017},
  volume 10627 of {\em Lecture Notes in Computer Science}, pages 240--250.
  Springer International Publishing, 2017.
\newblock \href {http://dx.doi.org/10.1007/978-3-319-71150-8_21}
  {\path{doi:10.1007/978-3-319-71150-8_21}}.

\bibitem{Klein1967}
Morton Klein.
\newblock A primal method for minimal cost flows with applications to the
  assignment and transportation problems.
\newblock {\em Management Science}, 14(3):205--220, 1967.
\newblock \href {http://dx.doi.org/10.1287/mnsc.14.3.205}
  {\path{doi:10.1287/mnsc.14.3.205}}.

\bibitem{Lehmann:2017:SAW:3077839.3077843}
Sebastian Lehmann, Ignaz Rutter, Dorothea Wagner, and Franziska Wegner.
\newblock A simulated-annealing-based approach for wind farm cabling.
\newblock In {\em Proceedings of the Eighth International Conference on Future
  Energy Systems}, e-Energy '17, pages 203--215, New York, NY, USA, 2017. ACM.
\newblock \href {http://dx.doi.org/10.1145/3077839.3077843}
  {\path{doi:10.1145/3077839.3077843}}.

\bibitem{2013Lumbreras}
Sara Lumbreras and Andres Ramos.
\newblock Optimal design of the electrical layout of an offshore wind farm
  applying decomposition strategies.
\newblock {\em IEEE Transactions on Power Systems}, 28(2):1434--1441, 2013.
\newblock \href {http://dx.doi.org/10.1109/TPWRS.2012.2204906}
  {\path{doi:10.1109/TPWRS.2012.2204906}}.

\bibitem{online:NYS:OffshoreMasterPlan}
New York State Energy Research and Development Authority.
\newblock {\em New York State Offshore Wind Master Plan}, 2017.
\newblock
  \url{https://www.nyserda.ny.gov/-/media/Files/Publications/Research/Biomass-Solar-Wind/Master-Plan/Offshore-Wind-Master-Plan.pdf},
  \mbox{Accessed: 2018-08-15}.

\bibitem{Ouorou2000}
Adam Ouorou and Philippe Mahey.
\newblock A minimum mean cycle cancelling method for nonlinear multicommodity
  flow problems.
\newblock {\em European Journal of Operational Research}, 121(3):532 -- 548,
  2000.
\newblock \href {http://dx.doi.org/10.1016/S0377-2217(99)00050-8}
  {\path{doi:10.1016/S0377-2217(99)00050-8}}.

\bibitem{Radzik1994}
Tomasz Radzik and Andrew~V. Goldberg.
\newblock Tight bounds on the number of minimum-mean cycle cancellations and
  related results.
\newblock {\em Algorithmica}, 11(3):226--242, Mar 1994.
\newblock \href {http://dx.doi.org/10.1007/BF01240734}
  {\path{doi:10.1007/BF01240734}}.

\bibitem{SWB-343458624}
David~J. Sheskin.
\newblock {\em Handbook of parametric and nonparametric statistical
  procedures}.
\newblock A Chapman \& Hall Book. CRC Press, Taylor \& Francis, Boca Raton
  [u.a.], 5. ed. edition, 2011.

\bibitem{Valverde:2013:ISOPE:WindFarmLayout}
Pedro~Santos Valverde, Ant\'{o}nio J. N.~A. Sarmento, and Marco Alves.
\newblock Offshore wind farm layout optimization -- state of the art.
\newblock {\em Journal of Ocean and Wind Energy}, 1(1):23--29, 2014.

\bibitem{online:WindEurope2017report}
WindEurope asbl/vzw.
\newblock {\em Wind in power 2017}, 2018.
\newblock
  \url{https://windeurope.org/wp-content/uploads/files/about-wind/statistics/WindEurope-Annual-Statistics-2017.pdf},
  \mbox{Accessed: 2018-08-15}.

\bibitem{f6079fe0004111dab4d5000ea68e967b}
Menghua Zhao, Zhe Chen, and Frede Blaabjerg.
\newblock Optimization of electrical system for a large {DC} offshore wind farm
  by genetic algorithm.
\newblock In {\em Proceedings of NORPIE 2004}, pages 1--8, 2004.

\end{thebibliography}
